# Why China Succeeds: A Road to Prosperity

Qingjie XIA

# Introduction of the Author

Qingjie XIA（夏庆杰）, BA & MA in Economics (Renmin University of China), PhD in Economics (Bath University, England), Post-Doc (Nottingham University, England). Professor in Economics, School of Economics at Peking University. Editorial Board Member of Journal of "Economic Science《经济科学》". Having published in "Journal of Comparative Economics", "World Development", "Journal of Development Studies", "China Economic Review", "Journal of Chinese Economics and Business Studies", "China and World Economy", "China Economist", "Journal of Economic Research《经济研究》", "China Economic Quarterly《经济学季刊》", "Journal of World Economy《世界经济》", etc. Referee for Oxford University Press, Asian Development Bank, "Oxford Economic Papers", "Oxford Bulletin of Economics and Statistics", "Journal of Comparative Economics", "Journal of Development Studies", "World Development", "Southern Economic Journal".

# Contents

# Preface

Over the past decade and more, I have devoted my spare time amid teaching and empirical research to the study of history, political sciences, philosophy, and political economy. From this long-term study, I have formed a set of personal insights and published some academic papers. Now I have compiled all these articles into a monograph, which I hereby present to specialists, scholars and the general public for their valuable comments and corrections.

I used to admire the various hypotheses that scholars put forward. Later, in the course of exploring the origins of the Chinese civilization, I proposed a hypothesis of my own — what I call the "collectivism" hypothesis. Before the Xià (夏, 2070 BCE - 1600BCE) and Shāng (商, 1600 BCE -1046BCE) Dynasties, prehistoric peoples along the Yellow River's middle and lower reaches had already developed settled agricultural civilization, much like other great river valley civilizations. Yet the Yellow River's periodic flooding posed a constant, lethal threat, compounded by the fact that a breach anywhere upstream inevitably endangered tribes downstream—illustrated by the "barrel principle," where water escapes through the weakest point. No single tribe could withstand these floods alone. This shared vulnerability forced tribal chiefs to surrender parochial interests and gradually unite under common leadership, since effective flood control demanded unified command over manpower and resources across the entire river system—echoing Engels' observation that the state arises from society's irreconcilable internal contradictions. Historical progress was uneven: certain stronger tribes, with superior leadership, economies, and flood-fighting experience, took the lead in organizing relief efforts and absorbing weaker tribes. This process of centralizing authority paralleled the process of water control, gradually building joint institutions—military, fiscal, and judicial—while spurring advances in technology, logistics, and the training of capable administrators. Ultimately, the collective struggle against flooding likely catalyzed the emergence of a unified state in prehistoric China. This is the origin of my "collectivism" hypothesis.

Since the Western Zhōu (西周 1046BCE - 771BCE), China's tradition of rule by virtue (德) has centered on benevolence toward all people—rule by virtue and people-centered governance being two sides of one coin. New dynasties typically arose from reflection on their predecessors' failures: the Zhōu's emphasis on "virtue (德)" emerged from critiquing Shāng misrule. According to Sìmǎ Qiān (司马迁, 145BCE - about 86BCE), successive Zhōu chieftains—Hòu Jì (后稷), Gōng Liú (公刘), Ancient Duke Dǎnfù (古公亶父), Jì Lì (季力), and King Wén (文王)—each practiced benevolent governance, sharing benefits with the people and winning their genuine loyalty, so that when King Wǔ (武王) of Zhōu campaigned against Shāng, popular support enabled swift victory. This experience convinced Zhōu founders that Heaven's Mandate is not permanent but conditional: it favors the virtuous and abandons the wicked, belonging to all under heaven rather than one lineage. Rulers must therefore govern with humility and virtue to sustain their mandate. The Book of Documents 《尚书》 links virtue to nurturing the people's livelihood; Guǎn

Zhòng (管仲, about 723BCE-645BCE) ties it to benefiting all people; Lǐ Zéhòu and Xú Fùguān see it as impartial, selfless governance; Zhào Tīngyáng frames it as a strategy for winning hearts through selfless benefit-sharing. Xúnzǐ (荀子, about 313BCE- 238BCE) contrasted three modes of unification—virtue, force, and wealth—arguing only virtue sustains lasting kingship, while force weakens and wealth impoverishes. Confucius compared virtuous rule to the North Star, inspiring voluntary shame and submission rather than mere compliance. Mencius held that winning hearts means winning the realm, since people naturally gravitate toward benevolence. Thus King Wén, King Wǔ, and the Duke of Zhōu represent China's true exemplary sages, with Confucius and Mencius merely transmitting their legacy. Confucian thought centered on virtue was later designated by Emperor Wǔ (汉武帝, 156BCE- 87BCE, reign 141BCE- 87BCE) of the Western Hàn (西汉, 202BCE-8BC) as the sole official ideological standard of the Hàn Dynasty. This norm was adopted by all subsequent dynasties and endured until the Qīng Dynasty.

The Mandate of Heaven does not stay fixed in place; it settles, in the end, on those who govern with virtue. At the heart of virtue lies the fair and equal distribution of benefit to the people. The well-field system of the Western Zhōu embodied this principle of equitable sharing. So too did the equal-field system carried out under the Northern Wèi, the Suí, and the early Táng, which likewise extended fair benefit to commoners. And in the early years of the People's Republic of China, land taken from landlords and capitalists was redistributed equally among farmers—yet another instance of this same impulse to share gains fairly with the people. Although the rule by virtue practiced by the rulers of the Western Zhōu and later dynasties, together with the discourses on benevolent governance by later thinkers such as Guǎn Zhòng, Confucius and Mencius, stemmed from utilitarian motives, the implementation of it under the commandery-county system granted equal treatment in terms of material and land as human beings to inhabitants across each dynasty's territory. This constitutes one of the major institutional arrangements that sustained China's great unification over the past more than two millennia.

During the Western Hàn, Emperor Wǔ established the "jǔxiàolián (inspection-and-recommendation)" system, which inspected and recommended candidates on the basis of filial piety and integrity. To keep aristocratic clans from monopolizing the power of appointment, the imperial examination system was introduced beginning in the Suí and Táng dynasties and continued in use all the way through the Qīng. Both the inspection-and-recommendation system and the imperial examination system drew social elites into government service, laying a solid foundation for good governance. They also gave ordinary people a path to upward mobility, keeping society open and fluid. Most importantly, these two systems allowed the gentry class to realize their ideal of an integrated family-state structure. The implementation of the imperial examination system from the Sui and Tang dynasties through the Míng and Qīng dynasties guaranteed political openness of the country, which is why I describe the ancient and pre-modern China as an "open society".

Compared with India's caste system, which blocked social mobility; European feudalism prior to the 17th century, where the titles of feudal lords were inherited in perpetuity; Russian serfdom before 1861, where the the identity of a serf could not be changed; and the slave-official and conscription systems in Egypt and the Ottoman Empire before the 20th century, which tore families apart — China's imperial examination system, judging candidates solely on morality and talent, regardless of family origin, was far more humane, a quality that account for its remarkable longevity.

After Qín unified China, it put in place nationwide systems of standardized script, uniform cart-axle gauges, and consistent weights and measures across the entire territory. The imperial court adopted the Three Lords and Nine Ministers system: the Chancellor headed civil administration, the Censor-in-Chief handled supervision, and the Grand Commandant oversaw military affairs. Below them, the Nine Ministers each managed a specific domain—rites, finance, justice, civil affairs, and so on. All officials were appointed and removed by the imperial court rather than inheriting their posts, forming a body of professional civil servants. At the local level, the commandery-county system (郡县制) replaced the hereditary enfeoffment practiced before the Qín. Each commandery was headed by a governor appointed by the imprial court. Besides the Commandery Governor, there was the Governor Assistant, the deputy to the commandery governor, who assisted with civil administration and official documents and acted on the governor's behalf during his absence. The imperial court also appointed a provincial Military Governor, exclusively in charge of the commandery's military affairs, public security, criminal arrest and conscription. Not subordinate to the Commandery Governor, he answered directly to the imperial court to divide the governor's military authority. The Censorate of the imperial court dispatched Supervising Censors to each commandery as local inspectors. These officials oversaw the Commandery Governor, Commandery Governor Assistant, Commandery Military Governor and all local bureaucrats. They possessed the authority to impeach officials directly to the imperial court, serving as a check on regional administrative power. Below commandery, there were counties. At the county level, there existed a full hierarchy of officials whose institutional framework largely mirrored that of the commandery administration, all of whom were appointed and removed directly by the imperial court. These officials drew fixed salaries, held no fiefdoms of their own, and exercised their power under the constraints of central oversight. Later dynasties—the Hàn, Táng, Sòng, Míng, and Qīng—all maintained broadly similar administrative frameworks. From the Suí and Táng dynasties onward, officials selected through the imperial examination system came to fill civil service posts at every level of administration. Through the commandery-county system, all residents within its territory were guaranteed security, unified metrology and written script, and subject to uniform tax and corvée obligations — in other words, all inhabitants were treated with equal standing as subjects of the state and respected as human beings.

By contrast, the Roman Empire granted political rights only to native Romans, never extending such privileges to residents of its other provinces. It ruled its territories through military governance and never imposed unified writing systems or standardized tax, corvee, weights and measures across its domains. According to Mencius's theory, violent repression only provokes resistance, rendering such rule ultimately unsustainable. In short, the Roman Empire never functioned as an integrated, organic whole. This is one reason why Europe has never achieved political unity again since the empire's fall.

China's historical institutional innovations — collectivism, the centralized commandery-county system, rule by virtue, and the Confucian open society epitomized by the imperial examination system — yielded a momentous outcome: continuously provision of increasingly equitable treatment and dignity as human beings to all residents within its territory, while offering a growing number of fair avenues for social advancement to those who aspired it. The more humane a civilization's institutional arrangements are, the stronger the sense of belonging among inhabitants within the territory, the longer its governing and sustained stability.

Virtue means the equitable distribution of benefit to all people. Marxism-Leninism, at its core, likewise seeks the well-being of the broad proletariat. This convergence—between the ancient Chinese concept of virtue and Marxist-Leninist ideals—explains why Marxism-Leninism took such firm root in China. After the founding of New China, land seized from landlords and capitalists was redistributed equally among landless peasants. The assets of bureaucratic and comprador capitalists were confiscated outright, while factories and mines belonging to national capitalists were bought out through redemption and brought under state ownership, creating jobs for the assetless working class. The government built up, step by step, a complete education system spanning primary, secondary schooling, and higher education, along with a state-run academy system and a nationwide public healthcare network. With the setup of 156 heavy- and chemical-industry projects aided by the Soviet Union, the country built a relatively comprehensive industrial system. To safeguard national security, China developed the atomic bomb, the hydrogen bomb, and an artificial satellite—the "Two Bombs, One Satellite" program—becoming a nuclear power. After reform and opening-up in the late 1970s, the Communist Party of China drew on the strong organizational and executive capacity (state capability) forged during the revolutionary wars to "embed" market mechanisms onto the existing planned economy, creating a socialist market economy. This path drove industrialization and turned China into a global high-tech power, even in the face of suppression from the United States since the mid-2020s.

During a break from an academic conference held in Chengdu on November 16, 2024, I visited the Dūjiāngyàn (都江堰) Irrigation System near Chengdu, Sichuan Province. Built in 256 BCE over the course of nine years under the leadership of Lǐ Bīng (李冰), the Governor of Shǔ Commandery (蜀郡) of the Qín State during the Warring States Period, this hydraulic project has, for nearly 2,300 years and continuing to this day, supplied water for production and daily life to tens of millions of residents across the Sichuan Basin, including people of Chengdu. On the morning of December 5, 2025, President Xí Jìnpíng accompanied French President Emmanuel Macron on a special visit to Dūjiāngyàn in Sìchuān Province.

From January 14 to 28, 2026, I was fortunate to be included as a member of an investigating group from the Tiānjīn Municipal's Development and Reform Commission on an inspection tour in Gansu Province, reviewing the progress of paired assistance between eastern and western regions. What struck me most over the course of this two-week inspection trip was the story of the relocated residents from the Mǐn Mountains (岷山) who have now put down permanent roots in Xīnpíng (新坪) Village, Jìngyuǎn County(靖远县). In 2013, an earthquake struck the Mǐn Mountains. The Gānsù provincial government subsequently relocated a portion of the earthquake-stricken residents to Jìngyuǎn County. At the time, Jìngyuǎn County allocated land from a state-owned farm to these 595 relocated households. The government built new houses for each family, and each household needed to pay just over 10,000 yuan to move in. Later, the government invested a further 50,000 yuan per household to build agricultural greenhouses for growing cucumbers. The relocated villagers, however, were unfamiliar with cucumber cultivation, having grown traditional Chinese medicinal herbs back in Mǐn County, so the government brought in agricultural technical instructors to teach them cucumber farming hands-on. It also offered a 600-yuan incentive in the first year, along with guaranteed sales channels for their produce. Since then, the villagers have made cucumber farming their main livelihood. The cucumber seedlings were sourced from the Academy of Agricultural Scientific Research in Tiānjīn, and the 50,000-yuan subsidy per household was funded through the east-west paired assistance program of Tiānjīn - Gansu. We visited one relocated household of seven — an elderly couple, a young couple, and three children. Beyond their own 80-square-meter greenhouse, the family also leased additional greenhouses from other villagers, earning an annual income of 100,000 yuan. We then toured the cucumber trading market, where 2,000 tons of cucumbers change hands each day, with produce grown here shipped to Russia via the China-Europe Railway Express.

Through the poverty alleviation campaign carried out from 2015 to 2020, China eradicated absolute poverty by the end of 2020. To prevent households that had just shaken off poverty from slipping back into destitution, the Chinese government has maintained the policies and fiscal input in the targeted poor regions rolled out during the 2015-2020 poverty alleviation drive throughout the 14th Five-Year Plan period (2021–2025) and the ongoing 15th Five-Year Plan period (2026–2030). The east-west paired assistance mechanism launched in 1996 likewise continues to deliver tangible results. Township governments in the targeted poor regions conduct regular monitoring to identify households at risk of relapsing into poverty and extend targeted support to vulnerable families, including public welfare posts in rural areas, full coverage of expensive hospital stays and medical bills, interest-free loans for university students, and financial subsidies to families with no labor capacity, and the like.

From June 27 to July 2, 2026, I visit Ningxia and Chongqing by accompanying the “Zhi-Xing China” U.S. University Student Delegation. In Ningxia, we toured the Tengger Desert Photovoltaic Power Generation Project and learned about local desert control efforts. We saw an endless blue sea formed by rows upon rows of solar panels, alongside stretches of once-barren desert now covered with thriving grass and trees. Since the new century, China’s power generation from clean energy sources such as photovoltaic and wind power has developed rapidly. In 2025, electricity generated by clean energy accounted for 40% of China’s total power output, with photovoltaic and wind power each contributing 11%. A key advantage of photovoltaic and wind power is that solar and wind energy carry no fuel cost. Since the founding of the People's Republic of China, the Chinese government has placed great emphasis on desertification control, and to date effective management has been achieved in both the Tengger Desert and Mu Us Desert. In Xinjiang, ecological shelterbelts have been successfully built around the 330,000-square-kilometer Taklamakan Desert — meaning vegetation has been planted along the desert's perimeter to halt its outward expansion — and efforts to rehabilitate the desert continue to be scaled up. In Chongqing, we visited a waste-to-energy incineration plant with a processing capacity of 4,000 tons of garbage every day. Reports indicate that among China's 1,137 waste incineration power plants, many are now operating below their designed capacity because of insufficient waste supply.

The above is a brief introduction to the context of this book. This book offers a comprehensive political-economic analysis of China's historical development and its contemporary rise, tracing continuities between imperial governance structures and the institutional foundations of the modern Chinese state. The central thesis holds that China's distinctive trajectory—from ancient collectivist civilization to industrial and technological power—reflects a coherent institutional logic rather than a series of disconnected historical episodes. The analysis begins by locating the roots of Chinese "collectivism" in the recurring ecological challenges of flood control along the Yellow and Yangtze Rivers and in persistent nomadic threats, which together drove the evolution from the Western Zhōu enfeoffment system toward centralized bureaucratic rule under the Qín and Hàn dynasties. The adoption of Confucianism as state ideology and the subsequent development of the imperial examination system created what the book terms a "Confucian open society," in which governance was shared between emperors and examination-selected elites. This meritocratic mechanism, the book argues, prevented the entrenchment of hereditary aristocratic power and the concentration of land-based wealth, thereby sustaining centuries of political unification even as private land tenure expanded. Turning to the modern era, the book frames the Communist Party of China's institutional capability—rather than any single policy—as the decisive factor in national development. It traces a lineage from Soviet-style Five-Year Planning in the early PRC period, through the introduction of market mechanisms after 1978, to a distinctive hybrid model combining state direction with market forces. This "China Model" is credited with building manufacturing capacity, infrastructure, and, since 2012, a financial regulatory framework explicitly oriented toward long-term socialist development goals and technological self-sufficiency. A substantial portion of the book addresses China's trajectory relative to the United States, forecasting continued geopolitical and technological rivalry through roughly 2040, based on a claimed historical pattern of thirty-year cycles in international relations. It links this forecast to the expansion of Chinese high-tech enterprises and the Belt and Road Initiative as vehicles for South-South economic cooperation, projecting eventual Chinese breakthroughs in strategically contested technologies. The book further contends that China's post-1978 economic transformation should be understood not as a rejection of its earlier planned economy but as a continuation that unlocked latent capacities—administrative experience, industrial infrastructure, and a trained workforce—accumulated during the pre-reform decades. This argument extends into an examination of how China's "socialist market economy" emerged through incremental Party-led decisions rather than a predetermined blueprint, producing a diverse economic ecosystem of state, private, and foreign enterprises. The author positions this experience as a challenge to conventional Western developmental economics, which typically presumes that formal institutions must precede growth rather than co-evolve with it. The concluding chapter situates China's rise within broader global patterns of industrialization, poverty reduction, and great-power competition, citing contemporary events—including the 2021 withdrawal from Afghanistan and 2026 US-Iran tensions—as evidence of the limits of military power against unified nations. It closes by casting China as a stabilizing force for developing nations navigating industrialization amid ongoing risks of great-power conflict and nuclear instability. Overall, the book presents an integrated historical-institutional narrative connecting imperial statecraft, Communist Party governance, and China's contemporary global positioning, written from a perspective sympathetic to the achievements of the Chinese political-economic model.

The detail of my book is organized as follows. Chapter 1 of the book is titled "Political Economic Implications of the Historical and Philosophical China." This chapter attempts to explore what a "philosophical" China is like and its political economic implications in historical perspective since the Greats Yáodì, Shùndì and Yǔdì. The periodic floods of the Yellow River and the Yangtze River and the frequent nomadic incursion had created the "collectivism" character of the Chinese civilization. In order to effectively fight floods, control rivers, and resist nomadic incursion, the political system of "collectivist" China gradually changed from the enfeoffment system of the Western Zhōu Dynasty to the centralized power system or the system of commanderies and counties in the Qín and Hàn Dynasties. In order to maintain the unified ideology of the country, the Emperor Wǔdì of the Western Hàn adopted Confucianism and banned all other schools of thought. Under the imperial examination system, and from the Suí and Táng to Míng and Qīng dynasties, China carried out the recruitment of imperial officials from all elites who studied the Confucianism and passed the exams. Therefore, the emperors governed the country together with the elites collectively. This might be called a "Confucian open society". After the middle of the seventeenth century, China fell behind in the tide of industrialization and hence was continually beaten by the West and Japan. After having had completed its political, economic and social reconstruction in the process of being beaten, the "collectivist" Chinese people finally ushered in the birth of the PRC in 1949. Since its foundation, the PRC has successfully realized industrialization and today become a global high-tech power, and the construction of its "open society" has also caused revolutionary changes.

Chapter 2 is titled "The Code of China's Historic Great Unification: A Political-Economic Institutional Analysis." In agrarian societies, effective state control depended on a well-organized administrative apparatus and reliable land-based taxation. To meet these demands, successive Chinese dynasties progressively developed two foundational pillars: a political mechanism rooted in the imperial civil service examination system for recruiting officials, and an economic framework grounded in functional land and tax regimes. The civil service examination system served as a powerful instrument for regulating social mobility and preventing the concentration of landed wealth that might otherwise enable powerful elites to capture state authority and divert land tax revenues. While land tenure gradually shifted toward predominantly private ownership in response to population growth and agricultural realities, the institutionalization of the examination system ensured that land consolidation never gave rise to entrenched aristocratic clans strong enough to challenge central rule. This dynamic was instrumental in preserving national unity across centuries. In this light, regulated social mobility and well-calibrated land taxation institutions represent the twin political and economic foundations that underpinned China's enduring Great Unification throughout its long imperial history.

Chapter 3 is titled "How China Succeeds: The Formation and Leverage of 'State Capability.'" Confucius observed, "Do not worry about having no position; worry about the means to stand firm." Amartya Sen argued that capability deprivation is the root cause of individual poverty. By extension, a nation's poverty and underdevelopment stem equally from weak state capability. It follows, then, that building state capability must be the foremost priority for developing countries. China's formidable state capability is grounded in the leadership of the Communist Party of China (CPC). In the early years of the People's Republic, China adopted the Soviet Union's Five-Year Plan system as its foundational planning framework. Following the launch of reform and opening-up in 1978, the CPC leveraged its strong executive capability to successfully embed market mechanisms into the existing planned economy. Drawing on this hybrid foundation, China systematically built manufacturing capacity and national infrastructure through successive Five-Year Plans and targeted industrial policies. Since the 18th National Congress of the CPC in 2012, China has gone further—establishing, for the first time in human history, a regulatory and institutional framework that orients the banking and financial sectors toward serving the people and the long-term imperatives of socialist development. Through continued Five-Year Plans and strategic industrial policy, China has also managed to overcome market constraints in high-technology innovation, cultivating indigenous high-tech industrial capacity on a significant scale.

Chapter 4 is titled "Forecasting China-US Economic and Technological Competition." It generally takes about 30 years to see major changes in international relationships and global events as well as in China's domestic ones. Thus, after China overtook Japan as the world's second largest economy in 2010, the US' suppression of China will likely continue until about 2040. The capacity of the US to suppress China is based on its mastery of certain advanced technologies. Since the 18th National Congress of the CPC in 2012, the number of China's high-tech enterprises have surged, and China has also launched the "Belt and Road Initiative (BRI)" the following year, through which China acts as a leading or principal actor in the internationalization process aimed at promoting the economic development and cooperation of the Global South. More critically, since its founding in 1921, the CPC has successfully led China's revolution, nation-building, economic growth, and high-tech development. Therefore, there is reason to believe that by 2040 or so, China will have made a comprehensive breakthrough in "chokepoint" technologies and will be an international leader in unknown technological fields, and the internationalization led by China to promote the development of the developing countries will have made greater progress. By that time, the US will have nothing to hold on to even if it wants to suppress China.

Chapter 5 is titled "How the Correct Path of Economic Development Has Been Identified Since the Founding of the New China?" China's economic miracle since reform and opening up hinges primarily on its successful industrialization. Some scholars argue that township-based industrialization centered on light industries such as textiles and apparel constitutes the core of China's industrial revolution. This paper holds that China's economic miracle can be mainly attributed to the reform and opening-up policies steadfastly enforced by the Chinese government since the late 1970s. The essence of these policies lies in the continuous introduction of market mechanisms into an economic system dominated by a planned economy and SOEs. While the heavy industrialization strategy and rural collective economy pursued in the early years of the PRC failed to eradicate poverty among the Chinese population, pre-reform China had laid all the fundamental foundations for industrialization. These foundations include sound state governance and institutional frameworks, a unified domestic market, power and energy infrastructure, heavy industrial production capacity, transportation and communications networks, state experience in economic administration, and a workforce of hundreds of millions of educated personnel including engineers, to name a few. Reform and opening up unlocked the tremendous latent strengths accumulated since the founding of New China. The pivotal force that unleashed China's enormous potential reservoir was the integration of market mechanisms into its economic system. The triumph of China's industrialization, or the success of the China Model, since reform and opening up arises from the synergistic interplay of a proactive government, the state-owned sector, and the market economy.

Chapter 6 is titled "How was China's 'market economy' established?" This chapter examines the choices made by the CPC and the Chinese government — along with their consequences — in constructing a "socialist market economy" and carrying out the "opening-up policy" since reforms were initiated. Drawing on a review of decision-making from 1978 onward, the study argues that neither the socialist market economy nor the opening-up agenda could have taken shape without a succession of critical decisions formulated and firmly implemented by the CPC and the Chinese Government. Progress on both fronts gave rise to a broad range of market actors across the country — including township and village enterprises, private urban businesses, and foreign-invested companies — while state-owned enterprises provided the essential industrial backbone in sectors such as transportation, communications, and energy. Collectively, these forces drove China's industrialization and propelled it to become the world's second-largest economy and a global high-tech power. The lived experience of China's "institution-building through development" model poses a direct challenge to the prevailing Western economic assumption that sound institutional frameworks must precede — rather than evolve alongside — economic growth.

Chapter 7 is titled "Where Is the World Heading?—Examining Global Trends Through the Lens of Industrialization's History and Development." Judging from the global history of industrialization since the 18th century, the world today remains in an era of industrialization and informatization. Achieving industrialization and informatization still constitutes the primary means for people in developing countries to lift themselves out of poverty and attain prosperity. The "Kabul moment" in 2021 and the forced signing of an armistice agreement between the United States and Iran amid the US-Iran conflict in 2026 demonstrate that no matter how powerful a country is, it can hardly conquer weak nations and states that stand united. China's economic miracle has revealed universally applicable laws governing economic development in the age of industrialization. Nevertheless, global affairs are still dominated by great-power politics. Given the extreme destructive power of nuclear weapons, the instability plaguing major powers including the United States and Russia, and the ongoing Russia-Ukraine war, the likelihood of regional conflicts breaking out in the future remains relatively high. Disadvantaged impoverished groups across developing countries are in urgent need of escaping poverty and gaining access to relatively advanced national infrastructure, which serves as a major engine for global economic growth. An increasingly powerful China stands as a core force underpinning stability in Asia and the wider world.

The above is the main content of my book. Valuable comments and suggestions from all experts, scholars and readers are greatly welcome! The reason that I dare try to write papers on political economy today stems entirely from my undergraduate studies in the Department of Political Economy at Renmin University of China in the mid-1980s. During this period, the late Professor Yang MENG's lectures on Das Kapital laid a solid foundation for my theoretical thinking. I pursued my master's degree at the Department of International Economics, Renmin University. Under the supervision of the late Professor Houwen DU, I began publishing academic papers in journals including *Journal of World Economy* (《世界经济》). After graduating with my master's degree in the early 1990s, I joined the Department of International Economics at Renmin as a faculty member. During my tenure at Renmin, I was fortunate to enroll in the one-year Ford Program master's curriculum in economics held at Renmin. The program's module teachings were primarily delivered by economics professors from universities in the United States and the United Kingdom.

In the mid 1990s, I undertook an academic visiting fellowship in the Department of Economics at Salford University in the UK, where I collaborated with the late Professor Colin Simmons—an experience that marked the start of my academic research in modern economics. Subsequently, funded by a UK Overseas Research Scholarship and a scholarship from the University of Bath, I completed my PhD studies at the University of Bath at the beginning of the first decade of the new century, under the meticulous guidance of Professors Christopher Heady and Simon Appleton. This experience paved a smooth path for my research in modern economics, and basing my PhD thesis I published an academic article in *China Economic Review*.

In the early 2000s, I was employed as a research assistant at the Institute of Contemporary China Studies of the University of Nottingham in UK under the supervision of Professor Lina Song, Professor Simon Appleton, and the late Professor John Knight. This experience deepened my understanding in modern Western economics, and together with the aforementioned supervisors, I co-authored and published multiple papers in international academic journals. The gurgling waters of the River Avon in Bath, academic seminars at St Edmund Hall, University of Oxford, and the hours spent wandering Oxford's streets still linger in my dreams to this day.

I joined the School of Economics, Peking University, in the mid of the first decade of the new century. Because of a comfortable office, friendly, caring colleagues, and Peking University's timeless, elegant campus — steeped in history and stately grandeur — I have been able to engage in extensive reading in history, political science, philosophy, and economics alongside my teaching and research. Debates with fellow faculty members across academic fields in the Dining Hall for Professors, and talks with friends and students while strolling along the shores of Weiming Lake, have sparked endless imagination and reflection in me. In particular, conversations with Professors Sabina Alkire, Heping CAO, Hailong CHANG, Xiaoqiu CHEN, Yi CHEN, Yuzhui CHENG, Quan CHONG, Jianhua CUI, Quheng DENG, Wei DI, Songgen DONG, Zhiyong DONG, Jiang DU, Liqun DU, Yang DU, El Azami Hicham, Xiaojian FAN, Fang FANG, Honglin FANG, Min FANG, Jianhua FENG, Yukun FENG, James Foster, Jun FU, Song GAO, Zigang GE, Hanhui GUAN, Haiyan HAN, Yuhai HAN, Yu HAO, Xiaofeng HE, Bikang HUANG, Chun HUANG, Guitian HUANG, Liaoyu HUANG, Gerhard ILLING, Xi JI, Xiaotao JIANG, Xu JIN, Guoyu JU, Jinfeng KANG, Desheng LAI, Da LEI, Xiaoyan LEI, Anshan LI, Bo LI, Hong LI, Lianfa LI, Lixing LI, Qiang LI, Shaojun LI, Shaorong LI, Shi LI, Xin LI, Justin Yifu LIN, Wanlong LIN, Changgeng LIU, Chong LIU, Guoen LIU, Haifang LIU, Junwen LIU, Minquan LIU, Wei LIU, Wenxin LIU, Xiaoyan LIU, Yajun LIU, Yi LIU, Zhenya LIU, Yongtu LONG, Chuliang LUO, Gerald MBANDA, the late Sir James Mirrlees, Zhesheng OUYANG, Xinqiao PING, Xuezheng QIN, Zhihui QIN, Jianguo QIU, Amartya Sen, Beseri Geleta SEYOUM, Yan SHEN, Jianhuai SHI, Xinzheng SHI, Jin SONG, Fangxiu SONG, Jian SU, Changfu SUN, Hao SUN, Qixiang SUN, Lingyan SUO, Hua TIAN, Haiyuan WAN, Bo WANG, Dashu WANG, Jian WANG, Meiyan WANG, Shuguang WANG, Wen WANG, Wenan WANG, Xiao WANG, Xiaolin WANG, Yiming WANG, Yuesheng WANG, Alfons WITBEEN, Zhong WU, Fei WU, Weixing WU, Zenan WU, Zhenlei WU, Chen XIAO, Guoliang XIAO, Dandan XU, Xianchun XU, Liping XU, Zhijie YAN, Dengcai YANG, Rudai YANG, Shujie YAO, Yang YAO, Anpei YE, Jingyi YE, Linxiang YE, Xingjian YI, Chunhai YU, Jiantuo YU, Miaojie YU, Xiaodong YU, Cheng YUAN, Ye YUAN, Ximing YUE, Daojiong ZHA, Peng ZHAN, Yong ZHAN, Bo ZHANG, Deliang ZHANG, Haibin ZHANG, Guoyin ZHANG, Hui ZHANG, Xing ZHANG, Yaguang ZHANG, Yan ZHANG, Yunling ZHANG, Zheng ZHANG, Liuyan ZHAO, Yaohui ZHAO, Zhong ZHAO, Kai ZHENG, Wei ZHENG, Jianbo ZHOU, Rongguo ZHOU, Yongmei ZHOU, among others, along with attending their lectures and studying their publications, have provided me with endless intellectual inspiration.

In addition, Qi TANG offered valuable input during the writing of "Political Economic Implications of the Historical and Philosophical China" (chapter 1). Qi TANG and Xiangzhou LIU made substantial contributions to the writing of "The Code of China's Historic Great Unification: A Political-Economic Institutional Analysis" (chapter 2).

The writing-up of this book was supported by the Major Project of the National Social Science Fund of China (国家社会科学基金重大项目) entitled "Research on Household Wealth Distribution under the New Development Pattern (新发展格局下居民家庭财富分布研究)" (Grant No. 23&ZD045). I hereby express my sincere gratitude. Any errors and omissions are the sole responsibility of the author.

Qingjie XIA (夏庆杰)
School of Economics, Peking University, Beijing, China
Summer of 2026

# Political Economic Implications of the Historical and Philosophical China

## 1. Introduction

Western scholars such as Wittfogel[1] and Acemoglu[2] often characterize China's political system as "despotism," "dictatorship," "autocracy," or "authoritarianism." Daniel Bell, a comparatively neutral observer, instead describes it as "meritocracy."[3] Chinese scholars, for their part, tend to call it a centralized power system or the system of commanderies and counties (jùn xiàn system). Fukuyama, meanwhile, credits the Shāng Yāng (商鞅) Reform under Duke Xiào of Qín (秦孝公) around 350BC with producing the first modern state in human history, and argues that ancient China should therefore serve as a reference point for studying the development of political institution.[4] Acemoglu, by contrast, concludes that liberal democracy leads to prosperity and strength, while dictatorship inevitably result in the failure of the country—economic collapse and popular immiseration. Bell examined "meritocracy" in detail, and argues that China's version of it may constitute a viable alternative to Western liberal democracy. Notwithstanding these comments, for the period from the first Emperor Shǐ Huángdì (秦始皇) of the Qín (秦) Dynasty which unified China in 221 BC to the late Míng (明) and the early Qīng (清), China had been the most powerful country in the world in terms of political, economic and cultural development.[5] Only from the late Míng/early Qīng period to the Kuomintang's (国民党) defeat in 1949, did China fail to keep pace with the global technological development and the Industrial Revolution. The consequence was a century of poverty, backwardness, and repeated defeat at the hands of the West and Japan—from the Opium War of 1940 to the Chinese victory in the Resistance Against Japanese Invasion in 1945. It was precisely through this process of being repeatedly beaten that China rebuilt its political, economic, social and cultural system, culminating in the founding of the People's Republic of China (PRC) in 1949. In the seven decades since, China has achieved industrialization, and became the world's second-largest economy, and emerged as a global high-tech power. For a civilization with this record, labels such as "despotism," "dictatorship," "autocracy," "authoritarianism," or even "meritocracy" fail to capture its essential character. Similarly, the terms favored by Chinese theorists—"centralized power system" or "system of commanderies and counties" may not answer to the deeper question: what kind of country is China? To address "What China ultimately is" in philosophical terms, we must trace the very origin of Chinese civilization itself and undertake a systemic analysis and generalization of its political, economic, social, cultural, and historical trajectory. Only through such an inquiry can we arrive at a properly "philosophical" China in metaphysical sense and grasp its political and economic implications.

## 2. The Origin of China's "Collectivism" and the Formation and Changes of its Institutional Arrangements

### 2.1 The Origin of "Collectivism" in China

Lǎo Tzu (老子，571-471 BC), who lived in the late Spring and Autumn Period, wrote: "Man follows the earth, the earth follows the heaven, the heaven follows the Tao (道), the Tao follows the nature."[6] Therefore, as long as the "Tao" is followed, the state of "ruling by doing nothing" can be achieved. Lao Tzu's ideal social state is: "neighboring states face each other, chickens and dogs hear each other, and people will never communicate with each other until they grow old and die."[7] Sun Yat-sen Sun Yat-sen offered a starkly different diagnosis of the Chinese condition: "China's four hundred million people are like the same scattered sand. Is this natural?"[8] From the Qín and Hàn dynasties through the late Míng and early Qīng, China remained an agrarian society organized around dispersed smallholder farming. This raises the central question: how did a traditional smallholder society—one in which "transportation relies mainly on walking and communication relies on shouting"—come to develop "collectivist" characteristics?

Hobbes argues that in the state of nature, the condition of "everyone against everyone" produces massive loss of life and property, leading people to cede certain rights to a powerful sovereign—Leviathan—in exchange for peace and security.[9] People relinquish certain rights in order to preserve their lives. Zhào Tīngyáng contends that this hypothesis suffers from a fatal flaw: it contains no implicit premise of cooperation, and thus cannot explain how cooperation arises within the ruling group that constitutes the Leviathan in the first place. Having identified this contradictions in Hobbes's framework, Zhào Tīngyáng argues that Xúnzǐ (荀子，313 BC – 238 BC) was arguably the first thinker to address the problem of the original state—nearly two thousand years before Hobbes. Xúnzǐ begins from the premise that community precedes the individual, as captured in his maxim: "Human beings cannot live without communities."[10] It follows that cooperation within a community must predate conflict. Xúnzǐ further argues that the ancient sage-kings established rites (礼) to govern this cooperation: "Human beings are born with desires. When desires go unfulfilled, people inevitably seek to satisfy them. If this pursuit lacks measure and boundary, strife becomes unavoidable. Strife breeds chaos, and chaos brings destitution. Detesting this chaos, the ancient sage-kings instituted rites and righteousness in order to apportion resources and establish norms."[11] To prevent conflict from endangering life and property within the community, its sage-kings established norms of cooperative conduct among its members. Zhào Tīngyáng draws out the implications of this contrast: "Xúnzǐ postulates an innate propensity for cooperation, thereby circumventing the predicament inherent in Hobbes' reasoning. Nevertheless, Xúnzǐ's hypothesis cannot fully account for the problem Hobbes identifies. While Hobbes' framework struggles to explain how cooperation becomes possible within a group, it offers a compelling account of conflict under anarchy. The two theories are therefore complementary rather than competing, and can be integrated into a single, more comprehensive theory of the original state—one capable of explaining a wider range of political phenomena. This synthesis may be termed the Xúnzǐ-Hobbes Hypothesis: an original state characterized by internal solidarity within communities and warfare between them."[12]

It may be hypothesized that prehistoric China witnessed not only the "conflicts between individuals" and "intra-group cooperation" that Hobbes and Xúnzǐ theorized, but also two additional pressures: periodic flooding that struck every two or three years, and annual raids—plunder, arson, and killing—launched by nomadic people from the north. This raises two questions. First, did these periodic floods and nomadic incursions contribute to the formation of the state in their own right? Second, how might a state born of "periodic flooding and nomadic raiding" differ from one shaped solely by "interpersonal conflicts" and "intra-group cooperation"?

Before the Xià (夏，2070-1600BC) and Shāng (商，1600-1046BC) dynasties, the people of prehistoric China had already settled the central plain—the alluvial floodplain of the Yellow River—where they faced two recurring threats: periodic flooding of the river and yearly incursions from northern nomadic groups.[13] While it is impossible to date precisely when prehistoric China entered agrarian civilization, communities along the middle and lower reaches of the Yellow River had already taken up agriculture before the time of Xià and Shāng. Agriculture, unlike the nomadic practice of following water and pasture, is inherently a settled mode of life. The climate along the middle and lower Yellow River is temperate, and the alluvial plain formed by the river were well suited to cultivation and settlement—much as the Tigris-Euphrates basin in West Asia, the Nile Delta in ancient Egypt, and the Ganges basin in India supported comparable agricultural civilizations. However, the Yellow River flooded every few years, and each flood claimed many lives. Periodic flooding thus becomes the paramount security threat facing farming communities along the middle and lower reaches. Compounding this danger, the river flows from west to east into the sea, meaning that even a tribe with a well-maintained embankment remained vulnerable: a breach anywhere upstream or along a neighboring tribe's section would still bring floodwaters to communities downstream. No single tribe, in other words, could withstand the river by acting alone. The barrel principle illustrates this logic well: water escapes from a barrel's weakest point, just as floodwaters break through a river's weakest embankment, regardless of how well other sections are maintained. This structural vulnerability—and the massive loss of live and property it produced—compelled tribal chiefs and their members to set aside local interests and gradually unite against the shared threat of flooding. More significantly, managing the river as a whole required a unified command capable of coordinating manpower and resources across all affected tribes. Flood control, in other words, demanded that tribal chiefs cede authority to a common leader or joint governing body. To prevent flooding from endangering the lives of tribal members, the chief ceded authority to the faction leading the resistance against the flood. This dynamic echoes Engels's description of the state as arising from a society's need to resolve contradictions it cannot otherwise dispel — "a power, apparently standing above society," in his words, "that would alleviate the conflict and keep it within the bounds of order."[14]

Historical and social progress rarely unfolds evenly. Tribes along the middle and lower reaches of the Yellow River likely varied considerably in size and strength, and a small number of them may have advanced further toward statehood than their neighbors. These more advanced tribes would have had stronger leadership, more developed economies, more capable armed forces, and greater accumulated experience of flood control. When floods struck, such a tribe could mobilize quickly— and could extend that capacity to rescue weaker neighboring tribes as well. Under these conditions, the process by which tribes ceded authority to a common chief likely became intertwined with the practical work of flood control itself. As power became increasingly concentrated, the integration of tribal elites and members would have deepened correspondingly, gradually advancing the construction of shared leadership institutions and joint governing bodies—and, eventually, the core institutions of statehood: military, command, taxation, and judicial authority. The sustained effort of fighting and managing floods would also have driven technological and organizational advances—in construction tools and techniques, building materials, food supply, and broader logistics. Just as importantly, this process would have trained a cadre of leaders: figures who understood local conditions intimately and possessed genuine organizational and leadership capacity. Together, these developments would have propelled tribal alliances steadily towards unified statehood. In short, the sustained struggle against flooding may well have been a driving force behind the gradual formation of a unified state in prehistoric China.

Beyond the periodic flooding of the Yellow River, the Yangtze, the Huái, and other river systems, the yearly incursion of nomads from northern China posed another major security threat to the settled farming tribes of the middle and lower Yellow River in prehistoric times. As with flood control, resisting these incursions required a standing military organization under unified leadership, backed by reliable logistical supplies. Periodic flooding and recurrent nomadic raids thus worked together to reinforce a shared collective consciousness across every tribe and every one of its members: no tribe can stand alone and survive. Only by uniting under strong, centralized leadership could tribes successfully manage floods and effectively resist and deter the nomadic incursions. Through repeated cycles of flood control and defense against raids, the elites and members of tribes along the middle and lower Yellow River would likely have arrived again and again at the same realization: none of them could survive, let alone prosper, without unifying together. The path toward a common chief, a central authority, and ultimately a unified state was almost certainly not a smooth one. It would have been marked by constant struggles between unity and fragmentation, by the repeated collapse of joint tribal institutions, and by their equally persistent reconstruction. Yet each collapse and each reconstruction likely deepened the same underlying conviction: that no tribe could survive or prosper except through unity.

Once the common chief, the joint tribal authority, and its military, taxation, and judicial systems were in place, the central authority likely began to develop according to its own internal logic. This might have included the unified deployment of military and civilian power to control water and resist invasion; the promotion of agriculture, animal husbandry, handicrafts, and trade; the issuance of currency; the construction of central cities; the maintenance of public order; and the use of forceful means to suppress any challenge to the unified state. The central authority might also have relied on official titles, ranks, and enfeoffment to co-opt tribal nobles and leaders—binding them to the central government and thereby securing the support of tribal elites and reinforce central power. Finally, the growing strength and cohesion of this common authority in prehistoric China likely proved attractive to neighboring tribes, drawing some to join the emerging polity outright and inducing others to cultivate friendly relations with it. Each new tribe that joined or aligned itself further strengthened the development and stability of the central authority and the nascent state.

This account shares a common thread with two others: Hobbes's Leviathan, in which the “war of all against all” is resolved by ceding rights to a sovereign, and Xúnzǐ's hypothesis of intra-group cooperation secured through a sage-king's decree. All the three ultimately serve to preserve human life. Flood control and resistance against external threats not only made cooperation between separate communities feasible—they gradually integrated that “inter-group cooperation” into a single, unified collective. This may well be the origin of the collectivist tradition within Chinese civilization, one later distilled into such enduring principles as “pooling resources to accomplish major tasks”[15] and “when disaster strikes one region, aid arrives from every quarter.”[16] These principles, in turn, represent the primary drivers behind the emergence, evolution and maturation of collectivism in prehistoric China.

### 2.2 The Evolution of Power Center and Political Structure of the Ancient China

Effective collective action required a central authority capable of mobilizing the finances of all clans and tribes, with each clan and tribe obligated to obey its order unconditionally. Unified command and deployment, in turn, demanded a power center with an organizational system, a centralized taxation system, and an allocating property system. From the Xià (夏), Shāng (商) and Zhōu (周) dynasties onward, the Chinese state continuously constructed, refined, and optimized this collectivist central authority and its accompanying political structure.

The political structure of the Xià and Shāng dynasties remain only vaguely documented in the historical records. Therefore, the discussion of China's political structure can only begin with the Western Zhōu (西周, 1046 - 770BC). Before King Píngwáng (周平王) moved the Western Zhōu’s capital eastward from Hào Jīng (镐京，around Xī Ān 西安) to Luò Yì (洛邑，today’s Luò Yàng 洛阳) in 770BC, the dynasty’s political structure centered on a King who exercised power from the imperial court, while the relationship between the central authority and the localities took the form of a one-off enfeoffment. According to Mencius: “He brought the whole realm under his rule and enfeoffed seventy-one states. Of these, fifty-three were ruled by members of the King’s clan, yet no one in the realm accused him of favoritism.” [17] In practice, the king—as holder of the Mandate of Heaven—entrusted administrative and judicial responsibilities to trusted kinsmen, and granted land to senior royal family members and select marital kin bearing the title of “arch-lords” (侯 hóu). In exchange, these arch-lords pledged loyalty: they were obligated to pay homage to the king, submit to his will, offer tribute, and—crucially—bring their own troops and logistical support to fight on the king’s behalf in times of war. Their descendants would continue to enjoy the blessing of both the king and Heaven, along with the land and the labor of the people upon it. The king retained his own territory, population, and six armies; his power and influence derived from his authority to grant titles and enfeoff land to the nobility. Besides, the King commanded vastly greater military strength than individual vassal states, which exerted a deterrent effect upon the latter. Most importantly, the King held the authority to appoint key officials in the vassal states, including their supreme military commanders. The Book of Rites · Royal Regulations records: “In large states, all three senior ministers are appointed by the King”; “In intermediate-rank states, two of the three senior ministers are appointed by the King, and one by the state's own ruler.”[18] This institution is repeatedly corroborated in historical text such as the Zuǒ Zhuàn.[19] Bronze vessel inscriptions from the reign of King Mù of Western Zhōu likewise document that the Zhōu king issued an official mandate appointing Dòu Bì as Sīmǎ (military chief) of the state of Yú.[20] However, as more and more vassal states were enfeoffed over time, the king’s direct control over land, population, and military manpower steadily diminished—until eventually there was no land left to grant. This erosion caused imperial power and influence to drop sharply. Compounding the problem, the Western Zhōu's central-local relationship depended on bonds of blood kinship to hold the arch-lords in loyalty to the throne; but within a few generations, this kinship tie between the king and the great fiefdom-holding nobles inevitably grew attenuated. After the King Píngwáng's (周平王) move east, the King's ability to control the enfeoffed arch-lords effectively ended, marking the start of the Eastern Zhōu (东周) period. By the Eastern Zhōu, royal authority had become little more than a symbol. The fiefdoms evolved into de facto independent states, forming alliances and waging wars against one another at will. The enfeoffment system of the Western Zhōu was thus destined to fail as a means of preserving unity. A feudal China, evidently, could not sustain the demands and burdens that a genuinely "collectivist" China would require.

After King Píngwáng moved the capital east in the late eighth century BC, China entered the Spring and Autumn Period (春秋 770-476BC) and, later, the Warring States Period (战国 475-221BC). Although the Zhōu King nominally retained his title, the various dukedoms now governed their own territories independently, attacking and absorbing one another with increasing frequency. Economically, China remained in the Bronze Age through the end of the Spring and Autumn Period, and agricultural production was still organized around the well-field system (井田制).[21] Because farming tools made of stone and bone remained primitive, plowing, planting, and harvesting required the coordinated labor of multiple households; agriculture was consequently organized on a collective basis. The Book of Odes captures this arrangement: "May the rain fall first upon our public fields, and then upon our private ones."[22] Following the Spring and Autumn Period, China entered the age of iron tools and ox-drawn plowing—a transformation that, for the first time, gave individual households the capacity to farm independently. Against this backdrop of rapidly advancing productivity, the Shāng Yāng (商鞅) undertook his reforms of the Qín state (秦国) around the mid-fourth century BC. He abolished the feudal system, redistributed land directly to peasant households, imposed taxes on them accordingly, and introduced a legal code that applied uniformly regardless of social class. He replaced the hereditary aristocracy with appointed officials at every level of government, establishing a standing army, and instituted promotion based on military merit rather than birth. At its core, the Shāng Yāng Reforms replaced the feudal system and hereditary nobility with a centralized administrative state—and in doing so, represented ancient China's first attempt to construct what might be called an "open society." Politically, positions for civil and military officials throughout the state were opened to commoner scholars and low-ranking military officers alike. Economically, hereditary aristocratic estates gave way to small-scale household farming, with the central government now levied land taxes and corvée labor directly on individual farming households. Strengthened by these reforms, the state of Qín ultimately conquered all rival states and unified China in 221 BC.

Although the Qín Dynasty (秦, 221-207BC), which first unified China, did not last long, the centralized power system it established—the system of commanderies and counties (郡县制)—was carried forward and further refined in the subsequent Hàn (汉, 202BC-220AC), Suí (隋, 581-618AC), Táng (唐, 618-907AC), Sòng (宋，960- 1279AC) dynasties. Compared with the enfeoffment system of the Western Zhōu, this centralized system represented a fundamental transformation in China's political development. Under enfeoffment, the King lacked direct administrative authority over individual vassal polities. Under the centralized system, by contrast, the emperor could dismiss and appoint the governors of commanderies and counties at will. Relative to enfeoffment, the centralized power system performed three key functions. First, it stripped the hereditary aristocracy of its power to appoint government officials, transferring the authority to the central government. Second, it opened official appointments to non-aristocratic scholars—the first step toward what might be called an "open" political system. Third, it substantially strengthened the central government's capacity to administer state affairs directly. Under the Qín Dynasty, administrative authority ran through two tiers below the center—commanderies and counties—allowing the central government's directives to reach all the way down to the county level. By Max Weber's criteria, the centralized system of commanderies and counties established under the Qín Dynasty should properly be classified as a modern system of state governance:[23] one in which the appointment of official and the conduct of state administration were based not on blood ties or hereditary succession, but on merit of candidates as judged against established impersonal rules.

Through the commandery-county system, all residents within its territory were guaranteed security, unified metrology and written script, and subject to uniform tax and corvée obligations — in other words, all inhabitants were treated with equal standing as subjects of the state and respected as human beings. The Roman Empire, by contrast, though contemporaneous with the Qín and Hàn dynasties, granted political rights only to Roman citizens, while imposing military rule over the inhabitants of its other conquered provinces and governing them indirectly through local aristocrats in the latter's own language and customs. According to Mencius's theory, violent repression only provokes resistance, rendering such rule ultimately unsustainable (see next subsection of this chapter). Put differently, residents of Rome's provinces outside the Italian heartland were denied the standing accorded to citizens. After the fall of the Roman Empire in 476 CE, Europe was never reunified. Viewed from this perspective, the commandery-county system stands as one of the most critical institutions sustaining China's enduring unity over the past two millennia.

### 2.3 The Ideology of Ancient Collectivist China

Every nation-state seeks to establish its own mainstream ideology, and China is no exception. In order to confront yearly floods and nomadic incursions and safeguard the lives of its people, collectivist China consistently required full executive authority to mobilize manpower and resources nationwide at critical moments. This necessity, in turn, demanded that its citizens—elites above all—align themselves with the emperor's rule not merely in action but in thought and consciousness.

The ideological core of collectivism in ancient China rests on two closely related notions: rule by virtue and people-centered governance. Since the Western Zhōu Dynasty, China has sustained a tradition emphasizing rule by virtue, understood principally as the practice of extending benevolence to all people under heaven. In this sense, rule by virtue and people-centered governance are best understood as two sides of the same coin. The concept of virtue (德, dé) has occupied the center of Chinese politics and ethics ever since the Zhōu: "The lesson of Yīn (殷) is not far off; it lies in the reign of the Xìa rulers."[24] During dynastic transitions in ancient China, newly founded dynasties were typically most acutely conscious of the fallings of the regimes they had just overthrown. The recuperative policies of the early Western Hàn, for instance, were a direct corrective to the harsh taxation and forced corvée labor that had characterized the late Qín Dynasty. Similarly, the extensive human sacrifice and live burial practiced under the Shāng reflected a broader disregard for common lives among the Shāng ruling class. From this pattern, it can be inferred that the early Western Zhōu's emphasis on the concept of "virtue "(德) emerged from sustained reflections on precisely this failure of virtuous governance during the late Shāng.

According to the "Basic Annals of Zhōu (周本纪)" in Sīmǎ Qiān's (司马迁) Records of the Grand Historian (《史记》), successive chieftains of the Zhōu clan cultivated a tradition of sharing benefits with the people. Hòu Jì (后稷), the clan's ancestral founder who lived during the reign of Emperor Yáo (尧, traditional chronology: c. 2357 – 2256 BCE), "delighted in farming, surveyed land to assess its suitability, and sowed grain on fertile ground; all common people took him as their model."[25] Toward the end of the Xià dynasty, the Zhōu clan dwelled among the Róng (戎) and Dí (狄) tribes. Gōng Liú (公刘), then leader of the clan, "revived the undertakings of Hòu Jì, devoted himself to ploughing and sowing, and made good use of land suited for farming; he crossed the Wèi River (渭水) from the Qī (漆) and Jǔ (沮) Rivers to procure timber and other supplies."[26] As a result, the Zhōu people "those traveling had provisions, and those settled at home had stockpiled stores."[27] Gōng Liú's virtuous governance won him the heartfelt affection of his people: "The populace relied on his blessings; all common folk cherished him, and many migrated to place themselves under his protection and submit to his rule."[28] During the late Shāng dynasty, the Zhōu leader known as the Ancient Duke Dǎnfù (古公亶父) "revived the undertakings of Hòu Jì and Gōng Liú, accumulated virtue and practiced righteousness, and all clansmen held him in high regard."[29] When the Róng and Dí tribes grew powerful and frequently plundered and harassed the Zhōu people, Dǎnfù was forced to lead his clansmen to migrate from Bīn (豳地[30]) to Qíshān (岐山): "The entire population of Bīn, supporting the elderly and carrying young children, all moved to follow Ancient Duke Dǎnfù at the foot of Qíshān. When neighboring states heard of Ancient Duke Dǎnfù's benevolence, many of their people also flocked to submit to him."[31] Jì Lì (季历), leader of the Zhōu clan in the late Shāng Dynasty, "steadfastly upheld righteousness, and all feudal lords submitted to him."[32] King Wén of Zhōu (周文王), who ascended the throne after Jì Lì, "carried forward the undertakings of Hòu Jì and Gōng Liú, and followed the institutions established by Ancient Duke Dǎnfù and Jì Lì." Under his governance, within the Zhōu territory, "farmers yielded boundaries of their farmland to one another, and among the populace it was customary to defer to elders," "many shì (士, scholar-officials) thus went over to his side; all feudal lords turned to him," "He cultivated virtue and practiced benevolence, so numerous feudal lords defected from King Zhòu (纣王) of the Shāng the Tyrant and came to pledge allegiance to the Western Earl (西伯)."[33] The Zhōu clan's successive chieftains had, over generations, practiced virtuous governance and ensured that the benefits of rule were shared with the people. This legacy of benevolent leadership meant that when King Wǔ (武王) launched his campaign against the Shāng, both the Zhōu populace and neighboring feudal lords rallied to his cause in overwhelming numbers, enabling him to bring down the Shāng Dynasty in a single decisive military strike. On this, King Wǔ remarked: “I launch this campaign in the name of King Wén and dare not act on my own arbitrary will,” and “it is because my forefathers possessed abundant virtue that I, their humble descendant, inherit their meritorious legacy.”[34]

The rise of this tiny western state to overthrow the mighty Shāng realm profoundly shook King Wǔ of Zhōu and the Duke Dàn of Zhōu (周公旦), driving the founding rulers of the Western Zhōu Dynasty to realize the truth that “the Mandate of Heaven is not everlasting. Those who walk the path of goodness shall gain it; those who stray into evil shall forfeit it.”[35] That is to say, the Mandate of Heaven is transferable, and the legitimacy of political power is conditional. The realm does not belong to a single clan or lineage alone, but to all people under heaven. “Heaven shows no partiality; it only aids those of virtue. The hearts of the people are inconstant; they cherish only those who bestow benevolence.”[36] To sustain the Mandate of Heaven perpetually, a ruler must win the support of all people under heaven, which in turn requires governing them through virtuous rule. Moreover, rulers ought to maintain humility at all times. “The supreme Lord on High revoked the Mandate bestowed upon its eldest son, the great Shāng state. Though you, the King, have received this Mandate boundless in blessings, it also brings boundless cares. Alas, how can you fail to uphold reverence?”[37] When the Duke Dàn of Zhōu (周公) admonished King Chéng (成王), he also said: “Alas! Only our Great King Dǎnfù and King Jì of Zhōu knew how to restrain themselves and live in awe of heaven. King Wén humbled himself to labor at tasks of peace and farming. Gentle, virtuous and reverent, he cared for and protected the common folk, extending kindness even to widows and widowers. From morning till midday and late afternoon, he had no leisure even to eat, all for the sake of bringing harmony to all the people....” [38] The Book of Documents (《尚书》), a representative text of the Western Zhōu Dynasty, contains discourses on “virtue (德)”: “Virtue resides in sound governance, and governance consists in nurturing the people. Rectify virtue, make good use of resources, enrich the people’s livelihoods, and uphold harmony above all.” [39]

On the meaning of “virtue”, Guǎn Zhòng (管仲) stated: “When the ancient sage-kings seized control of the realm, theirs was a boundless great virtue! It refers to bringing material benefits to all things and people.” [40] Sage-kings built their enduring realms not through brute force or private wealth, but by securing popular recognition through public interests that benefit all under heaven. Lǐ Zéhòu (李泽厚) argues: “Virtue springs from an upright heart, and originally denotes equitable and impartial principles of governance and distribution.” [41] Xú Fùguān stated: “The concept of 'virtue (de)' in the early Zhōu Dynasty refers to impartial distribution in governance and integrity and selflessness within one’s inner heart.” [42] Zhào Tīngyáng remarks: “Virtue (de) is a political strategy for winning the hearts of the people. To love and nurture the common folk, to deliver benefits to them while refraining from personal greed, is how one gains popular support. When the people reap benefits, the ruler secures political authority.” [43]

On the political function of rule by virtue, Xúnzǐ (荀子) stated: “There are three ways to unify and bring other peoples under one’s sway: uniting others through virtue, through military might, and through wealth. He who unites others by virtue shall become a true king; he who relies on force shall grow weak; he who depends on riches shall fall into poverty. This truth has remained unchanged from ancient times to the present.”[44] Winning over the people through virtue enables a lasting royal reign. Oppressing the populace with armed might incites popular resistance, dooming the state to eventual decline. Maintaining rule by showering subordinates with material wealth will inevitably plunge the realm into hardship, for human desires are insatiable. Implementing virtuous governance yields numerous political effects. Confucius remarked: "Governing through virtue is like the North Star: it remains in its fixed place, and all other stars turn to orbit around it." "If you guide the people with decrees and restrain them with punishments, they will avoid wrongdoing yet lack a sense of shame. If you guide them with virtue and regulate them through ritual propriety, they will possess a sense of shame and wholeheartedly submit to virtuous order."[45] This means that rule by virtue wins the wholehearted obedience of the people, prompts them to exercise self-restraint, and eliminates any desire to rebel. Mencius (孟子) stated: "To win the people is to gain the realm... To win their hearts is to win the people." "The people gravitate toward benevolence just as water flows downward and wild beasts run toward open wilderness."[46] The support of the people is equivalent to political legitimacy. Under virtuous rule, the populace submit voluntarily and have no desire to defect. Xúnzǐ observed: "Whoever governs a state will attain true kingship if righteousness prevails, hegemony if trustworthiness prevails, and ruin if crafty schemes prevail." "Therefore, when those governing a state let righteousness overcome private gain, the people will grow attached to their superiors, delight in their ruler, and willingly lay down their lives to serve him."[47] Rule by virtue (righteousness) unites popular support, generates the strongest collective strength, and secures the most stable governance.

The true great Confucian scholars of China ought to be King Wén of Zhōu, King Wǔ of Zhōu and the Duke Dàn of Zhōu. Confucius and Mencius merely recounted the deeds of these three men and passed them down to later generations through teaching.[48,49] Although the rule by virtue practiced by the rulers of the Western Zhōu Dynasty, together with the discourses on benevolent governance by later thinkers such as Confucius, Guǎn Zhòng and Mencius, stemmed from utilitarian motives, the implementation of benevolent governance granted equal treatment as human beings to inhabitants across each dynasty's territory. This constitutes one of the major institutional arrangements that sustained China's great unification over the past more than two millennia.

## 3. Collectivist Confucian Open Society of the Ancient China

### 3.1 The Creation of Confucian Open Society under the Systems of Inspection and Recommendation, and the Imperial Examination

The openness of ancient Chinese politics—what might be called an "open society" in this specific sense—refers primarily to institutional arrangements designed to broaden the pool from which imperial officials were drawn. Precedents for recruiting officials from outside the hereditary aristocrats already existed during the Spring and Autumn and Warring States Periods; some of Confucius' own students, for instance, went on to hold political office. Yet the true starting point for the institutionalization of openness in ancient Chinese politics was the "inspection and recommendation system" (举孝廉, jǔxiàolián), initiated in 134 BC under the Western Hàn Dynasty. From a political standpoint, this system is best described as a form of "limited openness." The right to inspect and recommend candidates was reserved exclusively for existing imperial officials and the local gentry, meaning that candidacy for office depended on recommendation by the privileged and well-connected—not on open competition available to every commoner. Over time, this monopoly on recommendation hardened further into an entrenched system of vested interests: eligibility for office became effectively restricted to the descendants and relatives of officials and gentry families already in power, so that official appointments came to be dominated by a self-perpetuating network of prominent households.

From the collapse of the Eastern Hàn Dynasty (东汉, 25-220AC) to the reunification of China under the Suí (隋, 581 to 618 CE) and Táng (唐，618 to 907 CE), China endured nearly four centuries of political fragmentation. This period encompassed the Three Kingdoms era (三国, 220-280AD), the Western Jìn (西晋 265-316AD) and Eastern Jìn (东晋 317-420AD), and the Southern and Northern Dynasties (南北朝, the former 420-589AD, the latter 386-581AD). This prolonged fragmentation had many causes, but one significant factor was the long-term monopoly over official appointments held by a small number of eminent, powerful households. By the late Eastern Hàn, a handful of wealthy families—each controlling vast landholdings—had effectively become a de facto hereditary aristocracy. Their control over official appointments during this period might best be described as a "disguised hereditary system," one that substantially eroded imperial authority, hastened the dynasty's collapse, and ultimately fractured the country. This four-hundred-years period of disunity suggests several conclusions. First, the political arrangements underpinning "collectivist" China remained incomplete by the end of the Eastern Hàn. Second, the centralized power system, Confucianism, and the inspection-and-recommendation system, taken together, proved insufficient to preserve national unity. Third, only further institutional reform could prevent the country from fracturing again. Fourth, the capture of central government power by entrenched interest groups posed a direct threat to national unity. Finally, any institutional arrangement that permitted the reemergence of hereditary aristocracy would, sooner or later, plunge the country back into chaos.

To address the fatal flaw of the inspection and recommendation system, the Suí and Táng dynasties created an imperial examination system (科举, kējǔ): a nationwide competitive examination through which successful candidates recruited directly into the imperial bureaucracy. This innovation opened official recruitment, for the first time, to people from all walks of life—not merely the sons and relatives of officials and gentry families who had monopolized access under the earlier inspection and recommendation system. Following the Suí and Táng, the imperial examination system was continuously refined through the Sòng, Míng, and Qīng Dynasties, making an indelible contribution to official recruitment across successive reigns. Rather than appointing hereditary aristocrats, emperors of these dynasties could now, in effect, "draw on all the talent under the Heaven" —a practice reflecting a form collective governance shared between the emperor and a meritocratic scholar-official class. The subject matter of the imperial examinations likewise underwent successive reforms before ultimately settling on the Confucian canon known as the Four Books and Five Classics. Four Books comprised the Great Learning (大学), the Doctrine of the Mean (中庸), the Analects of Confucius (论语), and the Words of Mencius (孟子); the Five Classics included the Book of Songs (诗经), the Book of Documents (尚书), the Book of Rites (礼记), the Book of Changes (周易), and the Spring and Autumn Annals (春秋). By studying these texts and their associated commentaries, students throughout the country prepared themselves—both academically and ideologically—for possibility of official appointment. The choice of Confucianism as the core examination subject played a decisive role in establishing and sustaining the country's mainstream ideology: from the inspection-and-recommendation system of the Western Hàn through the abolition of the imperial examination system in 1905, generation after generation of students and scholar-officials played an indispensable role in maintaining and developing an official ideology grounded in Confucian thought. Compared with the inspection and recommendation system, the imperial examination system eliminated—at least at the level of institutional design—the tendency toward hereditary aristocracy in the appointment of state officials. Seen in this light, the continuous refinement of what might be called a "Confucian open society" under "collectivism" functioned as a powerful safeguard: one that resisted the reemergence of hereditary aristocracy and helped prevent the country from falling back into fragmentation.

Additionally, the implementation of the imperial examination system opened channels of upward social mobility for talented commoners, and afforded ample scope to aspirants in society who sought to "cultivate oneself, regulate the family, govern the state, and bring peace to the world." Compared with India's caste system, which blocked social mobility; European feudalism prior to the 17th century, where the titles of feudal lords were inherited in perpetuity; Russian serfdom before 1861, where the the identity of a serf could not be changed; and the slave-official and conscription systems in Egypt and the Ottoman Empire before the 20th century, which tore families apart — China's imperial examination system, judging candidates solely on morality and talent, regardless of family origin, was far more humane, a quality that account for its remarkable longevity.

Once a candidate passed the imperial examination—including the final court test personally overseen by the emperor—he earned the title of the imperial scholar "Jìn Shì (进士)" and become an imperial official. Newly appointed Jìn Shì would first serve at court in the Hànlín Academy (翰林院) or the Imperial Academy (国子监, Guózǐjìan), acting as court secretaries; after a certain period, they would be dispatched to serve as officials at the county level. New officials typically began their careers at the bottom, in county-level posts, where they accumulated practical experience in administrative affairs. An official who achieved outstanding results governing a county or commandery could then be promoted step by step, potentially rising all the way to the position of prime minister. This entire promotion pathway remained, of course, firmly under the control of the imperial court. This system stands in some contrast to modern Western countries, where people elect presidents or prime ministers from among candidates who may have no prior administrative experience whatsoever. Traditional Chinese political philosophy, by comparison, placed greater faith in the principle that "the great generals rise from the ranks of common soldiers, and the great prime ministers rise from the rank of county and prefectural officials."[50] This preference likely reflected the practical demands of "collectivist" ancient China, which confronted the flooding of the Yellow and Yangtze rivers and nomadic incursions virtually every year. Only experienced civilian officials and army generals, tested and proven at lower levels of government, could bring floods swiftly under control and repel invasions—and thereby save lives. All in all, compared with the so-called elected officials in the West, Chinese government officials at every level were bound more closely, ideologically, to the state's mainstream ideology, and had risen through direct administrative experience from the lowest ranks to the highest—an arrangement that corresponds closely to what the Western scholars describe as "politics of meritocracy."

Through the imperial examination system, the dynasties from the Suí and Táng through the Míng and Qīng achieved genuine openness in the official recruitment, drawing talent from all over the country into state service. Korea and Vietnam adopted and adapted this same examination system to recruit their own government officials from the tenth century onwards, and the resulting East Asian Confucian cultural circle stands as clear testimony to the influence of Confucian thought. Yet this system carried a significant cost. Because the imperial examination questions focused almost exclusively on the Confucian classics, it discouraged the study of natural sciences and, over time, impeded scientific and technological progress in imperial China—a limitation that became especially consequential as European countries industrialized through the eighteenth and nineteenth centuries. The European countries fought one another repeatedly across the period from roughly 1500 to 1900,[51] in order to prevail, they invested heavily in military-related scientific research aimed at producing more advanced guns and gun-powders. This state-funded research helped give rise to modern chemistry, physics, and the related manufacturing technologies—development that, in turn, laid the groundwork for industrialization. China, by contrast, enjoyed relative peace and security through the Míng and early Qīng, which left the state with little urgent incentive to fund comparative scientific research—a key reason it fell behind once industrialization took hold in England and the rest of Europe.[52]

### 3.2 The Economic Balance of the Ancient Collectivist Confucian Open Society

Before the end of the Míng and the beginning of the Qīng, "collectivist" China remained fundamentally an agrarian society, yet it stood as the most powerful country in the world across political, economic, social, and cultural dimension. The neighboring countries were relatively backward compared with China. In this international context, the central tasks facing "collectivist" agrarian China were twofold: feeding the people while continuously preparing for fighting flood and other natural disasters, and perpetually resisting nomadic incursion. Flood fighting and disaster relief required the construction and maintenance of river embankments, the establishment and operation of countrywide grain reserves to stabilize food prices, and the grain warehouses for famine relief. Resisting nomadic incursion, in turn, required a standing army, the construction of the Great Wall and other major fortifications, and canal systems to transport military logistics. Therefore, in addition to the expropriation of land taxes, "collectivist" China also required a large amount of military service and corvee labor every year. In other words, "collectivist" China had to strike a continual balance between "fighting floods and defending against invasion" on one hand and "feeding the people" on the other. When the burden of large-scale construction projects exceeded what the people could bear, it posed a direct threat to the stability of the country. The premature demise of both the Qín and Suí Dynasties illustrates this danger: the Qín fell in part due to the mobilization of millions of laborers to build the Great Wall, while the Suí collapsed following the Emperor

Yángdì's (隋炀帝) massive conscription of labor to construct the Grand Canal. Many Chinese dynasties nonetheless lasted for more than two hundred years— the Western Zhōu, Western Hàn, Eastern Hàn, Táng, Sóng, Míng, and Qīng—and hence the population had increased rapidly during each of these long-lasting dynasties. The rich and powerful were insatiable—the rich had thousands of miles of land, the poor had no place to stand—and taxation was heavy. These circumstances plus the natural disaster such as floods or pandemic, could trigger large-scale food shortage severe enough to bring down an entire dynasty: a pattern consistent with what is often termed the "Malthusian population trap."

### 4. Value of "Virtue First" and "Duties"

#### 4.1 The Social Mainstream of "Governing by Virtue" and Men's Duties

The classics from the Zhōu Dynasty reveal how the idea of virtue and morality, as the ideological foundation of the country, gradually expanded to become the mainstream of Chinese political thought. One classical source describes the Shāng approach to governance this way: "People in Shāng Dynasty respect the gods, lead the people to do things for the gods, put ghosts first and then rituals, punish first and then rewards, respect but not close, (when politics and religion are declining) its people become dissolute and unquiet, striving for strength and victorious without being ashamed."[53] In other words, Shāng society relied on religious rituals to maintain social cohesion. The early rulers of the Western Zhōu, by contrast, pursued a markedly different and more energetic approach: they "practice benevolence and righteousness, respect the elders, and be kind to the young; be courteous to the sages, forget lunch and rest in order to receive the scholars, so the scholars turned to the Zhōu one after another"[54] and "never forget the benevolence of the ancestors, inherit and carry it forward. Only by complying with the mandate of heaven all the time can we obtain more blessings from the heaven."[55] Wáng Guówéi observed at the beginning of the twentieth century: "The most dynamic change of the Chinese politics and culture happened at the end of Shāng and the beginning of Zhōu."[56] Since the Zhōu Dynasty, virtue and morality had become the internal cohesive force binding collectivist society together, rising to serve as an essential foundation upon which a ruler could claim—-and retain—the Mandate of Heaven, and thereby lead the collective. The key to maintaining a "collectivist" society lies in cultivating the unity of purpose among its peoples. The so-called "the country becomes strong when the people unite, whereas the country gets weak when the people are not united;"[57] and "there will be more private interest groups when a state is in chaos, there will be more violence when the state army is weak."[58] The thinkers of the turbulent Spring and Autumn and Warring States Periods clearly realized that the inconsistency between society's moral ideals and the actual rules governing behavior would inevitably produce widespread disorder. In the process of normalizing the people's behavior, both the deterrent force of severe penalties and the cultivating force of virtue and moral education played significant roles in the long history of China. However, given the high social cost of unrest that the Legalists' harsh punitive system tended to provoke, the dynasties following the Western Hàn increasingly turned to Confucianism's emphasis on virtue and moral education as their primary strategy for maintaining social order. In the Four Books and Five Classics and other core Confucian texts, moral behavior was systematically codified as ritual propriety (礼, lǐ)— with the understanding that "rituals and social etiquette to a state are what the scale is to weight, what the ink-line is to crookedness and straightness, and what compasses and squares are to circles and squares."[59] The state rulers, accordingly, were expected to follow rites and etiquette, embodying virtue as a model of entire country, and in doing so guide society toward the shared ethical norms that held collectivist society together.

Under this system of virtue-based rule, the people were to be enlightened and educated through "ritual propriety" (礼, lǐ), cultivating a shared sense of "righteousness" (义, yì) that the people would internalized as genuine belief. This, in turn, established a moral framework governing the standards of conduct expected of each social class. In a collectivist society with the top ruler at its center, the promotion of virtue and morality was thus closely bound up with the maintenance of political rule itself: placing virtue first helped ensure that governmental orders and directives were implemented properly, and helped the ruler retain the Mandate of Heaven. In addition to the ruler's own moral conduct, however, an effective moral system under collectivism also depended on the social mobility across classes. The social mobility—from the bottom to the top—was addressed, in large part, by the Confucian open society under the imperial examination system. Downward moral enlightenment—diffusing virtue from the ruling class to the population at large—depended instead on the ruling class actually embodying virtue and moral role as a visible model for others to follow. Therefore, "What Confucius called "teaching", the knowledge that it focuses is moral knowledge."[60] Virtue and moral standards were disseminated through the principle of "education regardless of class"—the idea that the moral concepts essential to social stability under collectivism should reach the entire population, not merely an elite. The imperial examination system reinforced this dynamic directly: it stressed virtue and moral standards alongside practical competence, and rewarded both moral behavior and scholarly learning. Consequently, this system unified moral cultivation, social mobility and collectivist governance into a single coherent structure—completing the integration of collective concepts and open society dominated by the Confucianism. This integration ensured that, within the cultural context of collectivism, successive Chinese dynasties consistently pursued what was pointed out in the Book of Documents: "The ruler must first cultivate himself in prosperity and virtue, so that the members of his own clan and extended kin will draw close to him; from there, the neighboring clans can be brought under good governance; and ultimately, even the most distant states can be unified;"[61] and "the nine major regions have achieved consensus, and the places where the rivers bend were inhabited. The mountains of the nine regions were connected by wooden roads, and the river flowed smoothly through the plains. Dams had been built for the large lakes in the nine regions, and the whole country had been unified."[62] This integration reflects a consistent pattern across Chinese dynasties: within a collectivist cultural framework, rulers pursued governance as a process of concentric expansion — beginning with self-cultivation, extending outward to kin, then neighboring regions, and ultimately achieving unification of the realm.

In the West, "Life is dear, love is dearer. Both can be given up for the sake of freedom."[63] In China, people value "cultivate the self, bring order to the clan, govern the state, and bring peace to all under the heaven,"[64] which refers to a man's responsibilities and obligations toward himself, his family, his country and society. Regarding "cultivate the self", Confucian classics contain numerous discussions. A disciple of Confucius remarked, "An educated gentleman may not be without strength and resoluteness of character. His responsibility in life is a heavy one, and the way is long. He is responsible to himself for living a moral life; is that not a heavy responsibility? He must continue in it until he dies; is the way then not a long one?"[65] A disciple of Confucius, the favorite Yán Huí, enquired what constituted benevolence. Confucius answered, "To discipline oneself and return to ritual propriety constitutes benevolence."[66] Benevolence (仁，rén) originates from human inner sentiments, first manifested in familial affection: "filial piety and fraternal respect constitute the root of benevolence."[67] It expands outward from the love between fathers and sons, as well as between elder and younger brothers. Rather than an externally imposed set of rules, benevolence is an innate moral capacity within human beings. Confucius also articulates the two dimensions of the principle of zhōng-shù (loyalty-and-reciprocity): "Wishing to establish himself, one establishes others; wishing to attain success himself, one enables others to attain success;" [68] and "Do not impose upon others what you do not desire for yourself." [69] Together they constitute the method to practice benevolence (rén). In *the Great Learning*, a Confucian classic, emphasized on self-cultivation "From the Emperor down to common people, without exception all take self-cultivation as the fundamental."[70] Confucian classics converge on self-cultivation as a lifelong, arduous moral responsibility grounded in innate benevolence (rén)—rooted in filial and fraternal affection, disciplined through ritual propriety, and enacted through the reciprocal principle of zhōng-shù—standing as the universal foundation of moral life for all persons, from ruler to commoner.

On responsibilities and obligations toward the family, the state and society, Confucian classics offer far more discussions. A disciple of Confucius enquired what constituted a wise and good man, Confucius answered, "He cultivates himself so as to give rest to all the people."[71] Zǐgòng said: "Suppose someone bestows extensive benefits upon the people and brings relief to the multitude — what would you make of him? Could he be called benevolent?" Confucius replied: "This is far more than benevolence; this is sageliness! Even Yáo and Shùn found it difficult to achieve." [72] Mencius said: "When in poverty, one cultivates one's own virtue; when in success, one brings succor to all under Heaven."[73] As Xúnzǐ · Zi Dao states: "Follow the Dao rather than the ruler; follow righteousness rather than one's father."[74] This maxim indicates that moral principle takes precedence over unconditional personal loyalty to superiors or kin. Confucian classics extend moral obligation outward in concentric circles—from filial piety as the root of benevolence, through self-cultivation aimed at benefiting the people, to the sage-like aspiration of relieving all under Heaven—while ultimately subordinating personal loyalty to ruler or father to the higher authority of the Dao and righteousness themselves. In Ku Hung-Ming's words, these Confucian rules of conduct are called "honor," or "a code of honor," or "the law of gentlemen."[75]

### 4.2 Cultivate the Value of "Virtue First" in Eulogizing of "Virtue" and Condemning "Evil Behaviors"

To ensure the survival of "virtue and morality first" as core value within a collectivist society, the moral standards established by the great deeds of the sages and classical texts were reinforced through the careful commendation of historical figures and folklores. Every civilization reveres its heroic deeds and condemns its villains: the praise of virtue and the condemnation of evil behavior function as a manifestation of social order in their own right. Additionally, each civilization tends to develop its own distinctive moral standards and vocabulary. The virtue and morality that gradually established over the course of Chinese long history clearly reflect the spirit of the Confucian open society operating within a collectivist framework. Across this unbroken civilization continuity, Chinese values—or "value orientation" —were stabilized through the ongoing process of praising great deeds and condemning evil ones, generating both the restraining force of public opinion and positive incentive models for social conduct.

Under collectivism, the measure of virtue was whether an act served the collective interest. Therefore, within the unitary framework shaped by early flood control and resistance to nomadic incursions, virtue and moral standards possessed an inherent unity running from rulers and officials at the top down to the common people. The deeds of great monarchs praised across millennia include "the legendary rule of the Greats Yáodì (尧帝) and Shùndì (舜帝)," "the Great Yǔdì (禹帝) controlled the waters, and did not enter his home for three times when passing by,"[76] "the Duke Dàn (周公旦) in the early Zhōu Dynasty worked day and night to maintain and improve the unity of the country,"[77] "Qín Shǐhuángdì (秦始皇) unified China," "the rule of Emperors Wéndì (汉文帝) and Jìngdì (汉景帝)" in the early Western Hàn, and "the rule of Emperor Tài Zōng (唐太宗)" in the early Táng Dynasty, "the rule of Emperors Kāngxī (康熙) and Qiánlóng (乾隆) in the Qīng Dynasty" and so on. Each of these periods allowed the people of the country to live in relative peace and security, with lighter corvee burdens and fewer worries over food and clothing. Since the beginning of modern era, similar praise has extended to more recent leaders and their achievements—commonly summarized in phrases such as "Máo Zédōng enabled China to stand up," "Dèng Xiǎopíng made China prosperous," "Xí Jìnpíng has made China strong." The heroic efforts to preserve Chinese people's lives—resisting foreign invasion, defending national sovereignty and territory, and recovering occupied land—came to define the virtues attributed to national heroes for 1840 to 1945, and later to the leadership of the CPC and PRC. Where a leader succeeded in preserving peace, security, territory, and sovereignty, that leader was understood to embody the virtues meant to shine through the annals of history. At the same time, the virtue and morality of ordinary people mattered equally, with the defense of collective honor and interest regarded as a form of great righteousness. Popular sayings such as "The river harnessed by Wāng Jǐng (王景, an imperial official in the early Eastern Hàn Dynasty) lasted a thousand years," and "the Yellow River become clean once a Saint come into being"[78] reflect how water-control projects vital to the lives and livelihood of millions were praised by both official history and popular folklore. Official records and the folklores heaped praise on the individuals who rose from humble origins through the channels of the Confucian open society: "Mencius's mother moved three times in order to provide him with good education," "the four-year-old Kǒng Róng (孔融) knew of giving his pear to the elder," "Scholar Sū Qín (苏秦) in the Warring Period studied extremely hard" and the likes are being regarded the personal virtue in terms of "cultivating oneself and elevating one's family in all aspects." Samuel Huntington said: "The Chinese civilization is the oldest one in the world. The Chinese have a clear understanding of the uniqueness and achievements of their civilization."[79]

Since ancient times, collectivist China has functioned as a tightly integrated country, one in which top leaders and officials at every level were expected to "care for the people," "be honest and upright," "levy less and lighten corvee labor," "be fair and just," "reward and punished clearly," and "take the lead," and the like. Only through this kind of "virtue and morality first" conduct, the argument goes, could government orders be properly carried out, natural disasters be brought under control, and foreign invasions be repelled through national collective endeavor at critical moments, and China's continuity ultimately be sustained.

Just as collectivist China praised virtue, it ruthlessly condemned evil behavior. For thousands of years, the Chinese people have reserved their deepest condemnation for the rulers who "destroying the country and its people"—the monarchs who led rotten life, letting traitors holding key position, torturing the loyal, and ruthlessly exploit the people—as well as the traitors, and foreign aggressors who killed the Chinese people and occupied the Chinese territories. Rulers commonly cited as embodiments of such misrule include the desolate and cruel King Jiéwáng (夏桀王) in the late Xià, King Zhòuwáng (商纣王) in the late Shāng, the Second Emperor of Qín, Emperor Yángdì (隋炀帝) of Suí, and the like. Not only did "evil conduct" corrupt the social atmosphere, but more importantly, caused the collapse of the dynasty. The top ruler's indulgence and licentiousness, government officials' "corruption and betray of the law," "not rewarding the good and not punishing evil," "favoring cronies" and the like, would inevitably lead to deterioration of public morality, less people obeying the laws and decrees, the dignitaries stealing the public wealth, the lives of the people being miserable, no one willing to fight for the country, and finally the collapse of the dynasty. The most important reason that several dynasties rose soon and fell fast in the long Chinese history is the moral corruption of the emperor and government officials at all levels.

For thousands of years, the Chinese people have despised those who betray the nation and seek to split the country. Frequently cited examples include: "Ān Lùshān (安禄山) and Shǐ Sīmíng (史思明) Rebellion in the late Táng Dynasty," "Shí Jìngtáng (石敬瑭) Ceding Yōuyún (幽州) Sixteen Prefectures to the Khitan (契丹) at the end of Táng Dynasty," "Emperors Huīzōng (宋徽宗) and Qínzōng (宋钦宗) of the Northern Sòng Dynasty being captured and prisoned in the north," "the traitor Qín Huì (秦桧) in the Southern Sòng Dynasty," "the scholar Hóng Chéngchóu (洪承畴) of the late Míng Dynasty (who claimed that 'the Emperor's grace is like the sea and the ministers' loyalty should be as firm as a mountain') surrendered to the Qīng Dynasty," "ceding Hong Kong to Britain after the 'Opium War' in the late Qīng Dynasty," "continually ceding land and paying indemnities to the Western and Japan by the Empress Dowager Cíxǐ (慈禧太后) and Prime Minister Lǐ Hóngzhāng (李鸿章)," "ceding Taiwan to Japan," "the Japanese and Russian war in Liáoníng and the government of the late Qīng Dynasty ceded Lǚshùn (旅顺) Port to Japan," and the like. Traitors, in pursuit of temporary safety and personal gain, voluntarily cede border passes, islands and entire territories to foreign invaders, directly shrinking the nation's living space. Once vast tracts of land fall out of national jurisdiction, their loss is enshrined in treaties and consolidated through prolonged foreign immigration and rule. Even if the country regains its national strength in later generations, it can hardly recover these lands in full, leaving permanent territorial regrets. The many unequal treaties of the late Qing Dynasty involving territorial cessions serve as a prime example: vast swathes of land in Northeast and Northwest China were stripped from the national map, pushing the border lines far inward. People living on the ceded lands were instantly reduced to second-class subjects under alien rule, their household registrations, land, and property no longer protected by the law of their own country. Invaders imposed discriminatory policies and extorted heavy taxes, plundering grain, gold, silver, and mineral resources, leaving common people who toiled year-round unable to secure adequate food and clothing. To entrench their occupation, the invaders pursued policies of cultural erasure: destroying historical records and classics, banning education in the native language, distorting local history, and forcing people to abandon their traditional customs. Those who resisted faced mass massacres; villages and towns were burned to the ground, countless families were torn apart, and generations endured the humiliation of subjugation under foreign rule.

The barbarian deeds of the imperialist are as below. "The Opium War of 1840," "The Anglo-French coalition burned down the Old Summer Palace," "the Eight-Power Allied Forces invaded Beijing," "the Japanese invasion of China marked by 'the September 18th Incident in 1931,' 'the July 7 Incident in 1937,' and 'the Nánjīng Massacre,'" and the like. China's century of humiliation in modern history was brought about entirely by imperialist aggression: the nation shackled by unequal treaties, its territory carved up, its people mired in destitution, its culture devastated, and countless compatriots subjected to enslavement. This agonizing chapter of suffering has forged a shared national sentiment among all Chinese people. From the bottom of their hearts, the people detest imperialist invaders and firmly stand against all hegemonic and aggressive acts.

**5. Recreation of Collectivist China and its Open Society in the Age of Industrialization**

From the seventeenth century onward, the West—pioneered by England—entered the age of industrialization. In this unprecedented tide of the industrial revolution and scientific advancement, China lagged behind. Compared with the West of that era, China suffered from political corruption, limited educational reach, technological and industrial backwardness, and widespread poverty—conditions attributable in part to an imperial examination system that tested only Confucian classics but not mathematics, physics, chemistry, engineering, and the like. As a consequence, China gradually lost the capacity to resist foreign aggression from the West and Japan. "Collectivist" China thus faced the formidable task of rebuilding its social, political, economic, and cultural institutions from the ground up. After a series of major historical upheavals— repeated invasion and slaughter at the hands of the Western powers and Japan, the end of dynastic rule, the rise of party politics, the joint resistance of the Japanese Invasion by the Kuomintang and the CPC, and finally the founding of the PRC in 1949—collectivism China had finally succeeded in reconstructing its core institutions. In an international environment still dominated by Western powers, the newly reconstituted PRC subsequently fought a series of wars: the Korean War (the War to Resist US Aggression and Aid Korea), the self-defense against India in 1962, and the resistance to US aggression of Vietnam.

Having completed the "recreation" of old China, the PRC went on to construct its revolutionary "open society". The CPC, by its very nature, is a political organization representing the broad masses of working people who were exploited and oppressed by imperialism, feudalism, and bureaucratic capitalism before the founding of PRC; many of its early leaders themselves rose from modest social origins. As far as political participation is concerned, any lawful Chinese citizen is entitled to apply for joining the CPC and becomes a member of it, and anyone has the opportunity to become a staff member of the Chinese government or even a senior leader of the country. Therefore, the PRC's politics is open to all Chinese citizens. The PRC has also created a series of modern public services that are open to all Chinese people, especially the working people: nine-year compulsory education system, higher education system, public healthcare, retirement pension, gender equality, ethnic equality, religious freedom, and so on. During the period from the founding of PRC to the eve of the reform and opening-up in the late 1970s, the building of open society has been transformed from simply "absorbing elites into politics" of the dynastic China to "guaranteeing people's livelihood," "enhancing human capital," and "ethnical equality," guided by the "people-centered" socialist framework. The CPC's own regulations bar it from accepting donations from individuals or organizations outside the CPC, a rule insulates the Party's decision making from influence of outside interest groups. However, the Chinese government made some mistakes in the first 30 years of the PRC. The Great Leap Forward (1959 – 1961) precipitated a severe famine, and the Cultural Revolution (1966 – 1976) brought profound disruption to the country: higher education admissions were suspended for roughly a decade, national scientific research came to a standstill, and the broader economy suffered extensive damage.

Since the reform and opening up in the late 1970s, the construction of "collectivist" China and its "open society" has entered an unprecedented new stage. "Reform and opening up" refers, broadly, to two parallel tracks: domestic economic reforms and opening up to the outside world. The domestic "openness" reform measures include: distributing land to rural households on the basis of long-term leases (replacing collective farming with the household responsibility system); allowing private enterprises to emerge and develop in both urban and rural areas; permitting rural surplus laborers to work in cities; substantially expanding enrollment in colleges and universities; and abolishing the agricultural tax that the rural households had paid since the beginning of the Chinese Civilization; rebuilding the new rural cooperative medical insurance system; eradicating poverty through the "rooting out poverty campaign", among other measures. Opening up to the outside world, meanwhile, encompassed: allowing foreign-funded enterprises to invest in China through joint ventures or wholly-foreign-owned enterprises; sending students and scholars abroad for study; joining the World Trade Organization; hosting international events such as the Asian Games and the Olympic Games; creating multinational institutions such as the Shanghai Cooperation Organization and the Asian Infrastructure Investment Bank; and launching the Belt and Road Initiative, among other efforts. Since the reform and opening up, collectivist China has successfully industrialized, become the world's second largest economy, emerged as a global high-tech power, and finally escaped from the curse of the "Malthusian population trap."

Taken together, the political structure of "collectivism" China evolved from the enfeoffment system of the Western Zhōu, through the fragmentation of the Spring and Autumn and the Warring States periods, to the centralized commandry-county system established under the Qín and Hàn. Subsequent dynasties largely retained this centralized structure. Such an advanced political system of the Qín and Hàn did not appear in other major civilizations in the same period in the world. Under successive dynasties, the imperial governance could only reach the county level—not the grassroots. In addition, under the political framework spanning the Qín and Hàn through the Míng and Qīng, the hierarchies were strict, and identification with the national ideology remained largely confined to officials, scholars, and gentry class. In other words, imperial authority and its ideology never fully penetrated down to the laboring masses. Kuomintang China's governance during its continental era was not much different from previous dynastic politics, mainly to safeguard the interests of bureaucratic capitalists, comprador capitalists, and land-owning gentry class.

What really makes the fundamental difference is the anti-aggression war and the people's revolution against bureaucratic capitalists led by the CPC. Not only had the cadres and army of the CPC never invaded the interests of laboring masses, but also distributed land to them—first through rent and interest reduction during the eight years of the Kuomintang-CPC cooperation against Japanese invasion (1937-1945), and later through full land reform—gaining the sincere support of hundreds of millions of toiling masses. On this foundation, the CPC effectively mobilized the bottom laboring masses of China's vast rural areas and drove the Kuomintang regime, despite its roughly eight-million-strong army, to Taiwan in just three years (1946-1949). After the founding of the PRC, the most thorough land system reform in human history was implemented: the land of landlords and capitalists was confiscated and redistributed equally among hundreds of millions of working people, eliminating large-scale private control over land as the principal means of agricultural production. Thus, the deprivation of the core productive means – land – was rooted out. Alongside this, the PRC built out a governance structure reaching well below the traditional county level. Beyond the central, provincial, prefectural, and county governments, it set up township government (each of them governs ten to twenty villages), along with Party branch and a villagers' committee in every village. During the period of rural collectivization—from the Great Leap Forward in 1958 until the eve of the reform and opening up—townships were reorganized as people's commune, villages became production brigades, and each brigade was further divided into production teams. In urban areas, sub-district offices and residents' committees were established under the municipal and district governments. Taken as a whole, the PRC's system of governance reached, in a way earlier dynasties never achieved, into nearly every household in the country.

Most of the political, social, and economic policies implemented since the founding of the PRC drew on the Marxism-Leninism, which was closely connected to the equalitarian ideals of the French Revolution in 1789. The ideas of equality of the Enlightenment culminated in the French Revolution of 1789-1830. The French Revolutionary, whose revolutionary regime issued the Declaration of Human Rights, emphasizing the equality of human beings alongside the legal protection of private property. This is the so-called "the second generation of human rights".[80] The equality concept triggered revolutionary struggles in European countries against feudal autocracy, the rise of parliaments representing the interests of the bourgeois interests, the labor movement against capitalist exploitation, and the colonial people's resistance to colonial oppression and pursuit of national independence. The anti-aggression and anti-oppression revolution led by the CPC in China is also an important part of this broader global current of revolutionary movements among the oppressed people and nations since the French Revolution in 1789.[81] In the earlier period of the PRC, China put more emphases on the second generation of human rights, i.e., implementing the most thorough equality measures, such as equal distribution of land in rural China, universal access to the nine-year compulsory education and the medical care, gender equality, and the like. Since the economic reforms of the late 1970s, by contrast, China has given more consideration of the first generation of human rights, such as allowing private enterprises, foreign investment, returning arable land to rural households on long-term lease, rural-urban migration, and the development of labor and stock market.

**6. Conclusions**

Both the Yangtze River and the Yellow River originate from the west Qīnghǎi-Tibet Plateau, flow through the middle of China, and finally merge into the Pacific in the east coast of China. The annual flooding disasters of the Yangtze River and the Yellow River make it vital for China to have a strong leadership to mobilize and coordinate the residents of various tribes to fight the flood. Similarly, the yearly northern nomadic incursion forced the regime of North China's Central Plain to establish a standing army to defend the border. Encountered floods and invasion, no one would be able to survive without depending on his/her country — the collective — which is the fundamental reason of the creation and existence of the collectivist China.

Collectivist China experienced the “feudal system” of the Western Zhōu Dynasty, the separatism of the vassals during the Spring and Autumn and Warring States periods, and finally moved towards the centralized power system or the system of commanderies and counties in the Qín and Hàn Dynasties. This political system has continued to exist in modern China. In order to enhance the collectivist China, all the dynasties of China had continually carried out the construction of “open society”. The main carrier of it is the imperial admittance of officials by civil examination system. From the Qín and Hàn dynasties to the late Míng and early Qīng dynasties, China — a “collectivist Confucian open society” — had always been the most powerful and most developed and civilized country in the world in terms of political civilization and economic development.

The ideology of pre-Qín China began with the formation of the concept of de (virtue 德). Western Zhōu texts like the Book of Documents emphasized “virtue” as the foundation of legitimate rule, understood by later thinkers—Guǎn Zhòng, Lǐ Zéhòu, and Xú Fùguān—as impartial, selfless governance that delivers genuine benefit to the people rather than serving the ruler's private interests. Confucian and Xúnzǐ’s philosophy further elaborated this into a political theory: Confucius likened virtuous rule to the North Star commanding natural deference, while Mencius and Xúnzǐ argued that winning popular hearts through benevolence and righteousness—rather than force or wealth—produces voluntary submission, genuine shame-based self-restraint, and lasting political stability. Ultimately, this tradition holds that only rule by virtue generates the deepest collective loyalty, as it alone can inspire people to attach themselves to their ruler so completely that they would willingly give their lives in his service.

Throughout Chinese history, successive dynasties worked continually to build a more "open society" as a way of strengthening collectivist China. The Qín Dynasty abolished the feudal, hereditary system in favor of centralized power, implemented through the commandery-and-county system: the emperor now directly appointed officials at every level, replacing the hereditary aristocracy. To broaden the pool of candidates for office, the Western Hàn introduced a system in which local officials and gentry recommended candidates to the emperor. Overt time, however, this inspection-and-recommendation system degenerated into a tool by which a handful of powerful clans controlled official appointments, effectively becoming a new hereditary aristocracy in disguise. Following the collapse of the Eastern Hàn in the early third century CE, China splintered into nearly four centuries of division, not reunified until the Suí Dynasty in the late sixth century. To prevent official appointments from once again being monopolized by a few eminent families, the Suí pioneered the imperial examination system—a more open and competitive mechanism that selected officials on the basis of merit demonstrated through examination performance. In this light, collectivist China was sustained and reinforced not by rigid hierarchy, but by the ongoing expansion of meritocratic openness.

In order to "control floods" and "resist nomadic incursion" more effectively, "collectivist" China required a unified ideology for the entire country to a greater degree than other civilizations. As early as the Spring and Autumn Period, Confucius advocated "restoring the Rituals of Zhōu"—that is, returning to a state in which the whole country recognized only a single legitimate "leadership". Under the rein of Emperor Wǔdì of the Western Hàn, Confucianism was elevated as the sole orthodoxy, while all other schools of thought were suppressed. From this point on, Confucianism gradually became both the core doctrine of the imperial examination system and the guiding ideology of the entire empire. Unlike the Western world, which has emphasized liberty since the Renaissance, traditional Chinese Confucian philosophy has long placed greater emphasis on responsibilities and obligations. Confucian tradition centers on "cultivate the self, bring order to the clan, govern the state, and bring peace to all under heaven," a framework describing a person's expanding responsibilities from himself outward to family, state, and society. Self-cultivation is portrayed as a lifelong, weighty moral burden grounded in innate benevolence (rén), which begins in filial piety and fraternal affection, is disciplined through ritual propriety, and is practiced through the reciprocal principle of zhōng-shù—what you do not wish for yourself, do not impose upon others; when one wishes to attain success oneself, one helps others attain success. This self-cultivation is not merely personal but foundational and universal, applying "from the emperor down to common people" as the root of moral life. From this inner foundation, Confucian obligation radiates outward in concentric circles: the cultivated man serves to give the people rest, aspires toward the sage-like ideal of relieving all under Heaven, and—per Mencius—cultivates virtue in poverty while aiding the world in success. Crucially, this moral framework ultimately places allegiance to the Dào (道) and righteousness above unconditional loyalty to ruler or father, constituting what Ku Hung-Ming termed a code of honor or "the law of gentlemen."

Wittfogel's central claim is that water control civilization gave rise to "despotism" as a system of governance, and on this basis he classifies China — a typical water control society— as an instance of "Oriental Despotism." China's recurrent flooding meant that residents along its major rivers could not survive without collective water-control efforts; this fact can itself be abstracted as “collectivism”. In ancient China, controlling flood disasters and repelling frequent nomadic incursions required establishing nationwide, unified command over both water-management agencies and military forces. The emergence of “collectivism” as a response to these crises is therefore, in essence, the unification of civil and military authority—precisely what Wittfogel, viewing it from the outside, characterized as “despotism”. Acemoglu, for his part, argues that despotism or autocracy entails the unchecked extraction of resources from the ruled by a small number of ruling elite, and that such extractive systems ultimately end in the defeat and disintegration of this autocratic state. What Wittfogel failed to account for is that, since the Reforms of Shāng Yāng in the mid-fourth century BCE, “collectivist” China had already begun building an “open society” through its political and economic institutional arrangements, as discussed above. Although China fractured into nearly four centuries of division following the fall of the Eastern Hàn, it was reunified under the Suí and Táng, and the continued construction of a more “open society” across the Suí, Táng, and Sòng dynasties made China the world's most powerful state from the Suí-Táng era through the late Míng and early Qīng. By contrast, following the fall of the Roman Empire in 476 CE, Europe descended into division and the “Dark Ages” of the medieval period, not beginning to emerge from feudal servitude until the Italian Renaissance of the early fourteenth century.

After having gradually set up a series of “openness” institutional arrangements, “collectivist” China's institutional construction had been continuously improved. However, the patrimonial system in Chinese dynastic politics still resulted in child or weak emperors without any administrative experience, which in turn caused imperial power to fall into the hands of vested groups or eunuchs, poor state governance and even the periodic collapse of the dynasty. In addition, the imperial examination system took the Confucian classics as the core content, and did not examine natural sciences and engineering. More importantly, the peace and security enjoyed by China during the period of Míng and the early Qīng Dynasties did not generate enough urgency for the state to fund scientific research to manufacture powerful guns and gun-powders, which happened in the European countries that continually fought with each other during the period from 1500 to 1900. Because of this, since the late Míng and Early Qīng period, China was ignorant in science and technology, backward in industrial development, economically impoverished, and out of date in the era of the western industrial revolution that initiated in the middle of the seventeenth century.

Due to its backward education, technology, industry and economy, China was humiliated by the West and Japan in every possible way from the Opium War in 1840 to the victory of the Anti-Japanese War in 1945. During this period, China changed from dynasty politics to party politics, and from a patrimonial system to an “open” selection system for the top leaders of the country based on meritocracy, and completed the reconstruction of its political, economic, social and cultural system, the symbol of which is the founding of the PRC led by the CPC in 1949. The core purpose of the CPC is to “serve the people wholeheartedly” and pay special attention to serving the toiling masses. It should be said that this is the most collectivist declaration of “openness” and “inclusiveness.” To this end, a series of more thorough and “open” political, economic and social institutional arrangements have been created and established, such as the abolition of private land ownership, the setup of a nine-year compulsory education system, public medical services and other public services. These policy measures manifested the second generation of human right theory, i.e., “human beings born equal.” The “reform and opening up” started in the late 1970s greatly accelerated the process of collectivist China's construction of “open society”, allowing free flow of labor, the establishment of non-state-owned enterprises and the investment of foreign-funded enterprises, and proactive integration into the process of economic globalization, and the like. These reforms reflected the first generation of human right theory in term of market economy. The rapid advancement of the “open society” has enabled collectivist China to successfully realize industrialization in about 40 years after reform and opening up, surpassing Japan to become the world's second largest economy in 2010. The years since the CPC's 18th National Congress in 2012 have witnessed a transformation in China's technological standing. What was once a country striving to close the gap with more advanced economies has emerged as a global high-tech power, its industries advancing across a broad and expanding frontier of innovation.

China's historical institutional innovations — collectivism resulted from saving lives from yearly recuring floods, the centralized commandery-county system embodied by providing security and imposing uniform taxes, corvees, metrology and written script to its people, rule by virtue characterized by equally and fairly distributing material benefits to its inhabitants, and the Confucian open society epitomized by the imperial examination system — yielded a momentous outcome: continuously provision of increasingly equitable treatment and dignity as human beings to all residents within its territory, while offering a growing number of fair avenues for social advancement to those who aspired it. The more humane a civilization’s institutional arrangements are, the stronger the sense of belonging among inhabitants within the territory, the longer its governing and sustained stability.

The history of China's anti-imperialist and anti-aggression since 1840 implies that only by maintaining a strong political, economic, social and cultural environment can China remain invincible in international political and economic competition. The history of the Chinese civilization since the Greats Yáodì, Shùndì and Yǔdì warns that China must hold onto the core of “collectivism” and strengthen “collectivism” through continual “openness” institutional construction. In other words, “collectivism” and “open society” are the core elements for China to remain strong.

Notes:

[1] Wittfogel, 1957.
[2] Acemoglu & Robinson, 2012, p.450-484.
[3] Bell, 2015.
[4] Fukuyama, 2011, p.25-30 and 99.
[5] Maddison, 2007, p.1.
[6] Lao Tzu, Chapter 25. "人法地，地法天，天法道，道法自然。"Tao is the basic concept of Taoism. The name "Taoism" itself comes from the word "tao". Generally speaking, tao can be translated by "way", "principle", "method". (https://www.laotzu.info/tao.html, entered on 5th of June 2021)
[7] Lao Tzu, Chapter 80. "邻国相望，鸡犬之声相闻，民至老死，不相往来。"
[8] Sun Yatsen, 1916, p. 316. "中国四万万之众如同一盘散沙，此岂天生而然耶？"
[9] Hobbes, 1651, p.112-124. Fukuyama, 2011, p.26-27.
[10] Xúnzǐ · The Kingly System (《荀子·王制》), "人生不能无群"
[11] Xúnzǐ · A Discussion of Rites (《荀子·礼论》), "人生而有欲，欲而不得，则不能无求。求而无度量分界，则不能不争；争则乱，乱则穷。先王恶其乱也，故制礼义以分之。"
[12] Zhao, 2016, p. 9–11.
[13] According to Deng (Deng, 1937, p. 47), there would be a severe flood every three and half years for the period from 1766 BC (since when there was the recorded history in China) to 1937.
[14] Engles, 1884, p. 229.
[15] Xí Jìnpíng, 2014. ("我国社会主义制度能够集中力量办大事是我们成就事业的重要法宝。我国很多重大科技成果都是依靠这个法宝搞出来的，千万不能丢了！")
[16] Xí Jìnpíng, 2020. ("武汉和湖北是疫情防控阻击战的主战场，武汉胜则湖北胜、湖北胜则全国胜。一方有难，八方支援。我们举全国之力实施规模空前的生命大救援……")
[17] Xúnzǐ · The Efficacy of Confucian Scholars. (《荀子· 儒效》, "兼制天下，立七十一国，姬姓独居五十三人，而天下不称偏焉。")
[18] 《礼记・王制》："大国三卿，皆命于天子"、"次国三卿，二卿命于天子，一卿命于其君。"
[19] In Zuǒ Zhuàn · Duke Xī, Year 12, when Guǎn Zhòng declined the ritual honours due to a senior minister, he remarked: "The two guardians appointed by the Son of Heaven, the Guo and Gao clans, are present." This explicitly records that the Guo and Gao clans in the State of Qí were "guardian ministers" appointed by the Zhōu King. Precisely because they were "mandated senior ministers" of exalted status, Guan Zhong, though a key minister under Duke Huán of Qí, could only rank below them. (《左传・僖公十二年》提到，管仲辞让上卿之礼时，说"有天子之二守国、高在"-19。这明确记载了齐国"国氏"和"高氏"是周天子任命的"守臣"-。正因他们是"命卿"--19，地位尊崇，所以管仲虽为齐桓公重臣，也只能位居其下。).
[20] 豆闭簋（guǐ）：西周穆王时期青铜器，铭文记载周王册命"豆闭"为"俞邦司马"（即诸侯国"俞"的军事长官）.
[21] The well field system (Jǐng Tián Zhì), was a Chinese land redistribution method existing between the ninth century BC (late Western Zhōu dynasty) to around the end of the Warring States period. Its name comes from Chinese character 井 (jǐng), which means 'well' and looks like the # symbol; this character represents the theoretical appearance of land division: a square area of land was divided into nine identically-sized sections; the eight outer sections (私田; sītián) were privately cultivated by serfs and the center section (公田; gōngtián) was communally cultivated on behalf of the landowning aristocrat. (Wikipedia)
[22] The book of odes, Xiaoya, Datian (《诗经・小雅・大田》"雨我公田，遂及我私。").
[23] Fukuyama, 2011, p. 21.

[24] The Book of Songs · Great Odes · Dang (《诗经·大雅·荡》"殷鉴不远，在夏后之世。")
[25] Sīmǎ, Qiān (BC91). "好农耕，相地之宜，宜谷者稼穑焉，民皆法则之。"
[26] Sīmǎ, Qiān (BC91). "复修后稷之业，务耕种，行地宜，自漆、沮度渭，取材用。"
[27] Sīmǎ, Qiān (BC91). "行者有资，居者有畜积。"
[28] Sīmǎ, Qiān (BC91). "民赖其庆，百姓怀之，多徙而保归焉。"
[29] Sīmǎ, Qiān (BC91). "复修后稷、公刘之业，积德行义，国人皆戴之。"
[30] This was the ancient homeland of the pre-Zhōu people, covering present-day Binzhou and Xunyi in Shaanxi Province (the tablelands north of the Wei River in the middle reaches of the Jing River). Gong Liu, ancestor of the Zhōu clans, moved here from Tai. His son Qingjie formally founded the state of Bin. Later, pressured by the Rong and Di tribes, Gugong Danfu led the entire clan southward from Bin to the Zhōu Plain at the foot of the Qi Mountains. (先周故地，范围即今陕西彬州、旬邑一带（泾河中游渭北台塬）。周人先祖公刘自邰迁居于此，其子庆节正式立国于豳；后古公亶父遭戎狄逼迫，全族自豳南迁岐山周原。)
[31] Sīmǎ, Qiān (BC91). "豳人举国扶老携弱，尽复归古公于岐下。及他旁国闻古公仁，亦多归之。"
[32] Sīmǎ, Qiān (BC91). "笃于行义，诸侯顺之。"
[33] Sīmǎ, Qiān (BC91). "遵后稷、公刘之业，则古公、公季之法"。"耕者皆让畔，民俗皆让长。" "士以此多归之"，"诸侯皆向之"，"修德行善，诸侯多叛纣而往归西伯。"
[34] Sīmǎ, Qiān (BC91). "言奉文王以伐，不敢自专。""以祖先有德，臣小子受先功。"
[35] The Great Learning quotes "Announcement to Kang" from the "Zhōu Documents" section of The Book of Documents. (《大学》引用《尚书・周书・康诰》"'惟命不于常。'道善则得之，不善则失之矣。").
[36] The Book of Documents · The Books of Zhōu · Charge to Cai Zhong (《尚书・周书・蔡仲之命》"皇天无亲，惟德是辅；民心无常，惟惠之怀。")
[37] The Book of Documents · The Zhōu Documents · Announcement to Shao (《尚书・周书・召诰》"皇天上帝，改厥元子兹大国殷之命。惟王受命，无疆惟休，亦无疆惟恤。呜呼！曷其奈何弗敬？")
[38] The Book of Documents · Against Idleness (《尚书・无逸》"呜呼！厥亦惟我周太王、王季，克自抑畏。文王卑服，即康功田功。徽柔懿恭，怀保小民，惠鲜鳏寡。自朝至于日中昃，不遑暇食，用咸和万民。")
[39] The Book of Documents · Counsels of the Great Yu (《尚书·大禹谟》"德惟善政，政在养民。正德、利用、厚生、惟和。")
[40] Guǎnzǐ · Speech on Hegemony, Chapter 23 (《管子・霸言第二十三》"夫先王取天下也，术术乎大德哉！物利之谓也。")
[41] Li, 1985, p. 20.
[42] Xu, 2001, p. 32.
[43] Zhao, 2007.
[44] Xúnzǐ · Discussion of Military Affairs (《荀子・议兵》"凡兼人者有三术：有以德兼人者，有以力兼人者，有以富兼人者。以德兼人者王，以力兼人者弱，以富兼人者贫。古今一也。")
[45] The Analects · Governing (《论语・为政》"为政以德，譬如北辰，居其所而众星共之。"" 道之以政，齐之以刑，民免而无耻；道之以德，齐之以礼，有耻且格。")
[46] Mencius · Li Lou I (《孟子・离娄上》"得其民，斯得天下矣…… 得其心，斯得民矣。" "民之归仁也，犹水之就下、兽之走圹也。")
[47] Xúnzǐ · Kings and Lords (《荀子・王霸》"用国者，义立而王，信立而霸，权谋立而亡。""故用国者，义胜私利，则民亲其上、乐其君，而轻死为之用。")
[48] Han Yu, 1987, p. 238. "Yao (尧) passed this Dao to Shun (舜); Shun passed it to Yu (禹); Yu passed it to Tang (汤); Tang passed it to King Wen, King Wu and the Duke of Zhōu. King Wen, King Wu and the Duke of Zhōu passed it to Confucius, and Confucius passed it to Mencius. … All those who came before the Duke of Zhōu ruled as sovereigns, so their institutions and governance could be put into practice. All those who came after the Duke of Zhōu served as ministers and common scholars, so their teachings were preserved and handed down through generations." ("尧以是传之舜，舜以是传之禹，禹以是传之汤，汤以是传之文、武、周公，文、武、周公传之孔子，孔子传之孟轲。……由周公而上，上而为君，故其事行；由周公而下，下而为臣，故其说长。")
[49] The Analects · Shu Er. The Master said: "I transmit but do not innovate; I am truthful and delight in antiquity. Privately I compare myself to our Lao Peng." Lao Peng was a virtuous senior official of the Shang Dynasty,

who delighted in recounting ancient tales. (《论语・述而》。子曰："述而不作，信而好古，窃比于我老彭。" "老彭"，商殷贤大夫，好述古事。)

[50] Hán Fēizǐ · The Prominent Schools. (《韩非子・显学》"宰相必起于州部，猛将必发于卒伍。")

[51] Kennedy, 1987, p. 1-540.

[52] Wēn, 2022, p. 341-482

[53] "The Collection of Historical Rite Documents" of "Book of Rites" (《礼记・表记》"殷人尊神，率民以事神，先鬼而后礼，先罚而后赏，尊而不亲")

[54] "The Record of the Zhōu Dynasty" of "The Record of the Grand Historian" (《史记周本纪》"笃仁，敬老，慈少…日中不暇食以待士，士以此多归之")

[55] "The King Wén Wáng" of "The Book of Songs" (《诗经・文王》"无念尔祖，聿修厥德。永言配命，自求多福")

[56] Wang, 1917, p.302.

[57] "On Military" of "The Xún Zǐ's Book" (《荀子・议兵》"民齐者强，民不齐者弱。")

[58] Shang Yang, 390–338 BC, p. 155. ( "国乱者，民多私义；兵弱者，民多私勇。")

[59] "Commentaries on the Classics" of "Book of Rites" (《礼记・经解》"礼之于正国也，犹衡之于轻重也，绳墨之于曲直也，规矩之于方圆也。")

[60] Chen, 2005.

[61] "Great Yáo's Documents" of "Book of Documents" (《尚书・尧典》"克明俊德，以亲九族；九族既睦，平章百姓；百姓昭明，协和万邦").

[62] "Great Yǔ's Documents" of "Book of Documents" (《尚书・禹贡》"九州攸同，四隩既宅。九山刊旅，九川涤原，九泽既陂，四海会同").

[63] That famous quatrain is actually a free Chinese rendering (by the translator Yin Fu, 1929) of a line from Sándor Petőfi's 1847 poem *Szabadság, szerelem* ("Liberty, Love").

[64] The Great Learning, *The Book of Rites*. ( 《礼记•大学》"修身、齐家、治国、平天下。")

[65] The Analects · Taibo.（曾子曰：″士不可以不弘毅，任重而道远。仁以为己任，不亦重乎？″《论语•泰伯》）

[66] The Analects · Yan Yuan.（《论语•颜渊》，"克己复礼为仁"）

[67] The Analects · Xue Er. (《论语・学而》，孝悌也者，其为仁之本与。)

[68] The Analects · Yong Ye. （《论语・雍也》，"己欲立而立人，己欲达而达人"）

[69] The Analects · Wei Ling Gong. （《论语・卫灵公》"己所不欲，勿施于人。"）

[70] *The Great Learning*. (《大学》"自天子以至于庶人，壹是皆以修身为本。")

[71] The Analects · Xian Wen.（《论语・宪问》" 修己以安百姓。" )

[72] The Analects · Yong Ye.（《论语・雍也》子贡曰："如有博施于民而能济众，何如？可谓仁乎？"子曰："何事于仁，必也圣乎！尧舜其犹病诸。"）

[73] Mencius · Jin Xin Shang.（"穷则独善其身，达则兼济天下。"《孟子・尽心上》）

[74] Xunzi · Zi Dao.（"从道不从君，从义不从父。"《荀子・子道》）

[75] Ku, 1915.

[76] Mencius · Teng Wen Gong Part One. (《孟子・滕文公上》"禹八年于外，三过其门而不入")

[77] Records of the Grand Historian · House of Zhou of Lu. (《史记・鲁周公世家》"我文王之子，武王之弟，成王之叔父，我于天下亦不贱矣。然我一沐三捉发，一饭三吐哺，起以待士，犹恐失天下之贤人。")

[78] "*Qionglin for Young Learners*, Volume One, *Geography*" (《幼学琼林・地舆》"王景治河，千载无恙". "圣人出，黄河清")。

[79] Huntington, 1997, p. 1.

[80] Jiàng, 2021. According to Jiàng, the core of the first generation of human rights theory was freedom, mainly for the bourgeoisie to oppose feudal lords and church rule. As a result, the European countries occupied and enslaved the colonies on a large scale, as well as severe wealth inequality.

[81] Ibid.

**The Code of China’s Historic Great Unification: A Political-Economic Institutional Analysis**

## 1. Introduction

Acemoglu and Robinson, in their book Why Nations Fail, categorize countries across the globe into two types: democratic and autocratic. They further formalize his conclusion into a rigid formula: democracy generates prosperity and advancement, while autocracy breeds corruption and backwardness.[1] Following Acemoglu’s logic, if China were an autocratic state, it would have collapsed repeatedly and descended into Balkanization (fragmentation) long ago. Nevertheless, dating back to the Xìa (夏), Shāng (商) and Zhōu (周) dynasties, and especially since Emperor Qín Shǐ Huáng (秦始皇) unified China in 221 BC, the vast, populous Chinese territory has remained a unified state for most of its recorded history. As Maddison notes: “China has consistently been the world’s largest political entity… a status it retained until the 15th century. China outpaced Europe in technological sophistication, the exploitation of natural resources, and the governance of expansive territories. It was only in the subsequent three centuries that Europe gradually pulled ahead of China in per capita income, technology and scientific capacity.”[2] Since the founding of the People's Republic of China (PRC), especially since the reform and opening-up, China has rapidly transformed from a poor and backward agrian country into the world second-largest economy and a global high-tech power.

For thousands of years, despite repeated transformations in China's social structures and successive shifts in its systems of governance, the primacy accorded to achieving and upholding “Great Unification” has remained unchanged across its political history. Even before the emergence of a unified state, China bore legends of the Yellow Emperor (黄帝), Yáo (尧), Shún (舜) and Yǔ (禹), ruling chieftains who effectively governed society and delivered public goods including flood control, calendar formulation and military defense. During the Xìa (夏), Shāng (商) and Zhōu (周) dynasties, when a unified state gradually took shape, the suzerains of regional polities and tribes strove unceasingly to expand their spheres of authority. Historical records preserve such passages as: “Of old was Tāng (汤) the Successful; from the Dí (氐) and Qiāng (羌), none dared but bring tribute, none dared but come to do homage as vassals”,[3] “Tāng (汤) dwelt in Bó (亳), King Wǔ (武王) in Hào (镐)… All under heaven formed one whole, and all feudal lords stood as ministers,”[4] and “What is meant by ‘the royal first month’? It signifies Great Unification.”[5] Therefore, tracing the origins of Great Unification, “prehistoric China” prior to the Xìa (夏) and Shāng (商) dynasties undoubtedly forged its foundational framework.[6] From the Xìa (夏), Shāng (商) and Zhōu (周) periods onward, and particularly after Emperor Qín Shǐ Huáng (秦始皇) unified China in 221 BC, the vast populous Chinese realm remained a unified state for most of its recorded history. What precise factors, then, have sustained China’s enduring unity? By contrast, following the collapse of the Roman Empire in 476 CE, numerous insightful European thinkers yearned for unification, yet Europe only achieved a union of sovereign states—the European Union—after the end of World War II. Historically, most sprawling empires that fractured struggled to reunite ever again. The successive political regimes of China's different dynasties, however, repeatedly succeeded in restoring national unity through effective institutional adaptations.

The formation of China's "Great Unification" is a natural product of historical development — a comprehensive institutional system of politics and culture, determined by and derived from the economic base. This institutional system was not something rulers imposed and established by force. Rather, it emerged out of the interplay between fundamental natural conditions and the productive forces of the time, taking shape through the repeated interactions and contestations among countless actors across all strata of society. Institutional construction tailored to basic socioeconomic conditions constitutes the core logic sustaining the fundamental structure of society. This holds true for China's millennia-long history of Great Unification. Time and again, periods of division eventually yielded to reunification, followed by more prosperity and bigger national strength. Behind this trajectory lie not only innate geographical and historical factors but also institutional designs capable of accommodating diverse social classes and groups within the country — these are the historical codes of China's Great Unification. Based on this understanding, this chapter seeks to address two core questions: Does China possess an inherent set of mechanisms that sustain national unity? If so, what exactly are these mechanisms? This study aims to decipher these codes through an investigation of the rise and fall of successive Chinese dynasties and the transformations of political and economic institutions throughout history.[7]

**2. What underlying mechanisms have sustained China's national unity?**

Historical records documenting the political, social and economic systems of prehistoric China as well as the Xìa (夏) and Shāng (商) dynasties remain sparse and unclear. For this reason, any inquiry into the underlying mechanisms that sustained China's national unity can only begin with the Western Zhōu (西周) Dynasty. Founded by King Wén of Zhōu (周文王) and King Wǔ of Zhōu (周武王), the Western Zhōu governed its realm through the enfeoffment system. This was a permanent hereditary system of enfeoffment, under which fiefdoms could be passed down in perpetuity through the descendants of the ruling lords. Vassal states bore formal obligations: to pay homage and present tributes to the King of Zhōu (周王) at regular intervals, and to lead their respective armies to defend the royal court in times of border incursions. The King's military power far exceeded that of any individual vassal state, giving him a decisive deterrent advantage over them. More significantly still, he retained the power to appoint the vassal states' key officials, including their top military commanders.[8] After enfeoffing the realm, the King of Zhōu (周王) retained his own royal domain, subjects and army.[9] As the King of Zhōu continued enfeoffing more vassal states, the territory and population under his own direct control grew smaller and smaller. In 770 BC, King Píng of Zhōu (周平王) relocated the capital eastward to Luòyì (洛邑), marking the start of what history would come to know as the Eastern Zhōu (东周) Dynasty. From this point onward, the King of Zhōu (周王) retained only symbolic sovereignty, while individual vassal states functioned as de facto independent polities. This ushered in the Spring and Autumn (春秋) and Warring States (战国) periods — an age defined by hegemonic rivalry, territorial annexation and institutional competition among feudal states.

In the mid-4th century BC, Duke Xiào of Qín (秦孝公) appointed Shāng Yāng (商鞅) to launch a sweeping set of reforms, whose core measures included "abolishing the well-field system (井田制), dismantling the field boundary ridges, and rewarding military merit."[10] The reforms overturned the age-old practice of hereditary aristocrats monopolizing civil official and senior military posts: from then on, all civil officials were appointed directly by the Duke, while military commanders earned promotion solely on the basis of combat achievements. Land private ownership was legalized, and centrally appointed officials were tasked with collecting tax. Bolstered by its advanced political, military and economic institutions, the Qín state went on to unify China. The unified Qín (秦) Empire adopted the centralized commandery-county (郡县) system, a framework that would be largely inherited by every Chinese dynasty that followed. This is why Máo Zédǒng (毛泽东) wrote that "All successive dynasties followed the political and legal institutions established by the Qín."[11] It is evident that the centralized prefeture-county system of the Qín and Hàn dynasties was far superior to the Western Zhōu enfeoffment system in sustaining China's prosperity and unity. Some might argue that the Western Zhōu enfeoffment system was better suited to the economic order that prevailed before the emergence of private, self-sufficient smallholder agrarian economies. By contrast, feudalism in Europe did not begin to disintegrate until the Industrial Revolution in the mid seventeenth century.

While fully inheriting the political, military and economic systems of the Qín Dynasty, the Western Hàn (西汉) established the Chájǔ sytemm of inspection and recommendation along with the Imperial Academy, to select and train government officials, and also formulated an official state ideology encapsulated in the policy of "deposing the hundred schools and revering only Confucianism." Private land ownership throughout the Western and Eastern Hàn (汉) Dynasties gave rise to a stark divide: "the wealthy owned thousands of mile of farmland, while the poor did not even have an inch of ground to stand on." The inspection and recommendation system gradually fell under the control of powerful aristocratic clans — gentry blocs formed over generations by families who had produced officials one after another. These same clans were also major landowners. By the late Eastern Hàn Dynasty, they had come to hold the central government hostage and had seized control of state power. The Eastern Hàn collapsed in the year of 220. National reunification was not achieved again until the Suí (隋) Dynasty at the end of the 6th century. The Suí (隋) and Táng (唐) dynasties largely adopted the equal-field system (均田制) inherited from the Northern Wèi (北魏) Dynasty and the militia system founded under the Western Wèi (西魏) Dynasty, which secured a steady supply of both soldiers and tax revenue. Since the old inspection and recommendation system had allowed aristocratic clans to monopolize official recruitment and appointment, the Suí abolished it and established a nationwide civil examination system for selecting government officials in its place.

When a civilization develops to a certain stage, statehood inevitably emerges, sustained by a set of hard-core institutions. These institutions encompass the appointment of civil and military officials at every level from the central to local governments, the mechanisms for recruiting, promoting and demoting officials, systems of production and taxation, as well as systems of conscription and military manpower. From the Qín and Hàn dynasties to the Suí and Táng dynasties, institutional foundations of China's golden-age dynasties were essentially finalized. In other words, the key mechanisms that sustained China's unity had by then became fully evident. At their core lay the centralized commandery-county system under which the central government uniformly appointed civil and military officials; the equal-field system practiced in the Northern Wèi, Suí and Táng dynasties; military manpower systems centered on the militia system, with rewards for military exploits granted on merit; the imperial examination system for selecting officials; and an official ideological framework centered on reverence of Confucianism. These constituted the overt political and institutional mechanisms that preserved national unity throughout ancient Chinese history. Nevertheless, no dynasty could sustain unity and prosperity indefinitely through coercion and harsh governance alone. Put differently, the political, economic and social institutions of any dynasty or state had to avoid arousing widespread resistance among the people. Beyond safeguarding national defense, personal safety, laws and social order, such institutions had to ensure the ordinary people could make a living and still hold out hope for the future. As an old Chinese saying goes, food is the paramount concern of the people.[12] Karl Marx remarked: "The point of the bayonet becomes as limp as a wick when confronted with acute economic problems."[13] From the Qín and Hàn dynasties through the Míng and Qīng dynasties, China remained largely an agrarian society, with land as the core factor of agricultural production. Land and taxation systems accordingly bore on more than the livelihoods of the ordinary people alone — they also bore on the rise and fall of the state itself.

To give people a sense of hope means that a dynasty's institutional arrangements allow for relatively rational social mobility. The inspection and recommendation system of the Hàn dynasties, and the imperial examination system used from the Suí and Táng through the Míng (明) and Qīng (清), fostered extensive social mobility of this kind. Many talented individuals of humble origins were able to attain important official positions at the imperial court by way of the imperial examinations. Guaranteeing people's livelihoods and enabling orderly social mobility together constitute the political mechanisms that sustained national unity throughout historical China.

### 3. The Imperial Examination System Constitutes the Core Political Mechanism Sustaining National Unity

When discussing the imperial examination system, many Chinese people are quick to dismiss it outright. They argue that scholar-officials under this system only devoted themselves solely to ancient Chinese classics such as the Confucian canon in order to pass the provincial and national examinations and secure official posts at court, remaining wholly ignorant of modern natural sciences like mathematics, physics and chemistry, modern technology, as well as of shifts in international politics and economics. It was precisely because of these drawbacks that the Qīng government abolished the imperial examination system in 1905 and replaced it with a modern university system aimed at cultivating modern talents versed in natural science, technology, and theories of international economics and politics. At that time, the Qīng Dynasty had already embarked on its modernization drive. Born out of small-scale self-sufficient agrarian economy, the imperial examination system could not meet the demands of industrial economic development. An industrial society required vast numbers of scientists, doctors, engineers, lawyers, senior managers, sales staff, journalists, and government officials versed in science and technology, along with a well-educated industrial workforce. Clearly, the imperial examination system, centered as it was solely on Confucian classics, was incompatible with the wave of industrialization that began in the 18th century. Yet even though it inevitably stepped down from the stage of history, the imperial examination system made tremendous contributions to sustaining China's successive unified empires.

Before the onset of the Industrial Revolution in the 18th century, nearly all nations across the globe remained agrarian societies. Prior to the Míng and Qīng dynasties, China had long functioned as an agrarian civilization centered on small-scale peasant natural economy organized around individual households. Under this peasant economy, household farming demanded little in the way of new technology, since the cultivation techniques handed down from earlier generations sufficed to sustain livelihoods. Consequently, for the most of common people, pursuing formal schooling was an unprofitable endeavor unless their goal was to pass the imperial civil service examinations. Farming-and-scholar families educated their sons chiefly to excel in these examinations, secure official positions at court, bring glory to their ancestors, and win noble titles and privileges for their wives and children. From the state's perspective, governing an agrarian society required virtually no knowledge of natural sciences. By contrast, ancient Chinese Confucian classics—centered on the core virtues of benevolence, righteousness, propriety, wisdom, trustworthiness, loyalty to the sovereign, and devotion to the nation—held far greater significance. It was the very nature of the small-scale peasant natural economy that dictated the scope of what the imperial civil service examinations tested.

The imperial civil service examination system originated in the Suí and Táng dynasties and lasted for more than 1,300 years until the close of the Qīng Dynasty. If we include the inspection and recommendation system of the Western and Eastern Hàn dynasties, this mechanism for selecting officials from the general populace by the way of examinations endured for over 2,000 years. Given that this system of selecting officials persisted across two millennia of Chinese history, it is worth examing the reasons behind its longevity, its rationality and merits, its core operating logic, as well as its functions and contributions. Under the imperial examination system, sayings prevailed such as "all occupations rank low, only scholarship stands supreme,"[14] and "within books lie mansions of gold, within books dwell beauties fair."[15] In hopes of passing the provincial and court examinations, bringing honor to their ancestors and winning official rank, aristocratic clans, wealthy merchant families, and especially farming-scholar households poured bulk of their resources into educating their sons. Following the syllabus set by the imperial examination system, boys studied ancient Chinese Confucian classics—the Four Books and Five Classics[16]—along with calligraphy, in hopes of one day succeeding in the examinations and fulfilling the age-old aspiration: "A farmer's boy at dawn, a courtier before the emperor by dusk."[17] In contrast to the hereditary aristocratic system that governed civil and military appointments before Shāng Yāng's reforms of 350 BCE, court officials after the Suí and Táng dynasties were recruited chiefly through imperial examinations, and official positions were no longer hereditary — only the throne itself remained hereditary. In other words, talented scholars from impoverished backgrounds could secure official posts at the imperial court by excelling in the civil service examinations. Consequently, for more than 1,300 years from the Suí-Tang era through the Míng and Qīng dynasties, the imperial examination system enabled and sustained steady, orderly upward mobility across China's social strata.

Chinese people have possessed a clan consciousness since ancient times, and "bringing glory to one's ancestors" has been a lofty ideal passed down through generations. Many prominent clans — the Kǒng (孔) and Qián (钱) clans among them — have endured continuously for thousands of years. Numerous families compile their own genealogy records, with their ancestral lineages traceable all the way back to the Three Sovereigns and Five Emperors.[18] From the Xià and Shāng dynasties through the Míng and Qīng dynasties, China remained a settled agrarian society, with household farming as the economic foundation of the entire nation. Clans sharing the same family name lived together in single village or across several neighboring villages, jointly performing sacrifices to their shared ancestors. Because agricultural yields depended entirely on weather and thus fluctuated unpredictably from year to year, mutual assistance among clan members and neighbors in times of hardship served as a crucial social survival mechanism within the smallholder peasant economy. The Chinese tradition holds ancestor worship and respects for elders in high regard. The system for inheriting land and property within families further reinforced filial piety at a material level. The system of inspecting and recommending "filial and incorruptible" candidates in the Western and Eastern Hàn dynasties, as its name suggests, selected individuals of virtue and talent who demonstrated deep respect for elders, allowing them to study at the Imperial Academy and go on to serve as court officials. The imperial civil service examination system established under the Suí and Táng dynasties also made filial piety a prerequisite qualification for examination candidates. In other words, the Chinese regard personal achievement as an accomplishment shared by the entire clan, and an individual's rise in social standing as honor bestowed upon all kin. The imperial examination system provided a path of upward mobility for ordinary people who aspired to govern the state and bring peace to all under the heaven, allowing individuals to advance in social status through moral character and scholarly competence alone, rather than through their clan's wealth or status. Accordingly, the imperial examination system organically wove together the concepts of "family" and "state" for the Chinese people, sustaining the core national sentiment captured in the maxim: "Cultivate the self, bring order to the clan, govern the state, and bring peace to all under the heaven."[19] Compared with India's caste system, which blocked social mobility; European feudalism prior to the 17th century; Russian serfdom before the October Revolution; and the slave-official and conscription systems in Egypt and the Ottoman Empire before the 20th century, which tore families apart, China's imperial examination system — judging candidates solely on morality and talent, regardless of family origin — was far more humane, a quality that account for its remarkable longevity.

The imperial examination system adopted Confucianism as its standard of assessment and promoted the benevolent and public-spirited dimensions of Confucian teachings. Human nature is multifaceted: it encompasses virtuous traits such as diligence, ambition, a disposition toward goodness, compassion, benevolence, and concern for all under heaven, yet it also inevitably habors selfishness and other base inclinations. Confucianism champions humanity's innate drive toward self-improvement and moral goodness, while condemning selfishness and vice. Confucian maxims widely familiar to the Chinese people—including "Every common man bears responsibility for the rise and fall of the nation,"[20] "benevolence, righteousness, propriety, wisdom and trustworthiness,"[21] "Do not do unto others what you would not have them do unto you,"[22] "A noble man helps others achieve their virtues, not their vices,"[23] "The noble man seeks harmony yet refuses blind conformity,"[24] "The ruler should act as a ruler ought, the minister as a minister ought, the father as a father ought, the son as a son ought,"[25] and "the world belongs to all"[26]—almost universally propagate values of upholding social order, embracing goodness while rejecting evil. Since Emperor Wǔ (汉武帝) of the Western Hàn Dynasty implemented the policy of "rejecting hundred schools and revering only Confucianism," Confucian learning has been enshrined as the official state ideology, and Confucian ethical codes governing individuals and families have become the code of conduct for society as a whole. Both the inspection and recommendation system of the Hàn dynasties and the imperial examination system established after the Suí and Táng dynasties took Confucianism as the benchmark for selecting and promoting officials. In other words, the imperial examination system, the Confucian ideological and behavioral norms, together with the system of centrally appointed officials under the commandery-county administration together served, to a considerable extent, to cultivate and expand humanity's virtuous, aspirational tendencies.

From the imperial court's perspective, the imperial civil service examination system enabled the government to recruit social elites from every stratum into the state apparatus through impartial scholarly assessments. Governance led by such talent, rather than hereditary aristocrats, was inherently more efficient. The system also bolstered the legitimacy of imperial power and significantly strengthened social stability. Had the hereditary aristocracy system persisted without the imperial examinations, the most dynamic, creative, capable and potentially disruptive elites across society would have largely remained outside the governing system — a situation bound to trigger social unrest. The court examination, which selected Jìn Shì (进士) degree holders under the imperial examination framework, evaluated candidates not only on their academic proficiency but also with an eye toward regional and provincial representation. Jìn Shì (进士) selection thus incorporated political calculations as well, aimed at upholding regional equity and representativeness within the imperial examination system, and at safeguarding social stability and territorial integrity.

From the Hàn era through the Míng and Qīng dynasties, most major dynasties in Chinese history upheld Confucianism as their dominant ideology and the bulk of examination questions under the imperial examination system centered on practical application of Confucian teachings. Driven by the immense influence of these imperial examinations, successive generations of scholar-officials from the Hàn through the Míng and Qīng dynasties devoted themselves to the arduous study of Confucian classics, including the Four Books and Five Classics. It can be said that the enduring elite stratum of scholar-officials fostered by the imperial examination system constituted the core force responsible for transmitting Confucian thought across generations.

The imperial examination system consisted of progressive ranks, from the lowest to the highest: tóngshēng (童生 candidates have not passing any examinations), xiùcái (秀才 successful county examination graduates), jǔrén (举人 successful provincial examination graduates), and Jìn Shì (进士 successful candidates from the court examination). Scholars who passed the official state-run imperial examinations were conferred these academic titles along with corresponding privileges, such as exemption from corvée labor and the right to stand rather than kneel before local officials. The institution under which imperial authority in ancient China rarely reached below the county level meant that gentry holding imperial examination degrees naturally became as the leaders of rural society. This degree-holding gentry, loyal to the throne and devoted to the nation, in turn strengthened social cohesion and stability. Across successive dynasties, the imperial examination system evolved into the established means of selecting both officials and gentry. Following the collapse of the enfeoffment system based on hereditary aristocracy prevailing in the Xià, Shāng and Zhōu dynasties, successive experimental official recruitment systems were tried in the Qín, Hàn, Wei (魏) and Jin (晋) dynasties: the military merit nobility system, the inspection and recommendation system, and the Nine-Rank Evaluation System (九品中正制) respectively. All of these institutions were originally designed to enable effective social mobility. The military merit nobility system collapsed under the excessive resource depletion caused by constant warfare. The inspection and recommendation system eventually degenerated into a hollow formality, as powerful aristocratic clans it nurtured came to monopolize political power. The Nine-Rank Evaluation System likewise soon fell into the trap of favoring established official clans. Yet the Confucian vision of social order did not fail. It was only with the emergence of the imperial examination system—and its rise as the dominant channel for selecting officials—that orderly social mobility was achieved, forging the core political code underpinning national unity.

**4. Land and taxation systems constitute the core economic foundation for national unification**

From the Xià and Shāng dynasties through the Míng and Qīng dynasties, China remained an agrarian society: commoners sustained themselves by farming the land, while state relied entirely on levying taxes on land and corvee labor on its people. The steady flow of land tax revenue— the lifeblood of governance in agrarian civilizations—into the imperial treasury was an essential precondition for maintaining national unity and stability. Such revenue was needed not only to cover massive military and administrative expenditures, but also to preserve the imperial court's upper hand in its shifting balance of power against nobles, powerful clan lineages and landowners, who likewise commanded large farming populations of their own. To prevent flooding, every household had to contribute corvée labor each year, organized by the government to build river dikes. In addition, guarding the frontier and repairing defensive fortifications also required large numbers of laborers. For China's prosperous, unified dynasties throughout history, the core economic underpinning of great unification lay in well-crafted land, taxation and corvee systems, working in tandem to secure the central government's governing authority.

For newly established central governments seeking to secure control over land, insituting a system of public or royal land ownership was the most natural institutional choice. As Friedrich Engels put it, "All civilized nations start out with communal ownership of land."[27] Pre-Qin society, especially during the Western Zhōu Dynasty, was based primarily on the well-field system (井田制). Under this arrangement, each well-shaped plot of farmland contained a central public field surrounded by private fields allotted to individual households, and all peasant families within an aristocrat's fief were required to work on the public field first before attending to their own private plots. As the "Great Fields" ode in The Book of Songs (《诗经·大田》) puts it: "May rain fall first on our public fields, and then extend to our private plots." The entire harvest from public fields went to the feudal lords, and even the private fields were subject to a tribute levy on part of their field. Neither type of field conferred full ownership; peasants held only the right to use the land.[28] The Zhōu Dynasty operated under the enfeoffment system, in which feudal states in turn carried out subinfeudation across multiple tiers. The relationship among enfeoffed lords, their fiefdoms, and the population residing within those territories remained relatively fixed.[29] Those called "guó rén" (国人) — town dwellers — lived within walled cities; they farmed in times of peace and took up arms as soldiers in times of war. The "yě rén" (野人) —country-dwellers — lived on the outskirts of cities; they performed corvée labor, such as cultivating public fields or rendering labor rent, but were exempt from military service.[30] Given a relatively low level of productivity at the time, institutional arrangements such as "communal cultivation by multitude" and "joint tending of the public fields" ensured that the King of Zhōu (周王) maintained effective control over societies' resources.

By the late Western Zhōu Dynasty, or at the latest by the Spring and Autumn Period, the emergence of iron farm tools and ox-drawn ploughs had made it possible for individual households to farm relying on their own labor alone — a marked contrast to the earlier age of bronze implements without ox-ploughing, when farming required several households to pool their labor with multiple people hauling a plough together to till the soil.[31] Since newly reclaimed private land was fell outside the tax reach of feudal states, people's enthusiasm for opening up private farmland was greatly stimulated.[32] As individual households gained capacity to cultivate new lands on their own, the well-field system collapsed. The feudal states, no longer under the control of the King of Zhōu's court, warred endlessly among themselves, and land annexation among states and among feudal lords broke through the restrictions the enfeoffment system had placed on landholding. As a result, the public fields had once belonged the King of Zhōu's court were gradually appropriated as private property by the feudal states and the rising landlord class, and the old rule that "land shall not be sold" was replaced by the practice that "land may be bought and sold".[33] The collapse of rituals and disintegration of ethical norms throughout the Spring and Autumn Period marked the onset of the central authority's loss of control over land. The King of Zhōu (周王), the nominal commen overlord, steadily declined in power, after the dismemberment of Jìn (晋) into Three States during the Warring States Period, his status was reduced to little more than that of an ordinary feudal lord.

Amid fierce rivalry centered on farming and warfare, feudal states began exploring new approaches to land governance, recognizing newly reclaimed private farmland and taxing it accordingly. The State of Qí (齐) led the way in this fiscal reform: Duke Huán of Qí (齐桓公) appointed Guǎn Zhòng (管仲) as prime minister, who in 686 BCE introduced the policy of "taxation graded by land quality", under which land rents were collected according to soil fertility grades, and peasants were granted permanent rights to cultivate the land.[34] This policy greatly spurred peasant labor, vastly expanded the acreage under cultivation, rapidly boosted

Qí's national strength, and secured its hegemony. Other feudal states followed suit: in 594 BCE, the State of Lǔ (鲁) implemented the "first land tax" reform, which Gōngyáng Commentary on The Spring and Autumn Annals (《春秋·公羊传》) interprets as "taxation levied per measured mǔ of land" — that is, tax was charged on every mǔ of land under cultivation — a measure further intensified class stratification in the period that followed.[35] In 548 BCE, the State of Chǔ (楚) enacted the policy of "adjusting levies based on household income," setting commoners' economic earnings as the benchmark for tax assessment.[36] In 538 BCE, the State of Zhèng (郑) introduced the "military levy by qiū (丘) unit of land" system, collecting military taxes on the basis of the qiū administrative unit of land.[37] Qín (秦) carried out its reforms relatively late: in 408 BCE, it adopted the "first grain land rent" system, imposing a fixed grain tax proportional to land area, similar in sprit to Lǔ's (鲁) first land tax. Later, Shāng Yāng's (商鞅) reforms brought further changes, "reorganizing field allotments and dismantling the boundary ridges that had divided farmland".[38]

Beyond the measure already mentioned, the core provisions of Shāng Yāng's Reforms also included "the registered land and housing system", and "requirement that commoners report their actual land holdings". "Reorganizing field allotments" involved the state reclaiming land held under the enfeoffment system as state property, then redistributing it to working peasants. Unlike under the old system, peasants were no longer required to return this land to the state; the plots were assigned to them permanently as private holdings and could be freely bought and sold. Such land fell within the category of state-allocated land.[39] Scholarly debate has long persisted over whether these allocated lands were publicly or privately owned, but since the excavation of the Shuìhǔdì Qín bamboo slips (睡虎地秦简) in Yúnmèng (云梦) in 1975, the view that state-allocated land constituted state property has gained growing acceptance. Scholars such as Yuan Lin,[40] Du Shaoshun,[41] Li Ruilan,[42] Jiang Chun[43] argue that the land allocation system formed the fundamental economic institution of Qín following Shāng Yāng's reforms. Under this framework, land granted alongside noble ranks awarded for military merit could not be passed down to descendants as private property; upon the recipient's death, the granted land reverted to the state and land transactions were prohibited. By contrast, scholars including Gao Min[44], Shi Weiqing[45], Shao Hong[46] hold that while the state land allocation system remained dominant in this era, private land ownership existed alongside it, and this later allocation system differed fundamentally from the land allocation practices of the well-field system era.

In addition, the State of Qín implemented the Named Land and Residence System, under which land and housing were awarded based on military merit. It stipulated that "social rank and nobility shall be clearly stratified in hierarchical order, and the allocation of titled land, residences, servants, concubines and apparel shall be determined by household rank." The Statute on Military Ranks from the Yúnmèng Qín Bamboo Slips (睡虎地秦简) decreed that "those serving in the army shall be assessed and rewarded in proportion to their meritorious service"— with "rewards" here referring to grants of land and residences scaled to to military achievement: the greater the merit, the more land and housing received.[47] In the 31st year of the Emperor Qín Shǐ Huáng's reign (秦始皇 216 BCE), an edict known as "Ordering Commoners to Report Their Own Land Holdings" was issued, requiring landowners and peasant landholders to voluntarily register the acreage of their lands, on the basis of which the state would levy taxes. Scholarly debate persists over the exact relationship between these two decrees. Many scholars argue that the Named Land and Residence System depended on the state's possession of vast tracts of uncultivated wasteland, and by the Emperor Qín Shǐ Huáng's reign (秦始皇), population growth had shrunk the land under state control, intensifying the pressure of distributing state-allocated plots—hence the promulgation of the edict mandating self-reported landholding. Yang Zhenhong[48], by contrast, holds that the Named Land and Residence System was precisely the official term for the state land allocation system prevailing under the Qín Dynasty, constituted the fundamental economic institution of the era. On this view, the edict "Ordering Commoners to Report Their Own Land Holdings" represented neither an overhaul of the Qín Empire's land regime nor the nationwide establishment of private land ownership, but was simply a nationwide land survey and registration campaign carried out after the Emperor Qín Shǐ Huáng (秦始皇) unification of China. Nevertheless, land institutional reforms spanning the Spring and Autumn (春秋) and Warring States (战国) periods, and Qín (秦) dynasty set in motion a gradual transition from public to private land ownership. Following Shāng Yāng's reforms, private land ownership gained formal state recognition.[49]

During the Pre-Qín period, the public land system underpinned the fundamental unity of the state. Institutional competition among warring states in the Warring States Period compelled regional dukedoms to recognize private land ownership in order to mobilize resources for war. When the population-to-land ratio was low, the public land system could directly sustain the governing capacity of political powers; by contrast, free land transactions and land reclamation under a private ownership regime enabled more efficient allocation of resources. Abundance of state-owned public lands in the Qín Dynasty and early Western Hàn secured the state's administrative capacity, while the emergence of private land owership boosted agricultural output and facilitated population recovery. Nevertheless, the inherent transferability of private land, combined with the accumulation of wealth and political resources, readily gave rise to severe land concentration. Three forms of land ownership coexisted in the Western Hàn: state-owned land, landlord-owned land, and smallholder peasant-owned land.[50] Even so, private land ownership gradually became the dominant social norm. "When transferring land property rights, both parties shall enter into a contract inscribed on iron tally slips with red characters as legal proof" — and even wasteland reclaimed into cultivated land would, "following the practice of land sales, have its transaction recorded through a written contracts and a commemorative stone stele".[51] Against the backdrop of the growing prevalence of private land ownership, a massive wave of land annexation emerged during the reign of Emperor Wǔ (汉武帝) of the Western Hàn. As an ancient record puts it, "the wealthy own stretches of farmland extending endlessly, while the poor do not possess an inch of ground to stand on," and the rank-based land allocation system collapsed completely. Later, surges of land annexation recurred under the Emperors Chéng and Āi of the Western Hàn (汉成帝和汉哀帝), far more severe than the earlier wave, drastically exacerbating social conflicts — "Wealthy magnates, officials and commoners amassed fortunes worth tens of millions, while the destitute sank deeper into hardship."[52] As land annexation grew increasingly severe, Wáng Măng invoked ancient precedent to restore the well-field system in an attempt to break the grip of powerful landed clans over the state's economy and politics — a set of policies came to be known as Wáng Măng's New Policies. Nevertheless, Wáng Măng(王莽) and his New Policies regime swiftly collapsed under the united resistance and rebellion of the landlord class inherited from the Western Hàn.[53] As rampant land annexation that plunged the populace into destitution, the Emperor Guāngwǔ of the Eastern Hàn (汉光武帝) issued the Land Survey Decree in 39 CE, aiming at tallying cultivated land acreage and household populations through land measurement so as to boost state fiscal revenue. It was, however, forced to be suspended amid fierce resistance from powerful local landowners and their attached peasants.[54]

Unchecked land annexation under the private land ownership system fueled undercurrents of social unrest and upheaval from the late Western Hàn onward. After the early Western Hàn ushered in a period of peace and stability, the Qín practice of conferring noble ranks based on military merit gave way to the inspection and recommendation system, which recruited talents through categories such as virtuous and upright scholars, experts in Confucian classics, legal specialists, and filial and incorrupt officials. The original intent of this system was to curb the power of meritorious aristocratic clans, forcing them to "yield ground to classical and legal scholars selected through recommendation and official summons,"[55] so as to enable social mobility and prevent land from accumulating in the hands of powerful hereditary clans. Over time, however, "civilian gentry lineages large and small leveraged their family status and influence to secure official posts at all levels in priority... seizing wealth and prestige by virtue of political authority."[56] This tendency matured fully under the Eastern Hàn, giving rise to prominent aristocratic ministerial clans including the Yáng clan of Hóngnóng (弘农杨氏) and the Yuán clan of Rŭnán (汝南袁氏). In the early Western Hàn, the state still held vast reserves of land, and land annexation had not yet spiraled out of control; under these conditions the inspection and recommendation system functioned effectively, and the central government secured prosperous governance on the strength of steady tax revenues drawn from land. In the late Western Hàn, the rise of powerful literati clans distorted the system of "selecting talents via inspection and recommendation" into a practice of "recruiting officials by family lineage." Fed by monopolized political resources, land inevitably became further concentrated in the hands of these aristocratic lineages, and the central government, deprived of effective control over land distribution, had no choice but to raise taxes levied on the dwindling rank of independent peasant landholders — even resorting to selling official titles and ranks to sustain state operations, which made the dynasty's collapse all but inevitable. The Nine-Rank Evaluation System established under the Cáo Wèi (曹魏) regime thereafter "displayed an explicit bias in favor of literati aristocrats and became the dominant mechanism for official selection,"[57] culminating in the well-known state of affairs in which "no commoner of humble origins could attain top-tier official rank, while no member of the hereditary gentry ever filled a low-tier post."

During the Two Jìn (两晋) and the Northern and Southern Dynasties (南北朝), land annexation under the private land ownership system reached catastrophic levels. Although the Western Jìn (西晋) briefly unified the realm and, through the Land Occupation System, sought to cap the land holdings of powerful clans that had proliferated since the late the Eastern Hàn, "the disparity in land allotted to nobles, bureaucrats and ordinary commoners was enormous, with the highest tier possessing over seventy times the land of common households".[58] The political upheaval and weakened imperial authority that followed across the Northern and Southern Dynasties (南北朝) further fueled the rise of hereditary aristocratic lineages and regional warlord forces. High-ranking officials in the Southern Dynasties (南朝) "seized private farmlands and bullied the weak by virtue of their power". During the Liú Sòng (刘宋) reign, the Mountain Occupation Law was enacted — the first statute in Chinese history to legalize private ownership of mountain and marshlands. By the fall of the Chén (陈) state of the Southern Dynasties, "the state-registered taxable households numbered a mere 500,000, with a total population of only two million."[59] In the Northern Dynasties (北朝), meanwhile, powerful clan chieftains built fortified settlements to house their kin and retainers, monopolizing land and peasant labor while evading state taxes and military conscription. Against this backdrop, incessant warfare triggered a sharp population decline, leaving vast stretches of land fallow and wasteland widespread. In 485 CE, Emperor Tuòbá Hóng (拓跋宏) (Yuán Hóng 元宏) of the Northern Wèi (北魏) introduced a system of state-owned land and rolled out the Equal Field Allocation System, under which land was distributed according to labor capacity and household size. Males and females aged fifteen became eligible to receive an allocation of land, upon reaching seventy or death, households were required to return open farmland and hemp fields to the state, while mulberry orchards and residential plots were retained as hereditary holdings. The policy aimed to rectify the longstanding issue whereby "retainers attached to powerful aristocratic clans rendered labor and tribute solely to their clan masters, paying no taxes to the imperial government."[60] In the initial stage, the Equal Field System still allowed powerful clans to acquire extra land by exploiting the clause granting land entitlements to "servants and maids within regulated quotas." Under the Northern Zhōu (北周) and Táng (唐) Dynasties, however, this loophole was closed, and land entitlements for servants and maids were abolished entirely.[61]

The implementation of the Equal Field System marked the revival of the state-owned land regime — an institutional reform through which the central government sought to reclaim control over land and fiscal revenues from powerful landlord magnates. Alongside it, corresponding tax and military conscription systems were re-established. The tax reform introduced the New Rent-and-Tribute System, which tied taxation to land allocation: peasants granted state land were required to pay land rent and household tribute, under this arrangement, land taxes and poll taxes became nearly indistinguishable, both functioning essentially as rents owed to the state. Meanwhile, building on the Equal Field System, the Western Wèi (西魏) State grouped military households together for concentrated settlement and land allotment, exempting them from taxation, and thereby founded the Garrison Troops System (府军制). The Suí and Táng dynasties largely preserved the land allocation system and garrison troop institution. "Garrison soldiers were drawn from civilian households and folded with the Equal Field System, allowing peasants to combine farming with military service."[62]

The Equal Field System allowed the central government to exercise effective control household registers and land, contributing significantly to the unification and prosperity of the Suí and Táng dynasties. However, from the mid-Táng onward, the stock of state-owned wasteland continued to dwindle until there was no more land left to allocate. Combined with widespread land annexation and disordered household registration, the system became unsustainable. Not only did "cultivated land fall short of what was needed for distribution — leavling every household with insufficient allotted land" — but the government was also forced to break land parcels into smaller, scattered plots[63] that were often not contiguous with one another. This severely dampened farmers' enthusiasm for production and reduced farming efficiency.

With the collapse of the Equal Field System, the central authority lost its stable source of tax revenue from land-allotted peasants. The Táng dynasty's tripartite tax system — rent, tribute and corvée labor — fell apart entirely, and the supply of soldiers for the militia (garrison troop) system dried up completely. Although private landholding had existed alongside state-owned land throughout the early Táng,[64] once the Equal Field System collapsed, land increasingly came under the control of powerful landlords through annexation. In 780 CE, Emperor Dézōng (唐德宗) of the Táng abolished the old rent-tribute-corvée system in favor of the Two-Tax Law, under which "household levies were assessed according to a family's wealth-based tier" and "land levies were assessed according to the amount of cultivated acreage,"[65] with collections made twice a year, in summer and autumn. This reform comfirmed the legality of private land ownership, shifted the basis of taxation from population count to a property assessment, and marked the definitive end of the state-administrated land allocation regime under the Equal Field System.

The land and taxation systems built on the foundation of private land ownership and the Two-Tax Law remained essentially stable from the Sòng (宋) Dynasty onward. "The Sòng court did not impose any unified, nationwide restrictions on land transactions, and it declined to prohibit powerful landed families from acquiring land through annexation."[66] Ruling elites entirely abandoned proposals to revive either the well-field system or the Equal Field System, and later adjustments to land institutions were mostly limited to refinements in administrative practice and minor revisions to tax regulations. For example, the state carried forward earlier price stabilization mechanisms — such as grain price equalization and government market intervention — to keep small peasants households from falling into bankruptcy; it imposed monopoly taxes on commodities to restrain the accumulation of capital by wealthy merchants; and measures such as the Green Sprout Law, and Farmland and Water Conservancy Law were enacted to protect ordinary households from ruin.

The Míng and Qīng dynasties continued along the same trajectory. "The scope and volume of land transactions expanded dramatically, and property rights in land circulated far more freely than before."[67] Private land ownership grew more thoroughgoing during these dynasties, and property rights were more clearly defined than in any preceding era — developments that even gave rise to the permanent tenancy system, an arrangement that separated underlying land ownership (land title) from surface use rights (the right of cultivation), allowing the two to be bought and sold independently of one another. Under a system buildt on private land ownership, some degree of land annexation was was all but inevitable.

To secure a steady stream of tax revenue, central governments had to make moderate adjustments to their tax policies. Accordingly, the Míng dynasty implemented the Single Whip Law and the Qīng dynasty carried out the Poll Tax Merger into Land Tax reform, marking the last major overhauls of land taxation under China's imperial dynasties. "The Single Whip Law apportioned corvée obligations according to a combination of household population and landholding, converted these levies into silver payments, and collected them with autumn grain taxes."[68] Its core objective was to simplify the categories of taxation and curb corruption and revenue leakage in the collection process. The Poll Tax Merger into Land Tax reform went a step further, folding the poll tax—previously paid in silver—into land tax, so that total land acreage become the sole basis for taxation. This shifted a greater share of the fiscal burden onto large landholding households and boosted state revenue drawn from the them.

A review of the history outlined above reveals that ancient China's land and taxation systems moved through alternating phases of public and private land ownership before evetually settling into a framework dominated by private ownership. The ancient communal well-field system broke down under the pressure of rising productivity and population growth. The private land ownership system that succeeded it sustained the unified empires of the Western and Eastern Hàn Dynasties, but the accumulation of wealth and political influence in private hands eventually drove large-scale land concentration and gave rise to powerful landed clans. During periods of population decline and agricultural recession, the revival of the communal system—the Equal Field System—allowed the central government to exercise direct control over registered households and land. This institution endured for roughly three centuries, laying the groundwork for the flourishing of the High Táng (盛唐) era, only to collapse once the dynasty reached its very peak. After the Táng Dynasty, central authorities essentially abandoned all attempts to restore either the well-field or the Equal Field systems, confining institutional adjustments to minor refinements within the framework of private land ownership. Under systems of public land ownership, commoners were, in effect, tenants farming state-owned land. As Marx observed, "Where, as in Asia, the state stands over the people as landlord and simultaneously as the sovereign, rent and taxation coincide,"[69] under such a system, the government's tax base centered on individual households. Under private land ownership, by contrast, state taxation shifted its primary focus onto land itself, placing heavier fiscal burdens on landowners. This graduated approach to taxation allowed central governments to adapt to shifts in land institutions while still securing sufficient fiscal revenue to sustain the operation of state.

It can likewise be seen from the foregoing analysis that a dynasty's prosperity and territorial unification do not hinge on whether it adopted a system of public or private land ownership—as evidenced by the private land regimes of the Western and Eastern Hàns, and of the Míng and Qīng, on the one hand, and the public land system of the Suí and Táng on the other. As the core economic foundation for national unification, the rationality of a state's land system lies in its capacity to adapt to prevailing socioeconomic conditions and to establish a taxation framework that matches them. When registered households were scarce and unclaimed wasteland was abundant, states enforced public land systems, levying taxes directly on cultivators. Conversely, when rising populations depleted the reserves of vacant land, states were compelled to shift toward private land ownership and to tax on landholders instead. This raises a critical question: could dynasties afford to disregard land annexation under a system of private ownership? The lessons drawn from the chaotic eras spanning the late Hàn to the Northern and Southern Dynasties offer a clear answer: no. Economic resources and political power reinforce one another: landlords who grow wealthy through land annexation inevitably acquire commensurate political influence, and political influence, in turn, tends to translate back into further economic gain—with the same detrimental outcomes resulting either way. No taxation framework, however designed, can fully eliminate the persistent problem of "powerful landed clans exploiting ordinary peasants and quietly diverting fiscal resources away from the state." To avert the disastrous consequence of powerful aristocratic clans coming to dominate state governance, two viable solutions present themselves: enforcing public land ownership, or safeguarding sufficient social mobility to curb the excessive hereditary transmission of political authority and land resources. Robust political institutions can, in this way, mitigate the severe crises triggered by rampant land annexation. This same logic underpinned the historical progression from the inspection and recommendation system, through the Nine-Rank Evaluation System to the imperial examination system. By facilitating mobility across social classes, these institutions prevented land from accumulating in the hands of hereditary bureaucratic lineages—a development that would otherwise have erode the central government's tax base and administrative capacity. The success of this dual strategy underpinned the stability and prosperity of dynasties that practiced private land ownership from the Sòng dynasty onward. These regimes made use of private land tenure to strengthen incentives for agricultural production, built tax systems centered on land-based levies, and relied on the imperial examination system as a mechanism for selecting talent, thereby sustaining social mobility. This arrangement forestalled extreme land concentration, and with it, the proliferation of entrenched aristocratic clans that had plagued China from the Eastern Hàn onward. At this stage, the political and economic pillars sustaining a unified empire were fully and effectively integrated.

Emperor Taizong (唐太宗) of Táng, Lǐ Shìmín, once remarked, "Water can carry a boat, but it just as easily capsize it."[70] From the Qín and Hàn dynasties through the Míng and Qīng era, China remained, for most part, an agrarian society built on smallholder farming. The bulk of state fiscal revenue derived from grain taxes and corvée labor levied upon countless peasant households. Major dynasties in Chinese history, including the Western Zhōu, Western Hàn, Eastern Hàn, Táng, Sòng, Míng and Qīng—lasted, on average, roughly 250 years. In their early reigns, the rulers of these dynasties typically drew lessons from the collapse of preceding regimes. They voluntarily lightened taxes and corvée burdens to let war-torn nations recover, giving rise to golden ages such as the Rule of Wén and Jǐng (文景), the Zhēnguān (贞观) Era of Prosperity, and the Kāng-Qián (康乾) Flourishing Age.

Nevertheless, by the middle and late phases of each dynasty, rulers frequently gave themselves over to extravagance and pleasure-seeking. With corruption set from the top, officials at every level, civil and military alike, engaged in graft and preyed on the populace for personal gain, while the elite evaded taxation altogether. This forced repeated increases in the levies and labor burdens imposed on ordinary commoners, until the people could no longer bear the strain and the dynasty collapsed.

A small number of short-lived dynasties—most notably the Qín and Suí—collapsed primarily because of exorbitant taxes and excessive corvée labor imposed from the very founding of the dynasty. After unifying China in 221 BCE, the Qín embarked on massive construction projects and unending military campaigns: building the Epang Palace (阿房宫), the First Emperor's mausoleum (始皇陵墓), and the Great Wall, while simultaneously launching northern expeditions against the Xiōngnú (匈奴) and southern campaigns against the Bǎiyuè (百越) tribes. Roughly three million able-bodied men were conscripted for year-round corvée labor. Given the Qín's total population of around 20 million, this meant some 15 percent of the entire population was tied up in labor service. When male laborers ran short, women were conscripted for corvée work as well. This reckless overexploitation of manpower severely disrupted agricultural production and provoked widespread public outrage, and the Qín Empire ultimately collapsed amid waves of popular uprising.[71] After ascending the throne, Emperor Yāng of the Suí (隋炀帝) ordered the construction of palatial retreats across the realm and the Grand Canal (大运河), indulged in luxury and debauchery, and launched incessant wars. To fund these ventures, he imposed crippling taxes and mass conscription, drafting millions of laborers at a time for palace and canal construction. Emperor Yang squandered far more wealth and manpower than even Qín Shǐ Huáng (秦始皇) had.[72] The suffering populace rose in rebellion, bringing an end to Suí rule. The lessons drawn from the rise and fall of China's unified dynasties reveal that, alongside land systems, the severity of land taxes, poll taxes and corvée labor was a decisive factor in determining the survival of China's unified empires. Once taxes and labor obligations grew so oppressive that commoners were driven to revolt, the overthrow of the dynasty became all but inevitable. There existed a critical threshold for taxation and labor extraction: the point below which ordinary people could still sustain basic subsistence.

## 5. Conclusion

This chapter attempts to explore the origins of Chinese unification through the vicissitudes of dynastic rise and fall across Chinese history, and to analyze the political and economic institutions that sustained national unification for most of that history.

From the Xià, Shāng and Zhōu eras through the Qín, Hàn and Táng dynasties, the Chinese nation gradually forged a set of robust political and institutional frameworks designed to safeguard national unity—the collective stability of the realm. Having learned from the collapse of the Western Zhōu brought about by its system of enfeoffment, the Qín Dynasty carried forward the institutional achievements of Shāng Yāng's reforms: a centralized commandery-county system under which civil officials at every level of administration were appointed by the central government, and military officers were promoted on the basis of battlefield merit. The Western Hàn created three pivotal institutions: the inspection and recommendation system for selecting officials, the Imperial Academy at the capital for training civil servants, and the Confucian ideological framework embodied in the policy of "deposing the hundred schools and revering only Confucianism". To secure tax revenue and military recruits, Emperor Xiàowén (Tuòbá Hóng 拓跋宏) of the Northern Wèi (北魏) nationalized land and implemented the state land allocation system alongside the garrison-based militia system, both of which were largely inherited by the Suí and Táng dynasties. To break the grip of hereditary aristocratic clans over the inspection and recommendation system, the Suí and Táng established the imperial examination system, which selected officials on the basis of their performance in standardized written examinations. These institutions together formed the core political and economic mechanisms that sustained China's national unification throughout its history. Nevertheless, for a dynasty to achieve long-term stability and peace, it had to do more than address the basic livelihood needs of its people, such as food and clothing—it also had to give them grounds for hope.

The imperial examination system embodied the collectivist vision of self, family, and state captured in the Confucian maxim "cultivate the self, bring order to the family, govern the state, and bring peace to all under heaven". It achieved a harmonious alignment of incentives among individuals, families and the state as a collective whole, enabling orderly social mobility that spanned from ordinary commoners to high-ranking court officials. Under this system, multiple goals were fulfilled at once: individuals' and families' aspirations for upward mobility, the government's need to recruit talented people across the realm for effective governance, the maintainance of social stability, and the preservation of traditional culture. This is arguably the primary reason why the imperial examination system endured for more than 1,300 years from the Suí-Táng (隋唐) dynasties through the Míng and Qīng. The imperial examination system functioned as an inclusive institution that facilitated orderly social mobility. Sustaining inclusive and orderly mobility stands as the core political code behind a country's or society's prosperity, vitality and stable development. Accordingly, how to maintain orderly social mobility poses a fundamental challenge for any nation pursuing progress. The lessons drawn from the imperial examination system suggest that robust institutional safeguards are indispensable for such mobility to be sustained. Nevertheless, the flourishing of imperial examinations from the Sòng and Míng dynasties onward fostered the prevailing notion that "of all pursuits, none ranks above scholarship." Amid social stability and economic prosperity, a growing number of young people pursued education in the hope of joining the ranks of scholar-officials. Constrained by the agrarian natural economy of the Sòng, Yuán, Míng and Qīng dynasties, there were almost no alternative avenues for upward social mobility outside the examination track, resulting in a massive backlog and waste of human talent.[73] Moreover, the imperial examination system was, by its very nature, a product of the smallholder natural economy, and it proved unable to adapt to the demands of industrialization.

An investigation into the land systems of successive dynasties in Chinese history reveals that public and private land ownership alternated across periods, before evolving into a framework dominated by private ownership in later eras. A dynasty's prosperity and national unification did not hing on whether it adopted public or private land tenure, as demonstrated by the private land systems of the Western and Eastern Hàn and Míng and Qīng, on the one hand, and the public land regime of the Suí and Táng, on the other. As the core economic underpinning of national unity, the soundness of a dynasty's land system lay in its capacity to adapt to contemporary socioeconomic realities and to establish tax and corvée institutions to match. From the Qín and Hàn through the Míng and Qīng, China remained largely an agrarian society centered on smallholder farming. Dynastic fiscal revenue relied primarily on grain taxes and compulsory corvée labor levied on peasant households. Beyond land institutions, the burden of land taxes, poll taxes and corvée labor was also a decisive factor in the survival of China's unified empires. Once taxation and labor obligations became unbearably onerous and drove commoners to rebel, the dynasty's downfall would soon follow. There existed a critical threshold for tax and labor extraction: the point below which ordinary people could no longer sustain even basic subsistence.

The economic secret behind China's enduring unification lies in the fact that a state must do more than guarantee its people's security—it must also furnish the conditions and space for its citizens to survive and thrive. Since the 17th century onward, China fell increasingly behind Europe, the United States and Japan on the path toward industrialization. As a result, from the First Opium War in 1840 until the founding of the People's Republic of China in 1949, the Chinese nation endured a century of aggression, slaughter and oppression at the hands of Western powers and Japan. The founding of New China brought peace and stability to hundreds of millions of Chinese people who had long suffered humiliation under Western imperialist forces. Nevertheless, rapid population growth, combined with the nation's continued reliance on agrarian land use—its inability to escape the Malthusian Trap—left large swathes of the population mired in poverty prior to Reform and Opening-up. Following the Reform and Opening-up, industrial parks sprang up across the country. Providing industrial and commercial land to enterprises free of charge or at low cost was one of the key drivers of China's rapid industrialization. At the same time, explosive industrial growth generated massive demand for labor, creating job opportunities for hundreds of millions of rural workers who had previously lacked employment, and lifting large numbers of people out of poverty. In other words, successful industrialization allowed China to escape both poverty and the Malthusian Trap. Compared with any preceding period in Chinese history, Reform and Opening-up not only lifted the Chinese people to a level of moderate prosperity but also greatly expanded social mobility. Beyond government service, social elites in China could now pursue respected careers in business, academia, scientific research, white-collar professions, engineering and technical fields, medicine, journalism, performing arts, among others. Stalin once formuated: "the essential features and requirements of the basic economic law of socialism as folows: securing of the maximum satisfaction of the constantly rising material and cultural needs of the whole of society by the continuous expansion and improvement of socialist production on the basis of advancing techniques."[74] Yet the lessons of Chinese history demonstrate that, beyond meeting people's material and cultural needs, a nation must also provide ample room for orderly social mobility across fields such as governance, commerce, and the advanced intellectual professions.

Western-style democratic regimes and non-Western regimes differ merely in the organizational forms of government; form does not equate to substance. The criterion for judging whether a country is successful ought to be the living conditions of its ordinary people, rather than the structure of its government. For the citizens of any nation, their core demands come down to security, subsistence and development. Thomas Hobbes argued that human beings rely on the state to safeguard their personal safety. Adam Smith held that individuals pursuing their own maximum self-interest would, collectively, bring prosperity to the nation as a whole. Amartya Sen argued that the root cause of poverty in underdeveloped countries lies in the deprivation of people's capabilities—that is, their human capital. John Mearsheimer contended that, from a liberal perspective, divergent individual conceptions of a fulfilling life can give rise to mutual violence, making a state necessary to preserve peace.[75] When Western philosophers deliberated on political, economic and social issues, they centered their analysis on human beings within the state, not the institutional form of a country's government. It is evident, then, that debating governmental structures alone, while ignoring human beings and their fundamental needs for safety, survival and growth, puts the cart before the horse. Developed Western nations deliver security, affluence, and social mobility to their citizens;

modern China, likewise, provides its 1.4 billion people with public safety, moderately prosperous living standards and orderly social mobility. In this respect, Western developed countries and China share the same underlying goal: satisfying people's basic needs. Against the Western slogan of "liberal democracy", China has its own corresponding traditional of moral and ethical precepts rooted in collectivism, including "harmony without uniformity,"[76] "do not do to others what you would not want done to yourself,"[77] "cultivate the self, bring order to the family, govern the state, and bring peace to all under heaven,"[78] and "the world is for all."[79] The West's so-called universalist slogans demand that states and societies supply public goods to individuals. China's traditional morality and ethics, by contrast, define the responsibilities each person bears toward the self, the family, and the state—that is, toward the collective. From Mearsheimer's perspective, liberalism cannot cultivate a sense of communal belonging among individuals, nor can it unify society, in the way that nationalism can. This stems chiefly from liberalism's singular focus on individual rights at the expense of personal duties and obligations. Liberalism also risks eroding the bonds that sustain social cohesion, ultimately undermining the foundations of societal unity.[80]

In summary, if a country's residents are guaranteed personal safety, enjoy a prosperous life, and have ample opportunities for upward mobility in politics, business, senior management, and academia, the nation will naturally flourish culturally and advance technologically. Meeting citizens' fundamental demands for security, subsistence, and development constitutes the ultimate goal of state governance. Only on this foundation can individuals realize the Confucian ideal: " cultivate the self, bring order to the family, govern the state, and bring peace to all under heaven." The legacy of the imperial examination system, which prevailed from the Suí-Táng dynasties through the Míng-Qīng, offers a vital lesson: achieving effective incentive compatibility between individuals, the state, and the dominant national ideology, together with orderly social mobility, forms a grand institutional system critical to a dynasty's long-term stability and peace. Such a system must therefore be safeguarded by robust state institutions.

Notes:

[1] Acemoglu & Robinson, 2012, p.43–44, 317.

[2] Maddison, 2007, p. 1.

[3] Book of Songs · Odes of Shāng. (《诗经·商颂》“昔有成汤，自彼氐羌，莫敢不来王，莫敢不来享。”)

[4] Xúnzǐ · On Military Affairs. (《荀子·议兵》“汤居亳，武王居镐…天下为一，诸侯为臣”)

[5] The Gongyang Commentary on the Spring and Autumn Annals, Duke Yin, First Year. (《春秋公羊传·隐公元年》“何言乎王正月？大一统也。”)

[6] Chapter 1 of this book, Section 2.

[7] Beyond politics, economy, military affairs and social systems, language, culture, customs, shared historical classics and collective narratives exert an inestimable impact on the continuous restoration and consolidation of China’s great unification.

[8] Section 2.2 of Chapter 1 of this book.

[9] Lǐ Xuéqín holds the view that: “The Six Divisions were centered on Zōngzhōu in the western territory, hence known as the 'Western Six Divisions'; the Eight Divisions were centered on Chéngzhōu in the former Yin lands, thus called the 'Yin Eight Divisions', also referred to as the 'Chéngzhōu Eight Divisions'. The 'Six Divisions' and 'Eight Divisions' mentioned in Western Zhōu bronze inscriptions can only be explained by the integration of military and administrative systems of that era. The 'Six Divisions' and 'Eight Divisions' denoted not merely military troops, but also the rural districts that supplied the soldiers.”（李学勤认为：“六师以宗周为中心，在西土，所以称为‘西六师’；八师以成周为中心，在故殷之地，所以称为‘殷八师’，又名‘成周八师’。“西周金文的‘六师’、‘八师’只能以当时军事制度与行政制度的合一来解释。‘六师’‘八师’不仅指军队，也通指出军的乡”（参见 Li, 1987).）

[10] Records of the Grand Historian · Biography of Shang Yang (Sīmǎ Qiān). (“有军功者，各以率受上爵”（奖励军功，二十等爵制，第一次变法，前 356）; “为田开阡陌封疆，而赋税平”（废井田、开阡陌，第二次变法，前 350）)

[11] Máo Zé Dōng, 1973, p. 361.

[12] Records of the Grand Historian, Volume 97, “Biographies of Lì Shēng and Lù Jiǎ” (Sīmǎ Qiān). (“王者以民人为天，而民人以食为天”)

[13] Marx, 1972, p. 317.

[14] Wang, North Sòng Dynasty, p.3. (“万般皆下品，惟有读书高。”)

[15] Zhao, North Sòng Dynasty, p. 1268. (“书中自有黄金屋，书中自有颜如玉。”)

[16] Four books include: The Great Learning(《大学》), The Doctrine of the Mean (《中庸》), The Analects (《论语》), Mencius (《孟子》). Five Classics include: The Book of Songs / The Book of Odes (《诗经》), The Book of Documents / Shàngshū (《尚书》), The Book of Rites (《礼记》), The Book of Changes / I Ching (《周易》), The Spring and Autumn Annals (《春秋》) .

[17] Wang, North Sòng Dynasty, p.4. (“早为田舍郎，暮登天子堂。”)

[18] Three Sovereigns indicating: Suíren who invented fire by drilling wood (燧人氏：钻木取火), Fuxi who created the Eight Trigrams, established marriage rituals, and formulated rites and music (伏羲氏：创八卦、造嫁娶、制礼乐), Shennong (Yan Emperor) who tasted hundreds of herbs and taught people agricultural farming (神农氏（炎帝）：尝百草、教农耕). Five Emperors include: Huángdi / Yellow Emperor(黄帝), Zhuanxu (颛顼), Ku (帝喾), Yáo (尧), Shùn (舜).

[19] The Great Learning. (《大学》“古之欲明明德于天下者，先治其国；欲治其国者，先齐其家；欲齐其家者，先修其身…… 身修而后家齐，家齐而后国治，国治而后天下平。”)

[20] Record of Daily Knowledge, Volume 13, “Zhengshi” (Gu Yanwu). (顾炎武《日知录》卷十三《正始》“有亡国，有亡天下。亡国与亡天下奚辨？曰：易姓改号，谓之亡国；仁义充塞，而至于率兽食人，人将相食，谓之亡天下。…… 保国者，其君其臣肉食者谋之；保天下者，匹夫之贱与有责焉耳矣。”)

[21] Mencius, “Gaozi Part A". (《孟子・告子上》“恻隐之心，仁也；羞恶之心，义也；恭敬之心，礼也；是非之心，智也。仁义礼智，非由外铄我也，我固有之也。”)

[22] The Analects, “Wei-ling-kung". (《论语・卫灵公》子贡问曰：“有一言而可以终身行之者乎？” 子曰：“其恕乎！己所不欲，勿施于人。”)

[23] The Analects, “Yen-yüan”. (《论语・颜渊》“子曰：“君子成人之美，不成人之恶。小人反是。””)

[24] The Analects, “Tzu-lu”. (《论语・子路》子曰：“君子和而不同，小人同而不和。”)

[25] The Analects, “Yen-yüan”. (《论语・颜渊》齐景公问政于孔子。孔子对曰：“君君，臣臣，父父，子子。”)

[26] Record of Rites, “Li Yun”. (《礼记・礼运》“大道之行也，天下为公。”)

[27] Engels, 1878, p. 513.
[28] Chen, 2007.
[29] Wei, 2003.
[30] Yu, 1987 and Jin, 1965.
[31] Cao, 1989 & Liu, 2016.
[32] Liu, 2006.
[33] Han, 1959 & Li, 2013.
[34] Li, et al., 2018.
[35] Zou, 2015.
[36] Zhang, 1994.
[37] Li, 2013.
[38] Yu, 1987.
[39] Yang, 2001.
[40] Yuan, 1983.
[41] Du, 1984.
[42] Li, 1985.
[43] Jiang, 1987.
[44] Gao, 1981.
[45] Shi, 1986.
[46] Shao, 1992.
[47] Wang, 1996.
[48] Yang, 2003.
[49] Zhang, 2014.
[50] Liu, 1978 & Ge, 1999.
[51] Zhao and Chen, 2006, p. 18.
[52] Zhu, 2004.
[53] Zhong, 1992.
[54] Zhang, 2012.
[55] Yan, 2009, p. 22.
[56] Yan, 2009, p.82.
[57] Yan, 2009, p.137.
[58] Jiang, 1992.
[59] Zang, *et al*. 2020.
[60] Zhao & Chen, 2006, p. 26.
[61] Zhao & Chen, 2006, p. 30-31.
[62] Zhang, 1989.
[63] Zhao & Chen, 2006, p. 35.
[64] Yang, 1962; Zhao, 1994 and Liu, 2005
[65] Wang, 1963.
[66] Zang, et al. 2020.
[67] Shi, et al. 2006.
[68] Fan, 1982.
[69] Marx, p. 894.
[70] Essentials of Governance in the Zhēnguān Reign: On Governance. (《贞观政要·论政体》贞观六年，太宗谓侍臣曰：“看古之帝王，有兴有衰，犹朝之有暮，皆为蔽其耳目，不知时政得失……水能载舟，亦能覆舟。”)
[71] Editorial Committee for A History of Chinese Peasants’ Burden , 1991, p. 35-37.
[72] Editorial Committee for A History of Chinese Peasants’ Burden , 1991, p. 264-266.
[73] Wang, 1987.
[74] Stalin, 1952, p. 31.
[75] Mearsheimer, 2019, p. 62–64.

[76] The Analects, Book XIII, Zi Lu. (《论语・子路》子曰：“君子和而不同，小人同而不和。”)

[77] The Analects, Book XII, Yan Yuan. (《论语・颜渊》, 仲弓问仁。子曰：“出门如见大宾，使民如承大祭。己所不欲，勿施于人。在邦无怨，在家无怨。”)

[78] The Great Learning. (《大学》“古之欲明明德于天下者，先治其国；欲治其国者，先齐其家；欲齐其家者，先修其身…… 身修而后家齐，家齐而后国治，国治而后天下平。”)

[79] The Book of Rites, Li Yun. (《礼记・礼运》, 昔者，仲尼与于蜡宾…… 大道之行也，天下为公，选贤与能，讲信修睦。)
[80] Mearsheimer, 2019, p. 118–122.

## How China Succeeds: The Formation and Leverage of "State Capability"

### 1. Introduction

The great ancient Chinese thinker Confucius observed, "Do not worry about having no position; worry about the means to stand firm."[1] What Confucius meant is that one need not fear the absence of title or status, but should instead fear whether one truly possesses the knowledge, skills, and moral integrity required to be worthy of a position and to stand on solid ground. In essence, a person must first be educated in order to acquire the competencies needed to earn a living and participate meaningfully in society. Amartya Sen, the Indian-born Nobel laureate in Economics, identified the root cause of persistent poverty not in material deprivation alone, but in the deficit of individual "capability." His insight was that escaping poverty requires the state and society to ensure accessible education and healthcare for disadvantaged groups—a framework he formalized as the "capability approach" to poverty eradication in the developing world.[2] In Sen's view, "Development should be evaluated by the expansion of people's capabilities — that is, their real opportunities to do and be what they value — rather than by income, resources, or utility alone."[3] Sen pointed out the harm of the utility theory of neoliberal economics to developing countries, to this he said: "Utilitarianism is not only an inadequate basis for the evaluation of social welfare, it is also a misleading guide to policy. It ignores the diversity of human beings and the importance of substantive freedoms." [4]

Scholars have conducted extensive research on the concept of "state capability". Max Weber argued that a state is an organization that legally monopolizes the use of force within a specific territory and effectively enforces rules, and a rational bureaucracy serves as the organizational foundation of state capability.[5] Skocpol defines state capability as a state's ability to carry out and realize its own goals and will. She emphasizes state autonomy, meaning a state can formulate and implement policies independently of social interest groups.[6] Migdal argues that the state possesses the capacity to shape social order and reshape societal objectives. He maintains that the state has four core capabilities: First, extractive capability: to levy taxes and mobilize resources. Second, penetrative capability: to extend bureaucratic governance down to grassroots communities. Third, regulatory capability: to standardize market operations and social conduct. Fourth, distributive capability: to redistribute resources and deliver public welfare.[7] Wang and Hu pioneered a systematic analytical framework for studying China and also put forward four core state capabilities. The first is extractive capability, referring to fiscal revenue collection and resource mobilization. The second is regulatory capability, which covers macroeconomic management and social stability. The third is legitimating capability, including ideological construction and national identity building. The fourth is coercive capability, tasked with maintaining social order and national security.[8] Fukuyama defines state capability as the government's ability to formulate and implement policies and deliver public goods, with the quality of the bureaucratic system at its core.[9] In addition to the above discussions on state capability, we hold that a state should also possess capabilities in social infrastructure development and provision, manufacturing, education and scientific research, as well as high-tech industries.

Compared with developed countries, developing nations lag significantly in high-tech industries, manufacturing, education and research, healthcare, military production, and government administration, among others. Borrowing from the frameworks of Confucius and Amartya Sen, this gap can be understood as a profound deficit in “state capability” — encompassing weak organizational, high-tech industrial, manufacturing, educational, healthcare, military, and administrative capacities. In short, developing countries face capability deficits across virtually every dimension of national life. It follows, therefore, that if these nations are to achieve meaningful development, the foundational imperative is to systematically build and strengthen state capability across each of these domains.

China's foremost advantage lies in being a socialist country under the leadership of the Communist Party of China (CPC). The CPC is a Marxist political party founded on two pillars: Lenin's theory of party building — whereby individual party members subordinate themselves to the organization, lower levels defer to higher ones, and the Party as a whole answers to the Central Committee, a principle known as democratic centralism — and Lenin's experience leading the Russian Bolsheviks, with the primary mission of opposing imperialism and feudalism. Forged through 28 years of arduous struggle against both domestic and foreign enemies, the CPC developed an exceptionally disciplined organizational and political culture, tempered by the demands of prolonged warfare. China's greatest fortune in the 20th century was that it owns the CPC — a party of formidable executive and organizational capability — and its powerful party-led military. These foundations, combined with the socialist “Five-Year Plan” system and the socialist principle of “pooling resources to accomplish major tasks,”[10] have served as the central instruments through which China has systematically built and enhanced state capability.

This article primarily explores the origins of the CPC’s “strong execution and organizational capabilities,” examines how socialist China used the CPC capabilities to develop “heavy industrial capability,” “embedding market into its planned economy,” “national development planning capability,” “high-tech industrial capability,” and thereby building China into a global high-tech socialist power.

## 2. The origins of CPC's “strong execution and organizational capabilities”

On July 1, 2021, in his speech at the ceremony marking the centenary of the CPC, General Secretary Xí Jìnpíng pointed out: “After the Opium War of 1840, China was gradually reduced to a semi-colonial, semi-feudal society. The country endured intense humiliation, the people were subjected to great pain, and the Chinese civilization was plunged into darkness. The Chinese nation suffered greater ravages than ever before. From that time on, national rejuvenation has been the greatest dream of the Chinese people and the Chinese nation. To save the nation from peril, the Chinese people put up a courageous fight. As noble-minded patriots sought to pull the nation together, the Taiping Heavenly Kingdom Movement, the Reform Movement of 1898, the Boxer

Uprising, and the Revolution of 1911 rose one after the other, and a variety of plans were devised to ensure national survival, but all of these ended in failure. China was in urgent need of new ideas to lead the movement to save the nation and a new organization to rally revolutionary forces. With the salvoes of Russia's October Revolution in 1917, Marxism-Leninism was brought to China. Then, as the Chinese people and the Chinese nation were undergoing a great awakening and Marxism-Leninism was becoming closely integrated with the Chinese workers' movement, the CPC was born. The founding of a communist party in China was an epoch-making event, which profoundly changed the course of Chinese history in modern times, transformed the future of the Chinese people and nation, and altered the landscape of world development. Since the very day of its founding, the Party has made seeking happiness for the Chinese people and rejuvenation for the Chinese nation its aspiration and mission. All the struggle, sacrifice, and creation through which the Party has united and led the Chinese people over the past hundred years has been tied together by one ultimate theme — bringing about the great rejuvenation of the Chinese nation."[11] The above quotation of President Xí's speech effectively weaves China's modern history into a single coherent narrative, from its founding day, the CPC has anchored every struggle, sacrifice, and act of creation to a single overarching mission: the great rejuvenation of the Chinese nation.

### 2.1 Efforts to Save China from 1840 to 1921

After the defeat in the Opium War in 1840, the Westernization Faction in the Qīng Dynasty, represented by Zēng Guófān (曾国藩), Lǐ Hóngzhāng (李鸿章), Zuǒ Zōngtáng (左宗棠) and Zhāng Zhīdòng (张之洞), launched the Westernization Movement in the hope of saving the country through industrial and military development. The Westernization Faction advocated the ideas of "learning the advanced skills of foreign countries to resist foreign aggression"[12] and "the Chinese tradition as the core, the Western knowledge as the means"[13] — the latter capturing a conviction that Western technology could be selectively adopted without disturbing the fundamental order of the Chinese civilization. In practice, this meant pursuing Western-style manufacturing and military development — striving for "powerful ships and formidable guns" — while leaving the Qīng Dynasty's political structure entirely intact. It was modernization confined to the surface, grafted onto a decaying institutional core. The fatal limits of this approach were exposed in 1895, when the corrupt and enfeebled Qīng state and its military suffered a humiliating defeat at the hands of Meiji Japan in the First Sino-Japanese War. With that defeat, China's first sustained ambition toward industrialization was abruptly and decisively interrupted.[14] China's defeat in the First Sino-Japanese War of 1894–1895 precipitated a profound reckoning over the question of national salvation, giving rise to two distinct camps: the Reformist Faction, led by Kāng Yǒuwéi(康有为), and the Revolutionary Faction, led by Sūn Yat-sen (孙中山). Each offered a fundamentally different answer to the same urgent question — how to rescue China from decline and subjugation. The reformists, represented most prominently by Kāng Yǒuwéi(康有为) and Liáng Qǐchāo (梁启超), drew a crucial lesson from the defeat: China's backwardness was not merely technological or industrial, but systemic. The real source of weakness, in their view, lay in an ossified political structure that was no longer capable of meeting the demands of a rapidly changing world. Rather than overthrowing the Qīng Dynasty outright, they sought to reform it from within — preserving its overall institutional framework while modernizing its governance.[15] This conviction drove the Hundred Days' Reform of 1898, an ambitious but ultimately short-lived attempt to introduce political and institutional reforms into the Qīng system from above. In 1898, the Qīng court embarked on an ambitious program of top-down reform: appointing reformist figures to senior government positions, encouraging the establishment of private industrial and commercial enterprises, founding new-style schools to cultivate modern talent, commissioning translations of Western works, and promoting the dissemination of new ideas. This episode is known historically as the Hundred Days' Reform, or the Wùxū Reform. Yet the movement proved short-lived. By encroaching on the entrenched interests of the conservative faction centered on Empress Dowager Cíxǐ, it provoked a swift and decisive backlash — the reforms were suppressed just over one hundred days after they began, and their principal architects, Kāng Yǒuwéi and Liáng Qǐchāo, were forced to flee into exile abroad. The Qīng court subsequently adopted many of the very measures it had suppressed, but the concession came too late to alter the dynasty's trajectory. The First Sino-Japanese War had already laid bare the incompetence, institutional decay, and systemic corruption of the Qīng political order — and no piecemeal reform could repair what was fundamentally broken. Revolutionary pioneers such as Sūn Yat-sen drew the more radical conclusion: that China could not be saved by reforming the Qīng system from within, but only by dismantling it entirely.[16] If China was to resist imperialist domination and reclaim its sovereignty, the Qīng Dynasty had to be overthrown and replaced with a republic — a transformation not of policy, but of the political order itself. In late 1894, Sūn Yat-sen founded the Revive China Society (Xīng Zhōng Huì) in Honolulu, Hawaii, with the declared aim of expelling Manchu rule, restoring Chinese sovereignty, and establishing a unified republican government. A decade later, in 1905, this organization served as the nucleus around which a broader revolutionary coalition was formed: the Chinese Revolutionary Alliance (Tóng Méng Huì), a bourgeois revolutionary party whose guiding tenets expanded the original program to encompass four objectives — expelling Manchu rule, restoring China, establishing a republic, and equalizing land ownership.

Under the leadership of the Tóng Méng Huì, revolutionary martyrs mounted a succession of armed uprisings against Qīng rule. Among the most celebrated was the Huáng Huā Gǎng Uprising in Guǎngzhōu in April 1911, in which dozens of revolutionaries gave their lives. The cumulative momentum of these efforts culminated in the Wǔchāng Uprising of October 1911 — the decisive insurrection that toppled the Qīng Dynasty and gave birth to the Republic of China. Yet the revolution's promise proved difficult to sustain. The revolutionary party's organizational weakness and lack of cohesion, compounded by fierce power struggles among rival warlord factions, plunged China into an era of fragmentation, internecine conflict, and separatist rule. Rather than emerging as a sovereign and unified nation, China found itself consigned to the periphery of the capitalist-imperialist world order — its people subjected to the exploitation and oppression of foreign imperial powers, its national ambitions once again deferred.

In the early 20th century, and particularly in the wake of the Xīnhài Revolution of 1911, two principal factions emerged in fierce contention over the question of how to save China. The first, the Royalist Faction led by Kāng Yǒuwéi, rejected republicanism in favor of constitutional monarchy — arguing that China should retain the monarchical system and be governed according to the principles of Confucianism. In Kāng's view, the institutional continuity of monarchical rule, tempered by constitutional constraints, offered a more stable and culturally rooted path to national renewal than the republican experiment that the revolution had unleashed.[17] The second faction, the Total Westernization camp led by Hú, took an altogether different position: that China's salvation lay in wholesale adoption of the Western liberal democratic model, with the United States as its exemplar, and in the comprehensive Westernization of Chinese society, culture, and institutions.[18] The Royalist Faction fundamentally failed to grasp the true nature of the epochal transformation China had been confronting since the advent of Western industrialization in the 18th century — what might aptly be described as a change unprecedented in three thousand years of Chinese history. Before the 17th century, China had stood as the most economically powerful civilization in the world for 3,000 years. Yet in the wake of Western industrial ascendancy and the military humiliations that followed, China was progressively reduced — by 1840 and the decades thereafter — to the condition of a semi-colonial state: controlled, exploited, and subjugated by the imperial powers of the West. The Royalists, wedded to the inherited institutions of a pre-industrial order, offered no framework adequate to this civilizational crisis.

The Total Westernization Faction, for its part, saw only the luminous surface of the imperialist powers — their modernity, apparent civilization, and material prosperity — and concluded that China need only follow the Western path faithfully to eventually attain the same condition. What they failed to see was the foundation upon which that prosperity rested: the extraction of monopoly profits through the systematic exploitation and oppression of colonial and semi-colonial peoples across the globe. Nor did they reckon with China's actual position within this imperial order. By the time the Total Westernizers were advancing their program, semi-colonial China had already been drawn into the orbit of imperialist control — its national politics, strategic industries, defense and foreign affairs, finance, and monetary system all penetrated and subordinated to foreign imperial interests. Within this structure, China was consigned to a fixed and permanent role: a captive market for imperialist goods and an inexhaustible source of raw materials. In short, without a fundamental rupture from the imperial order, old China was structurally condemned to remain an object of exploitation and oppression — perpetually dependent, perpetually subjugated, and never truly sovereign.

### 2.2 Lenin's Theory of Imperialism and the October Revolution in the Soviet Union Pointed out a Path to Success for China

While Chinese intellectuals of the late 19th and early 20th centuries remained locked in debate over whether to restore the imperial system or embrace wholesale Westernization, Lenin, writing in 1916, offered a penetrating analytical framework that reframed the question entirely. In his view, the world was fundamentally divided between imperialist powers on one side and colonial and semi-colonial nations on the other — a structural division in which the former systematically exploited and oppressed the peoples of the latter.[19] From this diagnosis, Lenin drew two revolutionary conclusions: first, that the proletariat of all colonial and semi-colonial countries must unite in common struggle against imperialist exploitation and domination; and second, that the weakest links in the chain of imperialist rule were precisely the points at which a socialist state could be forged and established.

Lenin did not merely theorize — he acted. In 1917, he led the Russian Bolsheviks to overthrow bourgeois rule in the October Revolution and establish the world's first socialist state, translating his theory of imperialism into revolutionary practice on a world-historical scale. In the course of leading this struggle, Lenin also developed and refined the Bolsheviks' theory of party building and its organizational principles — a doctrine that would prove as consequential as the revolution itself. These principles rested on a clear hierarchy of deference: individual party members subordinate to party organizations; lower-level party organs subordinate to higher-level ones; and the entire party subordinate to the Central Committee. Binding this structure together was the doctrine of democratic centralism — a disciplined unity of will and action forged through open internal deliberation, followed by unconditional collective adherence to decisions once made.[20]

Inspired by the triumph of the October Revolution and with the organizational guidance and financial support of the Communist International (Comintern), dozens of progressive Chinese intellectuals convened in Shanghai in 1921 to found the Communist Party of China (CPC). From its inception, the CPC set itself two defining objectives: to oppose imperialism and feudalism, and to build China into a powerful socialist industrialized nation. The ideological debt to the October Revolution was profound and explicit. As Máo Zédōng observed: "The salvoes of the October Revolution brought us Marxism-Leninism. The October Revolution helped progressives in China and across the world to adopt the proletarian world outlook as the instrument for understanding the destiny of their country and reconsidering their own problems."[21] Under Máo Zédōng's leadership, the CPC drew on the political and organizational principles of the Communist Party of the Soviet Union, and undertook the creative task of adapting Marxism-Leninism to the Chinese contexts. Through this process, the CPC forged itself into a formidable political organization — one characterized by exceptional organizational capacity, ironclad discipline, and a heightened level of political consciousness unmatched by the revolutionary movements that had come before.

Since its founding, the CPC has traversed a series of monumental historical trials that forged its character and tempered its resolve. These include: the Northern Expedition of 1926, launched in cooperation with the Kuomintang (KMT); the KMT's catastrophic betrayal of the revolutionary alliance and its mass slaughter of Communist members in April 1927; the Nánchāng Uprising of August 1, 1927, led by Zhū Dé, Zhōu Ēnlái, and Hè Lóng, which marked the birth of the people's army; the Autumn Harvest Uprising later that year, led by Máo Zédōng; and the KMT's five successive campaigns of encirclement and suppression against CPC-led Soviet Areas between 1930 and 1934, each met by determined Red Army counter-offensives. Forced into a strategic withdrawal in October 1934, the Red Army embarked on what would become one of the most extraordinary feats of military endurance in modern history---the Long March. Fighting through relentless encirclement, pursuit, interception, and blockade by hundreds of thousands of KMT troops, the Red Army traversed the rugged terrain of Guìzhōu, Sìchuān, Gānsù, and Níngxià before reaching its destination in northern Shǎanxī — battered but unbroken, and more battle-hardened than ever.

In January 1935, the enlarged Politburo meeting convened at Zūnyì, Guìzhōu, marked a decisive turning point: it established Máo Zédōng's leading position within both the CPC and the Red Army, providing the revolutionary movement with the unified central leadership it had long needed. Between June and September of that year, a fierce intra-Party struggle against Zhāng Guótaō — whose separatist ambitions threatened to fracture the Red Army — played out at Màogōng in Sìchuān, testing the cohesion of the CPC at one of its most vulnerable moments. In October 1935, the Central Red Army completed its arduous march and arrived in northern Shǎanxī, where it joined forces with the local Red Army led by Liú Zhìdān, establishing the northern Shǎanxī Revolutionary Base Area as the new center of revolutionary operations. From this foothold, the Red Army successfully repelled successive KMT encirclement campaigns targeting the base area, consolidating its position in the northwest. In the first half of 1936, the Red Army crossed the Yellow River into Shānxī on the Eastern Expedition — a campaign designed not only to engage the enemy but to collect grain, raise funds, and expand the Red Army's ranks in preparation for the struggles ahead.

In September 1937, the CPC reached an agreement with the KMT government to form a united front against Japanese aggression.[22] Under this arrangement, the Red Army was reorganized as the Eighth Route Army within the National Revolutionary Army, fighting alongside KMT forces against the Japanese invaders. Under Máo Zédōng's strategic direction, the Eighth Route Army and the New Fourth Army pursued a distinctive approach: independent, self-reliant guerrilla warfare in mountainous terrain and sustained operations deep behind enemy lines. Through this strategy, the CPC established numerous revolutionary base areas in Japanese-occupied territories, extending its organizational reach and building the popular foundations that would prove decisive in the struggles ahead. After eight years of grueling resistance, China emerged victorious in August 1945 — not in isolation, but as part of the broader triumph of the global anti-fascist coalition. The War of Resistance against Japanese Aggression had exacted an enormous human and material toll, yet it had also transformed the CPC: the years of protracted warfare had hardened its discipline, expanded its mass base, and demonstrated, on a national scale, the extraordinary organizational and military capability that Máo's strategy had cultivated.

With Japan's defeat, Chiang Kai-shek's KMT swiftly moved to extinguish the CPC rather than consolidate the peace, launching a full-scale civil war against the CPC and its armed forces. The People's Liberation Army (PLA) under the command of the CPC, though confronting a KMT military of some eight million troops, prosecuted a series of decisive campaigns that progressively dismantled the Nationalist war machine. Central to the CPC's success was its land redistribution program, which granted farmland to the peasant masses and in doing so secured the active allegiance of China's vast rural population — transforming the PLA's struggle from a military contest into a genuine people's war. By 1949, on the eve of the founding of the PRC, Chiang Kai-shek's forces had been driven from the mainland entirely, retreating to Taiwan Island in irreversible defeat.

From 1927 through to the eve of the founding of the PRC, the CPC waged struggle on two simultaneous fronts. On the battlefield, it led the Red Army, the Eighth Route Army, the New Fourth Army, and ultimately the People's Liberation Army through successive phases of armed conflict — the Agrarian Revolutionary War (1927-1937), the War of Resistance against Japanese Aggression (1937–1945), and the Liberation War (1946–1949). In parallel, it directed an extensive network of underground revolutionary operations in enemy-occupied territories, working covertly in conditions of acute danger and repression. These underground operations proved indispensable to the CPC's ultimate success. They drew large numbers of revolutionary-minded youth from enemy-occupied areas into the CPC's ranks, steadily expanding its organizational base. They furnished the CPC with vital intelligence and material support. And they mobilized local populations within enemy-held territories, quietly building the broad social foundations upon which the CPC's eventual nationwide victory would rest.

Operating in a perilous environment where enemy and allied forces were inextricably intertwined — where a single misstep could mean arrest, torture, or death — underground work forged a cadre of revolutionaries of exceptional quality: acutely vigilant against enemy infiltration, unwavering in revolutionary conviction, disciplined in organization, and resolute under pressure. Among them were underground CPC members who, even when captured through betrayal and subjected to brutal torture, neither renounced their beliefs nor surrendered the CPC's secrets. The moral and organizational fiber demonstrated in these conditions was not merely admirable in the abstract — it represented a form of human capital of the highest order, one that would prove invaluable when the CPC turned from revolutionary struggle to the governance of cities and countryside in the years following the founding of the PRC.

In 1950, barely a year after the founding of the PRC, China was drawn into another major conflict. The outbreak of the Korean War, and particularly the dramatic reversal triggered by the United States-led amphibious landing at Incheon, brought American forces to the Chinese border — a strategic threat that Beijing could not ignore. China responded by dispatching the Chinese People's Volunteer Army across the Yālù River to resist U.S. aggression and aid the Democratic People's Republic of Korea. After more than three years of grinding and costly warfare, and equipped largely with weapons supplied by the Soviet Union, the Chinese Volunteer Army achieved what few observers had deemed possible: fighting the sixteen-nation coalition led by the United States to a military stalemate and restoring the pre-war boundary along the 38th parallel. The outcome announced to the world that the new China — impoverished, industrially underdeveloped, and barely a year old — possessed both the will and the organizational capacity to hold on its own against the most powerful military alliance on earth.

In 1964, the United States escalated its military intervention in Vietnam into a full-scale war of aggression. Once again, China found itself called upon to fulfill its commitment to resisting imperialist aggression — this time in support of the Communist Party of Vietnam. Drawing on its military organizational framework and the heavy industrial base it had built in the preceding decade, China provided sustained and decisive assistance that contributed to the withdrawal of American forces from Vietnam by 1973. Beyond these two theaters of external struggle, the same period witnessed a series of breakthroughs of historic significance on the strategic front. Under the leadership of the CPC, China successfully tested its first atomic bomb in 1964 and its first hydrogen bomb in 1967, and developed long-range ballistic missiles capable of delivering nuclear warheads. With these achievements, China entered the ranks of the world's nuclear powers — a transformation that fundamentally altered its strategic position and secured for it a measure of sovereign inviolability that no foreign power could henceforth ignore.

Since the reform and opening-up, the CPC has accumulated hard-won experience in driving rapid economic growth — developing a socialist market economy and integrating China into the global trading system on terms consistent with its own developmental imperatives. Since the 18th National Congress of the CPC in 2012, this experience has deepened further: China has navigated sustained strategic pressure from the United States, cultivated a new generation of globally competitive high-tech enterprises, and pioneered a distinctively Chinese model of international cooperation — one that offers an alternative framework to the Western-dominated norms that have long governed global engagement.

The CPC's long-term governance has served as a structural bulwark against the penetration of Western monopoly capital — most prominently the financial interests centered on Wall Street — into China's policy-making processes, strategic industries, and sovereign affairs. In this sense, the continuity of the CPC leadership has preserved something increasingly rare in the contemporary world: genuine, substantive independence — a sovereignty that is not merely formal, but real in its capacity to resist external domination and chart the nation's course according to its own developmental logic.

From its founding in 1921 to the establishment of the PRC in 1949, the CPC led the Chinese people and their armed forces through 28 years of unrelenting political and military struggle — confronting the reactionary KMT, prosecuting the eight-year War of Resistance against Japanese Aggression, and navigating the treacherous currents of underground revolutionary work. Forged in conditions of extreme adversity, the CPC and its armies developed a rock-solid organizational cohesion and an ironclad discipline that no comparable revolutionary movement of the era could match. Through these trials, Lenin's theories of party building were not merely upheld but tested, refined, and made authentically Chinese. After the founding of the PRC, the CPC accumulated a further layer of formative experience: resisting American strategic pressure, constructing the foundations of Chinese socialism, and sustaining an enduring commitment to the liberation and development of the Chinese people. Each chapter of this history — revolutionary struggle, national defense, and socialist construction — added to the institutional memory, organizational depth, and executive capacity that define the CPC's governance in the new era. It is the accumulated weight of this century-long experience, more than any single policy or leader, that underlies the CPC's formidable execution and organizational capability today.

### 3. From 1949 to 1977, China Leveraged the State Capability to Build up its Heavy Industry and Scientific Research and Development

Socialism arose in opposition to capitalism, imperialism, and colonialism. In New China, socialist construction aimed above all to end the exploitation and oppression that landlords, the bourgeoisie, and colonial powers had imposed on working people. After the PRC was founded, a planned socialist economy took shape gradually, modeled largely on the Soviet experience.

The early years of New China were marked by a transformation of historic proportions. Empowered by the executive strength, military reach, and mass mobilization capability forged through years of revolutionary war, the CPC-led government swept through the countryside with a program of land reform — seizing land from the landlord class and distributing it equally among the peasants, then steering agriculture toward collective organization. In doing so, it brought to a close a cycle of exploitation and oppression that had endured for thousands of years.

The transformation unfolded in cities as well. Enterprises held by the bureaucratic-comprador bourgeoisie were confiscated outright and absorbed into the state sector, while the industrial and commercial holdings of the national bourgeoisie were brought under public ownership through a more gradual process of compensated redemption. The result was the systematic dismantling of the economic structures through which the bourgeoisie had long exploited and oppressed the working class.

Step by step, New China constructed the institutional pillars of a modern society: a nationwide education system reaching from primary schools to universities, a web of scientific research institutions, and a public healthcare network that brought medical services to ordinary working people for the first time. These efforts not only widened access to education and healthcare but also nurtured the country's own capabilities in science, research, and learning. Against the backdrop of the Korean War, which broke out in 1950, and through the decades that followed, China harnessed the momentum of 156 Soviet-assisted heavy industry projects to lay the foundations of a socialist industrial economy.

Building heavy industry was a resource-intensive undertaking, demanding both advanced technology and enormous capital. On the technological front, New China drew on Soviet assistance while steadily cultivating its own technical workforce from within. Funding the endeavor, however, fell largely on the shoulders of agriculture: through the price scissors mechanism — farm prices held artificially low while industrial goods were priced high — and through state monopolies over the purchase and sale of major agricultural products, the rural sector effectively bankrolled industrial growth. In the context of New China's early years, redirecting agricultural resources toward rapid heavy industrialization was seen as an unavoidable necessity.

Grounded in its expanding scientific, educational, and industrial foundations, and empowered by the socialist planned economy's singular ability to marshal national resources toward a common purpose, New China embarked from the mid-1950s on an audacious push to master nuclear technology, rocketry, and space exploration. The results were transformative: a domestically replicated missile took flight in 1960, an atomic bomb shook the desert in 1964, a hydrogen bomb followed in 1967, and a satellite rose into orbit in 1970. With these achievements, China entered the ranks of the world's nuclear powers — capable of producing and deploying nuclear weapons — and in doing so, permanently closed the door on the nuclear blackmail and military intimidation that imperialist powers, past and present, had sought to wield against it.

The decades between New China's founding and the reform and opening up were a period of foundational construction. Exploitation was abolished, and the state built education systems from primary school through university, a public healthcare network, and a web of scientific research institutions — bringing services once reserved for the privileged within reach of ordinary working people. National security imperatives left China with little choice but to pour resources into heavy industry, financing that drive by redirecting surplus from the agricultural sector. The human toll, however, was considerable: up to 76 percent of rural population found themselves below the poverty line on the eve of reform and opening up.[23] Light industry, starved of investment and attention, fell well short of meeting the everyday needs of the population.

### 4. Since the Reform and Opening-up in 1978, China Has Leveraged its State Capability to "Embed" Markets into its Socialist Planned Economy

At its core, reform and opening-up was a response to a stark reality: China was poor, and it was falling behind. What followed was a carefully managed process in which the Chinese government began, step by step, to embed market mechanisms into the fabric of a socialist economy still organized around central planning. The logic behind this approach finds an echo in the work of Hungarian philosopher Karl Polanyi, who argued in The Great Transformation that markets do not arise from some natural order of things — they are institutional creations, deliberately constructed and embedded into existing economic arrangements through the active hand of the state.[24] Centuries of market dominance have worked a kind of historical amnesia on European and American societies. The fact that markets were once deliberately constructed and embedded into existing economic systems has faded from view, replaced by a quiet assumption that the market is simply the natural order of things. From this assumption flows a troubling corollary: that the market's destructive consequences — unemployment, environmental ruin, monopoly power, poverty, exploitation, and the deprivation of basic services like water, electricity, healthcare, and education in the world's slums — are simply the price of doing business, impossible to avoid. It is precisely this worldview that gave rise to American neoliberalism from the 1980s onward. China's situation, however, is fundamentally different. Where Western nations have lost the thread of how their markets came to be embedded — the memory worn away by centuries of familiarity — China's own experience of grafting market mechanisms onto a planned economy is recent, deliberate, and very much alive in the public consciousness.

In their research on embedding the market economy within China's economic framework, Zhang and Guo sorted out two core threads: the compatibility between public ownership and the market economy, and the relationship between the government and the market.[25] They explained how the market economy is integrated into the basic socialist system, drawing the core conclusion that the socialist market economy with the Chinese characteristics essentially represents an organic integration of the socialist system and market mechanisms. It upholds the dominant position of public ownership while leveraging the efficiency advantages of the market in resource allocation. Ren and Li analyze the gradual embedding trajectory spanning the planned economy, the planned commodity economy, and the socialist market economy, reveal the institutional evolution logic of "the market embedded within planning", and draw the conclusion: starting from a framework where planning played the leading role and the market a supplementary role, the market's function has continuously expanded, culminating in the market playing the decisive role in resource allocation.[26] From a historical perspective, Yang and Guo analyze how gradual reform has enabled the steady embedding of the market economy into the planned economy, thereby effectively averting social shocks and economic recessions triggered by radical reforms and minimizing the pains of economic transition to the greatest extent possible.[27] Xie and Kuang analyze the practical process of the gradual embedding of market mechanisms into the planning system in three stages: the planned commodity economy, the establishment of the market economy, and its improvement. This approach has not only guaranteed the continuity and stability of reform but also delivered sustained improvements in economic efficiency.[28] From the perspective of the New Era, Liu elaborates on the theories and paths for the deep embedding of market mechanisms into the basic socialist economic system. His conclusion holds that a high-standard socialist market economy system represents a mature form in which the market economy is fully, profoundly, and institutionally embedded into the basic socialist system. Upholding the socialist orientation while fully unlocking market vitality, it serves as the institutional foundation of Chinese modernization.[29] China's integration of market forces into its formerly planned economy has never been an unconditional embrace. Throughout the process, the government has pursued a carefully calibrated approach: drawing on the market's proven ability to stimulate economic dynamism while remaining alert to, and guarding against, its corrosive side effects. Underpinning this balance is a sustained state commitment to providing what markets routinely fail to deliver equitably — public services such as education, healthcare, social security, eradicating poverty, and the essential infrastructure of modern life, from water and electricity to gas supply.

The story of how China embedded the market into its planned economy is one of deliberate, staged expansion. It began in 1979, when the central authorities gave the green light to four Special Economic Zones — Shēnzhèn, Zhūhāi, Shàntóu, and Xiàmén. Within these designated enclaves, foreign investors could establish wholly owned enterprises and enjoy tax rates more competitive than even those in Hong Kong, making them powerful magnets for foreign capital. The experiment proved successful enough to prompt a bolder move: in 1984, fourteen coastal cities — among them Shànghǎi, Tiānjīn, Dàlián, Níngbō, and Guǎngzhou — were opened up to outside investment. The momentum continued into 1990, when Deng Xiaoping personally championed the opening of Pudong New Area in Shànghǎi, setting the stage for that city's dramatic rise. Running parallel to these urban reforms, rural China underwent its own quiet transformation around 1980, as the policy of granting household long-term leases over farmland took root across the countryside.

The return of educated youth from the countryside to China's cities created an urgent employment problem that could not be ignored.[30] The government's response, in 1980, was to open a new door: urban residents were granted permission to operate their own individual businesses — a modest but symbolically significant step toward accommodating private economic activity within the socialist system. The pace of liberalization quickened through the early 1980s. By 1983, rural households were free to launch their own businesses and bring on hired help, and a year later the rules reserving township and village enterprises exclusively for collective operators were swept away. In the years that followed — right up to the legalization of private enterprises in 1988 — it was the collectively owned township and village enterprises and urban collectives that shouldered much of the burden of reform and industrialization, serving as the primary engines of grassroots economic growth. Their success, however, owed much to an unlikely source of support. In an era when formal commercial credit systems were still embryonic, township and village governments stepped into the breach, putting their administrative authority behind these enterprises in the form of loan guarantees, debt repayment assurances, and endorsements of product reliability. That official backing — part credibility, part safety net — became the all-important key that unlocked markets and opportunities across the country, allowing township and village enterprises to flourish on a truly national scale.[31] Urban collective enterprises were sustained by the same institutional logic. On the ideological front, the report to the CPC's 13th National Congress in 1987 provided an important theoretical anchor, affirming that China remained at the primary stage of socialism — a framing that justified the gradual, experimental nature of the reforms underway. The pace of change accelerated from there: a constitutional amendment in 1988 gave private enterprises their first formal legal standing, opening the door to a sector that would grow to reshape the economy. Then, in 1993, the Third Plenary Session of the 14th Central Committee delivered what amounted to a definitive statement of direction, resolving to build a socialist market economy system and calling for the active development of markets for labor, capital, real estate, and beyond.

By the mid-1990s, widespread losses among state-owned enterprises (SOEs) prompted the central government to adopt a policy of closing, suspending, merging, or restructuring the majority of small and medium-sized SOEs, while retaining a limited number of strategically significant enterprises essential to national economic security and public welfare. SOE reform represented a pivotal juncture in China's broader economic transformation. Its advancement injected considerable dynamism into the economy: small and medium-sized SOEs that were closed, suspended, merged, or restructured no longer placed fiscal burdens on the central government, while those that were privatized became contributors to state revenue through taxation. In anticipation of accession to the World Trade Organization, the Chinese government undertook an extensive review and revision of its economic regulatory framework around 2000, bringing domestic rules and administrative regulations into conformity with international standards.

Since the 1990s, one of the central challenges China has wrestled with is how to build a banking and financial sector that genuinely serves socialist objectives and the wellbeing of all its people, even as market forces have been woven ever more deeply into the economic fabric. The stakes are high: whoever controls the flow of surplus capital effectively shapes the trajectory of the entire economy, steering growth toward whichever sectors receive investment. The American experience serves as a sobering reference point. In the 1980s and 1990s, Wall Street's financial interests — driven by the relentless pursuit of excess profit — systematically redirected capital away from domestic manufacturing and toward cheaper production bases in East Asia and China, leaving American industry hollowed out in their wake. Then, as the century turned, that same tide of financial capital crashed into the domestic real estate market, inflating a bubble of extraordinary proportions — one that ultimately burst in 2008, triggering a financial crisis whose tremors were felt across the globe.

In order to forestall the kind of industrial hollowing-out and real estate instability that had afflicted other economies, and to ensure sustained, dependable financial support for the real economy — particularly high-technology industries — China convened Central Financial Work Conferences in 2018 and 2023. The overarching aim of these gatherings was to lay the groundwork for a socialist financial system oriented toward the collective wellbeing of the Chinese people, and to align bank lending and financial investment with the country's long-term priorities for economic and social development as well as its industrial policy objectives. The principle that the financial sector must treat service to the real economy as its core mission has been a recurring emphasis of both the Central Financial Work Conference and the Central Economic Work Conference.[32]

China has moved to institutionalize its industrial protection goals through concrete regulatory frameworks. Through authoritative policy instruments such as the Guiding Opinions on Financial Support for New Industrialization and the Circular on Deepening Financial Services for the Manufacturing Sector to Boost the Advancement of New Industrialization, financial institutions are explicitly prohibited from extending financing, guarantees, or foreign exchange facilities to support the offshore relocation of manufacturing activities. More specifically, large-scale overseas investment, cross-border mergers and acquisitions, and asset transfers by manufacturing enterprises are subject to rigorous filing and approval requirements, authenticity verification, and scrutiny of capital origins. Strict controls are also applied against fictitious outbound investment and the diversion of production capacity and capital through overseas operations, with the explicit aim of preventing industrial hollowing-out.

In 2021, China took deliberate action to deflate its real estate bubble, with the broader goal of redirecting bank lending and financial investment toward the cultivation of new quality productive forces. Complementing this shift, central and local governments have established government-backed venture capital funds targeting high-tech sectors that align with national development priorities but carry elevated market risks that private capital is reluctant to absorb. State financial resources are also deployed to maintain fluctuation boundaries in stock and bond markets, thereby limiting the ability of financial oligarchs to manipulate these markets and extract wealth from ordinary investors. Taken together, these measures have gone a considerable way toward ensuring that China's banking and financial sector serves socialist objectives and the interests of the population as a whole, avoiding the distortions that arise in Western economies where financial monopoly capital exercises dominant influence over the financial system.

The success of the Chinese government's effort to embed market mechanisms into its economic system is reflected in a series of landmark achievements. China became the world's largest holder of foreign exchange reserves in 2006,[33] and its GDP surpassed Japan's in 2010 to make it the world's second-largest economy. It rose to become the world's largest trader of goods in 2013,[34] ranked second globally in trade in services in 2014,[35] and became the world's largest source of outward foreign direct investment in 2020.[36] In 2023, it overtook Japan to become the leading global exporter of automobiles,[37] and by 2025 its goods trade surplus had exceeded one trillion US dollars.[38] Since the 18th National Congress of the CPC, China emerges as a global high-tech power.

In contrast to the former Soviet Union and Eastern European countries, where economic transition entailed the wholesale dismantling of existing state institutions, China pursued a gradualist approach to reform. Rather than sweeping away the structures of the planned era, China allowed its state institutions, regulatory frameworks, and administrative authorities to adapt incrementally to the demands of a market economy and to assume constructive roles within it. The former State Planning Commission is illustrative: renamed the National Development and Reform Commission, it continues to perform core functions including the formulation and implementation of national plans, and the review and approval of major investment projects. Following the restructuring of small and medium-sized SOEs from the mid-1990s onward, the strategically significant enterprises retained under state ownership underwent a process of revitalization. These enterprises now bear primary responsibility for the construction and operation of major national infrastructure — spanning airports, ports, railways, power generation, telecommunications, and energy facilities.

Prior to market reform, national infrastructure projects were the exclusive domain of SOEs. The establishment of a market economy introduced competitive bidding processes, opening project tenders to participation by private and foreign-funded enterprises alike. Within the state apparatus, the career advancement of officials at all levels has been closely tied to the economic performance of their jurisdictions and their track record in delivering public works. Of equal significance is the continuity of political leadership: the CPC has remained in power throughout the reform and opening-up era, its fundamental governing objectives unchanged. What has evolved is the CPC's understanding of China's developmental context — by defining the country's current stage as the primary stage of socialism rather than a fully realized one, the CPC created the ideological space necessary to justify the introduction of market mechanisms and private enterprise.

From the founding of the Soviet Union through to the late 1970s, the contest between Soviet-led socialism and American-led capitalism played out across political, economic, and military dimensions. The scientific and military achievements generated by the Soviet five-year planning system, and the institutional capacity of socialism to concentrate resources toward strategic objectives, placed Western countries on the defensive in geopolitical competition. From 1980, however, the dynamic began to shift. The incoming Reagan administration embraced neoliberalism, rolling back regulatory constraints and injecting new energy into the American economy. The Soviet Union, by contrast, remained locked into a planned economic model that struggled to generate dynamism or deliver the sophisticated consumer goods its citizens increasingly demanded. These accumulated weaknesses ultimately proved decisive, contributing to the Soviet Union's defeat in both geopolitical and economic competition with the West. Since the late 1970s, China has pursued a policy of reform and opening-up that has drawn on the state's considerable executive and organizational capability to progressively integrate market mechanisms into its economic system, while preserving the five-year planning framework and the socialist institutional advantage of concentrating resources toward major national objectives. The outcome of this dual approach is that China has been able to harness the complementary strengths of both socialism and the market economy — a combination that has positioned it favorably in the arena of international political, economic, and military competition. In doing so, it has reinvigorated the global communist movement and ensured that China has not replicated the collapse of socialism that overtook the Soviet Union and the countries of Eastern Europe.

Even as the Chinese government has actively cultivated the expansion of market forces into virtually every corner of economic life, it has simultaneously transcended the limitations of the market by drawing on the socialist five-year planning system and the institutional capacity to concentrate national resources toward major objectives. This has enabled the state to provide public goods and services that markets characteristically fail to deliver — including poverty alleviation programs, healthcare, education, and a broader range of public services directed especially at vulnerable populations — as well as landmark infrastructure undertakings such as the South-to-North Water Diversion Project. Where markets have proven too slow to supply essential infrastructure, the state has stepped in to accelerate the development of transportation networks, logistics systems, and communications infrastructure. Industrial policy has been deployed additionally to incentivize research and development in advanced technology sectors where China seeks capabilities unavailable elsewhere.

**5. Gradually Forge the State Capability for Strategic Development Planning**

Ancient Chinese scholars often said: "Preparation ensures success, and unpreparedness leads to failure."[39] As early as 1955, Chairman Máo pointed out: "Our goal is to catch up with the United States and surpass it. The United States has a population of merely over 100 million, while our country has more than 600 million people; we ought to catch up with the United States... How many decades it will take depends on the efforts of all of us. It will take at least 50 years, perhaps 75 years — seventy-five years equals fifteen Five-Year Plans. Only when we catch up with and surpass the United States can we finally hold our heads high with relief."[40] In August 2020, General Secretary Xí Jìnpíng explicitly stated at a forum of experts on economic and social affairs: "Guiding economic and social development through medium- and long-term planning is an important approach for our Party to govern the country."[41] In April 2025, General Secretary Xí Jìnpíng pointed out at a symposium on economic and social development during the 15th Five-Year Plan period attended by officials from some provinces, autonomous regions and municipalities directly under the Central Government: "Formulating five-year plans in a sound scientific manner and implementing them in a sustained, successive fashion constitutes an important experience of our Party in national governance, as well as an important political strength of socialism with Chinese characteristics."[42] Timothy Geithner, US Treasury Secretary during the Obama administration, stated: "Plan beats no plan."[43] Although development economist Hirschman criticized rigid planning, he pointed out that in the context of development, well-thought-out planning is superior to the unplanned state of laissez-faire.[44] Professor Porter of Harvard University stresses that choosing a clear plan (a distinct strategy) is far better than drifting aimlessly in operational activities.[45]

Markets fail across a wide range of circumstances, and one of their most significant limitations is the inability to supply, in a timely and adequate manner, the public goods that socio-economic development requires. Large-scale infrastructure projects — high-speed railways, expressways, national power grids, power stations spanning hydropower, nuclear, thermal, wind, and photovoltaic generation, and telecommunications networks — cannot reliably be delivered by market forces alone, particularly in remote or less developed regions where commercial returns are insufficient to attract private investment. Markets are equally ill-equipped to provide the social infrastructure that a functioning society depends upon, including primary, secondary, and tertiary education, healthcare, social security, and poverty alleviation services — goods that must reach all members of society, and especially its most vulnerable. Nor, given that markets are fundamentally oriented toward profit, can they be expected to protect the integrity and completeness of a country's industrial system.

Reliance on market forces alone is insufficient to drive the advancement of high technology. Markets tend to be risk-averse with respect to high-tech research and development, and are generally reluctant to direct capital toward ventures whose returns are uncertain or long-deferred. They are equally indifferent to the quality and comprehensiveness of national education systems across primary, secondary, and tertiary levels, or to the scientific rigor of university curricula that form the foundation for cultivating engineers and scientists. Markets will not, of their own accord, establish academies or research institutions to conduct basic scientific inquiry. In the absence of long-term government planning, unchecked market forces will inevitably produce severe imbalances in socio-economic development and chronic undersupply of public goods. No market actor will volunteer to build highways, railways, or new energy infrastructure, operate hospitals and schools in impoverished regions, conduct scientifically necessary but commercially unviable research, or attend to the needs of the poor and marginalized.

The logic of advance planning is familiar at the household level: households budget for children's education, set aside time for parental engagement, and make provisions for retirement, housing, and major life events. A country operates on the same principle, though with incomparably greater complexity, making systematic forward planning all the more indispensable. At the national level, this entails allocating resources to major development initiatives, designating the government bodies responsible for their execution, and identifying the land, labor, and material inputs required. The seasonal rhythms of temperate climates offer another illustration: since winters and springs yield no harvests, people must accumulate sufficient food, clothing, and shelter in advance to survive these lean months — a practice that embodies both planning and preparedness against adversity. The failure to do so has historically meant starvation and exposure for the unprepared. This imperative of forward planning helps explain why organized state formations emerged early across the Northern Temperate Zone. In ancient China, the recurring floods that necessitated large-scale, coordinated flood control efforts predating even the Xià (夏 c. 2070 BC – c. 1600 BC), Shāng (商 c. 1600 BC – 1046 BC), and West Zhōu (西周 1046 BC – 771 BC) dynasties provided a powerful impetus for the development of centralized state power.[46] "Planning" is similar to "strategy", yet more detailed than the latter.

From the early seventeenth century through the end of the Second World War, armed conflict among European states erupted with remarkable regularity, at intervals of roughly three to five years. The imperative of military survival drove European royal houses to invest heavily and consistently in scientific talent, commissioning researchers over extended periods to advance the development of artillery and gunpowder technology. The scientific inquiry that grew out of this military patronage proved consequential far beyond the battlefield: it laid the intellectual and technical foundations of modern natural science and technology, and in doing so helped set the conditions for Europe's Industrial Revolution.[47] Scientific research rarely generates sufficient income to sustain the researchers who conduct it. Beyond governments, few enterprises are willing to fund scientific work whose commercial returns are uncertain or indefinite. It follows that sustained, long-term public funding is a foundational prerequisite for scientific and technological development. The historical underdevelopment of science and technology in pre-modern China can be understood in this light: the Qīng Dynasty (1644-1911) never acknowledged the necessity of financing scientific and technological research, while the Republic of China (1911-1949), beset by incessant warfare, was able to allocate only negligible resources to the same end.

In its early years, the PRC drew on the example of the Soviet Union — its senior socialist partner — by adopting the five-year planning system for national socio-economic development. The Soviet Union had pioneered this approach with its first Five-Year Plan in 1928, centering it on the development of heavy industry. The sustained fiscal investment channeled through successive Five-Year Plans yielded remarkable results: successful atomic and hydrogen bomb tests in 1949 and 1953 respectively, and the launch of Sputnik 1 — the world's first artificial Earth satellite — in 1957. That achievement demonstrated that Soviet scientific and technological capabilities had, in at least some domains, surpassed those of the United States, a revelation that reverberated through American political and academic circles in what became known as the "Sputnik Moment." The lesson drawn from this history is clear: scientific research and technological development cannot sustain themselves without consistent government fiscal support, which establishes them, in functional terms, as public goods.

Since its founding, China has completed fourteen five-year plans in succession, beginning with the First Five-Year Plan launched in 1953, with the fifteenth set to enter implementation in 2026. By the time the Third Five-Year Plan concluded in 1970, sustained and stable fiscal investment had enabled China to successfully develop its "Two Bombs, One Satellite" program and establish itself as a nuclear power. At the close of the 11th Five-Year Plan in 2010, China had overtaken Japan to become the world's second-largest economy and surpassed the United States as the world's leading manufacturing nation. By the conclusion of the 14th Five-Year Plan in 2025, China's GDP had reached approximately five times that of Japan, India, and Germany respectively. Xí Jìnpíng, the General Secretary of the CPC Central Committee, pointed out in his 2026 New Year message: "Science and technology have been deeply integrated with industries, and innovative achievements have emerged in large numbers. The development of large artificial intelligence models is advancing at a rapid pace, new breakthroughs have been made in independent chip research and development, and China has emerged as one of the economies with the fastest-growing innovation capacity."[48] China has additionally established leading global positions across a broad range of strategic sectors, among them power batteries, new energy vehicles, solar panels, large language models, quantum computing, rare earths, electromagnetic catapult technology for aircraft carriers, shipbuilding, high-speed rail, e-commerce, 5G communications, fighter aircraft, unmanned aerial vehicles, the application and manufacture of robotics, and the like. The breadth and depth of these achievements points to a further conclusion: the institutional arrangement of the five-year planning system itself possesses the fundamental attributes of a public good.

The formulation of a Five-Year Plan is a comprehensive, multi-tiered process. All ministries of the central government, together with the functional departments of provincial, municipal, and county-level governments, prepare individual plans detailing the initiatives they intend to pursue within their respective jurisdictions over the coming five years. County-level plans are submitted to prefecture-level authorities for consolidation, then forwarded to the provincial level for further integration, before being finally reported to the central government. These submissions are consolidated by the central department charged with drafting the plan, which evaluates, prioritizes, and where necessary eliminates proposed items before assembling a unified national draft. The draft is then circulated to all government departments, democratic parties, and local governments at every level for deliberation and comment. Feedback is incorporated into a revised draft, which is subsequently submitted to the Central Committee of the CPC for deliberation, amendment, and formal adoption. Once approved, all central departments and local governments translate their respective portions of the plan into action by preparing supporting budgets, designating implementing agencies, securing land and other resources, and developing detailed implementation roadmaps. Progress is monitored through inspections and evaluations conducted by the central authorities as departments and localities execute their assigned responsibilities. Upon the conclusion of each five-year cycle, all central departments and local governments are required to submit progress reports to the central authorities accounting for the completion of the plan components under their purview.

The priorities enshrined in each Five-Year Plan reflect the developmental imperatives of their respective periods. The First and Second Five-Year Plans, for instance, centered on the expansion of heavy industry, with the concrete objective of realizing and fully implementing across China the 156 industrial projects supported by the Soviet Union. The recently concluded 14th Five-Year Plan (2021–2025) addressed a markedly different set of ambitions: "during this period, China's economic strength, scientific and technological capabilities, and overall national power reached new heights, meaningful advances were made in the pursuit of Chinese modernization, and a solid foundation was laid for the next stage of progress toward the Second Centenary Goal."[49] The proposal for the 15th Five-Year Plan recently adopted by the central authorities calls for "strengthening original innovation and tackling key core technologies". It requires that "we improve the new nationwide system, adopt extraordinary measures, and advance decisive breakthroughs in tackling key core technologies in key fields including integrated circuits, industrial machine tools, high-end instruments, basic software, advanced materials and bio-manufacturing across the entire industrial chain."[50]

## 6. Leverage Industrial Policies to Break through the "Market Constraints for High-tech" and Foster "High-tech Industrial Capabilities"

We live in an era in which the most advanced capitalist nations deploy high technology as an instrument of geopolitical pressure against developing countries. Faced with this form of high-tech strangulation, developing nations have little choice but to pursue independent innovation and achieve original breakthroughs — what might be called the leap from zero to one. Yet this leap is an extraordinarily demanding one. Most transformative ideas and technologies originate in university laboratories, research institutes, or corporate R&D divisions; but the path from initial discovery to industrialization, and from industrialization to profitable commercial operation, is long and fraught with uncertainty. The most critical bottleneck along this path is securing investment. The United States, with its highly mature market economy, stands as the world's foremost model of high-tech innovation, home to a dense ecosystem of venture capital firms dedicated to funding early-stage technology enterprises. Venture capital operates on a clear logic: back nascent technologies with outsized profit potential, and when those technologies achieve commercial success — typically marked by a public offering — recover the principal and realize substantial returns. We define the class of new technological products and services that emerge solely through this market-driven investment mechanism as the "market constraints for high-tech."

A defining challenge for developing countries is their lack of competitiveness in high-tech product markets — both domestic and global — which remain largely dominated by advanced economies. This dominance is a direct reflection of developing nations' insufficient high-tech industrial capability. Given the near-absolute monopoly that developed countries hold over high-tech goods and services, it is exceedingly difficult for developing nations to cultivate such capability through market competition alone. Under these circumstances, governments in developing countries must leverage state power to actively build domestic high-tech industrial capacity. This may take several forms: funding qualified enterprises, universities, and research institutions to establish R&D centers targeting core bottleneck technologies; providing state venture capital, preferential tax treatment, and subsidized bank lending to firms manufacturing cutting-edge products and services; and offering consumer subsidies to stimulate demand for frontier high-tech goods. Together, these instruments constitute a coordinated industrial policy capable of doing what markets alone cannot. Once developing countries successfully cultivate high-tech industrial capability through such state support, they become positioned to break through what we have termed the "market constraints for high-tech" — transcending the technological ceilings imposed by market forces and enabling domestic enterprises to deploy newly developed technologies for commercial production and wider distribution. There is, moreover, an instructive irony at work: when advanced capitalist powers forbid high-tech goods and services to be exported to certain developing countries as instruments of geopolitical pressure, they inadvertently create market space within these targeted countries for domestically produced alternatives. State-backed industrial development thus not only builds capability — it generates the very commercial opportunity that allows domestic high-tech enterprises to take root and flourish.

Taking the development of electric vehicles and power batteries in China over the past more than thirty years as a case, we will analyze how China has leveraged industrial policies to boost the growth of the power battery and electric vehicle industries.

### 6.1 The Initial Stage of China's Power Battery R&D in the 1990s

The 1973 oil crisis sent crude oil prices soaring from around $3 to nearly $11 per barrel. This surge not only fueled a boom in small-displacement vehicle production but also pushed many countries to explore alternatives to petroleum as their primary energy source. In 1976, British scientist Stanley Whittingham pioneered research into lithium-based energy storage, laying the theoretical foundation for lithium-ion batteries and developed a rechargeable lithium battery with a voltage exceeding 2 volts. Although the battery suffered from severe safety flaws, and Whittingham is still recognized as the founding father of lithium batteries. In 1980, American scientist John Goodenough developed the lithium cobalt oxide battery, whose voltage reached 4 volts --- double that of Whittingham's original design. In 1985, Japanese scientist Akira Yoshino achieved a technological breakthrough. Building on Goodenough's research, he replaced pure lithium metal with safer lithium ions and invented the lithium-ion battery using carbon-based materials as the negative electrode. This greatly improved the stability of lithium batteries and established the basic framework of modern lithium-ion batteries.[51] In 1991, Sony launched the first commercially available lithium-ion battery, using lithium cobalt oxide as the cathode and carbon-based materials as the anode. Compared with previous generations of battery technology, it offered higher energy density, a longer service life and lighter weight.[52]

Given the rapid progress in global lithium battery research and the swift expansion of Japan's lithium battery industry, Qián Xuésēn wrote a letter to Zōu Jiāhuá, then Vice Premier of the State Council, in August 1992. He proposed that China's automobile industry skip the gasoline vehicle stage and moved directly to electric vehicles, so as to bypass Western patent barriers and achieve a leapfrog catch-up in new energy batteries.[53] To implement Qián Xuésēn's proposal, the State Science and Technology Commission led the launch of the Electric Vehicle Technology Research Program in 1992, investing 15 million yuan to tackle core technologies for new energy electric vehicles, with power batteries as the central research focus.[54] Meanwhile, the State Science and Technology Commission organized industry-university-research collaborations involving institutions such as Fudan University and Tianjin's 18th Research Institute to jointly pursue breakthroughs in power lead-acid battery research, leading to the birth of electric bicycles equipped with power lead-acid batteries in 1992. In 1994, the Ford-China Research and Development Fund sponsored a lithium-ion battery research project submitted by Researcher Chén Lìquán of the Institute of Physics, Chinese Academy of Sciences, with a grant of $30,000.[55] In 1994, the State Science and Technology Commission issued the Plan for the Promotion and Application of Electric Vehicles, which laid out development goals and roadmaps for technologies such as power batteries, and guided enterprises and capital toward battery research and development. It encouraged the early adoption of power batteries in low-threshold sectors such as two- and three-wheeled electric vehicles, using market feedback to drive technological optimization. Subsequently, the State Science and Technology Commission took the lead in setting up an expert group on electric vehicles to coordinate resources across universities, research institutes, automobile manufacturers, and battery producers. For example, Shengli Bus Factory joined forces with Westinghouse Electric Corporation of the United States, the Commission of Science, Technology and Industry for National Defense, and Beijing Institute of Technology to jointly advance integrated research and development of batteries, motors and electronic control systems for electric buses. In 1995, China successfully trial-produced its first trolley-free pure electric bus, the Yuanwang.[56][57] Continuing the energy storage materials (polymer lithium battery) project launched under the Seventh Five-Year Plan,[58] the lithium-ion battery basic research and pilot-scale testing received sustained funding throughout the 1990s under the Eighth Five-Year Plan and the 863 Program (National High Technology Research and Development Program). From 1993 to 1995, the Institute of Physics obtained an emergency grant of 800,000 yuan from its headquarter, the Chinese Academy of Sciences and built China's first pilot production line for lithium-ion batteries. In 1995, China's first Type A lithium-ion battery was developed, resolving key technologies for large-scale manufacturing.[59] The Chinese Academy of Sciences designated solid-state ionics and lithium batteries as key projects across three consecutive Five-Year Plans. In 1993, it coordinated research funding for lithium-ion battery development and connected research teams with industrial capital, facilitating the transition of technologies from laboratories to pilot testing and preliminary industrialization. State-backed research institutes such as the Tianjin Institute of Power Sources undertook long-term R&D on battery materials and manufacturing processes, building up technical expertise and talent reserves for the industry. In 1996, the Ministry of Science and Technology launched the Major Scientific and Technological Industrial Project for Electric Vehicles under the Ninth Five-Year Plan (1996–2000). During this plan period, the project completed the design of electric sedan concept cars, along with trial production and trial operation of converted electric vehicles. A national demonstration zone for electric vehicle operation tests was built between Shàntóu and Nán Ào Island to conduct on-site verification of battery and other related technologies. The prototype vehicles reached a top speed of 110 kilometers per hour, a driving range of 130 kilometers on a single charge, and an acceleration time of 8.5 seconds, matching the advanced standards of equivalent foreign models at the time.[60]

Amid the domestic and global boom in research, development, and manufacturing of power battery technologies, the General Research Institute for Nonferrous Metals of China founded Big Battery Co., Ltd. in Shēnzhèn in October 1993, with Wáng Chuánfú, a senior engineer from the institute, serving as general manager. The company embarked on efforts to commercialize lithium battery research outcomes. In February 1995, Wáng resigned from his post as general manager of Big Battery. He gathered a group of core technical staff as shareholders and established BYD Company Limited in Shēnzhèn. After thousands of rounds of trials, the team finally developed a viable mass-production process for lithium-ion batteries. In 2000, BYD began supplying mobile phone batteries to Motorola, marking an epoch-making milestone in China's lithium battery industry. As a qualified battery supplier for major handset manufacturers including Motorola and Nokia, BYD quickly grew into the world's largest lithium-ion battery producer outside Japan.[61] Around 2000, Japanese manufacturers including Sony, Panasonic, Toshiba, and Sanyo accounted for more than 90 percent of the global lithium-ion battery market.[62] In 1999, Zēng Yùqún, Chen Tanghua and Liang Shaokang resigned from SAE Magnetics, a Hong Kong-based hard-disk read-write-head manufacturer affiliated with Japan's TDK Group, co-founded Amperex Technology Limited (ATL), a company specializing in the production of polymer lithium-ion batteries.[63]

### 6.2 From 2001 to 2008, with Government Support, the Research, Development and Production of Power Batteries in China Entered a Take-off Stage

In 2001, preparations for the Beijing Olympic Games got underway, including a demonstration project for new energy vehicles featuring a fleet of 50 lithium-ion battery-powered electric buses. This project spurred R&D efforts among lithium battery manufacturers such as MGL in cathode materials and battery packs.[64] In addition, in September 2001, the Ministry of Science and Technology launched the Major Special Project for Electric Vehicles under the 863 Program of the Tenth Five-Year Plan. It established the "Three Verticals and Three Horizontals" R&D framework for electric vehicles: the three vertical lines referred to hybrid electric vehicles, battery electric vehicles, and fuel cell vehicles; the three horizontals covered multi-energy powertrain control systems, drive motors and power batteries. This initiative organized and launched large-scale R&D on new energy vehicle technologies in full swing.[65] During the Tenth Five-Year Plan period (2001–2005), the state invested 880 million yuan in the major electric vehicle project under the 863 Program. A group of technically competent automobile manufacturers and universities were brought together to form R&D partnerships and jointly pursue technological research and development along the roadmap outlined above. FAW, Dongfeng, Changan, and Chery focused on hybrid electric vehicles; SAIC Group collaborated with Tongji University to develop fuel cell power systems;[66] Tsinghua University, Beijing Institute of Technology, Foton Motor and Beijing Public Transport Corporation worked together to develop fuel cell and battery electric buses.[67] In 2004, the project designated Beijing, Tianjin, Wuhan and Weihai as the first batch of partner cities to facilitate the industrial application of newly developed new energy vehicle technologies. In 2007, the National Development and Reform Commission (NDRC) issued the Measures for the Administration of Market Access for New Energy Vehicle Products, which set out the thresholds and technical standards for mass-production new energy vehicles.[68] During the Eleventh Five-Year Plan period, the 863 Program's electric vehicle project further expanded in both capital investment and project scale, bringing in more cities and automobile manufacturers.[69]

With the support of CITIC Group in 2000, Professor Qi Lu from Peking University founded Zhongxin Guoan MGL (CITIC Guoan MGL) to develop lithium-ion batteries. In 2001, Qi Lu's research group achieved domestic production of cathode materials for small lithium-ion batteries, breaking China's long-standing reliance on imported battery materials for the first time. In 2003, Qi Lu developed a 100Ah power lithium-ion secondary battery in the laboratory for the first time. In 2004, the Beijing Municipal's Science and Technology Commission decided to install the 400Ah/400V battery pack developed by MGL on public buses for long-term testing, kicking off preparations to ensure electric bus operation for the 2008 Beijing Olympic Games.[70] Besides MGL, Peking University Pioneer jointly with the ATL's Power Battery Division, also secured several orders for Olympic buses.[71] In 2006, lithium-ion batteries for Dell laptops were recalled over defects in Sony-made cells, lithium-ion batteries for Lenovo laptops were also recalled over defects in Sanyo-made cells. In 2007, Nokia recalled batteries made by Panasonic (then Matsushita) for its mobile phones. The three major Japanese battery giants suffered heavy setbacks.[72] BYD made its IPO debut in 2002. In 2003, BYD acquired the former Xi'an Qinchuan Automobile Co., Ltd. and entered automobile manufacturing business. At the 2006 Beijing Auto Show, BYD unveiled its battery electric vehicle model F3e, which boasted a driving range of 300 kilometers. Equipped with the "world's first iron-based power battery ET-POWER", the F3e signaled that lithium iron phosphate batteries were on the verge of commercialization. Although John Goodenough discovered lithium iron phosphate ($LiFePO_4$), a battery material with a polyanionic structure, back in 1997, a long gap remained between this theoretical breakthrough and large-scale mass production with viable real-world applications. Numerous challenges related to material synthesis and modification processes remained unsolved. Liu Fei, an engineer at BYD, substantially improved the lithium battery technology pioneered by Nobel laureate John Goodenough. His refinements enabled mass commercial production of such batteries, securing a strategic upper hand for BYD in lithium battery manufacturing. In 2009, BYD acquired Midea Sanxiang Bus and obtained a bus production license, launching its production of K-series electric buses.[73] BYD launched its first all-electric vehicle, the BYD e6, in 2009, officially opening the door to the new energy vehicle market. The biggest highlight of this model was BYD's self-developed lithium iron phosphate battery, enabling a driving range of up to 300 kilometers.[74] Fifty electric buses equipped with lithium manganate batteries operated flawlessly throughout the 2008 Beijing Olympic Games with zero malfunctions.[75] The first-generation iPhone equipped with polymer lithium-ion batteries manufactured by ATL was launched in 2007. From then on, ATL finally ushered in a prosperous business period because of its polymer lithium-ion batteries, though it also faced fierce competition from South Korea's large conglomerates Samsung and LG. In 2010, ATL joined hands with BAIC Group and cathode material manufacturer Peking University Pioneer to co-found Pride Power through joint investment, launching its provision of power battery solutions for new energy complete vehicles.[76]

### 6.3 2009–2014 Marked the Phase of Large-scale Production of China's Power Batteries and Electric Vehicles Driven by Government Industrial Policies

In 2007, South Korea's LG secured an order from Hyundai Motor of South Korea. In January 2009, LG went on to win the contract to supply power batteries for General Motors' Chevrolet Volt electric vehicle in the United States. GM eventually put the Chevrolet Volt into mass production in 2010. In 2010, Japanese-funded enterprises captured half of the global lithium battery market, while South Korean companies accounted for 30%. In the global production share ranking of power battery manufacturers in 2012, LG of South Korea outperformed Panasonic and other competitors to claim the top spot, and Samsung SDI ranked seventh with a 2.2% market share.[77] To tackle the economic recession triggered by the financial crisis, the Obama administration of the United States launched a large-scale economic stimulus package in 2009, which included an extensive development program injecting billions of US dollars into the country's clean energy industry.[78] A123 established America's first power lithium-ion battery manufacturing plant in 2009, and later built supply partnerships with major automakers including General Motors and Fisker of the United States, as well as BMW of Germany.[79] Nissan and NEC had already established the joint venture AESC in 2007, which mainly engaged in the production of lithium-ion battery cells, modules and packs. At the end of 2010, Nissan launched its first all-electric vehicle, the Leaf.[80]

Against the backdrop of the booming global power battery and new energy vehicle industries, in January 2009, China's Ministry of Finance, Ministry of Science and Technology, National Development and Reform Commission, and Ministry of Industry and Information Technology jointly issued the Circular on Launching Pilot Demonstration and Promotion Work for Energy-Saving and New Energy Vehicles, kicking off the "Ten Cities, Thousand Vehicles" demonstration and promotion pilot program. Under this initiative, roughly 1,000 new energy vehicles were deployed annually in about 10 cities. Over three years, more than 30,000 new energy vehicles were rolled out nationwide.[81] Over the following ten-plus years, the central government injected more than 150 billion yuan of fiscal funds into the new energy vehicle industry.[82] In February 2009, the four aforementioned authorities issued the Interim Measures for the Administration of Financial Subsidies for the Demonstration and Promotion of Energy-Saving and New Energy Vehicles, laying a financial subsidy framework for the rollout of new energy vehicles. In March 2009, the General Office of the State Council released the Plan for the Restructuring and Revitalization of the Automobile Industry (2009–2011). For the first time, the document set forth a target for large-scale development of new energy vehicles: "Achieve mass production and sales of electric vehicles; upgrade existing production capacity to build an annual output capacity of 500,000 new energy vehicles, with sales of new energy vehicles accounting for approximately 5% of total passenger vehicle sales." To support this goal, the government planned to allocate 10 billion yuan and offer extensive low-interest loans to eligible enterprises, aiming to boost investment in new energy vehicles and other clean energy sectors. Furthermore, the Plan mandated the launch of numerous demonstration pilot programs for new energy vehicles. Deployment began with municipal fleets --- including urban public transit buses, sanitation vehicles, postal delivery vans, and taxis --- before gradually expanding to vehicles operated by commercial and private entities.[83] In June 2010, the Ministry of Finance, Ministry of Science and Technology, Ministry of Industry and Information Technology and National Development and Reform Commission jointly issued the Circular on Launching the Pilot Program of Subsidies for Private Purchases of New Energy Vehicles, marking the launch of China's industrial policy offering substantial subsidies for consumers buying new energy vehicles.[84] At that time, enterprises could receive a central government fiscal subsidy of 500,000 yuan for each large battery-electric bus sold, with additional subsidies provided by local governments. In October 2010, the State Council issued the Decision on Accelerating the Cultivation and Development of Strategic Emerging Industries, designating new energy vehicles as one of the seven strategic emerging industries. Plug-in hybrid electric vehicles and battery electric vehicles were identified as the core development orientations, charting a clear path for the advancement of power battery technologies.[85]

Around 2010, the global power battery industry was booming, and the Chinese government rolled out a series of supportive policies for power batteries, triggering an upsurge in lithium battery production across China. Founded in 2006, Hefei Gotion High-Tech seized the opportunity brought by government subsidies under the "Ten Cities, Thousand Vehicles" program. In 2010, it successfully supplied lithium batteries for Hefei Bus Route 18, which became the world's first public transit route fully operated by electric buses.[86]

In 2011, the National Development and Reform Commission and the Ministry of Commerce of China jointly revised the Catalogue for the Guidance of Foreign Investment Industries, stipulating that foreign capital shall not account for more than 50% of enterprises manufacturing key components of electric vehicles. As a Japanese-funded enterprise, ATL was forced to spin off its battery business for separate operation, giving birth to Contemporary Amperex Technology Co., Limited (CATL) in December 2011.[87] On May 28, 2011, the State Intellectual Property Office of China ruled that the core lithium iron phosphate patents filed by patent holders including Hydro-Québec of Canada were invalid, erecting a protective umbrella for the development of China's domestic lithium iron phosphate industrial chain. Since then, China's lithium battery market ushered in its first major boom. A host of lithium battery manufacturers rose rapidly, including Peking University Pioneer Technology, BTR New Energy Materials, Tianjin Stran, Gotion High-Tech, Shēnzhèn Waterma, and Tianjin Lishen (whose shareholder is Tianjin Institute of Power Sources). [88]

In March 2012, the state issued the Circular on Vehicle and Vessel Tax Incentive Policies for Energy-Saving and New Energy Vehicles and Vessels, stipulating that starting from January 1, 2012, vehicle and vessel tax would be halved for energy-saving vehicles and fully exempted for new energy vehicles.[89] In 2012, the Ministry of Finance and the State Taxation Administration further exempted new energy buses from vehicle purchase tax.[90] In June 2012, the State Council issued the Development Plan for the Energy-Saving and New Energy Vehicle Industry (2012–2020). According to the plan, China strived to achieve cumulative production and sales of 500,000 battery electric vehicles and plug-in hybrid electric vehicles by 2015. By 2020, the country aimed to build an annual production capacity of 2 million such vehicles, with cumulative production and sales exceeding 5 million units.[91] The central government allocated a special fund of 4 billion yuan to support implementation of the plan, with a focus on developing new energy vehicle models and their key components.[92] Between 2009 and 2016, the central government provided approximately 12.6 billion yuan (equivalent to around 1.8 billion US dollars) in fiscal subsidies for new energy vehicles.[93]

At the 2012 Beijing Auto Show, the BYD Qin plug-in hybrid sedan and the Denza EV all-electric concept car were exhibited simultaneously as new products showcasing updated technologies.[94] Denza was a joint venture between BYD and Daimler, launched as China's first brand dedicated to new energy vehicles. CATL started supplying power batteries to Brilliance BMW in 2012.[95] In October 2012, A123 Systems of the United States officially filed for Chapter 11 bankruptcy protection, delivering a heavy blow to lithium battery manufacturers.[96] Some lithium battery manufacturers in China withdrew from the market. Nevertheless, EVE Energy, a lithium battery producer, bucked the downward cycle. It formed a joint venture with Saina, a unicorn enterprise specializing in lithium iron phosphate materials, to enter the power battery sector.[97]

After launching the Roadster in 2008, Tesla weathered the U.S. financial crisis and achieved tremendous success with the Model S released in 2012. A large number of consumers grew interested in electric vehicles, drawing massive capital inflows into the new energy vehicle market. Starting in 2013, major Chinese cities rolled out plans to restrict gasoline-powered vehicles and scale up new energy vehicle deployment as part of efforts to control smog pollution. These policies fueled the expansion of the new energy vehicle industry and sparked a new growth cycle for power lithium batteries.[98]

The General Secretary Xí Jìnpíng pointed out in May 2014 that developing new energy vehicles is the only path for China to grow from a major automobile country to a powerful automobile nation.[99] Guiding Opinions of the General Office of the State Council on Accelerating the Promotion and Application of New Energy Vehicles, issued in July 2014, established a relatively comprehensive ecosystem for the development of new energy vehicles from six dimensions: building out charging infrastructure to meet the rising demand for electric vehicles; fostering innovative business models such as car-sharing and time-sharing rental of new energy vehicles; offering subsidies to stimulate electric vehicle consumption; expanding the application of new energy vehicles in public, municipal and corporate fleets; combating local protectionism; and raising consumer awareness. To support the implementation of the Guiding Opinions, the government successively rolled out policies exempting new energy vehicles from vehicle purchase tax,[100] procurement requirements for new energy vehicles by government and public fleets,[101] and incentive measures for charging infrastructure.[102] In 2014, China's electric vehicle market saw its first explosive growth. Major overseas power battery manufacturers---including Panasonic, Samsung SDI and LG Chem---rushed to to break ground on power battery plants in China.[103]

### 6.4 Strong Protection of China's Infant Industry — Domestic Power Batteries by the Chinese Government, 2015–2019

In January 2015, the state further stabilized and adjusted the subsidy policies for new energy vehicles, announcing that the subsidies would be extended until 2020 while signaling a gradual annual reduction in subsidies.[104] In March 2015, the Ministry of Industry and Information Technology (MIIT) officially released the Standard Conditions for the Automotive Power Battery Industry, establishing a "white list" mechanism for power batteries. Only new energy vehicles equipped with power batteries that meet the criteria and are included in the white list catalogue were eligible for new energy vehicle subsidies; vehicles fitted with batteries not on the catalogue could not receive any subsidies. The MIIT issued four batches of enterprise rosters complying with the Standard Conditions, covering a total of 57 battery manufacturers. The list featured well-known domestic power battery enterprises such as CATL, BYD, Gotion High-Tech and Tianjin Lishen. The barriers established by these Standard Conditions prevented foreign manufacturers---including Samsung SDI, Panasonic, LG Chem and South Korea's SKI Innovation---from expanding their businesses in China during the policy period. As a result, China, the world's largest power battery market, was remained effectively reserved for domestic manufacturers from 2015 to 2019.[105] Released in May 2015, Made in China 2025 put forward the strategic goal of "establishing a complete industrial and innovation system covering core components to complete vehicles, and bringing domestic-brand new energy vehicles in line with advanced international standards."[106] In 2015, China's production and sales volume of electric vehicles officially surpassed that of the United States. Since then, China has remained the world's largest market for new energy vehicles to this day.[107] In 2017, CATL claimed the top spot in the global power battery sales ranking for the first time, with BYD ranking third. The combined market share of the two firms reached nearly 30%. Chinese manufacturers including OptimumNano and Gotion High-Tech also secured positions among the world's top ten power battery suppliers.[108] In June 2019, China slashed subsidies for electric vehicles and abolished the white list policy that excluded foreign lithium battery manufacturers. This marked the end of the country's industrial policy designed to nurture its infant industry, demonstrating that Chinese lithium battery makers had gained substantial market competitiveness.[109] South Korean power battery enterprises with advantages in ternary lithium technology returned to China, only to find they could no longer shake the dominant position of Chinese power battery manufacturers.

Around 2015, the Chinese government set out to address the shortage of charging infrastructure, a major barrier constraining consumer uptake of new energy vehicles. The Guidance on the Development of Electric Vehicle Charging Infrastructure and the accompanying Guiding Opinions on Accelerating the Construction of Electric Vehicle Charging Infrastructure, both issued in 2015, together established targets to build 12,000 centralized charging and battery-swapping stations as well as 4.8 million distributed charging piles by 2020. These facilities were designed to cater to the charging demands of five million electric vehicles, alongside a national intercity fast-charging network rollout plan.[110,111] To support these targets and plans, the Ministry of Finance released incentive policies for charging infrastructure construction in 2016, under which each city could apply for a maximum of 120 million RMB to build charging piles.[112] In 2017, China issued the Medium- and Long-Term Development Plan for the Automotive Industry, guiding the gradual transition of the automobile manufacturing sector toward new energy vehicles. The Plan set targets that annual production and sales of new energy vehicles would reach 2 million units by 2020, and for new energy vehicles to account for over 20 percent of total new automobile production and sales by 2025.[113]

In mid-2018, China announced it would phase out foreign ownership caps in the automobile manufacturing industry, starting with special-purpose vehicles and new energy vehicles.[114] Foreign companies manufacturing electric vehicles are now allowed to produce automobiles in China without first establishing joint ventures with Chinese enterprises.[115] Just one month after the release of the new regulations, Tesla signed an agreement to build a Gigafactory in the Lingang Free Trade Zone of Shanghai.[116] Since Tesla entered China's market, it has played a positive catalytic role in the domestic new energy vehicle sector. Notably, the local component sourcing rate of Tesla Shanghai Gigafactory has exceeded 90%, greatly boosting the rapid development of the upstream and downstream industrial chains for China's new energy vehicles.[117] Afterwards, BMW and Great Wall Motors joined forces to produce electric vehicles under the Mini brand in China, through their jointly established venture, Spotlight Automotive.[118] Volkswagen of Germany regards China as its most important market worldwide and planned to invest 4 billion euros in China by 2020, around 40 percent of which would be allocated to research and development related to electrified vehicles,[119] it also plans to locate half of its global annual output of 4 million electric vehicles in China by 2028.[120] Toyota has also seized the opportunities brought by the development of electric vehicles and the transformation of the mobility industry to establish partnerships with BYD, CATL and Didi.[121,122,123]

In 2018, NIO became the first Chinese electric vehicle manufacturer to list on the New York Stock Exchange. Private brands among China's new generation of automakers---including NIO, Li Auto and XPeng---have drawn growing attention from the global automotive industry. Several of these Chinese EV startups have emerged as top fundraisers in global capital markets and are regarded as competitors to Tesla, the leading new energy vehicle manufacturer in the United States.[124]

### 6.5 Forge Ahead After Emerging as the Global Leader in Power Batteries and Electric Vehicles Post-2020

In 2021, the global installed capacity of power batteries reached 296.8 GWh, representing a year-on-year increase of 102%. CATL held an absolute leading position with a market share of 32.6%. LG Energy Solution ranked second with a 20.28% market share, followed by Panasonic at 12.2% and BYD at 8.8%. SK ON and Samsung SDI took the fifth and sixth spots, capturing 5.6% and 4.5% of the market respectively.[125] In 2022, CATL ranked first with a market share of 48.20%, followed by BYD (23.45%) and CALB (6.53%). All other power battery enterprises among the top ten held a market share of less than 5%.[126] In the past, China relied on imports for many components and raw materials used in battery manufacturing. By 2022, China had established the world's most complete and largest power battery industrial chain covering material research and development, battery production, recycling, and equipment support. The four core raw materials — cathode materials, anode materials, electrolyte and separators — have largely achieved domestic self-sufficiency. The domestic localization rate of lithium battery equipment exceeds 90%, while that of equipment for key production processes stands above 80%.[127] In 2023, China's automobile output and sales each exceeded 30 million units for the first time, overtaking Japan to become the world's largest automobile exporter. In 2024, both the output and sales of new energy vehicles surpassed 11 million units, accounting for over 65% of the global market share, ranking first worldwide for nine consecutive years.[128,129]

In 2020, BYD Company Limited launched its Blade Battery, boosting the volumetric utilization rate of traditional lithium iron phosphate (LFP) batteries by over 50% and substantially raising their energy density. Because of a series of technological breakthroughs, the driving range of LFP batteries has reached a level comparable to that of ternary lithium batteries.[130] The ternary lithium magazine battery launched by GAC AION in 2021.[131] The Kirin ternary lithium battery launched by CATL in 2022[132] has achieved a substantial improvement in safety performance. In 2024, CATL launched the Shenxing PLUS lithium iron phosphate battery. Through nano-level precise arrangement, the cathode material achieves a high compaction density standard, successfully overcoming three major industry challenges including driving range, fast charging and safety. This battery can deliver a driving range of over 1,000 kilometers on a single charge.[133,134]

In 2020, Tesla launched the 4680 large cylindrical battery (ternary lithium battery), realizing structural innovation.[135] In 2023, EVE Energy also rolled out large cylindrical batteries. The company has secured orders from BMW and built production plants in Shenyang and Hungary, which are now operational.[136] In November 2020, the State Council issued the New Energy Vehicle Industry Development Plan (2021–2035), which put forward requirements including enhancing the fundamental capacity of the lithium battery industry, accelerating the R&D and industrialization of solid-state power battery technologies, strengthening the development of the recycling system, and promoting the development of the full value chain.[137] In December 2021, the Ministry of Finance, the Ministry of Industry and Information Technology, the Ministry of Science and Technology, and the National Development and Reform Commission jointly issued the Circular on the Fiscal Subsidy Policy for the Promotion and Application of New Energy Vehicles in 2022, further improving the subsidy policies for new energy vehicles.[138] In 2024, the Ministry of Transport and nine other Ministries jointly issued the Several Measures on Accelerating the Improvement of Transportation Services and Safety Support Capacity for Power Lithium-ion Batteries of New Energy Vehicles, proposing to basically eliminate bottlenecks and pain points in the transportation of power lithium-ion batteries by 2027.[139] Starting from 2026, the preferential policy for new energy vehicle purchase tax will be adjusted from "full exemption" to “half levy”.[140] Currently, Japan, the Republic of Korea, the United States, the European Union and other economies are vigorously developing solid-state batteries.[141,142] At present, China's solid-state battery industry is still in a net investment phase, a situation that will persist for some time to come. Judging from the plans of major global enterprises, the research and development of solid-state batteries will be divided into three stages from 2026 to 2030: sample delivery in 2026, small-scale mass production between 2027 and 2028, and large-scale mass production after 2030.[143] The Suggestions of the Central Committee of the CPC on Formulating the 15th Five-Year Plan for National Economic and Social Development proposes to accelerate the development of clusters of strategic emerging industries including new energy, new materials, aerospace and low-altitude economy. It has drawn a blueprint and pointed out the direction for fostering new quality productive forces such as new energy over the next five years.

**7. Does state capability have limits?**

The planned-economy era represents an extreme case of China's exercise of state capability. For the three decades between the founding of the PRC and the launch of reform and opening-up, China operated under a planned-economic system: SOEs dominated the cities, collective agriculture prevailed in the countryside, and the state directed economic activity through administrative fiat. Because output bore little relation to individual effort—whether one worked diligently or idled made scarcely any difference to one's compensation—this system steadily eroded the work incentives of both workers and farmers. Reform and opening-up reversed this logic. Collective agriculture was dismantled, and rural land was contracted to individual households under long-term use rights. The state withdrew from most small and medium-sized enterprises, retaining direct control only over a limited number of large SOEs deemed strategically vital to the national economy and public welfare, which were then restructured under a modern corporate governance system. Private enterprise was permitted to develop, and foreign-invested firms were granted access to the domestic market. Under this socialist market economy, the state no longer commands production directly but instead guides non-public enterprises through industrial policy and economic incentives—reconciling state capability with market mechanisms rather than substituting one for the other. A country’s institutional arrangements should achieve compatibility between state capability and the incentive mechanisms of enterprises and individuals. The former Soviet Union, by contrast, failed to dismantle collective agriculture and divest from small- and medium-sized SOEs in time, and never succeeded in embedding market mechanisms within its planned economy. This failure to reform left it unable to keep pace in economic competition with the West, and the system ultimately collapsed under its own weight.

More than 2,500 years ago, the Chinese philosopher Lao Tzu (老子，571-471 BCE) observed that “governing a large state is like cooking a small fish.”[144] The implication is that state capability, however extensive, ought to be exercised with restraint: policy intervention should not overreach, and the inherent operating logic of society and the economy must be respected rather than overridden. India's experience over the past decade offers a cautionary illustration of what happens when this restraint is abandoned. Reports have repeatedly surfaced of a familiar pattern: India courts foreign-invested enterprises to establish operations, only to subject them to aggressive tax audits and fines once those firms become profitable. This after-the-fact extraction of value—rather than the transparent, predictable rule-setting that investment depends on—has driven a steady procession of foreign firms to withdraw from the Indian market, while prospective investors have grown increasingly wary of entering it at all. The lesson recalls Lao Tzu's warning: a state capable of far-reaching intervention still forfeits the trust essential to sustained economic development if it does not know when to hold back.

There is an old Chinese saying that goes, “however mighty the state, one addicted to war will surely perish.”[145] The United States is a negative example in this regard. During the Clinton administration in the 1990s, the United States had virtually no national debt at all. After the 2001 Afghanistan War, the 2003 Iraq War, the 2011 Libyan War, and the 2026 US–Israel–Iran War, the U.S. national debt has by now reached 40 trillion US dollars, while U.S. GDP in 2025 was only 30 trillion dollars. The United States now spends more each year servicing interest on its national debt than it does on its entire military budget. Since it has not deployed ground troops, the United States and Iran can so far only be said to be evenly matched in this war. The main reason for it is that the U.S. government had run out of money. The United States has overextended its national strength.

## 8. Conclusions

Confucius held that a person's foremost obligation is to cultivate their own capabilities. Amartya Sen, Nobel laureate in Economics, argues along parallel lines: the root cause of poverty lies not in material scarcity alone, but in the limited capabilities of impoverished groups. The same logic scales from the individual to the national: a country's poverty and backwardness are, at their core, a reflection of weak state capability. It follows, therefore, that for underdeveloped nations seeking to advance, the foundational priority is not the pursuit of wealth itself, but the deliberate and sustained construction of robust state capability.

China's robust state capability derives, above all, from the leadership of the CPC. To understand the origins of this capability, one must look to the CPC's own formative history — captured aptly by the celebrated phrase: “The salvo of the October Revolution brought Marxism-Leninism to us.”[146] Lenin's analysis of imperialism — its exploitation, oppression, and enslavement of colonial and semi-colonial peoples — combined with the triumph of the October Revolution, illuminated a path of anti-imperialist and anti-feudal struggle for China in the early 20th century. A cohort of progressive Chinese intellectuals, most prominently Lǐ Dàzhāo (李大钊), Chén Dúxiù (陈独秀), and Máo Zédōng (毛泽东), embraced Marxist-Leninist theory and the revolutionary model forged by the October Revolution. With the guidance and financial support of the Communist International (Comintern), they established the CPC in 1921. Central to Marxism-Leninism is Lenin's theory of party building, grounded in a set of organizational principles known as the Four Principles of Deference: individual Party members defer to Party organizations; the minority defers to the majority; lower-level Party organizations defer to higher-level ones; and all Party members and organizations defer to the National Congress and the Central Committee it elects. Binding these principles together is the doctrine of democratic centralism — a two-stage process in which open deliberation and democratic discussion precede the exercise of centralized authority. Once the CPC Central Committee reaches a decision through this process, the entire Party is bound to accept and implement it without exception.

From a historical vantage point, the 22 years spanning from the Nanchang Uprising and the Autumn Harvest Uprising of 1927 to the founding of the PRC in 1949 were defined by relentless struggle. The CPC endured successive trials — resisting KMT rule, prosecuting the War of Resistance against Japanese Aggression, and waging the Chinese People's War of Liberation — through both open armed conflict and brutal underground operations. These crucibles forged ironclad discipline and formidable organizational strength within the CPC and the people's army under its command. Through this experience, the CPC successfully applied democratic centralism to the socialism construction process in China, while also adopting and refining the Five-Year Planning system from the Soviet Union, thereby cultivating a robust capability for national development planning. From the founding of the PRC in 1949 to the launch of reform and opening-up in 1978, the CPC drew on this state capability — expressed through sound democratic centralism, organizational cohesion, and strict discipline — to build China's heavy industrial base, and establish its scientific research and development capacity.

Following the introduction of reform and opening-up in 1978, the CPC leveraged its strong executive capability to successfully embedded market mechanisms into China's planned economic system, and on that foundation cultivated substantial manufacturing and infrastructure development capacity. Since the 18th National Congress of the CPC in 2012, China has progressively established a regulatory and institutional framework that — for the first time in human history — orients the banking and financial sector toward serving the long-term imperatives of China's socialist development across political, economic, and social dimensions, as well as the well-being of the Chinese people as a whole. Backed by targeted government industrial policies in the post-2012 period, China has broken through high-tech market constraints and built world-class industrial capacity in power batteries and new energy vehicles. Today, China leads the world in the production and consumption of new energy power generation, power batteries, and new energy vehicles — surpassing Europe, the United States, and Japan across all three domains.

Market failures are pervasive across the developing world. Left to their own devices, markets cannot reliably deliver the material infrastructure — transportation, communications, energy, and power — that economic growth requires, nor can they construct, in a timely or comprehensive manner, the full institutional architecture of a modern society: systems of primary, secondary, and tertiary education; scientific research and technological development; healthcare; poverty alleviation; and social security. Beyond infrastructure and institutions, markets alone are ill-equipped to eradicate poverty, design long-term blueprints for economic and social development, or cultivate the high-tech industrial capacity that developing nations urgently need. Socialist China offers an instructive contrast. The CPC, drawing on its formidable organizational and executive capability — that is, its state capability — has treated each of these domains not as commodities to be allocated by the market, but as mega public goods to be deliberately supplied through state action. These include transportation, communications, energy, and power systems; education at every level; scientific research and technological development infrastructure; healthcare and social security schemes; eradicating poverty campaign (2015-2020); banking and financial services oriented toward the public interest; national venture capital; and, critically, high-tech industrial capacity. Where markets fall short, state capability steps in — and in China's case, does so at a scale and with a consistency that few developing nations have managed to replicate.

The CPC has further institutionalized national medium- and long-term planning for economic and social development. In concrete terms, this has involved the construction of comprehensive transportation, communications, energy, power, healthcare, and education systems serving ethnic minority regions in southwest and northwest China, as well as economically underdeveloped areas across the country. China has also formalized a structured mechanism of inter-regional collaboration and paired assistance — linking the more prosperous eastern provinces with the relatively underdeveloped central and western regions — as an enduring feature of its development architecture.Taken together, three interlocking foundations account for China's decades of rapid industrialization and modernization, as well as the explosive growth of its high-tech industries since the 18th National Congress of the CPC in 2012: the ample and sustained provision of mega public goods, made possible by sophisticated state capability; the institutionalization of Five-Year Plans as the central instrument of national development planning; and the vast, dynamic market economy that has taken shape since reform and opening-up. It is the interaction of these three elements — state capability, long-term planning, and market forces — that distinguishes China's development model and underlies its transformative economic trajectory.

The rise of China's new energy vehicle (NEV) and lithium battery industries offers a compelling illustration of how targeted government industrial policy can enable domestic enterprises to overcome technical barriers and achieve sustained growth. China's lithium battery research and development dates to the mid-to-late 1990s. In the period leading up to 2010, the Chinese government — through the 863 Program implemented across the Eighth through Eleventh Five-Year Plans — mobilized capable enterprises, universities, and research institutions to collaborate on the development of new energy and power battery technologies. In 2010, the government issued two landmark policy documents in succession: the Circular on Launching Pilot Work for the Demonstration and Promotion of Energy-Saving and New Energy Vehicles, known as the “Ten Cities, Thousand Vehicles” Program, and the Circular on Launching Subsidies Pilot for Private Purchases of New Energy Vehicles. By channeling fiscal subsidies toward NEV consumption, these measures delivered substantial and direct support to China's nascent NEV manufacturing sector. The “Ten Cities, Thousand Vehicles” Program represented the first genuine inflection point for China's electric vehicle and power battery industries. Although the vehicles covered were predominantly electric buses, sanitation vehicles, postal vehicles, and taxis operated by state-controlled companies, the program generated demand for tens of thousands of NEVs and provided manufacturers and battery enterprises with stable, predictable revenue streams. Equally significant was the signal it sent: the government's unambiguous commitment to the industry attracted a wave of new enterprises and talent into NEV and power battery manufacturing. From this foundation, a competitive industrial landscape — characterized by numerous players vying for position — began to take shape. The rise of China's new energy vehicle (NEV) and lithium battery industries offers a compelling illustration of how targeted government industrial policy can enable domestic enterprises to overcome technical barriers and achieve sustained growth (Tang, 2024). The “Ten Cities, Thousand Vehicles” initiative was guided by a deliberate strategy of avoiding the most competitive terrain and instead cultivating more manageable footholds — what might be described as “steering clear of the main thoroughfare and taking the side lanes.” In this analogy, the “main thoroughfare” represents the private passenger vehicle market for household use, while the “side lanes” denote government-procured assets: public buses, sanitation vehicles, postal vehicles, and fleets operated by state-controlled taxi companies. The logic behind this approach was sound. Electric buses and sanitation vehicles, as government procurement goods, operate within a more forgiving environment — technical faults, when they arise, carry limited public repercussions and can be addressed without triggering broader market consequences. Private electric passenger vehicles, by contrast, are acutely sensitive to public opinion. Any malfunction risks generating severe adverse sentiment, which could deal a crippling blow to an industry still in its infancy. By channeling initial deployment into lower-risk public sector applications, the government shielded the nascent industry from the full force of consumer scrutiny while it matured. Through this carefully calibrated industrial policy, the Chinese government succeeded in constructing a viable initial market for power batteries and new energy vehicles — one that generated real demand, real revenue, and real learning, without exposing the industry prematurely to conditions it was not yet equipped to withstand.

A further turning point came in 2011, when the State Intellectual Property Office of China ruled invalid the core lithium iron phosphate patents held by foreign patentees, including Hydro-Québec of Canada. This ruling effectively erected a protective umbrella over China's domestic lithium iron phosphate industrial chain, shielding it from foreign intellectual property constraints at a critical stage of development. In March 2015, the Ministry of Industry and Information Technology (MIIT) formalized this protective posture by issuing the Specifications for the Automotive Power Battery Industry and compiling an approved white list of qualified domestic power battery manufacturers. Between 2015 and 2019, Chinese electric vehicle manufacturers were required to source power batteries exclusively from domestic suppliers — a measure that decisively blocked Japanese and South Korean lithium battery companies from capturing China's rapidly expanding market. This period of deliberate state protection proved indispensable to the subsequent rise and global dominance of China's power battery sector. Alongside these industrial policy measures, China's sustained expansion of university enrollment since the late 20th century has generated a steady and substantial pipeline of engineering talent for high-tech research and development. The cumulative effect has been transformative: today, many of China's leading high-tech firms are able to employ upward of 100,000 engineers each, dedicating this enormous human capital base to continuous R&D at scale.[147]

The rise of China's new energy vehicle industry is chosen as an illustrative case precisely because of the starkness of its starting point: in the early 1990s, China had virtually no presence in either the NEV or power battery sectors. From that near-zero base, hundreds of privately owned enterprises — among them CATL, BYD, NIO, Xpeng, Li Auto, and EVE Energy — gradually emerged to lead both industries on the global stage. This trajectory is analytically significant: industries shaped by the interplay of government industrial policy and private enterprise tend to develop with considerably greater dynamism and sophistication than those driven by SOEs alone. That the Chinese government has succeeded in nurturing these industries from nothing into the world's largest represents a remarkable demonstration of its capacity to calibrate industrial policy with market forces — intervening decisively where markets fall short, while preserving sufficient competitive pressure to drive innovation and efficiency. The more than three decades of extraordinary growth in China's power battery and NEV industries offer perhaps the clearest available evidence of how a developing country can deploy state capability to help domestic enterprises break through high-tech market barriers and build world-class industrial capacity from the ground up.

Don’t you remember? For roughly a decade after the signing of the Copenhagen Accord at the 2009 Copenhagen Conference, European and American countries chanted slogans of “low carbon and emission reduction” every single day. Back in August 2021, Academician Dīng Zhònglǐ pointed out in the CCTV program Lǔ Jiàn’s Interview: “These Western countries are just making empty pledges. Do you really think they will deliver on emission cuts? Let’s wait and see. China is taking solid, genuine action on this front.” As Academician Dīng predicted, right after Donald Trump took the oath of office on January 20, 2025, the United States announced its withdrawal from the Paris Agreement for a second time, [148]scrapped the green energy policies rolled out by the Biden administration, and revoked mandatory regulations governing electric vehicles.[149] According to reports from Reuters, Agence France-Presse, Germany's Handelsblatt and other media in 2025, the European Commission officially announced in Brussels on December 16 that it would withdraw the original plan to completely ban the sale of new fuel-powered vehicles starting in 2035. In contrast, China, the world’s largest developing country, generated 3,500 terawatt-hours of electricity from clean energy in 2024, accounting for 35 percent of its total power output. The combined power generation from wind and solar energy hit 1,800 terawatt-hours in the same year, making up 19 percent of total electricity production.[150] Meanwhile, China ranks as the world’s largest producer and consumer of power batteries and new energy vehicles.

The international political and economic landscape is undergoing transformations of historic magnitude — a great restructuring unseen in a century. In this environment, only nations equipped with formidable state capability can respond swiftly and decisively to shifting global conditions, securing for themselves an enduring and unassailable position in the emerging world order. Nowhere is this competition more visible than in the intensifying rivalry between China and the United States across a broad front of strategic high-tech sectors: chip manufacturing, lithography machine development, large language model research, quantum computing, biopharmaceuticals, and advanced materials. The stakes of this contest are civilizational in scope, and its outcome will shape the global order for generations. China's own trajectory offers a compelling frame for assessing what lies ahead. It took approximately 30 years — from the founding of the CPC in 1921 to the establishment of the PRC in 1949 — to win political power. Over the following three decades, from 1949 to the launch of reform and opening-up in 1978, China successfully tested its first atomic bomb in 1964 and hydrogen bomb in 1967, and launched its first artificial satellite in 1970, cementing its status as a nuclear power and asserting its place among the great nations of the world. A further span of roughly 30 years — encompassing six Five-Year Plans, from the Sixth through the Eleventh — elapsed between the start of reform and opening-up and 2010, when China overtook Japan to become the world's second-largest economy. This rhythm of roughly 30-year developmental arcs — each marked by a qualitative leap in national capability — suggests a coherent pattern. Extrapolating from China's sustained economic momentum since reform and opening-up, and grounded in its robust state capability, expanding scientific and technological research capacity, and rapidly maturing high-tech industrial base, it is projected that within approximately 30 years from 2010 — another six Five-Year Plans — China will, by around 2040, match or surpass the United States across all of the aforementioned high-tech domains and less charted fields of high-tech innovations.[151]

The state capability must be balanced against market incentives and restraint. China's planned economy (1949–1978) demonstrated that centralized control without individual incentives erodes productivity, while post-reform China succeeded by combining state guidance with market mechanisms—contracting land to households, privatizing small enterprises, and retaining strategic SOEs. By exercising state capability in a calibrated manner, China has achieved industrialization and emerged as a global high-tech power over the more than four decades since reform and opening-up. The Soviet Union's failure to enact similar reforms contributed to its collapse. Lao Tzu's warning against overreach is illustrated by India's unpredictable taxation of foreign investors, which undermines trust and deters investment. Finally, the United States exemplifies fiscal overextension: repeated wars have driven national debt to unsustainable levels, exceeding GDP, weakening state capability through self-inflicted exhaustion.

Note

[1] The Analects · On Benevolence. (《论语・里仁》)
[2] Sen, 1979.
[3] Sen, 1999, p.3 and 18
[4] Sen, 1999, p.68
[5] Weber, 1919, p. 77-128; 1922/1978, p. 956-1005 and 1028-1031
[6] Skocpol,1985, p. 1-24
[7] Migdal, 1988, p. 4-5, 19-22 and 23-26.
[8] Wang and Hu,1993. p. 1–10, 23.
[9] Fukuyama, 2004, p. 12–14 and 35–42.
[10] Xí Jìnpíng, 2014. (“我国社会主义制度能够集中力量办大事是我们成就事业的重要法宝。我国很多重大科技成果都是依靠这个法宝搞出来的，千万不能丢了！”)
[11] Xí Jìnpíng, 2021.
[12] After the First Opium War, Wei Yuan advocated “learning the advanced techniques of foreign barbarians to resist them” (“师夷之长技以制夷”)(Wei, 1842).
[13] Following the defeat in the Second Opium War (1856–1860) and the signing of the Treaty of Peking, Feng Guifen finished his work Protests from the Jiaobin Studio in 1861, in which he argued: “We shall take China’s traditional ethical norms and moral teachings as the foundation, supplemented by the methods through which various foreign nations achieve prosperity and strength.” (“以中国之伦常名教委原本,辅之以诸国富强之术”) (Feng, 1861).
[14] Since the Meiji Restoration in 1868, Japan also learned from European countries and achieved industrialization. As a latecomer in industrialization, Japan also followed the bad habit of European countries of occupying colonies on a large scale, and attempted to occupy Korea and China as colonies. In the Sino-Japanese naval battle in 1894, Japan defeated the navy of the Qīng Dynasty in China. In 1895, the Qīng government was forced to sign the humiliating “Treaty of Shimonoseki” with Japan, ceding the Liaodong Peninsula, Taiwan Island and all its affiliated islands (including the Diaoyu Islands), and the Penghu Islands to Japan. The defeat in the Sino-Japanese War of 1894-1895 also led to the bankruptcy of the Westernization Movement of the Qīng Dynasty, and China's first process of industrialization miscarried.
[15] From 1895 to 1898, Kāng Yǒuwéipointed out in his memorials to the Qīng Emperor: “Looking at all nations across the globe, those that reform grow strong, while those clinging to old traditions perish.” (“观大地诸国，皆以变法而强，守旧而亡。”) “Observing the trends of all nations: those capable of reform survive; those that refuse to reform are doomed to fall; thorough reform brings prosperity, while partial reform still leads to ruin.” (“观万国之势， 能变则全，不变则亡，全变则强，小变仍亡。”)( Kang, 1898, p. 212) “I have examined the root of China’s decline and weakness; the accumulation of countless maladies all stems from the estrangement between the ruling system and the people.” (“尝考中国败弱之由，百弊丛积，皆由体制尊隔之故”。) (Kang, 1898, p. 219). Liáng Qǐchāo meanwhile argued: “The foundation of reform lies in cultivating talented people; cultivating talent lies in establishing schools; establishing schools lies in reforming the imperial examination system; and the ultimate prerequisite for full success in all undertakings is to restructure the official bureaucracy.” (“变法之本，在育人才，人兴，在开学校；学校之立，在变科举；而一切要其大成，在变官制”。) (Liang, 1896, p. 13).
[16] In the wake of China’s defeat in the First Sino-Japanese War, Sūn Yat-sen drew the following conclusion: “The Qīng Dynasty may be likened to a building on the verge of collapse, its entire framework fundamentally and

thoroughly rotten." ("满清王朝可以比作一座即将倒塌的房屋，整个结构已从根本上彻底地腐朽了"). "Any support for the tottering Qīng royal house is doomed to failure." ("对目前摇摇欲坠的满清王室的支持，那么注定是要失败的"). "A new, enlightened and progressive government must replace the old one ... and the obsolete Qīng monarchical system shall be transformed into the 'Republic of China.'" ("必须以一个新的、开明的、进步的政府来代替旧政府，... ...，把过时的满清君主政体改变为 '中华民国'。" ). (Sūn Yat-sen, 1904, p. 67-68).

[17] During the Hundred Days' Reform, Kāng Yǒuwéiargued that the fundamental path for China to achieve prosperity and strength through institutional reform lay in "establishing a parliament to communicate the sentiments of the common people." ("设议院以通下情。") (Kang, 1898, p. 150.). Kāng Yǒuwéimaintained that "Confucius is truly the spiritual leader of China." ("孔子实为中国之教主"). Centered on morality, Confucianism rests on human ethics rather than divine theology, making Confucius "an unparalleled spiritual leader of a truly civilized age, unmatched anywhere under heaven." ("真文明世之教主，大地所无也"). Therefore, he held that upholding Confucianism and pursuing political reform were not contradictory: "Politics and religion shall each occupy their own sphere, advancing side by side like two wheels; they run parallel without conflict and reinforce one another to achieve mutual success. The nation's power will expand, and the sacred Confucian teachings will flourish day by day. How greatly this will aid in honoring doctrines, encouraging learning, rectifying fallacies, and reforming social customs!" ("政教各立，双轮并驰，即并行而不悖，亦相互而相成。国势可张，圣教日盛，其于敬教劝学，匡谬正俗，岂少补哉?") (Kang, 1898, p. 281–283). After the fall of the Qīng Dynasty, Kāng Yǒuwéiremained a monarchist who opposed republicanism and continuously plotted to restore Puyi (溥仪) to the throne. In 1917, Kāng Yǒuwéicollaborated with Zhang Xun (张勋) to stage a restoration coup, reinstalling Puyi as emperor. The uprising was soon crushed by Duan Qirui (段祺瑞), Premier of the Beiyang Government, ending in total failure.

[18] Hu Shi argued that compared with Western civilization, Chinese civilization "lags behind in everything: not only in material machinery and political institutions, but also in morality, knowledge, literature, music, art, and physical fitness." ("百事不如人，不但物质机械上不如人，不但政治制度不如人，并且道德不如人，知识不如人，文学不如人，音乐不如人，艺术不如人，身体不如人") (Hu, 1930, p. 1-16). In his 1929 English essay Conflict of Cultures in China (published in the China Christian Yearbook), Hu Shi put forward the slogans of "Wholesale Westernization" ("全盘西化") and "Wholehearted Modernization" ("全力现代化"). (Hu, 1929).

[19] Lenin, 2015.

[20] Li, 2023.

[21] Máo Zédōng, 1949.

[22] In 1931, Japanese imperialism occupied China's three northeastern provinces. In 1937, it staged the Marco Polo Bridge Incident, launching an all-out war of aggression against China thereafter. This marked the second time Japanese imperialism disrupted China's industrialization drive led by the Kuomintang. Japanese imperialism would not allow semi-colonial old China to attain political and economic independence. World War II was a war fought among imperialist powers, as well as between imperialist powers on one side and the Soviet Union and semi-colonial countries on the other. In 1941, Japan launched a surprise attack on the U.S. naval base at Pearl Harbor in Hawaii, prompting the United States to declare war on Japan. In the end, the fascist powers of Germany, Italy and Japan, which sought to redivide the world map, were defeated by the Allied Powers headed by the United States and the Soviet Union, with China as a member. Pursuant to the Yalta Agreements, the Soviet Union sent troops into Northeast China in August 1945 and drove out the Japanese invaders from the region. Meanwhile, the Eighth Route Army and the New Fourth Army led by the CPC, together with Kuomintang troops, jointly expelled Japanese invaders from Chinese territory.

[23] Ravallion and Chen, 2005.

[24] Polanyi, 2001/1944, p. 16.

[25] Zhang and Guo, 2022.

[26] Ren and Li, 2024.

[27] Yang and Guo, 2022.

[28] Xie and Kuang, 2021.

[29] Liu, 2025.

[30] To address the shortage of urban employment, the government called on educated urban youths to go up to the mountains and down to the countryside. This policy was mainly implemented between 1968 and 1977, with a total of approximately 17 million educated youths dispatched to rural areas nationwide. The send-down policy for educated youths was abolished in 1978, triggering a large-scale return of educated youths to cities.

[31] Wen, 2016, p. 201-209.

[32] State Administration of Financial Regulation, Ministry of Industry and Information Technology, National Development and Reform Commission, 2024.

[33] China forex reserves become the world's biggest (Reuters) https://www.chinadaily.com.cn/business/2006-03/28/content_554267.htm
[34] Li Jiabao (2014). China now the world's top trader. https://www.chinadaily.com.cn/bizchina/2014-03/02/content_17316163.htm
[35] http://cn.chinagate.cn/reports/2015-05/15/content_35581288.htm
[36] Wang Wenbo (2021). In 2020, China's outward foreign direct investment (OFDI) flow ranked first in the world for the first time. Economic Reference News. September 30th, 2021. https://www.jjckb.cn/2021-09/30/c_1310218724.htm (王文博.(2021). 2020 年我国对外直接投资 流量规模首次位居全球第一. 经济参考报. 2021-09-30)
[37] http://www.china.com.cn/txt/2024-02/01/content_116978766.shtml
[38] https://www.bbc.com/news/articles/c9wx1v84rzyo
[39] The Doctrine of the Mean (Book of Rites). (《礼记・中庸》)
[40] Mao Ze Dong, 1955, p. 98-109.
[41] Xí Jìnpíng, 2021.
[42] Xí Jìnpíng, 2025.
[43] Geithner, 2014, p. 280-290.
[44] Hirschman, 1958.
[45] Porter, 1996, p. 61-78.
[46] Wittfolfel, 1957; Xia and Tang, 2022.
[47] Wen, 2022, p. 341–482.
[48] Xí Jìnpíng, 2025.
[49] Jiang, 2025.
[50] The 20th Central Committee of CPC, 2025.
[51] https://www.nobelprize.org/prizes/chemistry/2019/popular-information/
[52] Liu, Yu, 2022.
[53] Qian, Xuesen, 1992.
[54] In 1992, Qian Xuesen put forward a forward-looking proposal to the central authorities: China should skip the development path of fuel vehicles and directly develop new energy vehicles. (1992 年钱学森向中央提出前瞻性建议：跳过燃油车，发展新能源汽车 https://history.sohu.com/a/950836902_121161321)
[55] China's Lithium Batteries: Thirty Years of Turbulent Growth — Dedicated to All Veterans of the Lithium Battery Industry. September 5, 2024. (中国锂电池，激荡 30 年——敬献锂电老兵们。2024-09-05。https://baijiahao.baidu.com/s?id=1809336760135340342&wfr=spider&for=pc)
[56] Zhang, Xiyong, 2023.
[57] Two Decades of Tumultuous Progress: The Development History of Power Batteries. July 30, 2018. (《激荡 20 年动力电池发展史》。2018-07-30。https://www.glpoly.com.cn/article_2061.html)
[58] Han Yangmei & Liu Runan, 2024. According to the article, China launched the energy storage materials (polymer lithium battery) project under the Seventh Five-Year Plan of the National High Technology Research and Development Program (863 Program) in 1987. Chen Liquan served as the chief person in charge, with 12 research groups under his leadership.
[59] Yang Lu & Zhang Congzhi. (2025).
[60] Zhang Xiyong, 2023.
[61] A Review of BYD’s Power Battery Development Journey. September 28, 2018.（比亚迪动力电池发展历程回顾。2018-09-28。http://www.cbea.com/ldc/201809/798697.html）
[62] Liu Yu, 2022.
[63] Who Is ATL? The Global Leader in 3C Batteries. May 28, 2025. (谁是 ATL？3C 电池全球霸主。2025-05-28。https://baijiahao.baidu.com/s?id=1833359519159843533&wfr=spider&for=pc)
[64] Xiao Yisi, Wu Ziye, Tang Liuyang, 2023.
[65] Zhang Xiyong, 2023.
[66] China Association of Automobile Manufacturers (2004).
[67] State Key Laboratory of Automotive Safety and Energy, Tsinghua University. (2002).
[68] National Development and Reform Commission. (2007).
[69] Ministry of Science and Technology. (2007).
[70] Tang Liuyang. (2024).
[71] Yang Lu, Zhang Congzhi. (2025).

[72] Darlin, Damon (2006).
[73] A Review of BYD Power Battery Development History. September 28, 2018. (比亚迪动力电池发展历程回顾。2018-09-28。http://www.cbea.com/ldc/201809/798697.html)
[74] Has the development of global new energy vehicles spanned half a century longer than gasoline vehicles? A Review of 100 Years of New Energy Vehicles. January 4, 2023. (全球新能源汽车发展历程，比燃油车早半个世纪？回顾新能源 100 年。2023-01-04。https://mp.ofweek.com/nev/a456714076387)
[75] Xiao Yisi, Wu Ziye, Tang Liuyang. (2023)
[76] Qīng Lan. (2024).
[77] Liu Yu. 2022. LG Energy Solution embarked on lithium-ion battery development in 1995. It is the world's first company to realize mass production of ternary cathode materials. In 2007, LG Energy Solution rolled out the world's first mass-produced small-sized NCM523 batteries; in 2014, it achieved the world's first mass production of NCM811 batteries.
[78] Ball, J. (2019).
[79] A Brief History of Automotive Power Batteries. EV Observer. December 23, 2019. (车用动力电池简史。电动车观察家。2019 年 12 月 23 日。https://www.eeworld.com.cn/qcdz/hisic483678.html)
[80] The March of Power Batteries. July 28, 2020. (动力电池进击史。2020-07-28 https://m.thepaper.cn/baijiahao_8469504)
[81] Ministry of Finance. (2009).
[82] Yang Lu, Zhang Congzhi. (2025).
[83] State Council. (2009).
[84] Yang Lu, Zhang Congzhi. (2025).
[85] State Council. (2010).
[86] Yang Lu, Zhang Congzhi. (2025).
[87] Wang Linlin. 2022.
[88] Qīng Lan. (2024).
[89] Ministry of Finance. (2012a).
[90] Ministry of Finance. (2012b).
[91] State Council. (2012).
[92] ChinaNews. (2013).
[93] MIIT. (2019a).
[94] Qīng Lan. (2024).
[95] The Advance History of Power Batteries. July 28, 2020. (动力电池进击史。2020-07-28。https://m.thepaper.cn/baijiahao_8469504)
[96] Ibid.
[97] Qīng Lan. (2024).
[98] Jin Lingzhi, He Hui, Cui Hongyang, Nic Lutsey, Wu Chuqi, Chu Yidan, Zhu Jin, Xiong Ying, Liu Qian. (2021).
[99] Xí Jìnpíng. (2014).
[100] Ministry of Finance. (2014a).
[101] National Government Offices Administration. (2014).
[102] Ministry of Finance. (2014b).
[103] Xiao Yisi, Wu Ziye, Tang Liuyang. 2023.
[104] Ministry of Finance. (2015).
[105] Jin Lingzhi, He Hui, Cui Hongyang, Nic Lutsey, Wu Chuqi, Chu Yidan, Zhu Jin, Xiong Ying, Liu Qian. (2021).
[106] State Council of the PRC. (2015a). Made in China 2025 set clear technical targets for power batteries: by 2020, the cell-level energy density of power batteries should reach 300 Wh/kg, with cell cost falling below 1 RMB/Wh, system cost under 1.3 RMB/Wh, and a service life of 10 years. By 2025, the cell-level energy density of power batteries is required to exceed 400 Wh/kg, cell cost to drop to 0.8 RMB/Wh, system cost to decline to 1 RMB/Wh, while retaining a 10-year service life. Dong, J. P., Lai, X., Tang, H., & Yang, R. D. (2018).
[107] Qīng Lan. (2024).
[108] Ibid.
[109] Ibid.
[110] National Development and Reform Commission, National Energy Administration, Ministry of Industry and Information Technology, Ministry of Housing and Urban-Rural Development. (2015).
[111] State Council of the PRC. (2015b).
[112] Ministry of Finance. (2016).

[113] Ministry of Industry and Information Technology. (2017b).
[114] Ministry of Commerce. (2018).
[115] Foreign ownership caps (which required foreign automakers to form joint ventures with Chinese enterprises to invest and build factories in China, with foreign shareholdings capped at 50%; additionally, no more than two joint ventures could be established in China for the same category of complete vehicle products) were introduced in 1994 and remained in force until their abolition. Historically, these caps served as a protective umbrella for domestic Chinese automakers. Lifting such restrictions demonstrates the Chinese government's confidence in the technological capabilities of local manufacturers and unlocks greater investment potential from the global market.
[116] Ding, T. (2019, January 7).
[117] The Development History of Global New Energy Vehicles: Did They Emerge Half a Century Earlier Than Fuel Vehicles? A Review of 100 Years of New Energy Vehicles. January 4, 2023. (全球新能源汽车发展历程，比燃油车早半个世纪？回顾新能源 100 年。2023-01-04。https://mp.ofweek.com/nev/a456714076387)
[118] BMW. (2018, July 10).
[119] Volkswagen. (2019, November 21).
[120] Witter, F. (2018).
[121] Toyota. (2019a, July 25).
[122] Toyota. (2019b, July 17).
[123] Toyota. (2019c, November 7).
[124] Jin Lingzhi, He Hui, Cui Hongyang, Nic Lutsey, Wu Chuqi, Chu Yidan, Zhu Jin, Xiong Ying, Liu Qian. (2021).
[125] Did the History of Global New Energy Vehicles Start Half a Century Earlier Than That of Gasoline Cars? A Century-long Retrospect of New Energy Vehicles. January 4th, 2023. (全球新能源汽车发展历程，比燃油车早半个世纪？回顾新能源 100 年。2023-01-04。https://mp.ofweek.com/nev/a456714076387)
[126] Xiao Yisi, Wu Ziye, Tang Liuyang. 2023.
[127] Yang Zhongyang, Liu Jin. (2022).
[128] Peng Bo. (2026).
[129] Ren Ping. (2024).
[130] The competition over power battery technology routes remains intense. Securities Daily, April 8, 2025. (动力电池技术路线之争仍然胶着。《证券日报》2025-04-08。
https://gxt.fujian.gov.cn/zwgk/xw/hydt/xydt/202504/t20250408_6836067.htm)
[131] "Magazine Battery" vs. "Blade Battery": GAC AION Rescues Ternary Lithium Batteries. March 13, 2021. ("弹匣"迎战"刀片"，广汽埃安"拯救"三元锂。2021-03-13。
https://k.sina.com.cn/article_1497675775_5944b7ff001011a59.html)
[132] CATL Qilin Battery: From Zeekr Sedans to Xiaomi Hypercars, an All-Scenario Performance Disruptor — Uncovering the Technical Breakthroughs of the Qilin Battery Deployed on Six Flagship Models
Zhineng Auto, March 8, 2025. (宁德时代麒麟电池：从极氪轿跑到小米超跑，全场景性能颠覆者——揭秘麒麟电池在六大旗舰车型上的技术突破。芝能汽车。2025 年 03 月 08 日。https://finance.sina.com.cn/cj/2025-03-08/doc-inenwxxs9101272.shtml)
[133] The competition over power battery technology routes remains intense. Securities Daily, April 8, 2025. (动力电池技术路线之争仍然胶着。《证券日报》2025-04-08。
https://gxt.fujian.gov.cn/zwgk/xw/hydt/xydt/202504/t20250408_6836067.htm)
[134] You Xiaoying. (2024).
[135] 4680 Large Cylindrical Cells: A Potential Direction for High-End Lithium Batteries with Accelerated Industrialization. April 21, 2022. (4680 大圆柱：高端锂电池潜在方向，产业化发展加速。2022 年 4 月 21 日。https://guba.sina.com.cn/?s=thread&bid=21046&tid=97053)
[136] Eve Energy: Construction of Hungary Battery Plant Starts, Large Cylindrical Batteries to Supply BMW
The Paper, November 28, 2023, Official Account of The Paper. (亿纬锂能：匈牙利电池工厂已开工建设，大圆柱电池将供货宝马。澎湃新闻。2023-11-28。澎湃新闻官方账号。)
[137] Boston Consulting Group & China Society of Automotive Engineers. (2024).
[138] Analysis on the Development Trend of Power Batteries for New Energy Electric Vehicles. February 4, 2024. (新能源电动汽车动力电池发展趋势分析。2024-02-04。https://elbzl.com/h-nd-303.html)
[139] Lin Hongmei, Yuan Bo, Hu Xu. 2025.
[140] Li Zhengguang. (2025).
[141] You Xiaoying. (2024). According to the article, BMW has partnered with the U.S. firm Solid Power on solid-state batteries and plans to launch its first demonstration vehicle equipped with this technology by 2025. In January,

PowerCo SE, Volkswagen's battery subsidiary, also announced that its partner QuantumScape, headquartered in California, USA, had successfully developed a solid-state battery capable of retaining 95% of its original capacity after 1,000 charging cycles. The Financial Times reported that Toyota announced last year a "breakthrough" in its solid-state battery research and development. The company aims to halve the volume, cost and weight of its solid-state batteries without compromising performance. Toyota has set up a joint venture with Panasonic to focus on this technology and targets mass production by 2027. South Korea's Samsung SDI has built a pilot production line for solid-state batteries and also plans to achieve mass production by 2027.

[142] Yang Zhongyang, Liu Jin. (2022).

[143] Li Wenshan, Li Ting. 2025.

[144] Lao Tzu, Dao De Jing (Tao Te Ching). Chapter 60. ( "治大国，若烹小鲜。" )

[145] Sima Fa（The Methods of the Sima）, Chapter One: "Benevolence as the Root."《司马法・仁本第一》, "故国虽大，好战必亡。"

[146] Máo Zédōng, 1949.

[147] Reuters, May 31st, 2025, "Jamie Dimon says China not scared of U.S., cites 100,000 engineers for problem-solving". In the above paper Jamie Dimon said: "I just got back from China last week. They're not scared. There's a notion they're gonna come bow to America. I wouldn't count on that. And when they have a problem, they put 100,000 engineers on it and they've been preparing for this for years."

[148] https://content-static.cctvnews.cctv.com/snow-book/index.html?item_id=16105516189184692424

[149] Zhou, 2025.

[150] https://baijiahao.baidu.com/s?id=1852896325610911235&wfr=spider&for=pc)

[151] Xia, 2025.

# Forecasting China--the US Economic and Technological Competition

## 1. Introduction

As the old Chinese saying goes, "Thirty years east of the river, thirty years west of the river"[1] — meaning that fortune and circumstance shift dramatically over roughly three-decade cycles. Major global events and shifts in international relations have, broadly speaking, followed a similar rhythm across the twentieth century and into the present.

The major international and domestic events examined in this chapter are viewed primarily from a Chinese perspective, and include: the breakout of World War I in 1914, the end of World War II in 1945, the October Revolution in Russia in 1917; the founding of the Communist Party of China (CPC) in 1921; the establishment of the People's Republic of China (PRC) in 1949; China's entry into the Korean War in 1950; the breakdown of Sino-Soviet relations around 1960; China's reform and opening-up beginning in 1978; the establishment of diplomatic relations between China and the US in 1979; the dissolution of the Soviet Union in 1991; the US financial crisis in 2008; China's surpassing of Japan as the world's second largest economy in 2010; the Russian invasion of Ukraine in 2022.

## 2. The Thirty-Year Cycle in Global and China's Events since the Twentieth Century

### 2.1 The Thirty-Year Cycle in Global Events

Roughly thirty years separate the outbreak of World War I in 1914 from the end of World War II in 1945. Three years into WWI, in 1917, the Soviet Union — the world's first socialist state — was founded. Four years after the end of World War II, in 1949, the PRC, another major socialist power, came into being. The interval between these two founding events was likewise around thirty years. A further thirty years passed between 1950, when China and the US fought opposing sides in Korea war, and 1979, when the two countries formally established diplomatic relations. Having failed to win the Korean War in the 1950s and then become mired in the Vietnam through the late 1960s, the United States saw its position gradually erode from the peak of relative power it had had held at the end of World War II.[2] The Soviet Union, meanwhile, tested its first atomic bomb in 1949 and its first hydrogen bomb in 1953, then launched the world's first artificial satellite, Sputnik 1, in 1957 — an event widely regarded, at the time and since, as demonstrating Soviet technological parity with, or advantage over, the United States, and often remembered as the "Sputnik moment." Capitalizing on this relative shift, the Soviet Union pursued strategic expansion on a global scale throughout the 1960s and 1970s,[3] producing a broader dynamic in which the Soviet Union appeared to hold the initiative while the United States remained largely on the defensive. China, for its part, tested its first atomic bomb in 1964 and its first hydrogen bomb in 1967, becoming a nuclear power in its own right. By this point, China's principal external security concern had shifted to the Soviet Union, reflected in the breakdown of Sino-Soviet relations around 1960 and the armed border clashes of 1969. This shift created an opening for Washington, which sought to draw China into a loose strategic alignment against Moscow — laying the groundwork for the broader thaw in Sino-US relations. President Richard Nixon's visit to China in 1972 and the signing of the three Sino-US joint communiqués followed, culminating in the formal establishment of diplomatic relations in 1979.

A further thirty-year span connects the 1960 Sino-Soviet split to the Soviet collapse in 1991. The break with China weakened the Soviet Union's geopolitical position, and once Sino-US relations eased in the 1970s, the United States was able to abandon a two-front strategic posture concentrate its efforts on the Soviet Union alone.[4] During the wave of global market liberalization beginning in the late 1970s, the Soviet Union — under continued Communist Party leadership — failed to develop either a functioning market economy or an international competitive manufacturing base, leaving it unable to supply its population with consumer goods comparable to those available in Western market economies. The main reason for the disintegration of the Soviet Union was that its leaders voluntarily gave up the Communist Party's leadership of the country.[5]

Another roughly thirty-year interval runs from 1981, when Ronald Reagan came to power and began implementing neoliberal economic policy, to 2008, US financial crisis. The American financial regulatory system had, since the 1930s, rested substantially on lessons drawn from the Great Depression of 1929–1933, embodied above all in the Glass-Steagall Act (officially The Banking Act of 1933), which separated commercial banking from investment banking and barred commercial banks from using depositor funds to purchase securities. This framework helped sustain financial stability for decades. Beginning in 1981, however, successive US administrations gradually loosened financial-sector regulation, culminating in the Clinton administration's repeal of the Glass-Steagall Act in 1999 — a deregulatory trajectory that many economists regarded as one significant contributing factor, among several, to the conditions that produced the 2008 financial crisis.[6]

A final thirty-year span connects the 1991 Soviet collapse to the outbreak of the Russia-Ukraine war in 2022. Following the Soviet collapse, NATO expanded eastward in successive rounds. After Vladimir Putin came to power in 2000, Russia applied to join NATO several times but was not accepted,[7] even as NATO continued to deepen engagement with states along Russia's periphery. In 2008, Ukraine's and Georgia's moves toward NATO membership contributed to the outbreak of the Russo-Georgian armed conflict that the same year; in 2014, Russia's annexation of Crimea was followed by Russian-backed conflict in eastern Ukraine. As China's national strength has continued to grow, the United Sates has increasingly identified China as its primary strategic competitor and shifted greater military attention toward East Asia. This, coupled with Russia's status as a nuclear power, makes it so that the US and NATO would not dare to intervene directly if Russia were to make a move against Ukraine. Ukraine's moves toward seeking NATO membership had become more intense post-2014, ultimately contributing to the outbreak of the Russia-Ukraine conflict.

### 2.2 The 30-year Cycle in China's Domestic Affairs

With the support and funding of the Communist International under the Vladimir Lenin's leadership of, the CPC was founded in 1921, following the path created by Lenin's October Revolution of 1917 — one oriented toward anti-imperialism, anti-feudalism, and national independence. Under the leadership of the CPC, the Chinese people underwent the Land Revolution, the War of Resistance against Japan, and the Three-year War with the Kuomintang, ultimately driving out the Kuomintang forces led by Chiang Kai-shek to the island of Taiwan in 1949. Soon after, the CPC united the majority of China's territories and established the PRC. From the founding of the CPC in 1921 to the setup of the PRC in 1949, it took about 30 years.

The succeeding thirty years, from 1949 to 1978, was the buildup period of the PRC. Just after its founding, the Korean War broke out in 1950, and in order to "protect China," China sent its army to fight against the US in Korea.[8] With the support of the military industrial system of the Soviet Union, the Chinese army fought the US-led coalition to a stalemate, and stabilized the battle line between North and South Korea at the 38th parallel before the Korean War. Along with the Korean War, China carried out domestic land reform from 1950 to 1953, confiscating land and property from landlords and capitalists and redistributing it among the country's landless peasantry and urban poor. At the same time in the 1950s, a large-scale literacy campaign was carried out. A nine-year primary and secondary education system was set up, and a system of higher education institutions and state-owned scientific research institutes was established. During and after the Korean War, China received the 156 aided industrial projects from the Soviet Union (after 1960, when Sino-Soviet relations broke down, the Soviet Union withdrew its experts and stopped all the aided projects).[9] In the mid-1970s, with its own efforts, China finally built its own heavy industry system, realizing the "Two Bombs and One Satellite" and becoming a "nuclear power."[10] In 1964, the US launched an invasion of Vietnam, which was essentially an attempt to encircle China, and China carried out the "Resist the US, help Vietnam" campaign. In 1973, with the help from China and the Soviet Union, the Vietnamese army drove the US out of Vietnam by relying on China's newly established heavy industry and the Soviet Union's more advanced military industry. In the first thirty years of its existence, China mainly broke the military siege imposed by the US and maintained its territorial security. After becoming a "nuclear power," China not only won international respect, but it also became a major force for world peace. [11]

About thirty years passed between the Third Plenary Session of the 11th Central Committee of the CPC in December 1978 — when the party announced the "reform and opening-up" (改革开放) policy — and 2010, when China overtook Japan as the world's second largest economy. By taking advantage of the strategic rivalry between the US and the Soviet Union in the early 1970s and the US' intent to draw China in to constrain the Soviet Union, China normalized its relationship with the US and formed friendly ties with other Western countries. This laid the international foundation for the implementation of China's reform and opening-up. The period from the establishment of Sino-US diplomatic relations in 1979 to the collapse of the Soviet Union in 1991 provided a unique opportunity for China's economic development. After the collapse of the Soviet bloc in 1991, the US and its Western allies proudly proclaimed the triumph of "democracy" and "freedom" over "authoritarian rule." The small size of China's economy at that time had not yet drawn the attention of the US and other Western countries. Therefore, China used the decade between the collapse of the Soviet Union in 1991 and the US "war on terror" in 2001 to vigorously built a socialist market economy, and completed the most difficult part of China's reforms: the closure and merger of small and medium-sized state-owned enterprises (SOEs). China's WTO accession in 2001 further prompted a substantial overhaul of domestic regulations to align with international commercial norms. Over the subsequent decade — one in which US strategic attention was substantially absorbed by the "war on terror" — China's economy continued to expand rapidly, by 2010 it had both surpassed Japan to become the world's second-largest economy and emerged as the world's largest manufacturing country by output.

### 2.3 The Turning Point in the 30-year Cycle

Each of these thirty-year cycles, in this account, pivots around a significant turning point. The turning point between the end of World War I and the outbreak of World War II lay in the Treaty of Versailles imposed on Germany under the leadership of Britain and the United States after WWI. The treaty stripped Germany of its colonies, forced it to cede territory, pay massive reparations, and restrict its military strength, which fueled widespread resentment among the German people and which the Nazi Party later exploited in its rise to power, making Germany the epicenter of the war that followed. The broader postwar territorial settlement among Britain, France, Japan, and Italy was likewise widely seen as inequitable; dissatisfaction with this order helped drive both Japan and Italy toward territorial expansion and fascism. Inspired by the founding of the world's first socialist state, the Soviet Union, Marxism-Leninism began to spread across China. Later, with the assistance and funding from the Communist International (established in 1919), which supported colonial and semi-colonial nations and peoples in their struggle for liberation and independence, the CPC was founded in 1921. After leading the Chinese people through twenty-eight years of sustained wars and struggles, the Party established the PRC in 1949.

From the outbreak of the war between China and the US in Korea in 1950 to the establishment of diplomatic relations between the two countries in 1979, the turning point was the breakdown of Sino-Soviet relations around 1960, which laid the foundation for the reconciliation aimed at  the containing the Soviet Union. From the Sino-Soviet split in 1960 to the Soviet collapse in 1991, the turning point was the wave of global market liberalization beginning in the late 1970s — a shift that Soviet Union largely failed to keep pace with, leaving it at a growing structural disadvantage. From the adoption of neoliberal economic policy by the Reagan administration in 1981 to the US financial crisis of 2008, the turning point was the repeal of the Glass-Steagall Act by the Clinton administration in 1999, which allowed the reintegration of commercial and investment banking; this deregulation, combined with the broader use of short-term borrowing to fund longer-term lending across the financial system, is widely cited among the contributing causes of the 2008 crisis.[12] And from the Soviet collapse in 1991 to the Russia-Ukraine war of 2022, the turning point was the Russo-Georgian conflict of 2008 — caused by Georgia's attempt to join NATO — which many observers regard as marking the start of open confrontation between Russia and NATO.

A parallel set of turning points structures the domestic Chinese narrative. The turning point from the founding of the CPC in 1921 to the setup of the PRC in 1949 was Japan's full-scale invasion of China in 1937, which precipitated the joint resistance Japanese invasion by the Kuomintang and the CPC forces. The turning point from the founding of the PRC in 1949 to the Third Plenary Session of the 11th Central Committee in 1978 was China's emergence as a nuclear power, following successful tests of the atomic bomb in 1964 and the hydrogen bomb in 1967. The turning point from the reform and opening-up policy in 1978 to China's surpassing of Japan as the world's second largest economy in 2010 was the CPC's redefinition of China's development stage at its Thirteenth National Congress in 1987. There, the Party formally characterized China as being in the "primary stage of socialism" — a framework under which diverse forms of ownership were permitted to coexist alongside a state-owned economy retaining the dominant role. This ideological reformation helped unlock a new phase of economic dynamism in the years that followed.

### 2.4 Will China Surpass the US by 2040?

As China's economic rise became increasingly apparent around 2010, the United States came to view it with growing strategic concern. The Obama administration, which took office in 2009, implemented a "Pivot to Asia" rebalancing strategy—drawing down US commitments in the Middle East in order to concentrate greater attention on China. In the realm of trade policy, the administration championed the Trans-Pacific Partnership (TPP), an arrangement widely understood as designed to strengthen US-led regional trade and investment networks while leaving China outside them. The Trump administration, which took office in 2017, adopted a more confrontational posture under the banner of "America First," withdrawing from the TPP and numerous other international agreements while pursuing a broader strategy aimed at constraining China's rise. The administration publicly criticized Chinese trade practices—alleging technology theft, forced technology transfer from foreign firms operating in China, unfair subsidization of state-owned enterprises, and characterizing the Belt and Road Initiative as a form of "debt trap" diplomacy or "neo-colonialism" toward developing countries.

The administration also launched a trade war marked by substantially higher tariffs on Chinese goods, restricted exports of advanced semiconductors and associated manufacturing equipment to China, and moved to bar Huawei's 5G equipment from US networks and, through diplomatic pressure, from those of several allied countries. On matters of diplomacy and security, the administration leaned more heavily on Táiwān- and Xīnjiāng-related issues in its dealings with Beijing, deepened intelligence coordination through the Five Eyes alliance, and advanced the "Indo-Pacific Strategy," including the revival of the Quad grouping with Japan, India, and Australia. The Biden administration, taking office in 2021, largely maintained this trajectory: sustaining high tariffs on Chinese goods, coordinating more closely with allies on China policy, and adding the AUKUS security partnership (with the UK and Australia) to the Quad as a further element of its Indo-Pacific posture. The administration also intensified sanctions on Chinese firms,[13] pursued a more targeted "small yard, high fence" approach to restricting high-technology exports, and raised tariffs further in select sectors — including tariffs as high as 100% on Chinese electric vehicles.[14] In February 2023, the US secured expanded access to four additional military bases in the Philippines, positioned near the Taiwan Strait and South China Sea.[15] After taking office in January 2025, Trump launched tariffs against trading partners worldwide, and tensions with China escalated sharply—US tariffs peaked at 145 percent in April, while China retaliated by tightening export controls on rare earths, minerals it dominates globally, disrupting US automotive, semiconductor, and defense supply chains. A May 2025 truce lowered tariffs, but tensions flared again in October when China expanded controls ahead of a Trump-Xí summit, prompting further tariff threats. The two leaders met at APEC in Busan in late October, agreeing that China would suspend export controls for a year while the US eased tariffs and restrictions.

So how long will the US be able to suppress China? Applying the "thirty-year cycle" framework developed earlier in this paper, and counting from roughly 2010, one might project that the current phase of US pressure on China could extend to approximately 2040. China would by then have become the world's largest economy and a leading center of technological innovation across a range of critical "chokepoint" technologies. By that time, even if the US wanted to suppress China, it would have no means. Whether this scenario materializes, however, depends heavily on a more immediate and uncertain question: whether China can achieve genuine breakthroughs in these chokepoint technologies and establish a durable technological lead over the United States within the coming fourteen to twenty years.

### 3. Lightning Rise of China's High-technology Enterprises and the New Internationalization since the 18th National Congress of the CPC

Since the 18th National Congress of the CPC, China's high-tech industry has expanded rapidly, with 504,000 high-tech enterprises emerging — representing by the Beidou Navigation Satellite System, Huawei, BOE Technology Group Co., Ltd., Alibaba Group, Tencent Holdings Limited, Baidu, SF Express, DJI (Shenzhen DJI Innovation Technology Co., Ltd.), Contemporary Amperex Technology Co., Ltd. (CATL), BYD Company Limited, ByteDance, Hengrui Medicine, Mindray Medical International Limited, SenseTime, Longi Green Energy Technology Co., Ltd., China's high-speed rail, ChangXin Memory Technologies Co., Ltd., iFLYTEK Co., Ltd., Hikvision Digital Technology Co., Ltd., Xunlei Network Technology Co., Ltd., and Chenguang Biotechnology Group Co., Ltd, among others.[16] The Pearl River Delta and Yangtze River Delta have become two major hubs for global cutting-edge tech enterprises.[17] China leads the world in the research, development, and manufacture of electric vehicles, power batteries, and solar panels, remaining the largest producer and exporter of all three — achievements widely cited as milestones of “Made in China 2025” initiative. In 2018, the US started imposing severe sanctions on Huawei, cutting off access to 5G chips, Oracle’s enterprise resource planning (ERP) software and the Android operating system.[18] However, with the support of many domestic manufacturers, Huawei developed its own ERP in 2022 and developed a circuit board free of American components in 2023.[19] That same year, Chinese chipmakers achieved a breakthrough in 7-nm 5G chips.[20] In 2024, Huawei launched HarmonyOS 4.0, the world’s third largest mobile operating system alongside Android and iOS; the Huawei Mate 60, equipped with a domestically produced 7-nm 5G chip had been released the previous year, followed by the Mate 70 running HarmonyOS.[21] Chinese AI development accelerated sharply during this period as well. DeepSeek’s R1 model, released in January 2025, triggered a major selloff in US tech and semiconductor stocks — Nvidia along lost nearly $590 billion in market value in a single trading session — as investors questioned assumptions about US dominance in frontier AI. In April 2026, DeepSeek achieved compatibility with Huawei Ascend CANN ecosystem, enabling large-scale deployment of trillion-parameter large models independent of NVIDIA's CUDA environment, which broke NVIDIA's monopoly on intelligent chips. In July 2026, Moonshot AI Kimi K3, a 2.8-trillion-parameter open-weight model, produced a comparable market reaction, contributing to a roughly $470 billion selloff across US AI-related stocks and pushing the Philadelphia Semiconductor index into bear-market territory; commentators dubbed it a second “DeepSeek moment.” At the time of its release, independent benchmarks ranked Kimi K3 third globally in overall capability, behind Anthropic's and OpenAI's leading models, while leading on certain coding benchmarks—illustrating how rapidly the gap between Chinese and American frontier AI systems had narrowed, even as a meaningful gap persisted. These advancements suggest that US sanctions since 2018, intended to constrain China’s semiconductor and technology sectors, may in some respects have accelerated Chinese self-sufficiency[22] by closing off the domestic market to foreign competitors and creating space for domestic substitutes to mature.[23] Moreover, more than half of the new industrial robots installed globally are in China.[24] Today, China’s economy has shifted substantially from low value-added consumer goods at the beginning of the new century to capital-intensive and high-tech output, with its export mix increasingly dominated by mid-to-high-end mechanical and electrical products — mobile phones, computers, automobiles, EVs, solar panels, lithium batteries, large ships, railway locomotives, numerically controlled machine tools, drones, rare earths, graphene, and other intermediate inputs.

Unlike the US-led globalization of the late 1970s through the end of the twentieth century, a new phase of globalization — led by China, or in which China plays a major role — has accelerated since the turn of the century, and especially since the 18th National Congress of the CPC.[25] In June 2001, the Shanghai Cooperation Organization (SCO) was established by six countries—China, Kazakhstan, Kyrgyzstan, Russia, Tajikistan, and Uzbekistan—to promote counter-terrorism cooperation in Central Asia. In 2017, the SCO expanded for the first time, admitting India and Pakistan as its full members, and again in July 2023, when Iran joined. The BRIC organization, established by Brazil, Russia, India, and China in 2006, became BRICS with the addition of South Africa in 2011; the group established its New Development Bank, headquartered in Shanghai, in 2015. In 2024, BRICS underwent a major expansion with the addition of Saudi Arabia, Egypt, the United Arab Emirates, Iran, and Ethiopia, becoming known as BRICS+.

China launched the BRI in 2013, and more than 150 countries have since participated in some form. The Asian Infrastructure Investment Bank (AIIB), established in 2015 with China at its core, finances a portion of BRI-related infrastructure projects. China put forward the initiative of "joining hands to build a closer China-ASEAN community of destiny" in 2013, building on its 1996 accession to the Treaty of Amity and Cooperation in Southeast Asia. In November 2020, fifteen Asia-Pacific countries — the 10 ASEAN members together with China, Japan, South Korea, Australia, and New Zealand — signed the Regional Comprehensive Economic Partnership (RCEP), which entered into force in 2022 as the world's largest free trade area by population and GDP; Hong Kong, Sri Lanka, and Chile have since expressed interest to joining. On May 30, 2025, thirty-three countries signed the Convention on the Establishment of the International Organization for Mediation (IOMed) in Hong Kong, with representatives from 85 countries and nearly 20 international organizations, including the UN, in attendance. IOMed is billed as the world's first intergovernmental organization specializing in resolving international disputes through mediation, and is headquartered in Hong Kong. The process of internationalization aimed at promoting the development of the third world countries, in which China has taken a major role, is accelerating.

Among these frameworks, the BRI has borne the most fruits. From 2013 to 2023, China has signed more than 200 cooperation documents on the BRI with more than 150 countries and 30 international organizations.[26] Completed landmark projects include the China–Europe rail freight service; the Mombasa–Nairobi Railway in Kenya; the Addis Ababa–Djibouti Railway in Ethiopia; the China–Laos Railway between Laos and Yunnan; the Jakarta–Bandung High Speed Rail in Indonesia; the Pelješac Cross-Sea Bridge in Croatia; the Padma Bridge in Bangladesh; the Mekong River Bridge in Cambodia; the Port of Bire Dwarfos in Greece; the deepwater port of Lekki in Nigeria; the Gwadar Port of Pakistan; the Phnom Penh-Seaport Expressway in Cambodia; the Westport Expressway in Cambodia; and the Hanoi Metro Line 2A in Vietnam; among others. Further projects — including the Budapest–Belgrade High-Speed Railway, the East Coast Railway Project in Malaysia, the China–Thailand Railway Project, and the Kuantan Port Expansion in Malaysia — remain under construction. Huawei-supplied 4G networks account for 70% of the local 4G networks in African and Latin American countries, strongly supporting the development of local small and medium-sized enterprises and economic growth.[27] The BRI projects in Latin American countries include the highway traversing the jungles of Costa Rica, the railway between Bolivia and Argentina, the industrial park and container port in Trinidad and Tobago, the largest hydropower plant in Ecuador, and the first trans-oceanic optical fiber cable stretching from China to Chile, directly connecting Asia and South America.[28] By the end of 2022, the China–Pakistan Economic Corridor had brought a cumulative total of $25.4 billion in direct investment to Pakistan, created 236,000 jobs, and helped Pakistan add 510 km of new highways, 8,000 MW of electricity, and 886 km of the country's core electricity transmission grid.[29] The implementation of these projects has, firstly, led to a large number of jobs created in the host countries and, secondly, to an increase in China's exports.[30] Exports to countries participating in the BRI in 2025 accounted for

50.71 percent of China's total exports.[31] China's BRI has not only brought about a significant reduction in the poverty rate of the host countries but also greatly increased the income level of those countries.[32] Since 1995, trade among the Global South has increased by 14.1%, and South-South trade now accounts for a 25% share of international trade. Since the new century, the annual growth rate of trade among the Global South has been 9.8%, much higher than the 5.3% annual growth rate of global trade.[33] China's conception of international development is mainly through building "infrastructure" and realizing "connectivity" in developing countries under the BRI, with the aim of promoting the economic and social development of the Global South. China adheres to the principle of "Confucian Improvement," i.e., "to achieve what one wants to achieve, you must first help others achieve".[34]

Today, China is not the same as it was before the 18th National Congress of the CPC. In 2010, China just overtook Japan to become the world's second largest economy, and now China's GDP is more than four times that of Japan. In 2023, China overtook Japan to become the world's leading automobile exporter and overtook South Korea in shipbuilding output and exports. China has the Beidou satellite navigation system, which is on the same level as the US' GPS. China is now the world's second largest spacefaring nation, and its large airplane manufacturing sector has taken off. Furthermore, China has the world's first and most advanced high-speed rail transportation system, and its ports are equipped with 5G network cables on which self-driving trucks run—the most efficient in the world. China's network communication system is among the most advanced in the world and its e-commerce services are the most developed globally. In addition, China's number of annual college graduates is the largest in the world, and China's cumulative number of college graduates has reached 240 million.[35] With a middle-income class of about 400 million, China now owns the largest single consumer market of its kind globally, making it a major draw for both domestic and foreign manufacturers.

### 4. Advantages of the Socialist Market Economy with the Chinese Characteristics

China's greatest advantage is that it is a socialist country led by the CPC. The CPC was a Marxist party founded on Leninist principles of anti-imperialist and anti-feudalist, drawing directly on the experience of Lenin's Bolshevik leadership in Russia. The CPC has the experience of twenty-eight years of struggle against domestic enemies and the Japanese invasion prior to 1949, the extremely strict organizational and political disciplines forged during those war years, the effort to build the state amid the fights against the US empire after 1949, rapid economic growth following the construction of a socialist market economy and the integration into the global trading system after the reform and opening-up, sustained resistance to U.S. suppression in the post-Cold War period, the buildup of high-tech industry, and, since the 18$^{th}$ Party congress, a new phase of China-led internationalization. The CPC's long-term rule has made it impossible for the Western monopoly capital, led by Wall Street in the US, to influence the Chinese government's decision-making, control China's strategic industries, or interfere with China's internal and foreign affairs.

Compared with the US, China's advantage lies in that in addition to the market economy, it also has the central planning system, SOEs, and State-Owned Venture Capital (SOVC). China usually sets for five-, ten-, fifteen-, thirty-year plans for strategic high-tech sectors, backed by targeted industrial policy. China's recent breakthroughs in new-energy vehicles, power battery, and clean energy, on this account, reflect the cumulative effect of planning and subsidy decision made over the preceding three decades — supporting manufacturer R&D, subsidizing consumer purchases, mandating public-sector adoption of electric vehicles in municipal transit fleets.[36] Starting from August 1992, when Qián Xuésēn wrote to the central authorities proposing that China develop electric vehicles,[37] until June 2019, when the Chinese government phased out the white list policy for power batteries,[38] after over thirty years of planning and industrial policy support from successive administrations, China's new energy vehicles and clean energy products have ultimately secured the world's top position.[39] The Nobel Prize-winning Economist Professor Stiglitz (who served as chief economist in the Clinton administration and chief economist at the World Bank) recalled in an interview with the Financial Times in 2024: More than a decade ago, "I was in a meeting with the premier [Wēn Jiābǎo] where he told the car companies: you have to be electric within five years or you're out of here. China has made it clear it will be an EV country; we haven't." In the same interview, Prof. Stiglitz commented, "The underlying problem is that the US did not anticipate a rival such as China."[40]

SOEs and state-owned public sectors control strategic industries related to the lifeblood of the national economy, such as banking and finance, transportation, communications, energy and power, primary and secondary school, higher education systems, the system of the Chinese Academy of Sciences, water and power supply and gas supply systems, and medical and health-care systems, among others. State-owned construction firms also play a central role in implementing BRI infrastructure projects — railways, subways, bridges, highways, power plants, ports, and power grids — including higher-risk projects such as the China–Pakistan Economic Corridor and the Gwadar Port, which state planners judge better suited to SOE execution given the associated security considerations.

SOVC, meanwhile, is characterized by foresight, focusing on the future, planning-oriented investment with national strategic industries — directed at both SOEs and private enterprises — rather than the short-term return horizons typical of private venture capital.[41] Technological innovation is often divided into "chokepoint" technologies, where a specific capability gap must be closed, and open-ended innovation in less charted areas. State-owned-sector-led breakthroughs in "chokepoint" technologies are said to include the "Two Bombs and One Satellite" program of the 1950s-60s, and more recently the national space program, high-speed rail, the Beidou satellite navigation system, domestic large-aircraft project (including aircraft engines), ultra-high-voltage power transmission, automated port technology, LNG shipbuilding, among others. Successful cases of investment in high-tech projects of SOEs and private enterprises by SOVC as well as provincial and municipal SOVC include power battery projects, EV projects, solar panel projects, other new energy projects, 5G communication systems, and the 7-nm 5G chips. Of course, when SOVC invested in private enterprises which achieved technological innovation, the SOVC typically plays a catalytic rather than executing role. The Trump administration's restrictions on high-tech exports to China played a key role behind China's subsequent push for self-sufficiency in chokepoint technologies, including the establishment of a major state investment fund for advanced chip development.

Let us use the Héféi model to illustrate how SOVC has helped both state-owned and private enterprises develop high-tech products. In 2008, the Hefei municipal government's investment of CN¥17.5 billion helped stabilize display manufacturer BOE during its financial distress;[42] a large investment helped bring Lenovo's operations to the city in 2011, which in turn drew in several hundred firms in the laptop supply chain; [43] a 2017 joint venture with GigaDevice (兆易创新) to establish memory-chip maker Changxin Memory Technologies (CXMT), with the municipal government holding a majority stake; and a roughly CN¥7 billion investment in 2020 that helped rescue EV maker NIO from near-bankruptcy in exchange for a substantial equity stake, which is credited with contributing to NIO's subsequent recovery.[44] The year of 2023 saw Volkswagen invest CN¥7.5 billion to set up an EV R&D center in Héféi. In addition, electric car giant BYD and many other high-tech companies have settled in Héféi. Not all of Héféi's venture investments have succeeded, and some have resulted in losses. The Héféi municipal government has made huge returns from its successful ventures; its annual GDP growth rate was at a steady 8% from 2012 to 2022, and the disposable income per capita in Héféi now far exceeds the national average.[45]

Innovations in less charted territories, by contrast, is harder to plan and has instead emerged through competition among China's rapidly growing private sector since reform and opening-up — spanning 5G networks, 7nm 5G chips, memory chips, e-commerce (JD.com, Alibaba), cross-border e-commerce (Shein, Pinduoduo), logistics and delivery (SF express, Meituan), consumer drones (DJI), biotechnology, power batteries, solar panels, EVs, smartphones, and computing, among others. One of reasons for the Soviet Union's collapse was its failure to develop a functioning market economy, which left Soviet citizens comparing their living standards unfavorably against the West.[46] Much of China's recent private-sector innovations is attributed to the sheer scale of corporate R&D investment. Firms such as Huawei and BYD each employ well over 100,000 R&D personnel.[47] China's universities and research institutes are mainly engaged in fundamental research, and since the reform and opening-up, they have trained tens of millions of scientific and technological personnel for the society. This talent pool expanded dramatically following China's 1999 higher-education expansion policy: national college enrollment capacity, under 1 million in 1998, grew to more than 7 million by 2008.[48]

### 5. The US Approach to US-China Competition and Expected Outcomes

The first part of this chapter examined the coercive dimension of US policy toward China — tariffs, export controls, and sanctions. This section turns to a complementary track: the industrial policy measures the United States has adopted domestically to promote its own economic and technological competitiveness. China' economic success has led the US to abandon its "liberal" tradition of government non-intervention in the economy and to adopt a more active industrial policy approach with some structural parallels to elements of the Chinese model. The CHIPS and Science Act, signed by the Biden administration in 2022, aims to preserve US leadership in semiconductors and artificial intelligence by offering financial incentives for chipmakers — including firms from allied democracies — to invest in advanced chip production in the US, with individual awards to participating companies often reaching tens of millions of dollars or more. Under the Act, the federal government committed $11 billion to establish the National Semiconductor Technology Center (NSTC), intended to help the US maintain leadership in semiconductor technology standards, design, manufacturing, and semiconductor engineering talent.[49] In 2022, the Biden administration also signed the Inflation Reduction Act, which expanded tax credits for qualifying clean energy facilities and EVs.[50] In May 2023, the US Department of Commerce launched the Regional Technology and Innovation Hubs (Tech Hubs) program, backed by an initial $10 billion investment, to support regional development in fields including artificial intelligence, advanced computing, robotics, natural disaster resilience, advanced communications, biotechnology, data infrastructure, advanced energy, and next-generation materials.[51] Taken together, these measures suggest that, in the face of intensifying competition with China, US policymakers—across both major parties to varying degrees—have shown greater willingness to deploy the tools of industrial policy than the country's traditional free-market rhetoric might have predicted, even as officials generally continue to frame these interventions as consistent with, rather than a departure from, American economic principles.[52]

Even if the United States were to fully abandon its traditional preference for limited government intervention in the economy and adopt an industrial-policy approach resembling China's, the return of manufacturing to American soil would likely be a long and difficult process. This is due in significant part to more than three decades of industrial offshoring — roughly from Reagan's arrival in office in 1981 through the end of the Obama administration ended in 2016 — during which US manufacturing capacity was substantially hollowed out. Trump began actively promoting manufacturing revitalization upon taking office in 2017, but the results to date have been modest at best. The core obstacles are structural: decades of offshoring eroded the skilled industrial workforce, engineering talent pipeline, and manufacturing know-how the US once possessed, while the loss of supporting supplier networks and persistently high US labor costs make it difficult for manufacturing to compete domestically—absent very high tariffs on imported goods. Yet sufficiently high tariffs carry their own risks. If foreign producers were effectively locked out of the US market, other countries would have fewer opportunities to earn US dollars through trade, which over time could undermine the dollar's role as the world's primary reserve currency—a status that itself confers substantial economic benefits on the United States. Since taking office, the Trump administration has raised tariffs on trading partners worldwide, framing the policy in part as an effort to bring manufacturing back to the United States—but Trump's stated goal of reshoring large numbers of manufacturing jobs for American workers is likely to prove difficult to achieve at scale. Due to the fact that the automation of global manufacturing is accelerating, and it is an inevitable event for industrial robots to replace low-level repetitive labor.[53] The re-industrialization of the US will be based on using industrial robots as the main labor force.[54]

A more plausible outcome is that the US will succeed in reshoring production primarily in the relatively narrow set of advanced sectors where it retains a strong competitive or technological edge—such as semiconductors, artificial intelligence, and biopharmaceuticals—where high margins in these sectors could help offset elevated domestic labor costs and the capital costs of automation.

### 6. Conclusion

This chapter has traced a recurring pattern of roughly thirty-year intervals across major turning points in global affairs and modern Chinese history: from the beginning of World War I to the end of World War II; from the founding of the Soviet Union, the world's first socialist state, to the founding of New China, another major socialist state; from China and the United States going to war in Korea in 1950 to the establishment of Sino-American diplomatic relations in 1979; from the rupture of Sino-Soviet relations in 1960 to the collapse of the Soviet Union in 1991; from the Reagan administration's implementation of neoliberal economic policies in 1981 to the U.S. financial crisis of 2008; from the collapse of the Soviet Union in 1991 to the outbreak of the Russo-Ukrainian War in 2022; from the founding of the CPC in 1921 to the establishment of the PRC in 1949; from the PRC's founding to the launch of reform and opening-up in 1978; and from reform and opening-up to China's overtaking of Japan as the world's second-largest economy in 2010. Drawing on this recurring pattern as a heuristic rather than a strict causal law, this chapter suggests that the current period of intensified US pressure on China—particularly through technology embargoes and export controls—may, following the same rough thirty-year rhythm, extend to approximately 2040. The reasoning behind this projection is that by around 2040, China may have achieved breakthroughs across most or all of the "chokepoint" technologies currently subject to US restriction, while also establishing leading positions in unknown technological fields not yet well defined today. Under such conditions, the argument goes, US technological containment would lose much of its practical effectiveness, since the underlying capability gaps it depends on would have substantially closed. The central and genuinely uncertain question, then, is whether China can, within the roughly fourteen years between now and 2040, achieve the technological breakthroughs on which this projection depends, and take the lead in uncharted fields of technologies in the 15 years from now to 2040.

Since the 18th National Congress of the CPC, China's homegrown high-tech enterprises have risen to global prominence, positioning the country as a world leader in 5G communications, electric vehicles, power batteries, solar panels, robotics, drone, among others. Huawei achieved a significant breakthrough in domestic 7-nanometer chip manufacturing in 2023, releasing its Mate 60 smartphone with a domestically produced 7nm 5G chip that same year. In 2024, Huawei's HarmonyOS 4.0 came into existence alongside the Android and iOS. The emergence of China's DeepSeek in early 2025 and Moonshot Kimi K3 in July 2026 likewise demonstrated that Chinese AI development had narrowed the gap with, and in certain respects challenged, US leadership in the field—complicating assumptions underlying US technology-containment strategy. On May 7, 2025, amid the India-Pakistan conflict, Pakistan Air Force pilots flying Chinese-built J-10C fighter jets armed with Chinese PL series air-to-air missiles shot down two French-made Rafale fighter jets of the Indian Air Force, demonstrating the advanced performance of Chinese weaponry. In April 2026, DeepSeek completed the migration of its operations to a platform underpinned entirely by Huawei Ascend chips, which broke the monopoly held by NVIDIA in the GPU chip market. CXMT has grown into the world's fourth-largest DRAM manufacturer, breaking the monopoly of Samsung, Micron and SK Hynix in the DRAM sector; in July 2026, CXMT's market capitalization surpassed that of Industrial and Commercial Bank of China and Kweichow Moutai. From reform and opening-up in 1978 through the 18th Party Congress in 2012, China's economic and trade engagement with the world operated largely within the US-led postwar international system. Since 2012, however, China has increasingly pursued forms of internationalization in which it plays a leading or central role—including the expansion of the Shanghai Cooperation Organization, the growth of BRICS into BRICS+, and, most significantly, the Belt and Road Initiative, which more than 150 countries have joined, supported in part by the China-led Asian Infrastructure Investment Bank. Over roughly a decade, the BRI has produced a substantial record of completed and ongoing projects. Taken together, this record forms a continuous arc: from the CPC's founding in 1921, through the establishment of the PRC in 1949, the achievement of nuclear-power status through the "Two Bombs, One Satellite" program, industrialization following reform and opening-up in 1978, and the more recent rise of Chinese high-tech industry and China-led internationalization since 2012. Building on this trajectory, this chapter argues there is reasonable basis to expect that, over the next fourteen to twenty years, China will make further progress toward closing gaps in chokepoint technologies, achieve leading positions in unknown technological fields, and continue shaping a more multipolar international system. By that time, the US will not have any means to suppress China even if it wants to.

Given that both major US political parties have shown commitment to re-industrialization—particularly in advanced manufacturing—it is plausible, following the same thirty-year heuristic applied elsewhere in this chapter, that meaningful results from this effort may not fully materialize until around 2047, roughly thirty years after Trump took his office in 2017 and began attracting manufacturing industries to reshoring to the United States. By that point, US re-industrialization may have achieved significant progress, particularly in high-tech manufacturing sectors. At the same time, if current trends continue, China may have substantially closed the technological gap with the United States and, in a number of cutting-edge fields, moved ahead of it. By then, the US will no longer have any means to sanction China in the field of high technology. Thus, by about 2040, China and the US will have achieved a new equilibrium in the development of high technology, and each will need to learn from the other.

**Notes**

[1] Wu, 1749, p. 338. (“三十年河东，三十年河西。”)

[2] In his book, Kennedy (1989) traces how nations that overextend their military commitments—like the US through Korea and especially Vietnam—undermine their economic strength and enter national decline. He posits that post-WWII, the US reached its peak around 1945–1946 and describes how subsequent strategic overreach led to relative erosion by the 1970s.

[3] Byrne, 1996; Crocker, 1980; Dommen, 2002; Gaiduk, 2003; Garthoff, 1994; Gleijeses, 2006; Grau & Gress, 2002; Haslam, 2011; Keen & Hayes, 2012; Kinzer, 1991; Raghavan, 2013; Schmidt, 2013; Westad, 2005. During the 1960s and 1970s, the Soviet Union, amid the Cold War rivalry with the US, launched a series of global strategic offensives to expand its influence, challenge US hegemony, and promote its ideological and geopolitical interests. In Asia, the Soviet Union backed North Vietnam during the Vietnam War (1955–1975) with military aid, including weapons, advisors, and air defense systems (e.g., SA-2 missiles), helping counter U.S. aerial attacks and strengthening North Vietnam’s war effort. The Soviet Union also supported leftist governments in Afghanistan (beginning in the 1970s, culminating in the 1979 invasion), India (during the 1971 Indo-Pakistani War), and socialist movements in Laos and Cambodia. In Africa, the Soviet Union supported anti-colonial and socialist movements, such as the Popular Movement for the Liberation of Angola (MPLA) in Angola’s civil war (1975–2002), sending troops and weapons to counter U.S.-backed UNITA party and South African forces. The Soviet Union backed Ethiopia’s Marxist government (led by Mengistu Haile Mariam) during the Ogaden War (1977–1978) against Somalia, providing military aid and advisors. The Soviet Union established alliances with socialist regimes in Mozambique (under the FRELIMO party) and Guinea-Bissau, as well as with radical states like Libya under Muammar Gaddafi. In Latin America, the Soviet Union strengthened ties with Cuba after the 1959 Cuban Revolution, providing economic and military aid. The 1962 Cuban Missile Crisis (a critical Cold War flashpoint) reflected the Soviet Union’s attempt to place nuclear weapons in the Western Hemisphere, though it ultimately withdrew them. The Soviet Union also supported leftist guerrilla movements, such as the Sandinistas in Nicaragua (1960s–1970s) and the Salvadoran Farabundo Martí National Liberation Front (FMLN), to challenge US-backed authoritarian regimes.

[4] Hurst & Trubowitz, 2025.

[5] Grachev et al., 1995.

[6] Crawford, 2011.

[7] Mearsheimer, 2014.

[8] Academy of Military Sciences, 2020.

[9] Naughton, 2007; Zhang, 2009. “By the end of the 1950s, the Soviet Union had provided China with assistance in building 156 large-scale industrial projects, including heavy industry, defense, and infrastructure” (Naughton, 2007)

[10] “Two Bombs and One Satellite” (两弹一星) is a pivotal historical initiative in China’s modern history, referring to the country’s successful development of nuclear weapons (atomic bomb and hydrogen bomb), ballistic missiles, and artificial satellites from the 1950s to the 1970s. This program was a cornerstone of China’s efforts to strengthen its national defense, enhance its international standing, and lay the foundation for its aerospace and nuclear industries.

[11] Since the setup of the PRC, in particular since China became a nuclear power in the 1970s, East Asia has been getting more peaceful. Apart from the Korean War (1950–1953), China fought the following wars that brought about peace to China’s borderline and its neighbors. In the 1962 Sino-Indian War, China repelled Indian forces back behind the Line of Actual Control, ensuring stability along the Sino-Indian border since then (Maxwell, 1970). In 1969, following the breakdown of Sino-Soviet border negotiations in 1964, the Soviet Union provoked armed clashes along the border. Relying solely on its own strength, China withstood the pressure from the nuclear superpower Soviet Union and maintained stability along the frontier. In 1964, the US sent troops to support South Vietnam against North Vietnam, while China and the Soviet Union backed North Vietnam until the signing of

the Paris Peace Accords in 1973, which led to the withdrawal of US forces. In 1975, North Vietnam defeated South Vietnam, reunifying the country. In 1978, Vietnam invaded Cambodia in an attempt to dominate Southeast Asia. In 1979, China dispatched troops to Vietnam, thwarting its hegemonic ambitions. In May 2025, India launched a cross-border strike against Pakistan. However, thanks to Pakistan's deployment of Chinese-made weapon systems, which successfully shot down India's French-imported Rafale fighter jets, the India-Pakistan conflict was swiftly brought to an end. In this sense, China contributes to the world peace.

[12] Crawford, 2011.

[13] The Guardian, 2024.

[14] Gertz, 2024; The Guardian, 2024.

[15] Arugay & Storey, 2023.

[16] https://www.xinhuanet.com/info/20251231/c15b4b7a5030484eaa6c10ebab2f9e75/c.html. Xinhuanet, 2025 Dec. 31. Although many of these above-mentioned enterprises were set up before the 18th National Congress of CPC in 2012, the main technological breakthroughs have been made after 2012.

[17] Lin & Zhang, 2019.

[18] The Economist, 2024.

[19] Ibid.

[20] Ibid.

[21] Ibid.

[22] Yoon, 2025.

[23] Li & Gao, 2025.

[24] Simington, 2024.

[25] Li, et al. 2023; Zhao, 2021.

[26] Wang & Liu, 2024.

[27] Dahl, 2024; Goldman, 2023.

[28] Dahl, 2024.

[29] Jiang, 2023.

[30] Yang & Zeng, 2019; Jiang et al., 2022.

[31] National Statistics Bureau, Feb. 28, 2026.

[32] Zhang et al., 2023.

[33] Runde et al., 2024.

[34] Zhao Tingyang (2016) refers to Confucius' words: "夫仁者，己欲立而立人，己欲达而达人" (p. 71).

[35] Xi Jinping, 2025; Xinhuanet, 2022.

[36] As early as 2001 when the "863 Program" first proposed new energy initiatives, through dedicated investments in the 10th, 11th, and 12th Five-Year Plans' "863" projects, to the 13th Five-Year Plan which incorporated new energy vehicles into the national key R&D program, state-funded enterprises developing new energy projects have cultivated a large pool of talent for new energy vehicle development. This is precisely why new energy vehicle manufacturers have proliferated across the country in recent years.

[37] Qian 1992.

[38] In March 2015, the Ministry of Industry and Information Technology (MIIT) officially issued the Standard Conditions for the Automotive Power Battery Industry (汽车动力蓄电池行业规范条件) and formulated the power battery "white list" mechanism. Only new energy vehicles equipped with power batteries that met the requirements and were included in the "white list" catalogue were eligible for new energy vehicle subsidies; vehicles fitted with batteries not listed in the catalogue could not receive such subsidies.

[39] Shepherd, 2025

[40] Stiglitz, 2024.

[41] Ge et al., 2024.

[42] Chawla, 2023.

[43] Ibid.

[44] Ibid.

[45] Ibid.

[46] Yin, 2018.

[47] Huawei, n.d.; Wang, 2025.

[48] Xia et al., 2016.

[49] Shepardson, 2024.

[50] Department of Energy, 2024

[51] Economic Development Administration, n.d.

[52] Zhao, 2024.
[53] Han, 2022.
[54] Acemoglu & Restrepo, 2020.

**How the Correct Path of Economic Development Has Been Identified Since the Founding of the New China?**

## 1. Introduction

After China overtook Japan to become the world's second-largest economy in 2010 and emerges as a global high-tech power in 2020s, the whole world marveled at its economic miracle, with widely divergent interpretations put forward to account for the phenomenon. A multitude of factors appear to explain China's economic miracle since the launch of reform and opening up: for instance, the shift to independent household farming for rural residents, the expansion of Township and Village Enterprises (TVEs) as well as urban private businesses, the free mobility of labor, the existence of state-owned enterprises (SOEs), opening-up to the outside world, and the massive influx of foreign-invested enterprises into China. Nevertheless, many developing countries across Asia, Africa, and Latin America share these same conditions, yet they have failed to achieve rapid economic takeoff. Amartya Sen argues that the ample accumulation of human capital prior to China's reform and opening up served as the primary driving force behind China's economic success.[1] In contrast, Wén holds that township industrialization centered on light textile industries in the reform and opening-up era constitutes the core of China's successful industrial revolution.[2] To properly unpack this issue, it is first necessary to clarify the structural shifts in China's economy from the onset of reform and opening up to the present. Between 1978 and 2025, the share of the primary industry—encompassing agriculture, animal husbandry, fishery and mining—in China's GDP gradually declined from 27.7 percent to 6.7 percent. Conversely, the proportion of the tertiary industry climbed steadily from 24.6 percent to 57.7 percent over the same period. Meanwhile, the secondary industry's share fell from 47.7 percent to 35.6 percent, a substantial decline than the other two sectors but still a meaningful drop rather than a stable plateau.[3] Second, after adjusting for price factors, the added value of China's primary industry registered an average annual growth rate of approximately 4.4 percent from 1978 to 2025, far below the average annual GDP growth rate of 8.9 percent, while the average annual growth rate of value added from non-agricultural industries hit 9.8 percent.[4] It is evident from the above that China's economic miracle since reform and opening up hinges chiefly on the growth of industry and services — or more fundamentally, on the triumph of China's industrialization. This chapter intends to explore the drivers of China's successful industrialization from the perspectives of its politics, economy, society and state capability since the founding of the People's Republic of China (the PRC).

## 2. A Rational Interpretation of China's Economic Miracle: The Right Path of Industrialization

To eradicate poverty, China introduced the reform and opening-up policy following the Third Plenary Session of the Eleventh Central Committee in 1978. To address the severe shortage of food, the Chinese government first carried out reforms to the rural collective economy, granting farmers the right to use the rural collective-owned land and introduced the household contract responsibility system. This system greatly boosted agricultural productivity; grain output surged from 300 million tons in 1978 to 450 million tons in 1990.[5] Alongside the household contract reform, the government also permitted farmers to run own-account businesses, which led to the explosive growth of TVEs. By 1995, these enterprises employed 129 million people, and their total industrial output value accounted for 56 percent of the national aggregate.[6] The substantial rise in grain production and rapid expansion of rural industrial operations greatly lifted rural living standards, driving the rural poverty rate down to less than 20 percent by the late 1980s.[7]

In the early 1980s, the Chinese government allowed millions of educated youths who had been sent to the countryside to return to cities. However, urban state-owned and collective enterprises simply lacked the capacity to provide jobs for these millions of returnees, giving rise to severe employment and livelihood difficulties among them. To tackle this problem, the government permitted returning educated youths to engage in own-account businesses, thereby sanctioning private citizens to run profit-making commercial ventures in cities. In 1987, the government lifted restrictions on the number of employees that non-SOEs could hire. This marked the beginning of robust growth for non-state-owned businesses. Document No. 18 issued by the Shenzhen Special Economic Zone in 1987 legally affirmed the property rights of private enterprises. Throughout the 1980s, non-state-owned industrial enterprises grew twice as fast as their state-owned counterparts. By the mid-1980s, the industrial output value of non-SOEs made up 31 percent of the country's total industrial output value.[8]

Prior to the convening of the Third Plenary Session of the 14th Central Committee of the CPC in 1993, China had basically completed its primary phase of its industrial revolution, led by TVEs in rural areas and non-SOEs in cities.[9] TVEs and urban non-SOEs were the main participants in this industrial revolution, surviving entirely through market competition without disrupting agricultural production in the countryside. Amid steady growth in rural grain output, hundreds of millions of workers employed by TVEs (as well as urban non-SOEs) boosted their incomes by producing and selling light industrial goods such as textiles and garments. This constituted the primary driver behind the sharp decline in China's rural poverty rate throughout the 1980s. Before their own industrialization, Europe, the United States, Japan and other nations also suffered from widespread poverty, and they likewise lifted large populations out of poverty through the development of the textile industry.[10] Only the vigorous expansion of light textile industries with broad participation from ordinary working-class people can deliver large-scale and rapid poverty reduction.

One of the core secrets behind China's rapid completion of initial industrialization lies in the fact that administrative bodies at all levels of the Chinese government attached paramount importance to economic development. Administrative leaders at township and village levels even personally acted as entrepreneurs launching and expanding TVEs and guiding villagers to start businesses and achieve prosperity. This practice removed numerous institutional barriers standing in the way of rural industrialization. For instance, collective-owned enterprises under townships and villages could raise funds from villagers and secure bank loans in the name of township/village administrative organs, or of village collectives. When purchasing raw materials and selling finished goods, they also leveraged government credentials to boost market credibility. Unlike Western countries, China lacked a merchant class, credibility systems, and property rights institutions nurtured over centuries. Nevertheless, the rural collective economic administrative organizations established across the countryside after the founding of the PRC served as functional substitutes. The institutional innovation of setting up collective-owned industrial enterprises under the auspices of rural governments, and assigning village cadres to serve as enterprise managers, resolved financing risks, commercial risks, and diverse institutional obstacles hindering the growth of China's rural industries. This factor arguably constitutes another core secret underpinning China's swift accomplishment of initial industrialization.[11]

The vigorous expansion of primary industrialization, led by rural TVEs as well as urban private enterprises, in the 1980s generated effects across three dimensions. First, these entities posed fierce competition to many small and medium-sized SOEs. As a result, numerous loss-making small and medium-sized SOEs were forced to shut down, merge, restructure, or reorganize starting from the mid-1990s. Meanwhile, state-owned heavy industrial enterprises of critical strategic importance — covering energy, electric power, transportation and communications, aviation, railways, shipbuilding, banking and other sectors — underwent restructuring and institutional reforms to establish modern corporate governance systems. Second, the boom of rural TVEs and urban non-SOEs created massive demand for electricity, energy, transportation and communications, breathing new life into China's heavy industrial enterprises in these sectors. China's economic reform followed a gradualist approach. During the process of reform, opening-up, and market-oriented transformation, China did not adopt the comprehensive privatization policies implemented by the former Soviet Union and Eastern European countries. Consequently, the large number of state-owned heavy and chemical industrial enterprises established under the planned economy not only survived but also underwent restructuring and institutional overhaul. When China's economic expansion generated enormous demand for energy, power, transport and communications, these state-owned heavy industrial enterprises were revitalized and subsequently played a pivotal role in the country's heavy industrialization drive, providing the badly needed infrastructure.[12] Finally, the enormous purchasing power generated by the thriving TVEs and private businesses, alongside the resultant second wave of robust growth in heavy industry, attracted massive inflows of foreign capital into China.

### 3. How Was the Right Path of Industrialization Identified?

From the founding of the PRC to the eve of reform and opening up, China implemented a planned economy and a heavy industry-oriented development strategy, relying mainly on an industrialization model in which rural agriculture subsidized urban industrialization. Although this model delivered a certain degree of industrialization in the short run, the distribution systems of traditional SOEs and rural collective economies failed to link remuneration to work performance — urban workers and rural commune members received equal pay regardless of output. This severely dampened work incentives among both groups and dragged down the overall efficiency of the national economy. Meanwhile, the country's total population had nearly doubled prior to reform and opening up, resulting in acute shortages of daily necessities including food and consumer industrial goods. Per World Bank criteria, 76 percent of rural residents and 55 percent of urban dwellers lived below the poverty line before the launch of reform and opening up.[13]

Following the launch of reform and opening-up, China introduced the household contract responsibility system in rural areas, permitted rural collectives and individual farmers to establish TVEs for industrial and commercial operations, and allowed urban residents to set up individual and private enterprises for industrial and commercial activities. At the heart of these reform and opening-up measures lay the gradual introduction of market mechanisms into China's economic system dominated by central planning and SOEs. Specifically, after fulfilling their grain procurement quotas, farmers could decide what to produce and how much on their own. Similarly, rural collective and own-account enterprises, as well as urban private enterprises, operated on a self-responsibility basis for their own profits and losses, enjoying full autonomy over employment, production decisions, sales and other operational matters. These market-oriented reform and opening-up policies greatly galvanized the enthusiasm and initiative of the general public to create material wealth.

Judging from the industrialization trajectories of Europe, the United States, Japan, and China's own recent successful industrialization, pursuing a heavy industrialization strategy at the initial stage of industrialization is not a viable path. The reason is that such a development strategy demands massive fiscal subsidies from the state. One approach to sustaining this heavy industry strategy is to have agriculture subsidize industry, as practiced in the former Soviet Union and early China. The consequence was widespread poverty among the rural population engaged in farming. Another approach, adopted by some Latin American countries, is to rely on foreign debt. Over time, these nations inevitably become heavily indebted, crushed by crippling debt-repayment burdens, and are eventually forced to sell off state-owned assets to service their debts. In reality, implementing a heavy-industry-focused development strategy in underdeveloped countries is unsustainable.

By contrast, rural industrialization centered on light industries such as textiles, with massive participation from a country's grassroots population, represents the correct path of industrialization.[14] This model requires no state subsidies; hundreds of millions of ordinary people escape poverty simply by taking part in it, and in doing so generate tremendous purchasing power and market demand. This surge in demand, in turn, naturally paves the way for the subsequent advancement of heavy industrialization. Once launched, this type of industrialization sustains its own momentum — in other words, this industrialization model is sustainable.

The core of China's reform and opening-up lies in the introduction of market mechanisms, which allowed rural collectives and individual farmers to decide for themselves what to produce. In the early stage of reform and opening-up, a great many TVEs naturally gravitated toward textile and garment processing as their line of business, for a fundamental reason: food and clothing constitute people's most basic subsistence needs, yet the two differ sharply in one key respect. Food has low consumption elasticity — once people's basic food needs are met, there is little room for further growth in food consumption. Grain output, moreover, cannot expand indefinitely, given the fixed supply of land: in most Chinese provinces, each farming household is allocated less than one mǔ of land.[15] Reliance on agriculture alone can therefore guarantee only basic food and clothing, leaving farmers no path to sustained income growth.[16]

By contrast, textile and apparel products feature high consumption elasticity and relatively simple production techniques: spinning, weaving, and garment manufacturing can each be broken down into discrete steps, allowing output to expand without any inherent ceiling. Well-managed TVEs could therefore grow to substantial scales, which is why textiles became the industry of choice for initial industrialization.[17] The sector's labor-intensive nature adds a further advantage: with its strong capacity to absorb labor, it could fully utilize the surplus labor hours freed up by efficiency gains under the household contract responsibility system. The initial industrialization of Britain, the United States, and Japan all originated in rural textile industries. Rural industrialization of this kind yields two major outcomes. First, rural residents achieve rising incomes on a broad scale. Second, the extensive expansion of rural industry creates enormous demand for electric power, energy, transportation, communications, and machinery manufacturing, thereby triggering rapid growth in heavy industry.

It can be seen from the above analysis that initial industrialization must be a process accessible to the broad masses of rural grassroot residents. Ordinary people at the bottom of society escape the poverty trap by participating in initial industrialization, and this same participation naturally fuels robust development of heavy industry that follows. China's development experience shows that only industrialization centered on light industries such as textiles, with extensive participation from grassroot people, can deliver large-scale and rapid poverty reduction. It needs to be emphasized that, at the early stage of reform and opening up, rural collective and own-account enterprises chose to engage in light industries such as textiles and garments entirely through voluntary market decisions. In other words, operating textile and garment businesses was a spontaneous, market-driven choice made independently by rural collective and own-account enterprises, as well as urban private businesses. Europe, the United States, Japan, and other regions likewise had massive impoverished populations prior to their industrialization, and their residents lifted themselves out of poverty via rural industrialization of textiles and other light industries with broad participation from ordinary grassroot people — there is no alternative path. This aligns with one core tenet of Marxism: that the people are the primary creators of material wealth.[18] Any successful industrialization must enable extensive participation from the broad grassroot masses.

## 4. The Core Factors Behind China's Economic Miracle

### 4.1 Political Economy Theory of New China

What are the core factors behind China's economic success? The experience of established developed countries cannot fully account for China's achievements. One major reason is that the growth of traditional industrialized nations, including Japan, relied heavily on seizing colonial land, raw materials, and markets, as well as on oppressing, exploiting, and enslaving colonized populations. China's economic success, by contrast, stems entirely from a peaceful rise. The economic takeoffs of South Korea, China's Taiwan region, China's Hong Kong region and Singapore appear to follow a trajectory more similar to China's economic miracle, yet differences remain. For instance, neither South Korea nor post-war Japan maintains full independent national defense; both were positioned as anti-communist, anti-socialist bulwarks within the U.S.-led alliance system. Post-war development in China's Táiwān region and Singapore followed similar logic. Furthermore, these economies and territories are far too small in land area and population to be compared meaningfully with China. Some might argue that China has possessed robust political, social, and cultural traditions throughout its history—yet did the Míng and Qīng dynasties not have these same traditions as well, without achieving comparable economic success?

Then what is the key factor that other developing countries lack, that China lacked under the Qing Dynasty and the Kuomintang regime before the founding of the PRC, but that New China alone possesses? It is the leadership of the CPC and China's political, economic socialist systems. In other words, China possesses a government with formidable executive capability. Specifically, since the founding of the PRC, the CPC organization have spread across all sectors of society, enabling unified policy enforcement, robust implementation capacity, and an independent and powerful national defense. The CPC is a revolutionary party forged through long years of revolutionary wars, and the exceptional organizational discipline it cultivated during that period is unmatched by political parties elsewhere in the world. In the political, economic and social construction of New China after its founding, this high degree of organization and discipline has translated into unified government decrees and extraordinary executive power. Furthermore, the Chinese People's Liberation Army, tempered by prolonged revolutionary war and grown stronger ever since, constitutes the bedrock of China's powerful national defense. The PRC's unified policymaking, strong execution capability, and independent, robust national defense have together guaranteed a stable and orderly environment for the country's political, economic and social development. The successful Industrial Revolutions of Western powers — including Britain, Germany, the United States, and Japan — between the 17th and 19th centuries all rested on strong state power, mighty armies and fleets, and the plunder of colonies, colonial raw materials, and colonial markets. A large country like China, without strong state authority and solid national defense, would only have ended up being carved up by Western powers, with no prospect whatsoever of economic development or industrialization. Huntington explicitly argues that political order is a precondition for modernization, and that rapid social mobilization without strong institutions produces instability rather than development.[19] Tilly argues state-building in Europe was driven by the need to organize coercion and extraction, and that strong states were co-constitutive with capitalist economic development.[20] Peter Evans in his book "*Embedded Autonomy: States and Industrial Transformation*" also argues that the development of modern industrial societies hinges on a strong central government to maintain social order and stability.[21]

However, even before the reform and opening-up, socialist China under the leadership of the CPC already possessed strong executive capability and had achieved notable progress in political, economic, and social development. Even so, the national economy stood on the brink of collapse in the pre-reform era,[22] with a large majority of the rural population and much of the urban population still living below the poverty line. What, then, is the most distinctive feature of post-reform China when compared with other developing countries — and even with pre-reform China itself? We argue that it lies in the fact that most of the decisions made by the Chinese government on political, economic, and social development since reform and opening up have been sound. This raises a follow-up question: on what theoretical framework has China's proactive government drawn to make these correct policy choices? In other words, does New China under CPC leadership possess its own political economy theory — and if so, what are this theory's stance, viewpoints and methodology?

New China is a country led by the CPC, so the fundamental stance of New China's political economy theory is determined by the CPC's fundamental purpose: serving the people wholeheartedly, particularly the working masses, with no ulterior motives of any kind. The purpose of "serving the people wholeheartedly" is embodied in the PRC's major political, economic, and institutional arrangements, as well as in the formulation and implementation of its policies. For more than a hundred years, Chinese working people suffered exploitation and oppression under feudalism, imperialism and colonialism. Their core interests boil down to living and working in peace and contentment, freedom from cold and hunger, and access to education and medical care for their children. In other words, the core interests of the broad Chinese people can be summarized as national security, economic growth, and the development of social infrastructure including education and healthcare. These core objectives have thus become the central goals of New China's construction and development. Whether these issues can be resolved serves as the touchstone for testing whether New China adheres to its purpose of "serving the people wholeheartedly".

The revolutionary mentors left extensive writings on the purpose of socialism. After the October Revolution, Lenin stated: "After the proletariat seizes political power, its paramount and fundamental interest lies in expanding output and greatly developing social productivity."[23] Stalin likewise put forward the view that "We must ensure the maximum satisfaction of the constantly growing material and cultural needs of the whole society through the continuous growth and improvement of socialist production on the basis of advanced technology."[24] The revolutionary mentors' successors carried this thinking further. Dèng Xiǎopīng put it more plainly: "Poverty is not socialism."[25] "Only socialism can foster cohesion, resolve people's hardships, avert polarization, and realize common prosperity step by step... Development can only be achieved with stability. Only the leadership of the Communist Party can deliver a stable socialist China."[26] And: "The essence of socialism is to liberate and develop productivity, eliminate exploitation and polarization, and ultimately achieve common prosperity."[27]

It should be noted that the CPC's stance and viewpoints have not undergone fundamental change before and after the reform and opening-up drive. Nevertheless, at the dawn of reform and opening up, the CPC Central Committee thoroughly summed up the successful experience and lessons of failure from New China's first three decades, accurately grasped the historical experience and major trends of global political and economic development, and resolutely shifted the CPC's central work focus from class struggle to economic development.

The shift in the CPC's central work focuses also brought about a transformation in its working methods. Following the launch of reform and opening up, the CPC's specific working methods can be summed up as: seeking truth from facts, analyzing specific issues on a case-by-case basis, conducting investigations and research, carrying out pilot projects, and the rolling out proven practices nationwide. In developed Western countries such as the United States, Wall Street elites are, in most cases, appointed as Treasury Secretaries, while distinguished professors from academic circles are recruited as economic advisors and senior officials of the Federal Reserve. Wall Street elites inevitably prioritize the interests of Wall Street. Meanwhile, those renowned university professors rarely conduct field research in factories, mines and rural areas. Lacking first-hand knowledge of real-world conditions, they formulate policies purely based on abstract market theories conceived in their own minds. By contrast, China's economic advisors are mainly government civil servants who are well versed in the CPC and government's policies and guidelines. They regularly conduct grassroot research and are far less susceptible to interference from interest groups. When problems arise, they carry out fieldwork across numerous factories, enterprises and rural areas to identify root causes, and formulate solutions proceeding from specific realities.[28] Throughout this process, economic theories take a back seat to seeking truth from facts; such theories merely serve as a reference framework for analyzing concrete issues.

Since the late 1970s, the CPC Central Committee has unswervingly adhered to the basic policy of reform and opening up in economic construction, and has adopted a series of decisions conducive to developing a socialist market economy and building a great modern socialist country in light of national economic conditions. In the early 1980s, the household contract responsibility system was introduced in rural areas, and rural collectives and individual farmers were allowed to engage in industrial and commercial operations. Amid the employment pressure brought by the large number of educated youths returning to cities, urban self-employed businesses were permitted. In addition, there are the following theoretical innovations.

First, the theory of gradual reform. The theory of gradual reform is the core strategy of China's reform and opening up. It advocates that reforms be advanced in a step-by-step manner — starting with the easier issues before tackling the harder ones, expanding from pilot points to broader areas, prioritizing incremental changes while gradually adjusting the existing stock. Its practical pathways include the household contract responsibility system in rural areas, pilot special economic zones, the dual-track price system, and the progressive restructuring of SOEs. The role of this theory is threefold: it effectively reduces the costs and risks of reform, avoiding the severe social upheaval that "shock therapy" might cause; it accumulates experience through local experiments, providing a reliable basis for nationwide promotion; and it maintains social stability while ensuring sustained and rapid economic growth. The gradual reform theory has successfully achieved a smooth transition from planning to market, offered methodological guidance for perfecting the socialist market economy with Chinese characteristics, and is hailed as the "Chinese wisdom" in transition economics.

Second, the "dual-track" theory. The "dual-track system" specifically refers to the price dual-track operation mechanism implemented during the early stage of China's transition from a planned economy to a market economy (the 1980s). Its core essence lies in the parallel existence, for the allocation of the same means of production, of two systems: the state-mandated planned low-price distribution and the market-adjusted negotiated high-price system. On the functional level, this theory embodies both transitional and strategic characteristics: it served as a "stabilizer" for reform, using planned low prices to maintain basic operations and social stability while avoiding inflation that might result from shock therapy; it also acted as a "nursery" for the market, where above-quota output could be sold at high prices, stimulating enterprise vitality and the rise of the private economy. At the same time, the system bred malpractices such as "official profiteering" (rent-seeking by bureaucrats). As a pragmatic compromise of a specific era, the dual-track system successfully bridged the old and new systems, and by the 1990s, the tracks were merged, fulfilling its historical mission in the gradualist reform process.

Third, the theory of the primary stage of socialism. In 1987, the 13th National Congress of the CPC systematically elaborated the theory of the primary stage of socialism, pointing out that China had entered socialist society but was still in a specific primary stage characterized by backward productivity and underdeveloped commodity economy. The principal contradiction at this primary stage is between the ever-growing material and cultural needs of the people and the backwardness of social production. On this basis, the CPC established the basic line of "one central task, two basic points"[29] and formulated the "three-step" development strategy.[30] This theory became the fundamental basis for formulating correct policies and guidelines, providing a powerful theoretical weapon for reform and opening up and for socialism with Chinese characteristics. It scientifically defined China's historical orientation in its development, successfully bridged theory and practice, and vigorously propelled the process of reform and opening up.

Fourth, the theory of the socialist market economy. The 14th National Congress of the CPC in 1992 put forward the goal of establishing a "socialist market economy." The Third Plenary Session of the 14th CPC Central Committee in 1993 elaborated on the specifics of implementing this system, including the development of labor, capital, and real estate markets. The emergence of a labor market represented a major theoretical breakthrough, discarding the traditional socialist notion that labor was not a commodity. The essence of the socialist market economy lies in transcending the institutional barriers between planning and the market, establishing that both are means of economic regulation, and affirming that socialism can and should develop a market economy. This theory established the basic economic system in which public ownership remains the mainstay while diverse forms of ownership develop side by side, and it emphasized that the market plays a decisive role in resource allocation under the state's macroeconomic regulation. In practice, this theory has had far-reaching effects: it has greatly liberated and developed social productive forces, propelled China's sustained and rapid economic growth, and elevated its comprehensive national strength; it has driven reform of the ownership and distribution systems and stimulated the vitality of market entities; accelerated the reform of SOEs and the establishment of a modern enterprise system; facilitated China's deep integration into the global economic system; and ultimately enabled the people's livelihood to advance from subsistence to moderate prosperity in all respects. The theory of the socialist market economy is a key theoretical pillar of reform and opening-up, providing a scientific guide and fundamental principle for the economic development of socialism with Chinese characteristics.

Fifth, the theory of state-controlled financial sector and banking. As China's level of economic development has continued to rise, its exploration of socialist market economy theory has continued to advance. Since the 1990s, China has continuously explored how the financial sector can serve socialism and the people amid market-oriented reforms. Financial sector and banks hold society's surplus capital and play a decisive role in guiding industrial development. The lessons of industrial hollowing-out and the subprime mortgage crisis, both triggered by profit-driven financial capital in the United States, serve as a warning for China. To this end, China held the Central Financial Work Conferences in 2018 and 2023, successively clarifying the fundamental mission of finance to serve the real economy and building a socialist financial system. The state issued special documents to strictly restrict financing for manufacturing relocation overseas, standardize reviews of outbound investment, and guard against industrial hollowing-out. Starting in 2021, China took initiative to deflate the real estate bubble and channel capital toward strategic emerging and high-tech industries.

Meanwhile, government venture capital funds were established to support high-risk high-tech industries. State-owned financial capital has been leveraged to stabilize capital markets and curb disorderly profit-seeking behavior by capital. These measures have helped China avoid the drawbacks of financial monopoly seen in Western countries, enabling the financial sector to serve national development priorities and people's well-being.

Finally, using industrial policy. Since the launch of reform and opening up, China's application of industrial policy has been relatively successful. From 1992 to 2010, the State High-Tech Development Plan (863 Program) funded research and development on power batteries and electric vehicles. Between 2010 and 2015, to cultivate an electric vehicle market, public-fleets vehicles — including municipal service vehicles and postal vehicles in major cities — were replaced with electric vehicles. From 2015 to 2019, purchase subsidies were offered to private electric vehicles equipped with power batteries manufactured by Chinese enterprises. Through these three decades of industrial policies, China has secured the world's leading position in power batteries and electric vehicles.

### 4.2 The heavy industrialization development strategy before reform and opening up

From the First Opium War in 1840 to the founding of the PRC in 1949, after more than a century of brutal aggression and bloody oppression by imperialist powers, the Chinese people finally stood up under the firm leadership of the CPC. The newly founded PRC was left devastated, with countless tasks awaiting reconstruction. Faced with the covetous gaze of Western powers, New China's top priority lay in national defense security. Only by building an effective national defense capability capable of resisting foreign aggression could the country achieve all-round political, economic, and social development. The top priority of the PRC's economic development was therefore to establish its own heavy industrial system — covering steel, energy, transportation and communications, aerospace, military industry and other sectors — along with a national security system.

From the perspective of economic development stages, China's industrial development from the founding of the PRC to the launch of reform and opening up fell within the catch-up industrialization stage centered on technology learning. During this stage, the country primarily imported and mastered existing technologies with well-defined technical pathways. To achieve rapid industrialization, many underdeveloped countries tend to launch major industrial projects through SOEs or government-business joint ventures. Developing countries generally suffer from underdeveloped infrastructure, backward technology, capital shortages, and unsound banking and financial systems. Faced with such widespread underdevelopment, it became an inevitable choice for these nations to pool capital through state-controlled financial systems and to advance industrialization strategies through SOEs. From the standpoint of development economics, when private firms lack sufficient capital to invest in heavy industrial projects, SOEs stepping in to develop heavy industry essentially means the state itself acts as a venture capitalist. From an economic perspective, developing technology- and capital-intensive heavy industrial projects inevitably entails high costs for underdeveloped nations. Without state subsidies, the SOEs undertaking these large-scale industrial projects could hardly have sustained operations. Where, then, did these state subsidies come from before reform and opening up? China relied mainly on the planned economic system and the mechanism of using agriculture to subsidize industry to fund major industrial projects. Furthermore, drawing on the state's high credibility and institutional standing, the government was able to allocate competent technical personnel and managerial cadres to state-owned heavy industrial enterprises.

In the 1950s and 1960s, with support from the former Soviet Union, New China swiftly built 156 heavy industrial projects. These projects enabled the country to manufacture the basic industrial goods required for the national economy, and established a comprehensive infrastructure system covering transportation and communications, electronics, energy, aerospace, military industries, among other. In sum, since non-SOEs were incapable of developing high-risk, high-tech, and capital-intensive projects within a short timeframe, New China had no alternative but to rely on SOEs to advance heavy industrial construction. Moreover, the initial industrialization drive cultivated a large contingent of engineers, enterprise managers, operational administrators, skilled technicians, and other professionals for China. Entrepreneurs emerge through diverse channels, and working in large enterprises equipped with advanced technology undoubtedly offers a shortcut for potential businessmen to learn corporate operation and management. Coase pointed out that one of the core bodies of knowledge for entrepreneurs needs concerns pricing mechanisms, including raw material prices, labor costs, loan interest rates, product sales prices and land prices.[31] Employees in SOEs, especially managerial staff, gain access to all dimensions of corporate operational and management information: pricing data, internal administration, raw material supply chains, product distribution networks, as well as institutional ties with government authorities and various social stakeholders. In other words, SOEs also functioned as incubators for cultivating entrepreneurs.

### 4.3 Land Supply Institutional Arrangements in China's Miracle of Economic Growth

When exploring the miracle of China's economic growth, one issue has long been overlooked by the economics community: how land was supplied efficiently to sustain China's rapid economic expansion following the reform and opening up. William Petty famously stated that, "Labor is the father of wealth, and land is the mother of it."[32] Compared with agrarian societies, industrialized economies require far less land overall, which explains why many small-territory nations have developed into affluent industrial powers. Nevertheless, industrial development remains reliant on land: economic growth in industrialized societies is driven largely by investments of numerous entrepreneurs on land, meaning all investment activities must have a physical land carrier. Large manufacturing enterprises demand expansive factory premises, while the construction of infrastructure — including airports, power stations, ports, highways, and railways — consumes even larger tracts of land. A core tenet of mainstream Western economics holds that entrepreneurial investment hinges on well-defined property rights, particularly clear land property rights. Accordingly, during the take-off stage of economic development, a developing country's ability to supply construction land for industrial and commercial enterprises constitutes the primary prerequisite for realizing economic take-off. A transparent, well-documented system for registering land and housing property rights enables entrepreneurs to secure mortgage loans using their residential properties and other assets, and also facilitates debt recovery by creditors and judicial authorities. As Hernando de Soto observed, the single most important source of funds for new businesses in the United States is a mortgage on the entrepreneur's house. Capital can only be transformed into productive capital once it is placed in the hands of entrepreneurs.[33] China has a massive population yet faces severe shortages of arable land and construction field, and it lacks the fully delineated land property right systems seen in developed economies. Even so, throughout its period of rapid economic expansion, China managed to deliver timely and efficient land supply to industrial, commercial, and financial enterprises, laying a solid land foundation for the economic take-off in the 1980s and the sustained high growth that followed. This raises a critical question: how have China's land and real estate institutional arrangements adapted to the demands of economic growth since the launch of reform and opening up?

In the early 1950s, the PRC carried out a thorough land reform, confiscating land owned by landlords and capitalists. It introduced collective ownership of rural land at the village level and state ownership of urban land, with administrative villages serving as the carrier entity for collectively owned rural land. The household contract responsibility system implemented in rural areas after the reform and opening up essentially allocated the usufruct rights over village collective land to individual farmers, injecting tremendous vitality into the rural economy and greatly boosting production efficiency. Following the reform and opening up, urban state-owned land could be provided by the government to all types of industrial and commercial enterprises in exchange for fees, or sold to land developers for residential or commercial construction. Starting in November 1988, the government levied a land use tax on urban state-owned land as a type of local tax, with the revenue used mainly for urban construction and maintenance. Under the tax-sharing reform of 1994, all land transfer proceeds were designated as revenue for local governments. Collectively owned rural land can only be supplied to enterprises by governments at all levels acting on behalf of the state, after the government has expropriated it and converted it into state-owned land. The term of land use for industrial and commercial production purposes is 50 years. The Property Law of 2007 stipulates that the term of use for residential housing is 70 years. Since the duration of residential land use concerns the fundamental interests of hundreds of millions of households, such land-use rights shall be automatically renewed free of charge upon expiration of the 70-year term. Entrepreneurs holding housing property rights may use their residences as collateral to obtain bank credit.

Since the launch of reform and opening up, the Chinese government has consistently adhered to Dèng Xiǎopīng's guiding principle that "development is the absolute principle."[34] Put more plainly, the core policy objective of the Chinese government is to pursue economic growth, and the promotion of government officials is largely based on their performance in driving local economic expansion.[35] To drive economic growth, alongside SOEs, the government also needs to attract foreign investment and guide the development of domestic private firms. Accordingly, enterprises primarily obtain land for their business operations through government channels. To deliver robust economic growth, governments at all levels have streamlined land provision procedures for enterprises. Many local governments have even supplied construction land to investors free of charge in order to attract foreign capital. Additionally, numerous local governments have established industrial parks and economic development zones by allocating state-owned land to facilitate investment and reduce operational hurdles for businesses. Since the commercialization of urban housing in the mid-1990s, land for residential construction has mainly been supplied by governments at all levels via auctions of state-owned land-use rights. Revenue generated from these land auctions flows into local government fiscal coffers, creating strong incentives for local authorities to sell land. Between 2001 and 2010, land concession revenue as a share of local government's fiscal income surged from 16.6 percent to 76.6 percent.[36] Because of this institutional framework governing residential land supply, hundreds of millions of urban households have been able to acquire housing with full property rights in their own names. Through such institutional arrangements, the Chinese government has secured adequate land for corporate investment and mitigated excessive transaction costs that would otherwise stem from ambiguous land property rights. This corresponds to what Francis Fukuyama describes as China's "good enough" property rights regime.[37]

By distributing land-use rights to hundreds of millions of rural households in China, and by allowing farmers to make their own production decisions in response to market signals and to allocate their own labor between agricultural and non-agricultural work, a Pareto improvement was achieved. During the rural reform, the welfare of countless rural households was enhanced while virtually no group suffered losses. In Chinese cities, residential construction land is supplied through land auctions, enabling hundreds of millions of urban households to acquire property rights to commercial housing. Meanwhile, local governments at all levels generate substantial fiscal revenue, which can be invested in public undertakings including education, healthcare, social security and upgrades to public facilities. This also constitutes a Pareto improvement.

China's unique institutional arrangement for land property rights constitutes the primary prerequisite for China's economic take-off. Guided by the central government's goal of pursuing economic growth, local governments at all levels supply industrial and commercial land to private and foreign-invested enterprises by developing economic development zones, industrial parks, and other platforms. They also provide land for urban residential construction through auctions of residential land-use rights. Through rural land expropriation, collectively owned rural land is converted into state-owned land, releasing large tracts of land for the construction of public infrastructure such as highways, railways, power stations and ports. Relying on this government-led land supply mechanism, China has facilitated corporate investment, residential housing development, and the construction of public infrastructure. Since reform and opening up, the uniqueness of China's land institutional framework lies in its capacity to deliver an effective land supply for rapid economic growth without privatizing land — in other words, land ownership hasn't become bottlenecks for economic and social progress. This is arguably an unprecedented case in the global history of industrialization.

China has a large population relative to its limited land resources. Historically, land concentration in the hands of a small number of big landlords — leaving the vast majority of farmers without a plot to live and farm on — was the primary driver of extreme wealth polarization, peasant uprisings, and dynastic collapses in ancient China. This was also the core motivation behind the state-owned land systems and land allocation policies implemented during the Northern Wèi, Suí, and Táng dynasties. Nevertheless, rapid population growth meant that dynasties practicing land allocation systems, such as the Táng, eventually ran out of land to distribute in its later stages. As the landless population expanded, poverty became widespread. Coupled with natural disasters and social unrest, this ultimately contributed to the dynasty's collapse. Even after the founding of the PRC, which completely abolished private land ownership, rapid population growth[38] and the lack of viable livelihood options outside farming constituted the main cause of widespread poverty prior to reform and opening up, putting the country in the Malthusian trap.[39] After reform and opening up, however, local governments across China established industrial parks and supplied industrial land to enterprises of all kinds free of charge or at low cost — a key factor behind China's rapid industrialization. Meanwhile, fast-paced industrialization generated an explosive surge in demand for industrial labor, creating massive job opportunities for landless rural workers and driving large-scale poverty eradication. In short, successful industrialization allowed China to escape poverty and break free from the Malthusian trap.

Over the first two decades of the new century, the rapid expansion of the real estate sector fueled robust growth in China's economy. Local governments at all levels garnered massive fiscal revenues from land auctions, laying a financial foundation for improving people's livelihoods and upgrading urban infrastructure. In 2018, the United States launched a trade war against China and imposed an embargo on high-tech products such as chips and lithography machines, intensifying high-tech competition between the two countries. Against this backdrop, China's principal task shifted from sustaining high economic growth rates to boosting the development of high-tech industries. Accordingly, the Chinese government took the initiative to deflate the real estate bubble starting in 2021. This move was intended, for one thing, to avert a financial crisis triggered by a real estate meltdown like the one that hit the United States in 2007-2009. For another, it aimed to channel investment toward strategic emerging and high-tech industries. At that time, China already held a leading global position in electric vehicles, power batteries, and solar and wind power equipment, while its chip manufacturing industry was also taking shape. This meant high-tech industries were fully capable of replacing real estate as the new engine of economic growth. In hindsight, the 2021 policy to deflate the real estate bubble and redirect capital from the real estate sector toward high-tech industries has proven entirely correct. It has spurred vigorous development across solar power, wind power, power batteries, electric vehicles, semiconductors, and other high-tech sectors, propelling China to emerge as a global high-tech power.

### 4.4 Human Capital Preparation in China

In the early days of the PRC, the illiteracy rate in rural China exceeded 95 percent, while the enrollment rate of school-age children stood at only around 20 percent.[40] To reverse this situation, New China advanced education through two major approaches: launching mass literacy campaigns and implementing compulsory education. The literacy drive was framed as a key political task. Institutionally, a Central Committee for Literacy Work was established to guide the orderly rollout of literacy programs. Practically, a wide array of literacy classes tailored to adults and working people were introduced, including winter schools, night schools, and rapid character learning methods.[41] Since the reform and opening up, the Chinese government has safeguarded the delivery of compulsory education through the Constitution (the 1982 Constitution) and national education legislation (the Compulsory Education Law of the PRC enacted in 1986). Compared with developing countries such as India, one of China's most remarkable achievements since its founding — especially after reform and opening up — lies in human capital development. The nationwide popularization of nine-year compulsory education has drastically lifted the population's literacy rate and educational attainment, laying a solid human resource foundation for China's rapid economic growth following the reform and opening up. Amartya Sen, the Nobel laureate in Economics and Harvard University professor, highlighted this point in a speech delivered at Peking University on February 25, 2016: "It is clear that in China education has been not only a major force for social transformation but also a driving force of industrial transformation. I do not think China's success can be understood without linking its industrial revolution to its remarkable achievements in educational expansion. This is not a discussion about India, yet if there is a flaw in India's development thinking, it is this: industrialization cannot realistically take root in India before it cultivates a workforce that is well-educated and physically healthy." Over the 22 years from the restoration of the national college entrance examination in 1977 to 1998, Chinese universities admitted a cumulative total of 13.55 million students. In the subsequent 20 years of university enrollment expansion spanning 1999 to 2018, cumulative college admissions surged roughly sevenfold to approximately 107 million students. By 2023, the stock of university graduates in China reached 240 million.[42]

**5. The key to China's successful industrialization lies in the introduction of market mechanisms**

Prior to the reform and opening up, China had already laid the basic groundwork for industrialization. The reform and opening up served to unleash the tremendous momentum accumulated since the founding of the PRC. The catalyst Dèng Xiǎopīng deployed to unlock China's vast latent potential lay in the gradual introduction of market mechanisms into the country's economic system. Launched in the late 1970s, China's economic reform essentially entailed incrementally embedding market forces within a planned economy dominated by urban state-owned sectors and rural collective economies. In the early 1980s, the introduction of the household contract responsibility in rural areas, together with permission for TVEs and non-state urban businesses to develop, represented continuous steps toward incorporating market elements. In October 1984, the Third Plenary Session of the 12th CPC Central Committee clarified that China's economic reform was geared toward developing a socialist planned commodity economy based on public ownership. At the 13th National Congress of the CPC in 1987, it was clarified that China was in the primary stage of socialism, and it was pointed out that on the premise that the state-owned economic sector played the leading role, diverse economic ownerships were allowed to coexist. In November 1993, the Third Plenary Session of the 14th CPC Central Committee adopted the Decision on Several Issues Concerning the Establishment of a Socialist Market Economic System. From the mid-1990s onward, China carried out the shutdown, consolidation, merger, and transformation of loss-making small and medium-sized SOEs, guided by the reform principle of "invigorating the large while relaxing oversight of the small" (抓大放小). These measures restored SOEs to their proper functional roles. China's successful accession to the World Trade Organization in 2001 delivered two major outcomes: it facilitated foreign firms' access to China's domestic market and Chinese goods' entry into global markets, while also integrating international market forces into China's economy.

Over more than four decades of the reform and opening up, China has continuously introduced market mechanisms into its economic system once dominated by SOEs and collective agriculture, giving full play to the three major strengths of a proactive government, SOEs and non-SOEs, thereby accomplishing industrialization at a rapid pace. A landmark of China's successful industrialization came in 2010, when China surpassed Japan to become the world's second-largest economy.

Wén argues that successfully industrialized nations — Europe, the United States, Japan, as well as present-day China — have largely followed a similar industrialization trajectory. First, a strong government safeguards the domestic market and the overall environment for economic development, while also exploring international markets. Second, widespread preliminary development of rural industries dominated by light textile sectors converges to form a vast ocean of market economy. Third, massive demand generated after the initial success of industrialization — for power, energy, transportation, communications and machinery manufacturing — fuels the vigorous expansion of heavy industry.[43] Western developed countries spent over a century completing industrialization through extremely brutal means: massacring indigenous peoples, waging wars of invasion, plundering colonies, engaging in the slave trade and slavery, trafficking opium, and monopolizing international markets. By contrast, China accomplished industrialization within more than three decades amid peaceful development. From the perspective of the global international environment, China mainly leveraged the international trade, finance, and investment system established after World War II.

Furthermore, since the founding of the PRC — and especially after the launch of reform and opening up — China has put in place the following necessary preconditions for industrialization. First, industrial societies are primarily run by industrial and commercial enterprises and entrepreneurs; therefore, for a country to grow its economy, its political, economic, and institutional arrangements must permit the continuous emergence, survival and expansion of industrial and commercial enterprises and entrepreneurs under all forms of ownership. Second, industrial societies require a steady supply of employable labor, which necessitates the establishment of a labor market within the country's political, economic, and social institutional framework. Third, beyond blue-collar workers, industrial societies demand large numbers of technical and managerial staff — that is, white-collar employees. Industrialization thus compels a country to build a comprehensive system of universal education, as well as scientific and technological research, capable of cultivating and training managers, engineers, researchers, accountants, lawyers, and other professionals for industrial, commercial, and financial enterprises. Fourth, goods and services produced by industrial and commercial enterprises need outlets for sales, which calls for the existence and sound development of markets for goods and services. Fifth, to foster the innovative drive and capacity of enterprises and entrepreneurs, the state must effectively protect property rights over intellectual property and wealth created by businesses and innovators. Sixth, industrial societies rely on stable yet flexible financial and financing systems. From Schumpeter's perspective, credit constitutes the most essential factor of production for entrepreneurs and stands as the core driver of economic progress in industrial societies.[44] Only when household savings are channeled to entrepreneurs through capital markets and lending institutions can an economy break free of a simple circular flow and achieve rapid growth. Lastly, as argued in Section 4.1 of this chapter, sustainable economic and social development in a modern industrial society is only achievable when a country is governed by a strong central authority that maintains social stability and public order.

## 6. Conclusion

Since the 20th century, only a handful of underdeveloped countries and regions have successfully achieved full industrialization, namely Japan, the former Soviet Union, the Republic of Korea, Singapore, China's Taiwan region and China's Hong Kong region. Many economies once hailed as promising development stars, including Argentina, Chile, Mexico, the Philippines, Iran, Iraq, Turkey and Egypt, have long faded into obscurity. China is the first giant nation with a population of 1.4 billion to accomplish industrialization in human history.

There are diverse views on the explaining China's economic miracle. This chapter argues that its emergence is mainly attributable to the state capability of the CPC and the Chinese Government, together with the reform and opening-up policies the Chinese government has unswervingly implemented since the late 1970s. The essence of China's reform and opening-up policies lies in the gradual introduction of market mechanisms into an economic system once dominated by a planned economy and SOEs. In the early 1980s, TVEs, urban individual businesses, and private enterprises naturally opted to engage in light textile industries through voluntary, market-oriented operations. Hence, China's early industrial success at that time was largely a product of the development of the market economy. Large-scale, rapid poverty reduction can only be achieved through an industrialization process that allows the broad masses of ordinary people to participate extensively — there is no alternative path.

In the mid-to-late 1990s, China carried out thorough reforms of its economic system, with landmark measures including the establishment of a socialist market economic system; reforms to the banking, financial, fiscal and taxation systems to align them with a socialist market economy; the shutdown, consolidation, merger, and transformation of numerous loss-making small and medium-sized SOEs; and the restructuring of large SOEs of strategic significance to national economic security and people's livelihoods, aimed at establishing a modern corporate governance system. These restructured state-owned sectors achieved marked efficiency gains. Collectively, these reform initiatives greatly improved China's overall economic operating environment. Furthermore, the early-stage industrialization driven by the reform and opening up generated massive demand for machinery manufacturing and infrastructure, covering railways, highways, shipping, aviation, ports, airports, communications, power and energy, real estate development, and among others. To meet such robust demand, a large number of restructured and corporatized SOEs launched massive investment and expansion in machinery manufacturing and infrastructure construction, fueling a boom in infrastructure development. Another pivotal milestone was China's accession to the World Trade Organization (WTO) at the start of the new century, which triggered a massive inflow of foreign direct investment (FDI) throughout the first decade of the 2000s and delivered dramatic growth in China's import and export trade, ushering in a flourishing period for both foreign capital and cross-border trade.

Although the heavy industrialization strategy and rural collective economy adopted in the early days of the PRC failed to lift the Chinese people out of poverty, pre-reform China laid all the foundational prerequisites for industrialization. These included robust state governance and institutional frameworks, a unified national market, power and energy facilities, heavy industrial production capacity, transportation and communications infrastructure, state economic management expertise, and a workforce of hundreds of millions educated engineers and skilled laborers, among others. The reform and opening up served to unleash the enormous latent momentum accumulated since the founding of New China. The key catalyst that unlocked China's massive untapped potential lay in introducing market mechanisms into its economic system.

To gradually introducing market mechanism into China's planned economy, there has been a series of theoretical innovation since the late 1970s, such as the gradualism of reform, the dual-track system, the primary stage of socialism, the socialist market economy, the state-controlling of financial and banking sector, the successful use of industrial policies, among others.

Since the launch of reform and opening up, China's successful industrialization — or the success of the "China Model" — stems from the combined effects of a proactive government, the state-owned economy, and the market economy.

Notes:

[1] Sen, 1999, p. 41-43.
[2] Wén, 2017, p. 20-60.
[3] Data source: Self-calculated based on historical data from the National Bureau of Statistics of China, http://data.stats.gov.cn/easyquery.htm?cn=E0103.
[4] Ibid.
[5] National Bureau of Statistics of China, 1992, p. 346.
[6] National Bureau of Statistics of China, 1997, p. 360-380.
[7] Ravallion and Chen, 2007.
[8] Wu, 2009.
[9] Wén, 2017, p. 20-60
[10] Ibid.
[11] Ibid.
[12] Ibid.
[13] Ravallion and Chen, 2007.
[14] Wén, 2017, p. 20-60.
[15] One hectare is equivalent to 15 mu.
[16] Ibid.
[17] Ibid.
[18] Marx, 1875/1989, p. 81. "Labor is the source of all wealth and all culture."
[19] Huntington, 1968, p. 1.
[20] Tilly, 1992, p. 269.
[21] Evans, 1995, p. 323.
[22] Central Committee of the Communist Party of China, 1981.
[23] Lenin, 1921, p. 586.
[24] Stalin, 1951, p. 31.
[25] Dèng Xiǎopīng, 1984, p. 62-66.
[26] Dèng Xiǎopīng, 1990, p. 357-358.
[27] Dèng Xiǎopīng, 1992, p. 370-383.
[28] Yang, 2018.
[29] Dèng Xiǎopíng, 1989, p. 326. "The One Central Task" refers to taking economic development as the central task; "the Two Basic Points" refers to adhering to the Four Cardinal Principles and persisting in reform and opening-up. They constitute the core elements of the Party's basic line for the primary stage of socialism.
[30] Following the Third Plenary Session of the Eleventh Central Committee of the CPC, China's economic development strategy was broadly divided into three steps. The first step was to double the gross national product compared with 1980 to meet the people's basic needs for food and clothing. This task had been basically accomplished. The second step was to double the gross national product once again by the end of the century, enabling the people to live a moderately prosperous life. The third step was to raise the per-capita gross national product to the level of moderately-developed countries by the middle of the next century, ensure a relatively prosperous life for the people and basically realize modernization. We would then forge ahead on this basis.

[31] Coase, 1937.
[32] Petty, 1662/2011, p. 7-45.
[33] Soto, 2000, p. 11-35.
[34] Dèng Xiǎopīng, 1992, p. 377.
[35] Zhou, 2007.
[36] Xiang, 2014, p. 139-141.
[37] Fukuyama, 2011, p.25-30 and 99.
[38] China's population surged from 540 million in the early years of the People's Republic to 960 million in 1978.
[39] Thomas Robert Malthus (1766-1834), a British economist, pointed out in An Essay on the Principle of Population (1798) that population grows geometrically while subsistence resources only expand arithmetically, which inevitably leads to famine, wars and diseases. He called for decisive measures to curb the birth rate.
[40] Li, 2013.
[41] Asai Kayoko, Wang, & Liu, 1997; Peng, Yao, & Huang, 2016.
[42] Xí Jìnpíng, 2025.
[43] Wén, 2017, p. 20-60.
[44] Schumpeter, 1934, chapter 2 and 3.

## How was China's "market economy" established?

### 1. Introduction

What drove China to overtake Japan in 2010 to become the world's second-largest economy, and to emerge by the 2020s as a global high-technology power? Many would point to "reform and opening up" as the answer. But what does "reform and opening up" actually mean in substance? In my view, its essence lies in the gradual establishment of a "socialist market economy" in China. The decisions underpinning this process can be divided into two broad areas: first, the construction of a "socialist market economy"; and second, the opening up to the outside world, which allowed overseas capital to enter China and establish joint ventures, wholly foreign-owned enterprises, and other market entities, while progressively encouraging Chinese enterprises to venture abroad and engage with international markets since the turn of century onward. Through a chronological review of the extensive body of decisions made in the course of reform and opening-up, this chapter finds that a market economy could never have taken root in China without the CPC and the Chinese government overcoming enormous obstacles to formulate and resolutely implement decisions on building a "socialist market economy" and pursuing "opening-up." China's "socialist market economy" was therefore gradually established under the leadership of the CPC.

Over the more than 40 years since reform and opening-up, China — a large socialist developing country under the leadership of the CPC — has achieved industrialization, become the world's second-largest economy, and emerged as a global high-tech power through the construction of a "socialist market economy" and the pursuit of an "opening up to the outside world" policy. This has given developing nations across Asia, Africa, and Latin America reason to hope for their own economic development. China's economic reforms had no "blueprint" whatsoever. The policies and measures implemented between the rural household responsibility system introduced after the Third Plenary Session of the 11th CPC Central Committee and the establishment of the "socialist market economy" at the 14th National Congress of the CPC in 1992 were all formulated and carried out in response to the political, economic, and social conditions of the time — or, one might say, ideological constraints were broken through incrementally, step by step. China's real-world historical experience — advancing construction through trial and error, and formulating institutional frameworks incrementally as it builds a socialist market economy — has overturned the tenet of Western economics that sound institutional systems must be fully established prior to economic growth.

The success of China's economic development has also led the United States, Europe, and Japan to perceive the Chinese development model as a threat to their so-called "liberal democratic" development model. As a result, the U.S. government has designated China its "primary challenger" and "greatest strategic competitor," launching a trade war against China, imposing embargoes on high-technology products and services, playing the Táiwān and Xīnjiāng cards to interfere in China's internal affairs, and—building on the U.S.-Japan-India-Australia Quadrilateral Security Dialogue (QUAD)—establishing the tighter core grouping of AUKUS (Australia, United Kingdom, United States) to contain China. The Trump administration, which returned to power in 2025, unleashed a trade war on the entire world. Most countries except China capitulated to the United States; however, when China responded with rare earth export controls as a countermeasure, the United States was compelled to seek a settlement with China on trade matters. Looking back on the process by which China achieved industrialization, became the world's second-largest economy, and emerged as a global high-tech power through the establishment of a “socialist market economy” and the pursuit of an “opening up” policy, we can draw the following conclusion: only by continuing to uphold and refine the “socialist market economy” and the policy of “opening up to the outside world” can China break through the containment imposed by the United States and other countries and grow ever stronger.

Part Two offers a literature review on the development of China's socialist market economy. Part Three traces the major policy decisions through which the Chinese government progressively built the socialist market economy. Part Four examines the key decisions that shaped China's opening-up drive. Part Five assesses the achievements that have flowed from developing the socialist market economy. Part Six considers the contributions of additional factors to China's economic success. Part Seven draws conclusions.

## 2. Literature Review

Over the past decade or more, a substantial body of literature has emerged on the construction of China's “socialist market economy.” Zhu Zhixin and Yu Kang[1] approach the subject first from the standpoint of the theory of socialism's essential nature, analyzing the internal logic of Dèng Xiǎopíng's “socialist market economy” theory. Adhering to the essence of Marxism, Dèng Xiǎopíng held that “The essence of socialism is the liberation and development of the productivity, the elimination of exploitation and polarization, and the ultimate realization of common prosperity.”[2] They then proceed from the premise of China's primary stage of socialism—the most fundamental national condition—to identify the principal contradictions, core tasks, and basic objectives of that historical stage, as well as the path of economic structural reform, thereby bringing the ideal model of socialism back into alignment with objective reality and achieving an overall transformation of the socialist model. Finally, from a practical standpoint, they argue that the development of a market economy must be combined with the improvement of fundamental institutions. In 1992, Dèng Xiǎopíng observed: “It will probably take another thirty years for us to develop a full set of more mature and well-established institutions across all fields.”[3] On the question of how to handle the relationship between planning and the market, he stated: “If we handle this well, it will greatly boost economic development; if we mishandle it, things will go badly.”[4] Dèng Xiǎopíng also attached great importance to the development of science and technology, stating that “the tremendous growth of social productivity and the sharp rise in labor productivity rely mainly on the power of science and technology.”[5] On the subject of opening up to the outside world, Dèng Xiǎopíng argued that today's world is an open world, and that “invigorating the domestic economy and opening to the outside world are not short-term policies but long-term ones; they will remain unchanged for at least fifty to seventy years.”[6]

Yu Jincheng[7] argues that the theory of the “primary stage of socialism” provided the theoretical foundation for China's adoption of “market economy” methods. Confronted with the problem of widening poor and rich polarization and questions over whether reform should be deepened, the “Three Represents” thought resolved the difficulty of keeping reform moving forward by revising the nature and purpose of the CPC: “In terms of nature, it was adjusted from being the vanguard of the working class to being simultaneously the vanguard of the working class and the vanguard of the Chinese nation and the Chinese people; in terms of purpose, it was adjusted from serving the people wholeheartedly to representing the development requirements of China's advanced productivity, representing the progressive direction of China's advanced culture, and representing the fundamental interests of the overwhelming majority of the people.” This adjustment to the nature and purpose of the CPC provided the theoretical basis for upholding and deepening reform. In the face of severe rich-poor polarization, the concept of the “Scientific Outlook on Development” called for coordinating urban and rural development, regional development, socioeconomic development, and the harmonious development of humanity and nature, while also championing collectivist values. Its underlying purpose was for the government to intervene through macroeconomic regulation in the market's spontaneous tendencies—allowing the market to play its role in allocating resources, while at the same time having the government extend assistance to underdeveloped regions and vulnerable groups, so as to counter a pure dynamic of survival of the fittest.

Wei Xinghua and Li Xianlin[8] review the tortuous course through which China established the socialist market economy system, as well as the contributions of the older generation of revolutionaries to this process. In 1979, when receiving foreign guests, Dèng Xiǎopíng proposed that China too could develop a “market economy.” That same year, Chén Yún and Lǐ Xiānniàn put forward the theory of “planning as primary, market as supplementary.” Then in 1992, Dèng Xiǎopíng advanced his famous proposition: “A planned economy is not equivalent to socialism — capitalism also has planning; a market economy is not equivalent to capitalism — socialism also has markets.”[9] This definitively established the dominant position of the “socialist market economy” in China.

Wei Liqun[10] reviews Dèng Xiǎopíng's major statements on the market economy from the late 1970s to the early 1990s. Gu Yumin[11] argues that the development of “socialist market economy” theory and practice in China since reform and opening-up constitutes a revision, innovation, and advancement of Marxist theory on socialist construction. Gu Hailiang[12], Long Xuchao[13], Qian Lubo and Zhang Zhanbin[14], Xi Jianwu[15], Zhu Andong and Sun Jiemin[16], Zhou Wen and Liu Shaoyang[17], among others, have each reviewed the CPC's decisions regarding the construction of the “socialist market economy” since the Third Plenary Session of the 11th Central Committee of the CPC in 1978. The literature on China's “socialist market economy” is vast, and no attempt is made here to enumerate it exhaustively.

The studies reviewed above approach the establishment of China's “socialist market economy” primarily by examining the statements of major leaders since reform and opening-up, as well as the evolution of reports from successive CPC congresses and decisions of the CPC’s Central Committee plenums. They make virtually no mention of the government policies formulated and implemented in the wake of those congress reports and plenum decisions, nor do they address the promulgation and enforcement of the laws, regulations, and policies governing China's “opening up to the outside world.” Still less do they touch on the concrete facts and outcomes of China's economic growth and expansion since reform and opening-up, or on the other factors that have contributed to China's economic success in the post-reform era. It is precisely these aspects of the construction of the “socialist market economy" — absent from the existing literature reviewed above — that the present chapter sets out to examine.

### 3. Policy Decisions in the Construction of China's Socialist Market Economy

Prior to reform and opening-up, China pursued a heavy industry development strategy subsidized by rural agricultural production. The countryside operated under a collective economic system, while cities developed heavy industry through a planned economy built around SOEs, leaving no room for market forces to operate. Heavy industry is capital- and technology-intensive; however, in the 1950s and 1960s, China was a poor agrarian country beset by capital shortages and technological backwardness, and was therefore compelled to subsidize heavy industry through agriculture. With Soviet assistance, China undertook 156 heavy industry projects during this period. The Chinese countryside, when the PRC was just set up, was naturally extremely impoverished and underdeveloped. The price scissors policy — which artificially suppressed agricultural prices while inflating the prices of industrial goods — combined with the inefficiency of the rural collective economy, kept the vast majority of rural residents trapped in poverty from the PRC's founding right up to the eve of reform and opening-up. Rural poverty stemmed primarily from three factors: an excessively large rural population, agriculture's limited capacity to provide employment, and urban heavy industry's near-total inability to absorb rural labor — resulting in severe hidden unemployment in the countryside. Because the national development priority before reform and opening-up was heavy industry, resources allocated to light industry were limited, constraining urban industry's capacity to absorb workers; cities, too, suffered from an oversupply of labor, a situation that was the principal driver of the "sending educated youth to the mountains and countryside" movement. The underdevelopment of light industry had a twofold consequence: it could neither adequately supply urban and rural residents with everyday consumer goods—giving rise to a "shortage economy"—nor generate sufficient demand for the products of heavy industries such as transportation, communications, energy, electric power, and machinery manufacturing, leaving heavy industry's development without a sustainable source of momentum. Such was the state of China's economic development on the eve of reform and opening-up.

#### 3.1 Policy Decisions on China's Development of a Planned Commodity Economy

The convening of the Third Plenary Session of the 11th Central Committee of the CPC in December 1978 brought about a great "historic turning point" in China, launching the nation on its course of "reform and opening-up." The plenum called for shifting the central focus of the CPC and state work from "class struggle" to "economic construction," for "transforming the relations of production and the superstructure that are incompatible with the development of the productivity," and for resolutely "acting in accordance with economic laws, attaching importance to the role of the law of value, combining ideological and political work with economic means, and fully mobilizing the initiative of cadres and workers in production." Significantly, the resolution of the Third Plenary Session was the first to mention "attaching importance to the role of the law of value" and to call for mobilizing workers' enthusiasm through economic means — marking the first step toward a market economy. Reform was set in motion in the wake of the Third Plenary Session. In response to the difficulties faced by the countryside and factories in selling their output, governments at all levels worked to open up channels of circulation, permitting individuals and collectives to engage in long-distance trade and allowing commercial enterprises to expand their procurement channels. On the question of how to reform the countryside, opinions diverged widely from the central to the local level. Most areas of Anhui, Sichuan, Guizhou, and other provinces had adopted the systems of output quotas linked to work groups and household contract production by the end of 1980, while provinces such as Heilongjiang, Jiangsu, and Hebei did not begin implementation until after the issuance of the No. 1 Document in 1982.[18] The rural household contract responsibility system thus emerged through a bottom-up process that ultimately received endorsement from above.[19]

Urban reform proceeded through the gradual expansion of SOEs' autonomy in management and operations, with the “substitution of tax payments for profit remittances” (利改税) policy introduced for the SOEs in 1983. In the late 1970s, large numbers of educated youth who had been sent to the countryside returned to the cities, sharply intensifying urban employment pressures. To maintain social stability, the central government permitted returning youth to seek their own livelihoods. In 1981, the State Council promulgated the “Several Policy Provisions on Urban and Township Non-Agricultural Individual Economy,” breaking the restrictions on private business operations.

Following the introduction of the household contract responsibility system, capable peasants began engaging in individual non-farm businesses and even hired workers. The CPC Central Committee's “No. 1 Document” of 1983 gave formal endorsement to rural individual businesses and the hiring of labor. In March 1984, the CPC Central Committee and State Council forwarded the Ministry of Agriculture, Animal Husbandry and Fisheries' “Report on Opening Up a New Situation for Village and Township Enterprises” (Central Document No. 4), which approved renaming these enterprises “township and village enterprises (TVEs),” broke through the restriction that rural enterprises could only be run collectively, and extended recognition to enterprises established by individual households or by households in association. The document described TVEs as “an important component of diversified types of business, an important pillar of agricultural production, an important pathway for the broad masses of farmers to advance toward common prosperity, and an important new source of national fiscal revenue” — “an important force in the national economy and an important supplement to SOEs” — and called for these enterprises to be “treated on equal footing with SOEs and provided with necessary support.”

In 1981, the Resolution on Certain Questions in the History of the CPC Since the Founding of the PRC, adopted at the Sixth Plenary Session of the 11th CPC Central Committee, affirmed the direction of reform. It stated: “We must practice a planned economy on the basis of public ownership while letting market-based adjustment play an auxiliary role,” and stressed that “we must vigorously develop socialist commodity production and commodity exchange.” Drawing on experience gained from economic restructuring, the report to the 12th National Congress of the CPC in 1982 identified the framework “relying mainly on a planned economy with market-based adjustment as an auxiliary” as the fundamental one of China’s economic systems.

In 1984, the Third Plenary Session of the 12th CPC Central Committee adopted the Decision of the Central Committee of the CPC on the Reform of the Economic Structure. For the first time, the Decision put forward the proposition to "establish a planning system that consciously applies the law of value and develop a socialist commodity economy." It also stipulated that "we should pursue the policy of encouraging development by the state, collectives and individuals alike, and persist in developing diverse forms of ownership and modes of operation." The document further clarified that "on the whole, what we practice in China is a planned economy, namely a planned commodity economy, rather than a market economy entirely governed by market forces." In the terminology of the former Soviet Union, a "commodity economy" equates to a "market economy." For China, which had implemented a fully planned economy before 1979 and followed the principle of "relying mainly on a planned economy with market-based adjustment as an auxiliary" prior to the release of this Decision, the new formulation represented a major leap forward. Nonetheless, the Decision contained an additional caveat: "Under China's socialist conditions, labor is not a commodity, nor are land, mineral resources, banks, railways, SOEs, and all other state-owned resources."

It is evident that the traditional socialist notion that labor is not a commodity was still restraining the pace of reform and opening up. Even so, in the mid-1980s, the government steadily remove the range of products subject to mandatory planning, abolished the system of first-, second-, and third-tier wholesale stations to enable direct dealings between producers and distributors, introduced the contract responsibility system for SOEs along with the factory director accountability system under this framework, and permitted some enterprises to adopt the joint-stock system. Three of the CPC Central Committee's "No.1 Documents" issued between 1984 and 1986 authorized the development of shareholding cooperatives in rural areas, and allowed farmers to move to cities with their own grain rations to work, engage in business or set up enterprises. In 1985, the dual-track pricing system was proposed as a transitional approach for the reform of production material prices; the state monopoly over grain procurement was scrapped, while unified grain distribution was retained.

Prior to the convening of the 13th National Congress of the CPC, certain people in society took advantage of the campaigns against spiritual pollution and bourgeois liberalization to advance Leftist rhetoric that negated reform and opening up. In response to such rampant arguments, Dèng Xiǎopíng explicitly pointed out in his July 4, 1987 speech Two Basic Points of Our Country's Guidelines and Policies: "In pursuing modernization, reform and opening up, we face interference from both Left and Right tendencies." And "Interference from the Left stems mostly from entrenched habitual thinking. People have long grown accustomed to the old ways, making reform a difficult undertaking... Between Left and Right interference, the Left poses the primary threat. After the founding of the PRC, from 1957 to 1978, all our setbacks were brought about by Leftist mistakes."[20] Dèng Xiǎopíng's remarks set the political tone for the 13th National Congress of the CPC.

The report to the 13th National Congress of the CPC in 1987 elaborated on the primary stage of socialism: "The primary stage of socialism does not refer to the initial phase that every country inevitably passes through upon embarking on socialism. Instead, it specifically denotes a unique stage that China must traverse in building socialism amid backward productive forces and an underdeveloped commodity economy. The period spanning from the basic completion of the socialist transformation of private ownership of the means of production in the 1950s to the fundamental realization of socialist modernization will last at least a hundred years — and the entirety of this period falls within the primary stage of socialism. This stage differs both from the transitional period before a socialist economic foundation was consolidated and from the phase when socialist modernization has been fully accomplished. The principal contradiction confronting us at this stage lies between the people's ever-growing material and cultural needs and the backward social production." The formulation of the theory of the primary stage of socialism in the 13th National Congress report provided theoretical underpinnings for reform and opening-up as well as the development of a market economy, shielding the reform drive from Leftist interference. As early as February 6, 1987, prior to the convening of the 13th National Congress, Dèng Xiǎopíng remarked in a conversation with several major central leaders: "We used to follow the Soviet model and practice a planned economy. Later we advocated putting the planned economy in the dominant position, but we should no longer stick to that formulation."[21] Consequently, the report to the 13th CPC National Congress omitted the formulation of "taking the planned economy as the mainstay" and did not mention the term "planned economy" anywhere throughout the report. The report also put forward a new operational mechanism — "the state regulates the market, and the market guides enterprises" — which enabled the market-oriented reform drive to forge ahead on a sustainable footing. The guiding spirit of the report was quickly translated into concrete actions. In April 1988, the First Session of the Seventh National People's Congress adopted the Amendment to the Constitution of the PRC, adding the following provision to Article 11: "The State permits the private sector of the economy to exist and develop within the limits prescribed by law. The private sector of the economy is a complement to the socialist public economy. The State protects the lawful rights and interests of the private sector of the economy, and exercises guidance, supervision, and control over it." Shortly afterwards, the State Council issued regulatory documents including the Provisional Regulations of the PRC on Private Enterprises. This marked the first time the Chinese government recognized the private economy through constitutional provisions, triggering explosive growth of private businesses nationwide. Article 10 of the 1988 Constitutional Amendment stipulated that "The right to the use of land may be transferred according to law." This constitutional provision furnished legal grounds both for the prior transfers of state-owned land use rights and for the subsequent rise of the real estate market. At the same time, progress was made in developing China's financial market: the Shanghai and Shenzhen Stock Exchanges were formally founded and commenced operations in 1990.

### 3.2 Policy Decisions on China's Development of the Socialist Market Economy

In the three years following the political disturbances of spring and summer 1989, voices questioning reform and opening-up, opposing the market, debating whether reform and opening-up was "capitalist or socialist" in nature, and calling for a return to "taking class struggle as the key link" grew increasingly clamorous. It was ultimately Dèng Xiǎopíng's speeches during his inspection tour of southern China in early 1992 — reaffirming an unwavering commitment to reform and opening-up — that put an end to these debates and ensured the smooth continuation of China's reform and opening-up.

The report to the 14th National Congress of the CPC in October 1992 clearly pointed out that "The goal of China's reform of the economic structure is to establish a socialist market economy system, so as to further emancipate and develop the productive forces." And "The socialist market economy system we aim to build requires the market to play a fundamental role in resource allocation under the macro-control of the socialist state, enables economic activities to follow the law of value, and adapt to changes in supply and demand." Furthermore "In terms of the ownership structure, the public ownership economy, including the state-owned economy and the collective economy, shall constitute the main body, while the individual economy, private economy, and foreign-invested economy shall serve as supplements. Various forms of ownership will coexist and develop side by side over the long term, and enterprises of different ownership types may, on a voluntary basis, engage in joint ventures or other forms of cooperation." The report to the 14th National Congress achieved a major theoretical breakthrough by defining that "The goal of China's reform of the economic structure is to establish a socialist market economy system." From this point onward, China's socialist market economy system was formally established. The report removed obstacles for the subsequent reform and opening-up drive. By the end of 1992, the state monopoly over the purchase and marketing of grain was completely abolished.

In 1993, the Third Plenary Session of the 14th Central Committee of the CPC adopted the Decision of the Central Committee of the CPC on Several Issues Concerning the Establishment of a Socialist Market Economic System. Carrying forward the core proposition laid out in the Report to the 14th National Congress that "establishing a socialist market economic system means enabling the market to play a fundamental role in resource allocation under state macro-regulation," the Decision also called for developing the labor market, capital market, real estate market, and other markets. The development of the labor market represented a major theoretical breakthrough, discarding the traditional socialist notion that labor is not a commodity. The Decision also stressed that "we should guard against Rightist tendencies, but primarily prevent Leftist tendencies".

Following the Third Plenary Session of the 14th CPC Central Committee, China's reform entered a critical and decisive stage. Starting in the 1990s, small and medium-sized SOEs — long burdened with heavy social obligations — suffered widespread losses amid fierce competition from TVEs, urban private enterprises, and foreign-funded firms. Subsidies to loss-making SOEs imposed a crippling fiscal burden on governments at all levels, especially the central government. This predicament compelled the central authorities to resolve to close, suspend, merge, or redirect the business scope of chronically loss-making small and medium-sized SOEs, while transforming large SOEs of strategic importance to the national economy and people's livelihood through the establishment of a modern corporate system. This was an arduous round of reform: by the end of 2002, it had left nearly 30 million SOEs workers laid off. SOE reform can be described as the toughest nut to crack in China's reform and opening-up. Its completion between the mid-1990s and 2002 marked the resolution of the core and most challenging task of China's reform and opening-up, after which China's economy gained its full vitality.

### 3.3 Decisions of China on Improving a High-Standard Socialist Market Economic System

In the course of developing the socialist market economy system, shareholding system stands out as a pivotal issue. Commenting on the evolution of shareholding enterprises in China, Lì Yǐníng stated: “In August 1980, the Fushun Branch of the People's Bank of China acted as an agent for enterprises to issue stocks worth 2.11 million yuan. In 1982, Shenzhen Bao'an Company issued stocks to the general public. In September 1984, Beijing Tianqiao Department Store Co., Ltd., China's first shareholding commercial enterprise, was founded and issued three-year term stocks. In August 1986, Shenyang Trust and Investment Corporation pioneered over-the-counter trading services, facilitating stock and corporate bond transactions on behalf of clients. In 1988, the People's Bank of China allocated funds to set up 33 securities companies, while the Ministry of Finance also established a number of securities firms... Shenzhen was where China's stock market truly took shape. Back in 1988, the Shenzhen Special Economic Zone piloted shareholding restructuring for selected enterprises, designating five firms as experimental candidates for public stock issuance and listing.”[22] In early 1992, amid disputes over the shareholding system during his Southern Tour, Dèng Xiǎopíng pointed out: “We may observe, yet we must resolutely conduct trials.”[23] Dèng Xiǎopíng’s Southern Tour Speeches and the decision set forth in the Report to the 14th National Congress of the CPC on developing a socialist market economy together greatly accelerated the pace of shareholding restructuring. “In 1992, nearly 400 pilot shareholding enterprises were approved and established in cities across the country, bringing the national total of shareholding enterprises to more than 3,700. At the same time, the State Council authorized nine SOEs to restructure into joint stock companies and get listed in Hong Kong and other overseas markets.”[24] The 15th National Congress of the CPC, convened in Beijing in September 1997, ultimately affirmed the shareholding system. The report to the Congress stated: “Forms for realizing public ownership can and should be diversified. We should boldly employ all business and organizational forms that reflect the laws of socialized production. We must strive to discover forms for realizing public ownership that can greatly boost the development of productive forces. The shareholding system represents a form of capital organization for modern enterprises. It facilitates the separation of ownership from management rights and helps raise the operational efficiency of enterprises and capital. It can be adopted by capitalism, and socialism can also make use of it. We cannot generalize whether the shareholding system is public or private; the crux lies in who holds the controlling shares.” From then on, the shareholding system became the primary organizational form for both SOEs and private enterprises in China.

Over the decade from the 16th National Congress of the CPC in 2002 to the 18th National Congress in 2012, the Chinese government focused on establishing and improving various social security systems required for the socialist market economy, including medical insurance, subsistence allowances, urban and rural pension insurance, as well as the abolition of the agricultural tax. In November 2013, the Third Plenary Session of the 18th CPC Central Committee adopted the Decision of the Central Committee of the CPC on Major Issues Concerning Comprehensively Deepening Reforms, which pointed out that “the market should play a decisive role in the allocation of resources.” The Report to the 20th National Congress of the CPC, delivered in October 2022, stated: “We will build a high-standard socialist market economy. We will uphold and improve the basic socialist economic system, unswervingly consolidate and develop the public sector, and unswervingly encourage, support, and guide the development of the non-public sector. We will fully leverage the decisive role of the market in resource allocation and give better play to the role of the government.”

Since the 1990s, China has continuously explored a development path for the financial sector within the socialist market economy, with the core goal of guiding banking and financial services to uphold their socialist nature and serve the real economy and all people. The financial and banking sector controls society’s surplus capital, and its investment priorities directly shape the pattern of industrial development. Drawing lessons from U.S. financial monopoly capital, which relocated manufacturing industries and fueled a real estate bubble that ultimately led to industrial hollowing-out and the 2008 financial crisis in that order, China has steadily refined its financial governance system. China convened Central Financial Work Conferences in 2018 and 2023, clarifying that serving the real economy is the fundamental mission of finance, and establishing a financial system compatible with socialist development. At the same time, multiple top-level policies have been issued to strictly restrict financing channels for manufacturing relocation overseas. Stringent reviews are imposed on enterprises’ outbound investment and cross-border mergers and acquisitions to put an end to sham overseas investments and industrial capacity flight, thereby forestalling the risks of industrial hollowing-out. In 2021, China took proactive steps to deflate the real estate bubble and channel financial resources toward the development of new quality productive forces. In addition, governments at all levels have set up venture capital funds to provide targeted support for high-risk, strategic high-tech industries. State financial capital is leveraged to stabilize capital markets and curb malicious speculative trading by capital players. This array of measures has successfully avoided the pitfalls of financial monopoly prevalent in Western countries, fulfilling the core mandate of the financial sector to advance national development and benefit the people.

**4. The CPC and the Chinese Government’s Decision-making Process Concerning “Opening Up"**

As early as October 10, 1978, during a meeting with a press delegation from the Federal Republic of Germany, Dèng Xiǎopíng pointed out: “Now is the time for us to learn from advanced countries around the world.”[25] On January 17, 1979, in a conversation with Hú Juéwén, Hú Zǐ'áng, Róng Yìrén and others, Dèng Xiǎopíng stated: “We need to open up more avenues for our current construction drive. We may utilize foreign capital and technology, and overseas Chinese and ethnic Chinese abroad may also return to China to set up factories. Foreign capital can be introduced through compensation trade or joint ventures. We may start with sectors featuring fast capital turnover.”[26] Between 1978 and 1979, senior Chinese leaders paid frequent inspection visits to developed countries in Europe, America, Japan, and elsewhere. They witnessed the advanced technologies and prosperous economies of Western developed nations and deeply realized China’s backwardness in economic and technological development. Consequently, a consensus on opening up to the outside world was quickly reached among China’s top leadership, which was soon translated into concrete actions and legislation. On July 1, 1979, the Second Session of the Fifth National People’s Congress adopted the Law of the PRC on Chinese-Foreign Equity Joint Ventures. In July of the same year, the CPC Central Committee approved the establishment of four Special Economic Zones (SEZs): Shenzhen, Zhuhai, Shantou and Xiamen. The Law imposed several restrictions on foreign investment at that time: foreign-invested enterprises were required to meet a specified export sales ratio, balance foreign exchange revenue and expenditure on their own, and the establishment of wholly foreign-owned enterprises was prohibited. In addition, the general manager of a joint venture had to be appointed by the Chinese side. The SEZs enjoyed a host of preferential policies unavailable in inland regions. For instance, while only joint ventures were permitted inland, foreign investors could set up wholly foreign-owned enterprises within the SEZs. The enterprise income tax rate applicable to enterprises in the SEZs was far lower than that in inland areas, and even lower than Hong Kong's tax rate.

The report to the 12th National Congress of the CPC delivered in September 1982 pointed out: "Opening up to the outside world and expanding foreign economic and technological exchanges on the basis of equality and mutual benefit constitute an unswerving strategic guideline of our country." In early 1984, Dèng Xiǎopíng inspected three Special Economic Zones successively: Shenzhen, Zhuhai and Xiamen, and spoke highly of their construction achievements. Encouraged by the success of the four special economic zones, the CPC Central Committee decided in May 1984 to further open up 14 coastal port cities, namely Shanghai, Tianjin, Dalian, Qinhuangdao, Yantai, Qingdao, Lianyungang, Nantong, Ningbo, Wenzhou, Fuzhou, Guangzhou, Zhanjiang, and Beihai. Following the expansion of opening-up zones, the government resolved in 1985 to open the Pearl River Delta, the Yangtze River Delta and the Southern Fujian Xiamen-Zhangzhou-Quanzhou Triangle Area. The Law of the People's Republic of China on Wholly Foreign-Owned Enterprises was promulgated in 1986, which lifted the ban prohibiting foreign investors from establishing wholly foreign-owned enterprises in China. The Law of the PRC on Chinese-Foreign Contractual Joint Ventures was issued in 1988. In 1990, the Chinese government abolished two provisions: the requirement that senior management of Chinese-foreign equity joint ventures must be Chinese citizens, and the clauses governing the operation term limits of such joint ventures. The promulgation of the above laws governing foreign investment greatly improved the business environment for foreign investors in China. Driven by Dèng Xiǎopíng, the CPC Central Committee made a major decision in April 1990 to develop and open up Pudong in Shanghai.

After Dèng Xiǎopíng's Southern Tour Speeches in 1992, China's opening-up accelerated once again. The Decision of the CPC Central Committee on Several Issues Concerning the Establishment of a Socialist Market Economic System, adopted at the Third Plenary Session of the 14th CPC Central Committee in 1993, stressed that "we shall pursue all-round opening-up. We will continue to advance the opening-up of special economic zones, coastal open cities, coastal open belts, border areas, riverine areas and inland central cities, and give full play to the radiating and driving role of open regions." The resolution of the CPC Central Committee Plenum was quickly translated into concrete actions. In 1994, China overhauled the foreign exchange administration system, abolishing the dual exchange rate system featuring the coexistence of official quoted rates and foreign exchange swap rates, and unifying the official RMB exchange rate with the swap market exchange rate. The Regulations on the Administration of Foreign-Funded Financial Institutions were promulgated in 1994, permitting foreign investors to set up financial institutions in China including foreign banks, branches of foreign banks, the Chinese-foreign joint venture banks, foreign financial companies, and Chinese-foreign joint venture financial companies. The Law of the PRC on Certified Public Accountants, adopted in the same year, allowed foreign accounting firms to establish permanent representative offices in China and jointly launch Chinese-foreign cooperative accounting firms with local accounting firms. In 1995, the Interim Measures for the Administration of Chinese-Foreign Joint Investment Banking Institutions were issued, authorizing foreign investment banking institutions to establish Chinese-foreign joint investment banking institutions with Chinese financial institutions to conduct investment banking businesses. Also in 1995, China permitted the establishment of foreign-invested joint stock limited companies. Starting December 1, 1996, the RMB became convertible under the current account.

Starting from the mid-1990s, China carried out a host of reforms to prepare for its accession to the World Trade Organization (WTO). For instance, it substantially cut tariffs, and revised thousands of laws and regulations between 1999 and 2005 to align China's foreign-related legal framework with WTO rules. On December 11, 2001, China officially acceded to the WTO. This marked a qualitative leap in China's opening-up drive and full integration into the global market. Following WTO accession, China abolished both super-national treatment and discriminatory treatment for foreign-invested enterprises operating within its borders; in other words, it granted national treatment to such enterprises. Foreign-funded firms were allowed to purchase foreign exchange from the Chinese banks (instead of merely opening foreign exchange accounts with banks as previously required), and the provision mandating foreign-invested enterprises to submit their production and business plans to government authorities was repealed. In April 2004, the revised Foreign Trade Law of the PRC was adopted. Compared with the 1994 version, the new law introduced several major reforms: it permitted natural persons to engage in foreign trade operations, eliminated approval requirements for the right to import and export goods and technologies, and increased penalties for illegal acts and intellectual property infringements.[27] In 2006, the State Council issued the Provisions on Mergers and Acquisitions of Domestic Enterprises by Foreign Investors, allowing foreign investors to conduct cross-border mergers and acquisitions of Chinese enterprises. In 2008, China equalized the tax rates applicable to foreign-invested enterprises and domestic enterprises.

Unlike the opening-up in the early stage of reform and the one that relied on the US-dominated international system, the new type of internationalization led by China or with China as a core participant has gathered momentum since the new century, especially after the 18th National Congress of the CPC, fostering a diverse and inclusive landscape for international cooperation. China has taken the lead in establishing multiple key multilateral cooperation platforms. The Shanghai Cooperation Organization (SCO) was founded in 2001 and expanded its membership several times, centering on regional security and development cooperation. The BRICS mechanism was launched in 2006, continuously expanded and upgraded, with the New Development Bank headquartered in Shanghai set up to strengthen cooperation among emerging economies. In 2013, China put forward the Belt and Road Initiative (BRI), which now involves more than 150 countries. A host of landmark projects covering infrastructure, communications, energy and other sectors have been delivered, benefiting numerous nations across Asia, Europe, Africa, and Latin America. The Asian Infrastructure Investment Bank (AIIB) were inaugurated in 2015, filling the gap in global infrastructure financing. The Regional Comprehensive Economic Partnership (RCEP) was formally signed in 2020, creating the world's largest free trade zone and greatly advancing regional economic and trade integration. Among all these frameworks, the Belt and Road Initiative has yielded the most fruitful outcomes. Over the past decade and more, China has signed over 200 cooperation documents with various countries. Signature projects including China-Europe Railway Express, Jakarta-Bandung High-Speed Rail and Gwadar Port have been completed, while a large number of other projects are advancing steadily. The initiative has effectively boosted employment, infrastructure upgrading and income growth in host countries, while driving China's exports—exports to BRI partner countries account for nearly half of China's total foreign trade volume. The wider use of cross-border RMB settlements has cushioned the external economic shocks caused by US interest rate hikes, supporting steady economic growth in developing countries. At present, the Global South boasts robust economic and trade vitality, with its trade growth rate far exceeding the global average, gradually emerging as an independent force alongside Western developed economies. Rejecting hegemonism, China adheres to the vision of mutual benefit and win-win results, advances common global development through infrastructure connectivity, and practices the cooperative ethos of "if you want to stand firm yourself, help others stand firm too," pioneering a new model of international cooperation.

In 2018, a pilot free trade zone covering the entire Hainan Island was established. At the end of 2025, Hainan Free Trade Port fully launched its island-wide customs closure operation. The Foreign Investment Law was adopted in 2019, replacing the three outdated laws governing foreign investment. The law introduces the management system of pre-establishment national treatment plus a negative list for foreign investment, protects foreign investors' intellectual property rights, and prohibits mandatory technology transfer from foreign investors. In addition, multiple pilot free trade zones have been set up nationwide, and negative lists for foreign investment access have been formulated. China encourages foreign investment in high-tech industries, while an increasing number of high-energy-consuming and high-polluting industries have been included in the negative list.

### 5. Achievements of China's Socialist Market Economy

Before the reform and opening-up, China practiced a planned economic system. Through the price scissors between industrial and agricultural products — artificially suppressing the prices of agricultural produce while inflating those of industrial goods — agriculture subsidized urban industries to prioritize the development of capital- and technology-intensive heavy and chemical industries. The market economy was not allowed to function, meaning no non-public economic entities were permitted to exist. As a consequence, consumer light industrial goods were in severe shortage. 76% of the rural population lived below the poverty line,[28] and the entire national economy teetered on the brink of collapse.[29]

Reform and opening-up allowed the household responsibility system in rural areas, TVEs, urban private enterprises, and foreign-invested enterprises. The state offered no subsidies to rural households, TVEs, or urban private businesses. TVEs and urban private firms fully competed in the market, subject to survival of the fittest. The implementation of the household contract responsibility system injected tremendous vitality into the rural economy and greatly boosted production efficiency. Grain output surged from 300 million tons in 1978 to 450 million tons in 1990.[30] Given China's reality of a large rural population with limited farmland, the implementation of household responsibility system released surplus agricultural labor, which naturally shifted to non-farm work. Rural residents became self-employed non-farm workers, ran individual non-farm businesses, or migrated to cities for jobs. Local rural governments across many regions also launched collectively owned TVEs. All these rural economic activities operated in line with the laws of the market economy. The total output value of TVEs as a share of the national total industrial output rose from 24 percent in 1988 to 30.83 percent in 1991,[31] and further climbed to 56 percent in 1995, providing employment for 129 million people.[32]

In the early 1980s, the Chinese government began to permit the private enterprises in cities. Throughout the 1980s, the growth rate of private industrial enterprises was twice that of SOEs.[33] By 1984, the output value of non-state-owned industry accounted for 36 percent of the country's total industrial output value.[34] TVEs and urban private enterprises, which mainly manufactured light industrial goods, posed fierce competition to numerous small and medium-sized SOEs specializing in light industry. Due to the implementation of the household responsibility system and the rise of TVEs as well as urban private businesses, the supply of agricultural products and light industrial goods such as textiles and garments expanded substantially, enabling the government to gradually relax price controls on consumer goods. Zhū Róngjī, then Vice Premier of the State Council, reported that as of November 1992, "prices of all consumer goods had been fully liberalized. Only a small number of producer goods, mainly coal, petroleum, and steel, remained subject to government pricing, and even only for part of these products."[35] By the end of 1992, grain prices were basically deregulated, marking the initial formation of a market-based pricing mechanism.[36]

Starting from 1978, capital from overseas Chinese began to flow into China for investment, while a small amount of capital from Europe, the United States, and Japan arrived to establish equity joint ventures. China's actually utilized foreign direct investment (FDI) rose gradually from USD 916 million in 1983 to USD 4.333 billion in 1991. After Dèng Xiǎopíng's Southern Tour Speeches in 1992, large corporations from Europe, the US, and Japan started setting up joint ventures in China. The volume of actually utilized FDI surged from USD 11.008 billion in 1992 to USD 46.878 billion in 2001. By the end of 2001, more than 450 enterprises on the Fortune Global 500 list had conducted direct investment in China.[37] After China's accession to the WTO in 2001, foreign direct investment (FDI) ushered in a new boom. The actually utilized FDI climbed from USD 52.743 billion in 2002 to USD 111.71 billion in 2012, and further rose to USD 173.43 billion in 2021.[38] This growth was mainly driven by China's opening-up of the service sector, especially the financial industry, to foreign-invested enterprises following WTO entry. In 2013, FDI and foreign-invested enterprises accounted for 33% of China's gross domestic product, 27% of total employment, 47% of total exports, and 45% of total imports of the country.[39] Foreign-funded enterprises bring access to international sales and marketing networks and help local supporting suppliers upgrade their technological and managerial capabilities. Large multinational corporations usually attract a cluster of their supporting foreign suppliers to invest and build factories in China. Many major high-tech foreign firms have set up research and development (R&D) divisions in China. These R&D facilities cultivate local researchers and serve as benchmarks for domestic enterprises to learn from. In addition, advanced technologies introduced by high-tech foreign investors generate substantial technology spillovers. Foreign-invested enterprises have lifted the standards of products and services available on China's domestic market. The entry of foreign financial institutions and related service providers has elevated the sophistication of China's capital markets and enriched financial products. During the first two decades of reform and opening-up, a large number of talented professionals from domestic companies moved to foreign enterprises. The trend reversed in the subsequent two decades, with ample skilled personnel flowing back to domestic firms. This talent transfer facilitates the diffusion of knowledge and technology, driving technological upgrading, deepened institutional reform, and improved internal corporate governance among domestic enterprises.[40]

The vigorous development of TVEs, urban private sectors and foreign-invested enterprises exerted far-reaching impacts. First and foremost, they generated employment opportunities for hundreds of millions of rural and urban workers, lifting countless households out of poverty. The rural poverty rate dropped rapidly to below 20 percent by the late 1980s.[41] Secondly, the shutdown, consolidation, merger, and transformation of loss-making small and medium-sized SOEs carried out after the mid-1990s essentially represented the withdrawal of the state-owned economy from industries where the market could efficiently supply goods and services, making room for the private sector and the market economy. The number of SOEs fell from over 120,000 in the mid-1990s to fewer than 32,000 in 2004, a reduction of 74 percent.[42] Nearly all small and medium-sized SOEs that underwent shutdown, consolidation, merger, and restructuring were sold to private investors and turned into private firms, resulting in a dramatic marketization in China during the mid-to-late 1990s. Meanwhile, strategically important SOEs vital to national economic security and people's livelihood were reorganized and overhauled under a modern corporate system, focusing on supplying infrastructure such as transportation, telecommunications, energy and electricity to the market. Furthermore, the robust expansion of the socialist market economy generated massive demand for raw materials, machinery manufacturing, transportation, telecommunications, energy, and power. This not only boosted the growth of raw material industries but, more importantly, injected new vitality into SOEs specializing in machinery manufacturing, transportation, telecommunications, energy, and power. While the private sector could import advanced machinery and equipment from Western countries, demand for transportation, telecommunications, energy and power infrastructure had to be met by domestic suppliers. Unlike the pre-reform era, the development of these heavy industrial products was now underpinned by genuine market demand. What distinguishes China from other developing countries is its large contingent of heavy-industry SOEs engaged in transportation, telecommunications, energy, power and machinery manufacturing. As the flourishing private sector created enormous demand for heavy industrial goods and related services in these fields, China's SOEs swiftly ramped up production of such goods and services, establishing effective market supply within a short timeframe.

Since the new century, based on the information technology sector, a large number of high-tech manufacturing enterprises represented by Huawei, DJI, Chenguang Biotech, BYD and CATL have emerged in industries including the internet, smart hardware, artificial intelligence, new energy and pharmaceuticals. There are also internet giants led by Baidu, Alibaba and Tencent, as well as high-tech logistics enterprises represented by Taobao, JD.com, SF Express and Meituan. The number of high-tech enterprises in China reached 504,000 by 2025.[43] Since the 18th National Congress of the Communist Party of China in 2012, China has delivered a string of landmark technological achievements, including space stations, the Beidou

Navigation Satellite System, 5G communications, power batteries, electric vehicles, solar and wind power generation equipment, large aircraft manufacturing, 7-nanometer high-end chips and AI chips, Huawei's HarmonyOS, and the DeepSeek large language model built entirely on Huawei Ascend AI chips. These breakthroughs have broken the United States' monopoly over large AI models and NVIDIA's dominance in AI chips. Driven by the rapid growth of market entities—including private and foreign-invested enterprises—two internationally significant high-tech manufacturing hubs have taken shape in the Pearl River Delta and the Yangtze River Delta. Together with emerging manufacturing centers such as the Beijing-Tianjin-Hebei region, Xi'an, and Chengdu, these hubs have propelled China's industrialization, enabling it to emerge as the world's second-largest economy and a global high-tech power.

## 6. Contributions of Other Factors to China's Economic Success

### 6.1 Human Capital Development Before and After the Reform and Opening-up

When the PRC was founded in 1949, the rural illiteracy rate stood as high as 95%, and the national enrollment rate of school-age children was merely 20%.[44] To address this harsh reality, the Chinese government implemented two core initiatives: eliminating illiteracy and advancing compulsory education. The national illiteracy rate plummeted from 80% in 1949 to 38.10% in 1965,[45] fell further to 22.81% in 1982, and dropped to just 3.6% by 2015.[46] While launching nationwide literacy campaigns and establishing the compulsory education system, the PRC began building its higher education system and the Chinese Academy of Sciences starting in the 1950s. Nobel Prize-winning economist Amartya Sen observed that, in the decades between the establishment of the PRC and the launch of reform and opening-up, China had built a substantial human capital foundation that would later prove instrumental to its remarkable economic rise.[47] In the 22 years between the resumption of the national college entrance examination in 1977 and 1998, Chinese universities admitted a total of 13.55 million students. Over the two decades following the higher education expansion initiative launched in 1999 up to 2018, cumulative college admissions surged roughly sevenfold to approximately 107 million students.[48] By the end of 2023, the total stock of college graduates in China had reached 240 million.[49] China's consistent investment in education built a solid human capital foundation that underpinned the country's economic boom in the era of reform and opening-up.

### 6.2 The Development of Heavy Industry Prior to the Reform and Opening-up

When the PRC was newly founded, the whole nation was devastated and in urgent need of reconstruction. Faced with blockades and hostility from Western powers, the top priority of the new China was national defense security. Only by building an effective and capable defense force to fend off Western powers could economic, social and cultural development be realized. Therefore, the primary goal of China's economic development was to establish its own defense-related heavy industries, including steel, transportation and telecommunications, aerospace, energy and more. Defense industries largely fell under capital- and technology-intensive heavy industry categories. Nevertheless, the newly founded country suffered severe shortages of capital and technology. To tackle this challenge, China adopted the approach of relying on rural agriculture to subsidize urban industry. With assistance from the Soviet Union, it rapidly constructed 156 state-owned heavy industrial projects covering shipbuilding, steel, aerospace, electronics, energy, transportation and telecommunications, and military industries. The development of state-owned heavy industrial projects before reform and opening-up cultivated a large contingent of engineers, entrepreneurs, managerial personnel and skilled workers. The SOEs established in that era played a pivotal role after reform and opening-up, enabling China's economy to readily break through infrastructure bottlenecks in transportation, telecommunications, energy, power, machinery manufacturing and other sectors — hurdles that most other developing countries struggle to overcome.

### 6.3 Favorable International Environment Before and After the Reform and Opening-up

By the late 1960s, having failed to achieve victory in the Korean War and become mired in the Vietnam War, the United States witnessed a relative decline in its strength after reaching its peak immediately following World War II. Seizing the opportunity, the Soviet Union launched strategic offensives, and the global rivalry between the United States and the Soviet Union gradually took on a pattern of "Soviet offensive and U.S. defensive". For China, the successful development of the "two bombs, one satellite" greatly boosted national strength. The Sino-Soviet split starting from the mid-1960s culminated in armed border clashes between China and the Soviet Union in 1969, leaving the Soviet Union as China's primary security threat. These developments laid the groundwork for the rapprochement between China and the United States, which led to President Nixon's visit to China in 1972 and the signing of three Sino-U.S. Joint Communiqués, followed by the establishment of formal diplomatic relations between China and the United States in 1979. The thaw in Sino-U.S. ties also warmed China's relations with Japan, Western European countries, and Southeast Asian nations. The improvement of China's foreign relations created a relatively favorable external environment for the launch of reform and opening-up.[50] Throughout the reform and opening-up, China also took advantage of the U.S.-led international economic and trade system as well as the tide of economic globalization.

## 7. Conclusion: Theoretical Insights from China's Development of the Socialist Market Economy and the Reform and Opening-up

By reviewing the decision-making process and achievements of the CPC and the Chinese government in developing the socialist market economy and opening up to the outside world since the launch of reform and opening-up in 1978, we arrive at the following theoretical insights.

First and foremost, without the CPC's policy decisions and forceful implementation in developing the socialist market economy, it would have been utterly impossible to foster a market economy and pursue opening up to the outside world in China. From the First Opium War in 1840 to the victory of the War of Resistance against Japanese Aggression in 1945, China endured over a century of invasion, slaughter and atrocities perpetrated by imperialist powers including Europe and Japan. It was precisely amid this prolonged subjugation by Western powers that, under the leadership of successive generations of far-sighted Chinese intellectuals and political parties, China accomplished the recreation of its country and nation. Ultimately, under the leadership of the CPC, the PRC was founded in 1949. After the founding of the PRC, bolstered by the firm leadership of the CPC, China secured victories in the War to Resist U.S. Aggression and Aid Korea, the Vietnam War and the Self-Defense Counterattack against India, repelling armed threats from Western anti-China forces against the newborn republic. In the 1960s and 1970s, China successfully developed the "two bombs, one satellite", catapulting it into the ranks of global nuclear powers and vastly strengthening national defense capabilities. It can be said that from the founding of the PRC in 1949 to the eve of reform and opening-up in 1978, the people's military and defense industries built under the CPC's leadership gradually laid a solid foundation of peaceful international surroundings for the PRC. Since the launch of reform and opening-up, the CPC and the Chinese government have consistently led and pushed forward the development of the socialist market economy. Without the CPC's leadership, the socialist market economy could never have been established or expanded. The robust organizational capability and leadership of the CPC constitute the core secret behind all achievements the PRC has made since its founding. For the vast number of developing countries, China's development experience yields the primary lesson: a prerequisite for a country's economic growth is that it must possess a robust central and local government institutional structure dedicated to the well-being of the overwhelming majority of its people. Just as a capable person relies on a sound, developed brain, a nation first needs a powerful nerve center — a central government focused on national political, economic and social development, alongside a complete governance framework extending all the way down to grassroots communities. Moreover, such a government must place the interests of the people, particularly the toiling masses, at its very core.

Secondly, the development of China's socialist market economy since the reform and opening-up is mainly attributable to the growth and expansion of four major market forces. The household responsibility system implemented in rural areas in the early 1980s not only drastically boosted the output of agricultural products but also spurred the rise and expansion of TVEs, forming the first market force. The emergence and sustained growth of urban private economies throughout the 1980s constituted the second market force. The vigorous expansion of TVEs and urban private enterprises, which mainly manufactured light industrial goods, eliminated the chronic shortage of daily consumer goods. More importantly, they absorbed hundreds of millions of surplus rural and urban laborers, lifting countless households nationwide out of poverty and generating massive demand for daily light industrial products in turn. These enterprises produced goods for both domestic and international markets, earning substantial foreign exchange for the country. Since the new century, numerous private enterprises have grown into giant high-tech corporations, including Huawei, BYD, CATL, Tencent, Alibaba, DJI, among others. The privatization of more than 80,000 small and medium-sized SOEs that underwent shutdown, consolidation, merger and restructuring after the mid-1990s gave rise to the third market force. China basically completed the reform of small and medium-sized SOEs between the mid-1990s and 2002. The success of this reform marked China's resolution of the most arduous task in the overall reform of its economic system. Following the reform of small and medium-sized SOEs, the fiscal position of governments at all levels across China saw a fundamental improvement, for three primary reasons. First, governments at all levels were relieved of the fiscal burden of subsidizing loss-making SOEs. Second, the privatized small and medium sized SOEs began paying taxes to the government. Third, the massive influx of foreign-funded enterprises also contributed tax revenue. Most crucially, the reform of small and medium-sized SOEs fully invigorated China's economy. Lastly, starting from the mid-1990s, and especially after China's accession to the WTO in 2001, an increasing number of large corporations from developed economies in Europe, the United States, and Japan invested in China, forming the fourth market force. The influx of numerous foreign-funded enterprises exerted intense competitive pressure on domestic Chinese firms while cultivating a large pool of technicians and managers for China. The exemplary role of foreign enterprises, together with the large-scale return of technical and managerial talents who had worked in foreign firms, fueled the rise of indigenous Chinese enterprises. For developing countries at large, fostering diverse domestic private enterprises and opening the market to foreign investment to a moderate extent represents the primary path to boosting economic growth and development.

Thirdly, in the course of opening up to the outside world, the Chinese government has adopted a realistic approach to gradually expand the scope and scale of foreign investment in light of national economic conditions and affordability. This has prevented the collapse of domestic enterprises caused by excessively dominant foreign-funded firms. At the initial stage of opening-up, the policy of establishing special economic zones greatly facilitated foreign investment attraction and integration with the international community, while safeguarding the overall political, economic, and social stability of the country. While pursuing opening-up, China maintained controls over cross-border capital flows to prevent massive speculative inflows and outflows of international hot money from undermining exchange rate stability and overall economic growth. WTO rules prohibit industrial subsidies. By the time China acceded to the WTO in 2001, it had built a robust industrial base, most sectors of which no longer required protection or incubation. This indicates that joining the WTO when domestic manufacturing lacks competitiveness would deal a severe blow to the national economy and may permanently stifle its economic rise.

Fourthly, when reform and opening-up was launched in the late 1970s, China had no well-defined blueprint whatsoever for developing a socialist market economy. Instead, it adopted the approach of "crossing the river by feeling the stones", moving steadily to unlock the productive initiative of individuals, households, and enterprises. Representative measures included the household responsibility system in rural areas, expanded operational autonomy for SOEs in cities, and the authorization of urban private businesses. Some reforms were driven by practical realities. For instance, certain provinces took the lead in rolling out the household contract responsibility system at the early stage of reform and opening-up, which later received central government endorsement. The massive return of educated youth from rural areas created mounting employment pressure, forcing authorities to permit urban self-employment. After the introduction of rural household contracting, a large surplus of rural laborers shifted to non-agricultural private businesses, a practice subsequently sanctioned by the government. In the 1990s, small and medium-sized SOEs suffered widespread losses, yet the government lacked fiscal capacity to sustain subsidies, leaving shutdown, consolidation, merger and restructuring as the only viable option for these loss-making SOEs. Throughout the high-speed growth of China's economy since reform and opening-up, the country's socioeconomic institutional arrangements have been formulated and adjusted dynamically in response to socioeconomic development demands, the compatibility with mainstream ideology, and the tolerance threshold of the CPC and the Chinese government officials as well as the general public — in short, building new institutions while taking practical steps forward. China’s experience in developing a socialist market economy completely overturns the theoretical hypothesis of the Western institutional economics school, which claims that a country’s success in economic development hinges on its pre-existing institutional arrangements. Influenced by this school of thought, many developing countries attribute their underperformance in economic development to backward institutional frameworks. China’s experience of economic reform and rapid growth starting from the late 1970s reveals to the world the true secret of economic development: socioeconomic institutional arrangements are built incrementally in response to practical developmental needs, captured in the maxim “building institutions while advancing practice.” In reality, Western nations followed the same logic during their own industrialization. Later Western theorists merely repackaged their historical development experience under the rhetoric of “pre-existing institutional arrangements.” China’s development experience thoroughly exposes the fallacy propagated by Western institutional economics.

Fifthly, in the transition from a planned economy to a socialist market economy, China has not followed any Western economic theories or other doctrines. Instead, while adhering to the Four Cardinal Principles, it has crossed the river by feeling the stones and formulated laws, regulations, policies and measures realistically in light of the demands of developing a socialist market economy. China's successful economic transition once again bears out a hard-won lesson the CPC drew from its long revolutionary practice: whenever policies and measures are formulated mechanically according to dogma, undertakings suffer setbacks; conversely, whenever principles and policies are formulated by seeking truth from facts, undertakings achieve success. This is a critical point that developing countries must prioritize when drawing on China’s development experience.

Sixthly, unlike the “shock therapy (Big Bang)” economic reform programs adopted by Russia and Eastern European countries in the early 1990s, China pursued gradualist reform measures. Shock therapy inflicted widespread hardships on the populations of Russia and Eastern Europe, including mass unemployment, soaring prices, goods shortages and the collapse of social security systems. Russia’s economy has still not recovered to the levels achieved under the former Soviet Union. In contrast, China’s gradual reforms unfolded in stages. The household contract responsibility system was first rolled out in rural areas. Subsequently, urban self-employment, rural TVEs, and foreign-invested enterprises were permitted to operate nationwide. Once these market forces had grown sufficiently robust, the government carried out the shutdown, consolidation, merger and restructuring of loss-making small and medium-sized SOEs. Later, China fully integrated into the global economy by acceding to the WTO. Through the incremental development of a socialist market economy, China accomplished industrialization and successively overtook Japan to become the world’s second-largest economy in 2010, and emerges as a global high-tech power in the 2020s. As the saying goes, Rome was not built in a day. The relative merits of gradualist reform versus shock therapy are self-evident.

Seventhly, over the past more than 40 years from the late 1970s to the present, China, a socialist country, has achieved rapid economic growth, industrialization and become a global high-tech power through reform and opening up and the development of a socialist market economy. In the course of its industrialization, China has relied entirely on peaceful economic tools — market forces and state capability — rather than the violent and inhumane means of capital accumulation adopted by old-line Western powers during their industrialization, such as colonial plunder, the slave trade and the opium trade. China completed industrialization in roughly four decades by fostering a socialist market economy and pursuing opening-up, which dismantles the fallacy propagated by Western theorists that industrialization can only be realized under Western-style liberal democracy. Precisely for this reason, Western countries led by the United States regard China's industrialization model as a challenge to their own model. Accordingly, they have imposed high-tech embargoes on China, hyped up the Xinjiang and Taiwan questions to contain China, and formed political and military alliances to jointly counter China. The history of China's rapid industrialization proves that as long as China adheres to the socialist market economy and reform and opening-up, no hardships engineered by Western powers can defeat it. Western sanctions will only make China stronger and stronger.

Finally, China's industrialization and rise as the world's second-largest economy and a global high-tech power through developing the socialist market economy and advancing opening-up to the outside world signify two key points. First, this achievement represents not merely a Pareto improvement but also a Confucian improvement.[51] Second, to attain the grand goal of building China into a great modern socialist country by the middle of this century, China must unswervingly uphold the two fundamental magic weapons: the socialist market economy and opening-up to the outside world.

Notes:

[1] Zhu, & Yu, 2017.
[2] Dèng Xiǎopíng, 1992, p. 373.
[3] Dèng Xiǎopíng, 1992, p. 372.
[4] Dèng Xiǎopíng, 1992, p. 364.
[5] Dèng Xiǎopíng, 1975, p. 87.
[6] Dèng Xiǎopíng, 1982, p. 79.
[7] Yu, 2013.
[8] Wei, & Li, 2019.
[9] Dèng Xiǎopíng, 1992, p. 373.
[10] Wei, 2014.
[11] Gu, 2012.
[12] Gu, 2020.
[13] Long, 2014.
[14] Qian, & Zhang, 2018.
[15] Xi, 2018.
[16] Zhu, & Sun, 2020.
[17] Zhou, & Liu, 2020.
[18] Central Government Document No. 1 of 1982 pointed out: "Contracting work, contracting output and contracting all responsibilities mainly represent different methods for distributing labor fruits. The all-responsibility contracting system mostly adopts the model of 'handing over a fixed quota to the collective and keeping the surplus for oneself'; it abolishes work-point distribution, featuring simple procedures and popular among the masses." (1982 年中央 1 号文件指出："包工、包产、包干，主要是体现劳动成果分配的不同方法。包干大多是'包产提留'，取消了工分分配，办法简便，群众欢迎。")
[19] Zhao, 2022.
[20] Dèng Xiǎopíng, 1987, p. 248-249.
[21] Dèng Xiǎopíng, 1987, p. 203.
[22] Ma, G. C. (2023, March 7). Exclusive Interview with Li Yining (Part III) [Online interview]. (马国川专访厉以宁（之三）(http://www.hybsl.cn/beijingcankao/beijingfenxi/2023-03-07/76298.html。2023 年 4 月 20 日))
[23] Dèng Xiǎopíng, 1992, p. 373.
[24] Ma, G. C. (2023, March 7). Exclusive Interview with Li Yining (Part III) [Online interview].马国川专访厉以宁（之三）(http://www.hybsl.cn/beijingcankao/beijingfenxi/2023-03-07/76298.html。2023 年 4 月 20 日)。
[25] Dèng Xiǎopíng, 1978, p. 132-133.
[26] Dèng Xiǎopíng, 1979, p. 156-157.
[27] Jiang, 2019, p. 167–172.
[28] Ravallion, Chen, 2007.
[29] Central Committee of the Communist Party of China, 1981.
[30] National Bureau of Statistics of China, 1992, p. 346.
[31] Fan, 1993.
[32] National Bureau of Statistics of China, 1997, p. 360-380.
[33] Wu, 2009.
[34] Xiao, 2019, p. 152.
[35] Zhu, 1993.
[36] Xiao, 2018.
[37] Jiang, 2019, p. 122.
[38] National Bureau of Statistics of China. Foreign direct investment data. (国家统计局网站，https://data.stats.gov.cn/search.htm?s=外商直接投资。)
[39] Enright, 2016, p. 31-38.
[40] Jiang, 2019, p. 188-189.
[41] Ravallion, Chen, 2007.
[42] Naughton, 2007, p. 127.
[43] https://www.xinhuanet.com/info/20251231/c15b4b7a5030484eaa6c10ebab2f9e75/c.html
[44] Li, 2013, p. 5-13.
[45] Ding, 2019, October 25.
[46] He, & Ren, 2019, September 29.
[47] Sen, 1999, p. 41-49.
[48] Data on higher education enrollment cited above are sourced from annual datasets published on the official website of the National Bureau of Statistics of China, http://data.stats.gov.cn/workspace/index?m=hgnd; partial

figures are self-calculated by the author. (上述关于高等教育招生数据来源于中国国家统计局网站年度数据，http://data.stats.gov.cn/workspace/index? m=hgnd。 部分由作者计算得出。)

[49] Xí Jìnpíng, 2025.

[50] Xiao, 2022.

[51] Zhao, 2016, p. 71.

**Where is the World Heading?**
**—Examining Global Trends through the Lens of Industrialization's History and Development**

## 1. Introduction

The international political and economic landscape has undergone drastic shifts since the turn of the century. During Donald Trump's first term and following his second inauguration in 2025, the United States pursued an "America First" foreign policy, withdrawing from numerous international organizations, treaties, and agreements. Beginning in Trump's first term, his administration designated China as America's primary rival and initiated a trade war against it. The Biden administration retained this core framing, identifying China as the United States' foremost strategic competitor. Under the slogan "America Is Back," it rallied Europe, Japan, Australia, India and other partners to contain and balance China, seeking to curb its growth momentum. The U.S. rolled out a "small yard, high fence" strategy to stifle China's high-tech development, while simultaneously emulating China's industrial policies in an effort to revitalize its own high-tech sectors.

After returning to office in 2025, Trump imposed sweeping tariff hikes worldwide, pushing the average U.S. tariff rate on Chinese goods to nearly 75%. While most governments scrambled to manage the fallout, China stood prepared for robust countermeasures. Beijing not only announced reciprocal tariffs but also tightened export controls on seven rare earth elements—core raw materials indispensable to smartphones, fighter jets, and advanced equipment. Given China's controls of roughly 90 percent of global rare earth processing capacity, this move struck at the foundations of U.S. manufacturing and defense industries. Confronted with Beijing's leverage over critical minerals, the Trump administration quickly retreated, recognizing that tariff pressure could not offset China's dominance in key supply chains. Amid the global trade war Washington had ignited, China alone refused to yield. In the standoff's aftermath, the United States visibly softened its stance on core issues: it scaled back support for Taiwan and relaxed high-tech export restrictions. To avert further escalation, the Trump administration explicitly prioritized "strategic stability" with China in its 2026 National Defense Strategy.

Since 2012, mounting pressure from the United States and the West has, if anything, catalyzed a sustained surge of technological innovation and high-tech enterprise growth in China. The country has made remarkable strides across multiple domains — including power batteries, electric vehicles, photovoltaic and wind power generation, advanced semiconductors, large-language-model artificial intelligence, Beidou Navigation Satellite System, and large aircraft manufacturing — establishing itself as a global high-tech power. Yet the trade war launched by the United States has simultaneously rendered the global political and economic environment far more complex, severe, and volatile. Against this backdrop, widespread anxiety prevails about the trajectory of global development in the years ahead. By tracing global and Chinese political, economic, and social developments since the Industrial Revolution, this chapter examines the fundamental trajectory the world is likely to follow.

## 2. The Origins and Early Development of the Industrial Revolution

The roots of the European Industrial Revolution in the eighteenth century can be traced to the Italian Renaissance of the fourteenth century and the European Protestant Reformation of the sixteenth. At their core, these movements sought to break free from the brutal political and economic oppression of the Roman Catholic Church and feudal lords, and to accommodate the demands of the rising bourgeoisie's demand for unrestricted commercial activities. Christopher Columbus's arrival in the Americas in 1492 inaugurated the Age of Discovery, setting off a wave of maritime exploration through which European powers seized vast overseas colonies. This early economic globalization and the rapid accumulation of wealth greatly strengthened the European bourgeoisie and lent powerful impetus to the bourgeois revolutions. The Renaissance, the Reformation and the Age of Discovery together gave rise to the first generation of human rights theories grounded on Western liberalism. These doctrines chiefly opposed the autocratic rule and brutal oppression of the papacy and European feudal lords, and resisted state intervention in economic life in order to secure free-market operation.[1] Yet they also furnished ideological cover for the plunder of overseas colonies and subjugation of peoples in less-developed regions. The result was acute wealth polarization within European societies, alongside persistent poverty and backwardness in the colonized world. The spread of the Enlightenment thought across continental Europe partially fueled the American Revolutionary War. Subsequently, the egalitarian ideals advanced by Enlightenment philosophers, the example of the American Revolution, and social crises generated by stark internal wealth disparities together prepared the ground for the French Revolution of 1789. The revolutionary French government issued the Declaration of the Rights of Man and of the Citizen and enacted the Napoleonic Code, both of which affirmed the natural equality of human beings and extended legal protection to private property. The principle of inherent human equality soon ignited a wave of revolutionary movements across the continent. European states gradually adopted constitutional and parliamentary systems with elected representation to curb monarchical autocracy. At the same time, the oppressed working class launched mass movements against capitalists' exploitation, while colonized and semi-colonized peoples rose up in revolt to throw off colonial rule.

World War II was fought by nations seeking equality, freedom, and self-determination against the fascist aggression and colonial expansionism of Germany, Japan, and Italy. The victory of World War II marked a decisive repudiation of militarism and racial supremacy, and laid the institutional foundations for a new international order. Established in 1945 under the UN Charter, the United Nations — whose Security Council includes China, the United States, Russia (as successor to the Soviet Union), the United Kingdom, and France as permanent members — has served as a vital platform for international peace and development in the post-war era. International organizations under the UN framework — the International Monetary Fund, the World Bank and the General Agreement on Tariffs and Trade (later renamed the World Trade Organization) — provided critical venues for multilateral economic and trade negotiations. In the aftermath of World War II, a massive wave of national independence movements swept across Asia, Africa and Latin America, as colonial territories secured sovereignty through sustained struggle. The emergence of these independent states drastically elevated the influence of developing countries on the global stage. In 2025, however, the U.S. administration launched sweeping trade measures targeting countries worldwide, systematically dismantling the post-war international trading order built upon reciprocal tariff reduction and exemption, and imposing a coercive trade regime in which Washington unilaterally dictates the terms of commerce.

Since the Industrial Revolution in Britain in the eighteenth century, European countries, the United States, Japan, the former Soviet Union have successively achieved industrialization. From the 1960s through the 1980s, the Four Asian Tigers — the Republic of Korea, China's Taiwan region, China's Hong Kong region, and Singapore — emerged as newly industrialized economies. Since the reform and opening up launched in the late 1970s, China has realized rapid economic growth; by 2010, it surpassed Japan to become the world's second-largest economy and the largest manufacturing country, and has since established itself as a global high-tech power. Inspired by China's economic success since the early 1990s, India embarked on its own path of economic reform and development. Over the past three decades, India has emerged as a prominent new driver of global economic growth. Collectively, these developments demonstrate that the world remains in an era of successive industrialization, with an ever-growing number of nations achieving advanced economic development.

### 3. International Political and Economic Landscape and Its Evolution from the End of World War II to the Late 1970s

After World War II, a war-weary global populace, who had endured prolonged wars, cherished peace and development. Restrained by the UN Charter and shaped by the major powers, peace and development became the dominant trends in the post-war era. Nevertheless, ideological confrontation soon intervened. In 1949, the United States and its Western allies established the North Atlantic Treaty Organization (NATO), a military alliance targeting the Soviet Union and its socialist bloc. To counter this Western threat, the Soviet Union led its Eastern European satellites in forming the Warsaw Treaty Organization (Warsaw Pact) in 1955. Thereafter, the capitalist and socialist camps — including Europe, the United States, Japan, and the Soviet bloc — competed in economic growth and arms racing. However, mindful of catastrophic potential of nuclear warfare, the NATO and the Warsaw Pact maintained a precarious "cold peace" throughout the Cold War.

Economically, war-torn European nations devastated World War II urgently needed U.S. aid to rebuild. Through the Marshall Plan, Europe achieved a remarkable recovery and subsequent boom. The 1950s and 1960s marked a golden age of growth for Europe, the United States and Japan. Yet from the early 1970s, industrial overcapacity coupled with OPEC's newly assertive oil pricing power to drive global prices sharply upward. This oil shock largely triggered stagflation that plagued Western capitalist economies throughout the 1970s — the rare and damaging combination of sluggish economic growth and high inflation.

The Soviet Union, Eastern Europe, China, India, and numerous other Asian, African and Latin American nations also aspired to boost their economies. From the 1950s to the 1970s, socialist countries — including China, the Soviet Union, Eastern European countries, and Vietnam — achieved relatively robust growth under the planned economies. Alongside economic expansion, they scored remarkable achievements in national infrastructure: education, healthcare, transportation, communications, and energy. During this period, several socialist countries realized far superior human development indicators compared with other developing economies. The Soviet Union successfully tested the atomic bomb and the hydrogen bomb in 1949 and 1953, respectively, and launched the world's first artificial Earth satellite, Sputnik 1, in 1957. It proved that the Soviet Union had surpassed the US in scientific and technological development, and is therefore known as the 'Sputnik moment'. Taking advantage of the relative decline of the US, the Soviet Union seized the opportunity to launch strategic offensives on a global scale throughout the 1960s and 1970s. Furthermore, with assistance from the Soviet Union, China leveraged its capability of pooling resources to accomplish major undertakings to launch 156 state-owned heavy industrial projects.[2] Though the Soviet Union withdrew its technical experts after Sino-Soviet split of the early 1960s, China pressed ahead with heavy industrial development relying on its own scientific and technological capacity. It successfully tested an atomic bomb in 1964, a hydrogen bomb in 1967 and launched its first satellite in 1970. These accomplishments endowed the New China with a relatively robust industrial base and nuclear capability.

Nevertheless, socialist countries across the Soviet bloc and Eastern Europe prioritized heavy industry exclusively under their planned economic economies, neglecting light industry, resulting in chronic shortages of consumer goods and electronic products demanded by their residents. As for China at that time, excessive reliance on agricultural surplus extraction to subsidize industrial expansion led to low efficiency in rural collectives, shortage of daily manufactured goods, and widespread rural poverty. Other socialist countries faced comparable difficulties to varying degrees. These tangible shortcomings furnished Western developed nations with pretexts to criticize, even vilify, socialist systems.

Meanwhile, a vast number of Asian, African and Latin American countries were in urgent need of developing their economies and lifting their people's living standards. Having endured long-term exploitation and oppression by imperialists, many nations that gained independence from colonial rule lagged behind in the development of political, economic, social and cultural institutions; suffered from insufficient human capital; and lacked experience in state governance. In numerous newly independent former colonies, vast tracts of fertile land remained in the hands of former colonial powers, while their economic lifelines were controlled by multinational corporations. Over five centuries of the transatlantic slave trade from the fifteenth century to the late nineteenth, massive numbers of young and middle-aged Africans were trafficked to the Americas. This inflicted endless suffering on enslaved people and their families, while fostering mutual distrust and hostility among African tribes — rifts that persist to this day. Such divisions have hindered the formation of a unified national identity in countries ravaged by the slave trade, perpetuating economic backwardness and sluggish growth.[3] Furthermore, colonial powers deliberately sowed seeds of political and military conflict among territories before granting independence. As a result, many developing countries endured frequent coups, unstable governance, and cross-border conflicts. From the end of World War II through the late 1970s, developing countries generally posted slow economic growth, shortage of infrastructure, and large impoverished populations.

### 4. The International Political and Economic Development Landscape and Its Shifts Since the 1980s

Faced with the stagflation of the 1970s, Western Europe and the United States turned to neoliberal economics as a solution, centered on reducing government intervention and cutting taxes. As part of this shift, governments across Europe and America began selling off large swathes of state-owned enterprises (SOEs) and public utilities. In the 1980s, for instance, Margaret Thatcher's government privatized Britain's state-owned gas, rail, electricity, telecommunications, airline, oil, postal, and shipbuilding industries, disposing of nearly every state asset that could be transferred to private hands. By the time Tony Blair's Labor government took office in 1997, few state assets remained to privatize, so it ultimately auctioned off for 3G mobile network licenses for USD 35 billion. SOEs had long held a prominent position in French economy as well. During the 1980s and 1990s, the French government sold off more than 2,000 SOEs, largely withdrawing the state from manufacturing. Even so, it retained control of key enterprises, including Air France, Orange (France Télécom), Électricité de France, Engie (Gaz de France), La Poste, RATP (Paris Metro), and SNCF (French National Railway Company). The United States had comparatively few SOEs to begin with, yet its federal government still divested itself of Amtrak (the national passenger rail system), Conrail (the federal freight railroad), federal power utilities, and enriched uranium production facilities through the privatization of United States Enrichment Corporation. It also introduced measures outsourcing management rights for certain SOEs, public utilities, and certain government agencies to private investors — even some prisons and mints were contracted out private operation.[4] From the late 1970s onward, economic globalization and neoliberal economics became the dominant ideological framework in the United States. A core outcome of economic globalization was the unrestricted cross-border flow of capital: American corporations relocated investment and production to lower-cost countries in pursuit of higher profits, then shipped finished goods back to the U.S. market. While American capital holders and elites reaped enormous gains from this arrangement, it also drove accelerating deindustrialization the 1980s onward. Millions of American manufacturing workers lost their jobs, forcing them into lower-wage service work or unemployment — a shift that contributed to the erosion of the U.S. middle class.

In the United States and other Western countries, neoliberal economics also elevated financial capital to an unparalleled position within national power structures, with the 1999 repeal of the Glass-Steagall Act under President Bill Clinton standing as its most consequential milestone. In the wake of the Great Depression of 1929 – 1933, the U.S. government drew painful lessons and targeted what it identified as a root cause of financial crises: the commingling of commercial and investment banking. Congress responded by passing the Glass-Steagall Act, which barred commercial banks from engaging investment banking and investment banks from engaging in commercial banking. This legislation safeguarded the stability of the American financial sector for more than six decades. After the Glass-Steagall Act was repealed in 1999, commercial and investment banking resumed mixed operation in the United States. The dissolving boundaries between commercial banks, investment banks, various investment funds, private equity firms and insurance companies gave rise to sprawling, "too big to fail" financial conglomerates offering the full range of financial services under one roof, such as Citigroup. With financial risk oversight drastically loosened, many investment banks began funding long-term investments with short-term loans from the money market as a cost-cutting measure. To evade reserve ratio regulations, banks shifted enormous volumes of activities off their balance sheets, driving financial institutions to dangerously high leverage. High-risk financial instruments proliferated, while excessively long securitization chains left market participants unable to identify the underlying assets or gauge the risks embedded within them. Meanwhile, outsized salaries and bonuses became standard practice for senior financial executives, creating a pervasive principal-agent problem: executives captured all gains during booms, while shareholders and, ultimately, taxpayers absorbed the losses during downturns. This accumulation of systemic disorder culminated in the U.S. financial crisis of 2007 – 2009, which spilled over into a global financial meltdown and worldwide recession. More broadly, in weighting efficiency against equity, neoliberal economics that gained widespread traction in the early 1980s tilted decisively toward capital interests, at the expense of social equity.

As Western developed countries, led by the United States, unleashed waves of privatization and financial liberalization, economic reform movements simultaneously swept across China, the Soviet Union, and Eastern Europe from the late 1970s to the early 1980s. Reforms in the Soviet Union and Eastern Europe, however, ultimately ended in failure. This failure — compounded by the strain of the Soviet Union's overextended global ambitions — contributed substantially to the Soviet collapse in the early 1990s and the broader breakdown of the socialist bloc in Eastern Europe. In the wake of these sweeping political and economic transformations, Russia and other former Eastern Bloc countries hastily adopted shock therapy, selling off vast volumes of SOEs and other public assets in the hope of establishing a market economy virtually overnight. The results were catastrophic: shock therapy left Russia and its Eastern European counterparts mired in sluggish economic growth and severe hardship for ordinary citizens.

Since launching its reform and opening-up drive in the late 1970s, China has adhered to a gradual approach and a dual-track system, in which SOEs and non-state-owned sector, as well as planning mechanisms and market forces, coexisted side by side. In the early 1980s, without altering the collective ownership of rural land, China introduced the household contract responsibility system, which greatly boosted grain output. Surplus rural laborers were then permitted to migrate to cities for work, while the development of the private economy was encouraged in both urban and rural areas. In the mid-1990s, China embarked on the reform of its SOEs. Small and medium-sized SOEs unrelated to critical national industries or people's livelihoods were shutdown, merged, reorganized, or privatized, while SOEs in transport, communications, energy and other sectors vital to national security and public welfare were restructured into modern share-holding companies. At the same time, to attract foreign investment and integrate into the global economic and trading system, China revised its domestic economic policies, laws, and regulations to align with international trade rules, and formally acceded to the World Trade Organization (WTO) in 2001. Thereafter, foreign direct investment poured into China on a massive scale, while SOEs expanded rapidly through major infrastructure projects, building a sophisticated nationwide infrastructure network. Sustained rapid growth enabled China to overtake Japan and become the world's second-largest economy in 2010. Since the 18th National Congress of the CPC in 2012, China's high-tech industries have surged, transforming the country into a global high-tech power to this day.

During China's reform and opening-up in the 1980s and 1990s, greater emphasis was placed on economic efficiency. Since the turn of the century, the Chinese government has attached higher priority to equity, stepping up efforts to promote social fairness. Examples include the launch of the new rural cooperative medical care system and basic medical insurance for elderly and underage urban residents following the SARS crisis in 2003, the abolition of the agricultural tax in 2006, and the rollout of a basic old-age insurance system covering both urban and rural populations in 2009. Since the 18th National Congress of the CPC, China has waged a targeted poverty alleviation campaign, eradicating absolute poverty entirely by the end of 2020. To prevent China from repeating the manufacturing hollowing-out and the 2007-2009 global financial crisis experienced in the West — driven, respectively, by Wall Street capital's pursuit of deindustrialization abroad and by speculative excess in the real estate sector and financial sectors — China convened Central Financial Work Conferences in 2018 and 2023. These conferences mandated that domestic banks and financial institutions channel the bulk of their financing into high-tech real economic sectors aligned with the country's medium- and long-term development plans and supportive industrial policies, while strictly prohibiting the financial sector facilitating the relocation of manufacturing enterprises overseas. In 2021, China took proactive measures to deflate the real estate bubble and redirect investment toward the development of new quality productive forces.

As a country with a massive population, India is also worthy of close attention. After gaining independence in 1947, India largely pursued a state-led, planned economic development model, which resulted in sluggish growth and widespread poverty. In the early 1990s, prompted by a balance-of-payment crisis and drawing on the broader wave of market-oriented reforms sweeping the developing world, India launched its own economic reforms and introduced privatization policies. Over the past three decades, India has made remarkable economic progress amid its transformation toward industrialization and digitalization. Its economic output climbed from the world's 12th largest in 1990 to surpass the United Kingdom and France, and by 2025, India ranked as the world's fourth-largest economy. The economic takeoff of China and India — the world's two populous developing countries — has greatly encouraged underdeveloped nations across Asia, Africa, and Latin America in their own efforts to grow their economies.

### 5. China-U.S. Relations amid Profound Changes Unseen in a Century Since the New Century

Having overtaken Japan to become the world's second-largest economy in 2010, China has since grown into a global high-tech power. As a socialist country led by the CPC, which frames its governance around prioritizing the people's interests, China achieved industrialization in little more than 70 years. For Western countries led by the United States — whose systems are built around capitalism — China's trajectory is therefore often viewed as an outlier or anomaly. Wary that numerous developing countries might come to see China's development model as a viable alternative path to progress, Western nations led by the U.S. have, in recent years, ramped up political and economic pressure against China.

President Barack Obama, who took office in 2009, rolled out the "pivot to Asia" (also known as the Asia Rebalancing Strategy). In essence, the United States planned to draw down its involvement in the Middle East and shift its strategic focus toward Asia, to contain and counterbalance China's rise. In international economic and trade affairs, Obama argued that China must not be allowed to write the rules of global trade. He pushed forward the Trans-Pacific Partnership (TPP), seeking to build a U.S.-led trade and investment bloc that excluded China. Beyond its strategic aim of containing China, the TPP championed by the Obama administration also granted multinational corporations sweeping new powers, requiring participating nations to refrain from imposing restrictions on multinationals' investment and trade activities. Ultimately, however, Obama's TPP proved politically ill-timed. A key reason lay in the fact that, since neoliberal economics gained prevalence in the United States in the early 1980s, successive administrations had vested capital with excessive power and freedom at the expense of social equity and justice. Deindustrialization driven by Wall Street financial capital had already hollowed out America's middle class, and the wealth held by the richest 1 percent of U.S. households accounted for more than one-third of the total household wealth nationwide — a level of wealth polarization that fueled deep public backlash against further trade liberalization.[5]

Trump, who rode to power on the grievances of a hollowed-out middle class by campaigning against elites and Wall Street, pursued the "America First" strategy. He withdrew the United States from numerous international agreements and treaties, including the TPP, and moved aggressively to contain China's rise. Blaming China for many of America's domestic troubles, he leveled a series of accusations against Beijing: that China stole U.S. technology, forced foreign firms operating within its borders to transfer technology, subsidized SOEs to gain unfair competitive advantages, and that China's Belt and Road Initiative saddled developing countries with unsustainable debt and amounted to a form of neocolonialism. Meanwhile, the Trump administration launched trade disputes, sanctioned Chinese telecommunications firms, restricted exports of high-end chips and precision chip-making equipment to China, and pressured U.S. allies to bar Chinese-developed 5G communication systems. It repeatedly raised the Taiwan and Xinjiang issues as points of leverage against China, deepened intelligence cooperation on China through the Five Eyes alliance (FVEY, an intelligence bloc of five English-speaking nations: the U.S., Canada, Australia, the UK and New Zealand), rolled out the Indo-Pacific Strategy, and established the Quadrilateral Security Dialogue (Quad) grouping the U.S., Japan, India and Australia. After Democratic President Biden took office, he largely maintained the Trump administration's trade policies toward China while touting the slogan "America Is Back." Under this banner, the U.S. sought to rally fellow democracies to jointly counter China's influence. Building upon the Quad, Washington established AUKUS — a smaller core bloc consisting of the U.S., the UK and Australia — to deepen containment efforts, while further tightening restrictions on high-tech exports to China and expanding sanctions against Chinese firms. After winning a second presidential term in 2025, Trump launched trade wars against a broad range of countries worldwide. Most other countries moved to negotiate or accommodate U.S. demand, while China largely held its position in the ensuing standoff. U.S.-China tension shows little sign of easing in the near term.

Despite intense pressure from the United States, however, China's economy has not collapsed. Instead, more than 500,000 high-tech enterprises have emerged, led by firms such as Huawei, DJI, BYD, CATL, Tencent, Douyin, and DeepSeek, alongside the Beidou Navigation Satellite System.[6] Since Qián Xuésēn wrote a letter to the central authorities in 1992 proposing the development of electric vehicles, nearly three decades of consistent industrial policies have propelled China's power batteries and electric vehicles industries forward. By 2020, China had grown into the world's leading producer of power batteries and electric vehicles. In 2023, China achieved breakthroughs in 7-nanometer chip production. In 2024, Huawei launched Harmony OS, an operating system developed independently of Android and Apple's iOS. In early in 2025, DeepSeek's large AI model made global headlines, narrowing the perceived U.S. lead in artificial intelligence. In April 2026, DeepSeek achieved full operational success entirely based on domestically produced Huawei Ascend chips, ending NVIDIA's monopoly on high-performance chips. China deploys over half of the world's industrial robots.[7] On July 16, 2026, Moonshot AI released KIMI K3, a large language model that drew widespread attention for narrowing China's gap with leading U.S. models to a matter of weeks — a milestone that reportedly contributed to a sharp pullback in the market value of the major U.S. tech companies. Today, China ranks as the world's largest shipbuilder and ship exporter, as well as the leading manufacturer and exporter of electric vehicle, power batteries, and photovoltaic panels. Its large passenger aircraft program has achieved commercial success, and the Beidou Navigation Satellite System is now used by more than 100 countries worldwide. Chinese suppliers provide three-quarters of all 4G communication network equipment across Latin America and Africa.[8] China's total power generation from renewable energy reached 3.99 trillion kilowatt-hours in 2025, accounting for 38.3 percent of the country's total power output. Wind and solar power each exceeded 1 trillion kilowatt-hours for the first time, with combined wind and solar generation hitting 2.3 trillion kilowatt-hours and making up 22.1 percent of total electricity production.[9] China's export product structure has undergone a major upgrade: whereas light consumer goods dominated exports at the start of this century, high-tech products — including smartphones, computers, chips, photovoltaic and wind power equipment, power batteries, electric vehicles, large vessels, railway locomotives, rare earths, and intermediate microelectronic industrial goods — now take center stage, becoming indispensable inputs for industrialization in other countries. This resilience is reflected in China's trade surplus, which reached USD 1.2 trillion in 2025 amid a sluggish global economy.

Unlike its integration into the U.S.-led international order during the early years of reform and opening up, China's new mode of engagement with the world — one it now often leads or shapes through deep participation — has gathered pace since the turn of the century, and especially since the 18th National Congress of the CPC. China has both initiated and joined a range of multilateral cooperation mechanisms. The Shanghai Cooperation Organization (SCO) and BRICS have each expanded memberships multiple times, the Asian Infrastructure

Investment Bank (AIIB) has been established, and the Regional Comprehensive Economic Partnership (RCEP) has entered into force. Together, these mechanisms have created diverse platforms for international cooperation and energized cooperation among Global South countries. Among them, the Belt and Road Initiative (BRI), launched in 2013, stands out as the most productive. Over the past decade, China has signed more than 200 cooperation documents with over 150 countries and more than 30 international organizations, delivering a wide array of landmark infrastructure projects — railways, ports, bridges, energy facilities, and communication networks — across Asia, Africa, Europe and Latin America. Flagship projects such as the China-Pakistan Economic Corridor have brought substantial investment, created abundant jobs, and upgraded infrastructure in host countries, effectively reducing local poverty and raising incomes. The BRI carries both economic and strategic weight. In 2025, exports to Belt and Road countries accounted for half of China's total export volume, while cross-border RMB settlement has helped cushion developing countries against the economic shocks of U.S. interest rate hikes, supporting steadier growth in relevant countries. More broadly, emerging economies continue to gain strength, and the trade growth and global trade share of Global South countries keep rising — gradually forming a counterweight to the traditional dominance of Western developed nations. In contrast to the hegemonic intervention model long favored by Western powers, China's approach is guided by the ancient maxim "to establish oneself, one must first help others establishes themselves; to succeed, one must first help others succeed."[10] Centered on infrastructure connectivity, China continues to deepen international cooperation in support of shared development among developing countries, reshaping the framework for global collaboration in the process.

Since the outbreak of the Russia-Ukraine conflict in 2022, Europe has lost access to low-cost oil and natural gas from Russia, driving up both domestic production costs and living costs. This situation became even more acute following the U.S. war of aggression against Iran in 2026. Confronted with this fossil energy predicament, European countries are likely to place growing emphasis on developing green energy sources such as photovoltaic and wind power in the future.

Since the turn of the century, a large number of emerging market economies and developing countries have achieved rapid growth, accelerating the world's advance toward multi-polarity. Any analysis of China-U.S. relations must be situated within this broader historical backdrop of profound changes unseen in a century. So long as China avoids major strategic missteps and continues to strengthen itself through its own efforts, virtually no external force can shake a powerful of its scale and resilience. Therefore, China-U.S. relations are likely to continue to evolve along the path of peaceful competition for the foreseeable future.

### 6. Universal Laws Reflected by China's Economic Miracle and Its Inspirations for Global Development in the Complex and Volatile Global System

People still vividly remember New China's historic transformation: within an unprecedentedly short span of time, this populous, poor, and backward agrarian nation transformed itself into a global high-tech power. What universal laws of development, then, does China's rise reveal? And amid a complex and ever-shifting global landscape, what insights can China's achievements offer for the world's future direction?

First, just as a capable individual needs a strong, well-functioning mind, a country needs above all a strong command center — a central government dedicated to national development, backed by a complete administrative framework extending down to grassroots. This government must consistently place the interests of its people at the heart of all its work. The government led by the CPC embodies precisely this model: its fundamental purpose is to serve the people wholeheartedly, with no other agenda. Moreover, the CPC and the people's army under its leadership are a party and military forged through decades of domestic revolutionary struggle and wars of resistance against foreign aggression. They possess a distinctive ideological system, resilient and unbreakable willpower, and rigorous organizational structures and disciplines. Under the leadership of this steadfast party, New China secured victories in the War to Resist U.S. Aggression and Aid Korea, the War to Resist U.S. Aggression and Aid Vietnam, and the 1962 Sino-Indian Self-Defense Counterattack — triumphs that secured a peaceful environment conducive to New China's growth and advancement. By contrast, many developing countries continue to struggle from weak and ineffective governance. When such states uncritically adopt neoliberal economic prescriptions that hollow out state governance capacity, effective national administration becomes unattainable — let alone sound growth and development.

Secondly, build a thoroughly open society. Alongside granting the people political participation rights, the state must also secure economic rights for the general public — especially the working masses — such as land ownership and access to public services. A country can only thrive when its people prosper. After the founding of New China, the state carried out the most thorough land reform in human history, abolishing the feudal land ownership system that had long benefited the landlord class. This was followed by rural land collectivization and urban land nationalization. Without these measures, the large-scale development of infrastructure and industrial parks across vast stretches of land would have been impossible. To equalize access to social services, New China introduced nine-year compulsory education and public medical care available to all citizens. On this point, Nobel laureate economist Amartya Sen observed that even before reform and opening-up, New China had already laid a solid human capital foundation — one that would go on to fuel the country's economic takeoff in the subsequent reform era.[11]

Thirdly, robust state-owned research institutions and SOEs constitute a vital source of state capability. Just as a strong person needs well-developed limbs, these state research bodies and SOEs function as the limbs of a nation. Studies on the industrialization history of Europe show that major technological breakthroughs were driven by state fiscal investment, not by so-called "liberal democracy," and certainly not by ancient Greek science and logic. Likewise, the underdevelopment of science and technology in China prior to 1949 had little to do with insufficient scientific thinking in ancient China; rather, it resulted from a lack of sustained state financial support. The success of the "Two Bombs, One Satellite" program after the founding of New China was made possible precisely by the government's high prioritization of the effort and its massive state investment. In their early stages of development, European countries established numerous SOEs to advance national goals. Examples include the British and Dutch East India and West India Companies, state-backed entities operating in colonies that were even permitted to maintain armed forces. Without state funding, scientists could not sustain themselves through research alone.[12] New China's establishment of a national academy system and SOEs framework replicated the secret behind early European development. The key difference lies in the fact that New China possessed no colonies and never accumulated wealth through slave trading or opium trafficking. Instead, it forged its economic miracle by constructing a robust domestic infrastructure network and a massive market economy. SOEs supply goods and services that private firms are either unable to deliver or can only provide at a far slower pace, rather than replacing private enterprises entirely.

Fourthly, continuously improve the market economy and promote the growth of the private sector. A market economy relies on market mechanisms as the fundamental means of allocating economic resources. Since the Industrial Revolution in the 18th century, privately owned industrial and commercial enterprises, driven by market incentives, have gradually replaced small-scale peasant agriculture as the mainstay of national economies worldwide. While generating profits, these enterprises supply capital goods and consumer products to society, create jobs, and contribute tax revenue to the state. China's economic success rests on a steadily refined market economy. Even while strengthening and growing its SOEs, China has consistently fostered the development of the non-public economy. Establishing and perfecting market mechanisms, and encouraging the growth of private enterprise, remain core priorities for national development in the contemporary era.

Lastly, all of the political, economic, and social institutions established since the founding of New China — especially since the launch of reform and opening-up — have been progressively built and refined in response to China's own development needs. The reason why New China does not blindly defer to Western dogmas lies in a lesson drawn from the CPC's long experience in revolutionary and development practice: whenever policies were rigidly formulated by mechanically copying doctrinal texts, revolutionary and development efforts suffered setbacks and failures. On the contrary, formulating policy on the basis of seeking truth from facts has consistently underpinned victories in both revolution and national construction. The history of development and industrialization following the European Renaissance tells a similar story: the institutions and institutional arrangements forged during the industrialization of Europe, the United States, Japan, and other economies were built to continuously adapt to their own specific development demands. This stands in direct contrast to the claim of new institutional economists, who argue that a supposedly inherent "liberal democracy" of the West is the source of technological progress and economic prosperity.[13]

Within the complex and ever-shifting global system, China's development miracle offers a vital insight: development must remain the top priority. While universal laws govern development, no single blueprint applies to all. Only by crafting institutional arrangements tailored to their own national conditions and realities can countries achieve political stability, economic growth, and social peace and harmony.

### 7. Conclusion: Where is the World Heading?

According to Maslow's hierarchy of needs, human demands fall into distinct tiers: physiological needs (food, clothing, shelter, and transportation), safety needs (personal security and stable employment income), social belonging needs (friendship and connection), esteem needs, and self-actualization.[14] From the perspective of Nobel laureate economist Amartya Sen, the root cause of poverty in developing countries lies in insufficient human capital, stemming from inadequate access to education and public healthcare among impoverished populations.[15] In the era of industrialization and AI, a prerequisite for economic growth in developing countries is that the state and society's capacity to continuously supply a steady stream of scientists, engineers, senior managers, doctors, accountants, lawyers, teachers at all levels — primary, secondary, and university — along with technicians and skilled workers to support industrial advancement.[16] In short, the all-round development of human beings constitutes both the precondition and the ultimate goal of national development.

Now, in the 2020s of the 21st century, Western developed economies — including Europe, the United States, and Japan — remain the world's most advanced economic blocs, possessing robust education and scientific research sectors, a steady stream of high-tech products, sound infrastructure, and formidable military strength. Their citizens enjoy high-quality material lives, sound education, and access to premium medical and healthcare services. Nevertheless, the U.S. financial crisis of 2008-2009 fully exposed the pitfalls of the neoliberal economics that had prevailed in the United States since the early 1980s — an overemphasis on liberty, corresponding to what is sometimes called first-generation human rights. This framework enabled Wall Street financial capital to reap enormous profits by relocating manufacturing to low-cost Asian economies, while refusing to bear the costs of the resulting deindustrialization and hollowing out of the middle class, and disregarding social equity altogether. Worse still, U.S. politicians — from Trump to Biden to Trump — have shifted blame for the adverse consequences of America's own policies onto China. The political, economic, and social development of the United States since the 1980s demonstrates that Western-style liberal democracy cannot, on its own, guarantee social justice or prevent vulnerable groups from being marginalized. Nor it curbs the profit-driven behavior of capital. It has not stopped governments of so-called democratic countries from arbitrarily launching military aggression, strikes and occupations against other countries. What is more, Western liberal democracy cannot compel domestic politicians to engage in genuine self-reflection on flaws in their own government's policies, rather than scapegoating other nations. Ultimately, it offers no guarantee that developing countries adopting Western democratic systems will achieve sound economic growth, political stability, or the well-being for their people.

As the world's second-largest economy, China is home to roughly 500 million middle-class residents, concentrated primarily in the industrial hubs of the Pearl River Delta and Yangtze River Delta along the southeast coast, as well as major metropolises such as Beijing and Tianjin. Although China lifted its remaining roughly 100 million people out of poverty by the end of 2020, approximately 900 million Chinese still fall outside the middle-class bracket, alongside a sizable population at risk of falling back into poverty. In other words, China's economy retains substantial room for growth as it works to raise incomes, educational attainment, and medical standards for these 900 million citizens. China's economic takeoff stems from two sets of factors. Domestically, prior to reform and opening-up, New China made significant strides in building infrastructure spanning education, healthcare, transportation, telecommunications, and energy, while advancing heavy industrialization — laying a solid foundation. Then, economic reforms launched in the 1980s enabled the parallel growth of the state-owned and private sectors, alongside free mobility of labor. Externally, China's rapid economic expansion benefited from the relatively peaceful international political and economic landscape that emerged after World War II, as well as decades of trade, technological exchanges, and personnel interaction with developed economies — including the United States, Europe, and Japan — beginning in the 1980s. For the foreseeable future, China will continue to uphold its national policy of reform and opening-up, and will work to safeguard a peaceful order for global trade, technological cooperation, and people-to-people exchange. In addition, to uphold social equity, China will maintain a certain number of strategic SOEs that bear critical implications for national welfare and people's livelihoods. These entities help ensure a stable supply of public goods across the board, and in particular address the "last-mile" challenge common across developing countries — where vulnerable groups struggle to access effective public infrastructure services. They deliver goods and essential services that the market fails to provide, preventing the national economy from being held hostage by oversized private corporations and multinationals.

Since the outbreak of the China-US trade war, the United States has imposed embargos and blockades on high-tech exports to China. This has, in turn, opened up domestic market space for Chinese manufacturers. Over the past decade, a host of China's high-tech breakthroughs have emerged as a direct consequence of U.S. technology restrictions: the Beidou Navigation Satellite System, 5G communications, advanced 7-nanometer chips, HarmonyOS — an operating system independent of Android and iOS — power batteries, new energy vehicles, and industrial robots, among others.[17] The DeepSeek large-language model, launched in early 2025, broke the United States' monopoly over large-language models. On May 7, 2025, Pakistan shot down two French-made Rafale fighter jets belonging to India using Chinese fighter jets and air-to-air missiles, furnishing real-combat proof of the superior performance of Chinese weaponry. China's cutting-edge high-tech capabilities help sustain stability across South Asia. In April 2026, DeepSeek fully migrated its operating platform to one primarily powered by Huawei Ascend chips, breaking NVIDIA's monopoly on GPUs. In July 2026, Moonshot's Kimi K3 large-language model was released. This model narrowed the gap between China's large-language models and America's most advanced models to a matter of weeks, triggering a sharp drop in the US stock market. With another 15 to 20 years of development, China will undoubtedly stand shoulder to shoulder with the United States in the research and development. From the First Opium War in 1840 to China's victory in the War of Resistance Against Japanese Aggression in 1945 — a span of roughly one hundred years — China endured its weakest era in its 5,000-year history. During this period, China suffered brutal invasions and massacres at the hands of imperialist powers, including the European nations and Japan, while frequent wars ravaged much of Asia. Since the end of World War II, as the PRC has grown stronger and more prosperous, East Asia has increasingly become stable. China's strength and prosperity are therefore not only a necessary condition for the steady growth of the global economy, but also a sufficient and necessary condition for political and security stability in East Asia.

Driven by the examples of China emerging as a global high-tech power and India achieving robust economic growth, numerous developing countries across Asia, Africa, and Latin

America have begun to see the prospect of their own economic takeoff. The aspiration of billions of people in these developing nations to attain prosperity, access quality education and healthcare, and enjoy sound infrastructure remains, therefore, the central trajectory of global development. This raise a core question: how can billions of people in developing countries achieve affluence? The "Kabul Moment" — marking the U.S. withdrawal from Afghanistan in 2021 after two decades of occupation — together with the failed U.S. military aggression against Iran in 2026, appears to illustrate a clear truth: even the United States — the world's sole superpower, leading the globe in politics, economy, military strength, and technology, and backed by NATO, the world's most powerful military alliance, as well as the Quadrilateral Security Dialogue (QUAD) comprising the U.S., Japan, India and Australia — was unable to fully subdue impoverished, underdeveloped nations such as Afghanistan and Iran. This signifies that the fate of smaller, weaker countries will no longer be easily dictated by a single major power or a coalition of great powers in the future. Even major nations, then, must respect the right of less developed countries to govern themselves according to their own social, political, cultural, and religious traditions.

At present, numerous vulnerable groups in developing countries across Asia, Africa, and Latin America still struggle to secure basic needs — food, clothing, housing, and transportation — let alone achieve all-round development. How, then, can these underdeveloped countries realize growth? Evidently, the development paths taken by established economies such as Europe, the United States, and Japan are unlikely to be replicable. Meanwhile, the resources- and the environment-intensive development model pursued by some developing countries is no longer acceptable to the international community. Developing countries will therefore have to adopt green technologies and equipment featuring lower energy consumption and less pollution in production in the future. Global industrialization and AI will need to advance along a green, low-carbon trajectory. At the same time, stricter technical standards and higher resource and environmental requirements will push up production costs and final product prices, which in turn will suppress market demand and keep overall economic growth at a more moderate pace. Even so, impoverished communities in developing countries are in urgent need of escaping poverty. To this end, developing countries require a growing number of enterprises and entrepreneurs capable of generating ample job opportunities for low-skilled laborers. The development of light industries such as textiles and apparel is, to a large extent, well-suited to this stage of economic takeoff in developing nations. Furthermore, as labor costs rise in Asian economies, labor-intensive light industries including textiles and garments are set to shift to countries with lower labor expenses. Developing countries aspiring to boost their economies can seize this industrial relocation opportunity to fuel their own takeoff. Beyond attracting foreign investment and fostering domestic industrial and commercial activities, developing countries must also upgrade their national infrastructure — covering education, healthcare, transportation, telecommunications, and energy — while building up their human capital. In other words, alongside boosting economic efficiency, equal attention must be paid to social equity. Doing so will not only cultivate the human capital needed to support domestic industrial upgrading, but also foster the comprehensive development of people.

A review of global political and economic developments since the Industrial Revolution suggests that the world is likely to face continued instability in the years ahead. After Donald

Trump won the U.S. Presidency again in 2024, his administration launched trade wars against the countries worldwide; most countries, with China a notable exception, were forced to submit to U.S. demands. On January 3, 2026, U.S. forces captured Venezuelan President Nicolás Maduro in a military operation and flew him out of the country, and then reduced Venezuela to a quasi-colony. In February28, 2026, the United States and Israel launched joint strikes against Iran. Faced with Iran's resolute resistance, the fighting remains deadlocked. The Russia-Ukraine War, meanwhile, represents Russia's strategic pushback against NATO's eastward expansion. Following the rise to power of Sanae Takaichi, an ultra-right politician in Japan, Japanese militarism has shown signs of resurgence. As things stand, the extreme destructive power of nuclear weapons makes large-scale military conflict between major global powers highly improbable. Nevertheless, localized instability and armed clashes remain inevitable.

In terms of international economic, trade, and technological exchanges, the Trump-Biden-Trump administrations imposed export restrictions on China targeting advanced chips and lithography machines. After winning re-election, Trump moved to dismantle the postwar free trade system centered on tariff reductions, replacing it with a more coercive trade regime and launching trade wars against all countries. Developed economies including the United States, Japan and Europe have all sought to build more self-sufficient supply chains covering rare earth elements, power batteries, electric vehicles, and photovoltaic and wind power generation, while China has strived to shore up its own supply chains for high-end chips and lithography equipment.

By and large, over the foreseeable future, global politics and the economy will likely remain unsettled, driven by several factors: major powers such as the United States pursuing strategic retrenchment across the Atlantic and Pacific Oceans, the prolonged Russia-Ukraine War, and the resurgence of Japanese militarism. Against this backdrop, China is positioned to serve as a stabilizing force for global political and economic development. As large numbers of populations in developing countries across Asia, Africa, and Latin America still fail to secure adequate food and clothing, lifting people in underdeveloped nations out of poverty and helping these countries build national infrastructure will remain a primary driver of world economic growth. Industrialization will continue to be the core pathways for people in these nations to achieve prosperity.

## Notes

[1] Jiàng, S.G. (2021)
[2] Zhang, J. (2009)
[3] Nunn, N. (2008),
[4] Li J J, Shi B Y. (2006).
[5] Stiglitz, 2013, p.10-15.
[6] Xinhua Flash: China has over 500,000 high-tech enterprises. September 18, 2025. (新华社快讯：我国高新技术企业超 50 万家。2025 年 9 月 18 日。
https://www.xinhuanet.com/20250918/773d3ab1e667408ab5a611eda4b83cc1/c.html)
[7] Simington, (2024).
[8] Dahl, 2024., ("Already, up to 70 percent of Latin America's 4G-LTE cellular networks are supported by infrastructure from the Chinese tech giant Huawei"). Goldman, 2023, (According to one estimate, Huawei components comprise 70 percent of Africa's 4G networks.)
[9] China Hydropower Planning & Design General Institute, 2026.
[10] The Analects · Yong Ye. (《论语・雍也》"己欲立而立人，己欲达而达人。")
[11] Sen, 1999, p. 258-261.
[12] Wen, 2022, p.341-482.
[13] Ibid.
[14] Maslow, 1943.
[15] Sen, 1979; Sen, 1999, p.1-20.
[16] Xia, 2016.
[17] Li and Gao, 2025.

# Index